%% file: thesis.tex
\documentclass[12pt,twoside,longbibliography,nofootinbib,preprintnumbers]{puthesis}
\title{\LARGE{\lettrine[lines=1]{D}{iscovery} of magnetic} \\ \LARGE{topological crystals}}

\submitted{June 2019}  % degree conferral date (January, April, June, September, or November)
\copyrightyear{2019}  % year in which the copyright is secured by publication of the dissertation.
\author{{Ilya Belopolski}}
\adviser{M. Zahid Hasan}  %replace with the full name of your adviser
\department{Physics}

\usepackage{CJK}

\usepackage{verbatim}
\usepackage{longtable}
\usepackage{amsmath}

\ifdefined\printmode

\usepackage{url}

\else

\usepackage{hyperref}
\usepackage[dvipsnames]{xcolor}
\usepackage{cleveref}

\newcommand\myshade{85}
\colorlet{mylinkcolor}{violet}
\colorlet{mycitecolor}{YellowOrange}
\colorlet{myurlcolor}{Aquamarine}

\hypersetup{colorlinks,bookmarksnumbered,
  linkcolor  = mylinkcolor!\myshade!black,
  citecolor  = mycitecolor!\myshade!black,
  urlcolor   = myurlcolor!\myshade!black,
}

\fi

\usepackage{Zallman,lettrine}

\PassOptionsToPackage{svgnames}{xcolor}
\usepackage[object=vectorian]{pgfornament}
\usepackage{lipsum,tikz}

\newcommand{\sectionline}{
  \noindent
  \begin{center}
  {
    \resizebox{0.5\linewidth}{1ex}
    {{%
    {\begin{tikzpicture}
    \node  (C) at (0,0) {};
    \node (D) at (9,0) {};
    \path (C) to [ornament=88] (D);
    \end{tikzpicture}}}}}%
    \end{center}
  }

\usepackage{graphicx}   
\usepackage{graphicx}
\usepackage{dcolumn}
\usepackage{bm}
\usepackage{graphics}
\usepackage{afterpage}
\usepackage{float}
\usepackage{subfigure}
\usepackage{rotating}
\usepackage{multirow}
\usepackage{fancyheadings}
\usepackage{theorem}
\usepackage{moreverb}
\usepackage{euscript}
\usepackage{psfrag}
\usepackage{slashed}
\usepackage{mathtools}

\usepackage{listings}
\usepackage{color,xcolor}

\definecolor{deepblue}{rgb}{0,0,0.9}
\definecolor{deepred}{rgb}{0.85,0,0}
\definecolor{deepgreen}{rgb}{0,0.95,0}
\definecolor{mygray}{gray}{0.97}

\lstdefinestyle{python}{
  belowcaptionskip=1\baselineskip,
  breaklines=true,
  frame=L,
  xleftmargin=\parindent,
  language=Python,
  showstringspaces=false,
  basicstyle=\small\ttfamily,
  morekeywords={models, lambda, forms,True,False,None},
  keywordstyle=\bfseries\color{deepgreen!40!black},
  commentstyle=\itshape\color{gray},
  identifierstyle=\color{black},
  stringstyle=\color{deepred},
  rulecolor=\color{gray},
  backgroundcolor=\color{mygray},
}

\usepackage{amssymb,amsmath,latexsym,graphics, graphicx,epsfig,epstopdf,multirow,comment,hyperref,appendix,slashed,xcolor,afterpage, makecell} 
\usepackage{booktabs}
\usepackage{tabularx}

\newcommand {\be} {\begin {equation}}
\newcommand {\ee} {\end {equation}} 

\newcommand {\bes} {\begin {equation*}}
\newcommand {\ees} {\end {equation*}}

\usepackage{csvsimple}

\newcommand{\beq}{\begin{equation}}
\newcommand{\eeq}{\end{equation}}

\usepackage[perpage]{footmisc}

\usepackage{booktabs}
\newcolumntype{C}[1]{>{\centering\let\newline\\\arraybackslash\hspace{0pt}}m{#1}}

\usepackage{inconsolata}
\usepackage{aas_macros}

\makeatletter
\renewcommand{\@chapapp}{}% Not necessary...
\newenvironment{chapquote}[2][2em]
  {\setlength{\@tempdima}{#1}%
   \def\chapquote@author{#2}%
   \parshape 1 \@tempdima \dimexpr\textwidth-2\@tempdima\relax%
   \itshape}
  {\par\normalfont\hfill--\ \chapquote@author\hspace*{\@tempdima}\par\bigskip}
\makeatother

\usepackage{cyrillic}

\newcommand{\cyrit}{\fontencoding{OT2}\selectfont\textcyrit}

\makeatletter
\hypersetup{pdftitle=Discovery of Magnetic Topological Crystals,
        pdfauthor=\@author}
\makeatother

\ifodd 0

\renewcommand{\maketitlepage}{}
\renewcommand*{\makecopyrightpage}{}
\renewcommand*{\makeabstract}{}

\else

\abstract{\input{abstract.tex}}
\acknowledgements{\input{acknowledgements}}

\dedication{\cyrit{Moim roditelyam i moemu bratu Aleksandru.}}
\fi

\begin{document}

\makefrontmatter

\include{ch-theory/topological}
\include{ch-arpes/arpes}
\include{ch-TaAs/ch-TaAs}
\include{ch-heterostructure/heterostructure}
\include{ch-weylcriteria/ch-weylcriteria}
\include{ch-mwt1/ch-mwt1}
\include{ch-mwt2/ch-mwt2}
\include{ch-TaIrTe4/ch-TaIrTe4}
\include{ch-chiral/ch-chiral}
\include{ch-magnet/magnet}
\appendix
\include{ch-mainpubs/ch-mainpubs}
\include{ch-pubs/ch-pubs}

%\include{ch-nptfit/NPTFit}
%\include{ch-igrb/LMNS_arxiv_v2}
%\include{ch-darksky/fermi_darksky}
%\include{ch-clusters/fermi_data_clusters_arxiv_v2}
%\appendix 
%\include{ch-appendicies/fermi_darksky}
%\include{ch-appendicies/fermi_data_clusters_arxiv_v2}

% Make the bibliography single spaced
%\singlespacing
%\bibliographystyle{utphys}

% add the Bibliography to the Table of Contents
%\cleardoublepage
%\ifdefined\phantomsection
%  \phantomsection  % makes hyperref recognize this section properly for pdf link
%\else
%\fi
%\addcontentsline{toc}{chapter}{Bibliography}

% include .bib file
%\bibliography{ch-nptfit/NPTF,ch-igrb/fermi_igrb,ch-darksky/fermi_darksky,ch-clusters/fermi_clusters,ch-intro/introduction}

%\bibliography{../../master_bib}{}

\sectionline

\end{document}

%% file: abstract.tex
%\lettrine[lines=3]{T}{he} extraordinary phenomena of quantum mechanics have fascinated observers for more than a century. However, not content with merely being ``fascinated'', an ongoing obsession of scientists and engineers is to coax quantum properties to the macroscopic scale, where humans can enjoy them. Crystals are important in this pursuit, because the repeating atomic structure preserves quantum coherence over a length scale limited only by the quality of structural order. As Loren Pfeiffer says\footnote{In ``Ten nines purity'', a popular science video from Princeton Engineering, see:\\ \href{https://www.youtube.com/watch?v=0UhVGhg6GdA}{https://www.youtube.com/watch?v=0UhVGhg6GdA}.} in reference to ultra-pure films of crystalline gallium arsenide, ``We'd like to get the quantum mechanical coherence length to about ten to fifteen microns. If we could do that, it's enough to see the whole structure [of quantum mechanics] with visible light and a light microscope.''

\lettrine[lines=3]{T}{opological} phases of matter have established a new paradigm in physics, bringing quantum phenomena to the macroscopic scale and hosting exotic emergent quasiparticles. In this thesis, I demonstrate with my collaborators the first Weyl semimetal, TaAs, using angle-resolved photoemission spectroscopy (ARPES), directly observing its emergent Weyl fermions and topological Fermi arc surface states [{\it Science} {\bf 349}, 6248 (2015); {\it Physical Review Letters} {\bf 116}, 066802 (2016)]. Next, I consider structurally chiral crystals, which I argue are guaranteed to host exotic chiral fermions leading to giant topological Fermi arcs. I study the chiral crystals RhSi and CoSi and I discover high-degeneracy chiral fermions with wide topological energy window, maximal separation in momentum space and giant Fermi arcs [{\it Nature} {\bf 567}, 500 (2019); {\it Nature Materials} {\bf 17}, 978 (2018)]. I establish a natural relationship between structural and topological chirality, producing a robust topological state which we predict supports a four-unit quantized photogalvanic effect [{\it Physical Review Letters} {\bf 119}, 206401 (2017)].

Next, I discuss the first quantum topological superlattice [{\it Science Advances} {\bf 3}, e1501692 (2017)]. I study multilayer heterostructures of alternating topological and trivial insulators. The Dirac cones at each interface tunnel across layers, realizing a new kind of emergent superlattice, where the interfaces act as lattice sites and the Dirac cones act as atomic orbitals. Adjusting the stacking pattern offers unprecedented control of individual hopping parameters in the atomic chain. I realize a novel topological phase transition and I predict that this platform may allow particle-hole symmetry without superconductivity.

Lastly, I present the discovery of a room-temperature topological magnet [arXiv:1712.09992]. I study crystals of Co$_2$MnGa and I observe a topological invariant supported by the material's intrinsic magnetic order [{\it Physical Review Letters} {\bf 119}, 156401 (2017)]. In particular I observe topological Weyl lines and drumhead surface states by ARPES and, through a scaling analysis of the anomalous Hall transport response, I find that the large anomalous Hall effect in Co$_2$MnGa arises from the Weyl lines. I hope that my discovery of Co$_2$MnGa establishes topological magnetism as a new research frontier in condensed matter physics.

%% file: acknowledgements.tex
% !TEX encoding = GB 18030
\begin{CJK*}{GB}{gbsn}
\lettrine[lines=3]{I} { am} grateful and honored to have worked alongside so many talented people during my time at Princeton. I thank my wonderful Ph.D. advisor, Zahid Hasan, for his guidance and support during the long journey of my graduate career. He offered me tremendous academic freedom and essentially unlimited resources to carry out my work. If I did not always use that freedom wisely, at least it has helped me learn to find my own way.\\

I benefited greatly from the excellent support and mentorship of my groupmates. I thank especially Su-Yang Xu, who introduced me to angle-resolved photoemission in July 2012 when I went to my first beamtime at the Advanced Light Source (ALS) in Berkeley, CA. Since that time he has guided me every step of the way and he remains a treasured collaborator, invaluable mentor and cherished friend. I also thank Chang Liu, who I met that same summer in the thick of battle during an insane twin-beamtime that we were running at the Swiss Light Source and who first taught me the nuts and bolts of photoemission. I am also grateful to Madhab Neupane, who guided me writing my first beamtime proposals and helped me start my first independent projects. His great success as faculty at the University of Central Florida has inspired me to push my own boundaries in research. I thank Guang Bian for great discussions on molecular beam epitaxy and for hosting us at his apartment where we made ``topological dumplings'' (manuscript in preparation). I thank Nasser Alidoust for the camaraderie as we fought valiantly on many hopeless experiments, as well as Hao Zheng for his explanations on determining surface terminations in scanning tunneling microscopy and on many other topics. Thank you to Daniel Sanchez, close friend, co-conspirator and sounding board. I'm sure we'll find great opportunities to continue collaborating in the future! It has also been a privilege to work with the younger generation, especially Tyler Cochran who has been a great friend and collaborator and has already given me a new perspective on many aspects of photoemission. Thank you Songtian Zhang, for all those times you bought me ÀϸÉÂè\  and for the great discussions on detecting Majorana fermions by scanning tunneling microscopy. Lastly, thank you to Guoqing Chang for his collaboration and critical comments on all aspects of my manuscripts and Jiaxin Yin for his careful eye and unwavering attention to detail.

Beyond my immediate group there were so many friends in the Physics Department at Princeton who made my time wonderful. I thank Andrei Bernevig for making time to answer my questions on topological phases during the early years. I also thank Aris Alexandradinata for the great discussions and friendship as I studied the theory of topological invariants, and in particular the nauseating detail of our conversations on the Su-Schrieffer-Heeger model. I look forward to many close future collaborations. I also thank Titus Neupert for his wonderful mentorship first at the Les Houches summer school in 2014 and later as a close collaborator on many of my projects on Weyl semimetals. Also Lauren McGough, for hanging out at 5am on the fourth floor of Jadwin. And of course Farzan Beroz, who was actually the first person I met in my cohort, when we serendipitously shared a room at Nassau Inn during the Princeton visiting weekend. Farzan remained a dedicated co-procrastinator throughout our time at Princeton. I enjoyed the great dark matter detection discussions, and general lounging around, with Siddharth Mishra Sharma. I also wish to thank Junyi Zhang for the deep comments at $\sim 2$am on electron-phonon coupling, quantum spin liquids, topological invariants and (how could I forget!)\hspace{-0.08cm} Ó㳦½£. Thank you also to Justin Ripley, who has always been there to tell me about general relativity beyond minimal coupling when I needed it. I will also miss the late-night discussions on coherent states and superconductivity with Abraham Asfaw. Thank you to Mallika Randeria for all those times you had me over at your place, the quality time spent together talking about life, the universe and everything. Infinite thanks to Fedor Popov for enthusiastically highlighting all of my awkward Russian grammar mistakes. And, of course, thanks to Peter Czajka for pointing out that I didn't mention him in these acknowledgements.

My research was also supported by many collaborators at Princeton and beyond, especially for sample synthesis and first-principles calculations. I will not try to name all of these amazing people here, but they all taught me so much and I am so grateful to work with them. I thank all of the incredible staff scientists at the synchrotron ARPES beamlines that were crucial for our work. Vladimir Strocov of ADRESS at the Swiss Light Source (SLS) stands out for introducing me to the perils of photoelectron escape depth when estimating $k_z$ resolution, as well as for many other fascinating discussions. I thank Hugo Dil of COPHEE, also at the SLS, for his many insights on final state effects in spin-resolved photoemission. I must also mention Jonathan Denlinger of the ALS for his many enlightening conversations on surface preparation, core level spectral analysis, X-ray diffraction and quasiperiodic undulators. Thank you to Eli Rotenberg, Chris Jozwiak, Aaron Bostwick and Roland Koch also of the ALS for the world-changing Maestro beamline as well as the fascinating discussions on deflection mapping, nanoARPES, low-energy electron microscopy and photoelectron emission microscopy. I thank Moritz Hoesch of Diamond I05 for the careful and lucid discussions of energy and momentum resolution in ARPES. Thanks to Fran\c cois Bertran, Patrick Le F\`evre and Julien Rault for the fun times at Cassiop\'ee and the positive attitude even when I was struggling with my na\"ive molecular beam epitaxy experiments. I thank Elio Vescovo and Konstantine Kaznatcheev for the cutting-edge electron spectro-microscopy beamline at NSLS-II and for the nonsense conversations at $\sim 3$am. Lastly, I thank the Shik Shin group at the Institute of Solid State Physics in Kashiwa for the wonderful collaboration and especially Yukiaki Ishida for the fascinating comments on pump-probe/time-resolved ARPES, deathly spicy curry at CoCoICHI and all of that conveyor belt sushi. Let me also thank Sanfeng Wu for acting as second reader of this thesis as well as Duncan Haldane and Christopher Tully for serving on the committee for my final public oral examination.\\

During this journey, I could not have survived without my wonderful friends from Columbia, especially (but not only!)\hspace{-0.08cm} ³Â½õ·¼, ÕżκÌ, ¹ù×Ó¿¡\ and ÕçÊç·Ò. Also thank you to Jacob Andreas, for the great discussions on natural language processing in Berkeley. I also thank Alena Kotelnikova for all the fancy dinners at overpriced Parisian brasseries and for the (significantly less expensive) showings at the Th\'e\^atre du Ranelagh and the Apollo Th\'e\^atre. I thank Mary Ellen Keneally, Emma Routhier and Campbell Hutcheson, who always brought me so much joy whenever I heard from them. Lastly, I thank my whole family and especially Sasha for their love, support and patience as I pursue my dreams.
\end{CJK*}
\sectionline

%\lettrine[lines=3]{S}{ince} it is well-known that the Acknowledgements section is what people typically read in Ph.D. theses and since, at the same time, it is typically the case that ``there are far more people I would like to thank than there is room to name here'', we will take a different approach and present the key scientific results here in the Acknowledgements, and defer the the actual acknowledgements to the remaining 137 pages of the main text. Co$_2$MnGa crystallizes in the full-Heusler structure (Fig. 1a), with a cubic face-centered Bravais lattice, space group $Fm\bar{3}m$ (No. 225), indicating the presence of several mirror symmetries in the system. Moreover, the material is ferromagnetic with Curie temperature $T_\textrm{C} = 690$ K (Fig. 1b), indicating intrinsically broken time-reversal symmetry. This suggests that...
%\sectionline

%% file: ch-theory/topological.tex
\chapter{Topological phases of matter}
\label{ch:theory}

{\singlespacing
\begin{chapquote}{Becky G, \textit{Can't Get Enough}}
Jump on if you ready to go\\
Up, up and around the globe
\end{chapquote}}

\lettrine[lines=3]{T}{he} extraordinary phenomena of quantum mechanics have fascinated observers for more than a century. However, not content with merely being ``fascinated'', an ongoing obsession of scientists and engineers is to coax quantum properties to the macroscopic scale, where humans can enjoy them. Crystals are important in this pursuit, because the repeating atomic structure preserves quantum coherence over a length scale limited only by the quality of structural order. As Loren Pfeiffer says in ``Ten nines purity''\footnote{A popular science video from Princeton Engineering, see:\\ \href{https://www.youtube.com/watch?v=0UhVGhg6GdA}{https://www.youtube.com/watch?v=0UhVGhg6GdA}.} in reference to ultra-pure films of crystalline gallium arsenide, ``We'd like to get the quantum mechanical coherence length to about ten to fifteen microns. If we could do that, it's enough to see the whole structure [of quantum mechanics] with visible light and a light microscope.''

Also important in this pursuit is topology, which occupies itself with global properties of objects. In 1980 humans inadvertently discovered that topology can preserve quantum behavior up to the macro-scale, specifically for a sea of electrons confined to two dimensions and subjected to a large magnetic field. They found that the electrical resistance of this setup is \emph{exactly} quantized to a fundamental value---the von Klitzing constant---approximately equal to $25$ k$\Omega$. This value of the resistance is fixed by a topological invariant which, due to its global nature, is not easily modified by the disorder in the sample.

Gradually, physicists realized that in fact a variety of topological invariants appear in crystals, producing a zoo of topological phases of matter, each with their own attributes. I will discuss the topological invariants of several animals in this zoo, focusing on concepts relevant to the work in this thesis \cite{TKNN, Nobel_FDMHaldane, Review_HasanKane, ReviewTopology_XiaoGangWen, Review_Kitaev, ReviewSSH_Kane, Review_QiZhang}. Basically, I will focus on the Berry phase and Chern number, which are both integrals of the Berry connection $A(k) = i \langle k, n | \nabla_k | k, n \rangle$. I will not discuss $\mathbb{Z}_2$ insulator invariants or Pfaffians; invariants in the Bogoliubov-de Gennes formalism for topological superconductors; or any interacting classifications. My discussion will also focus predominantly on implications for ARPES experiments, where we are often most interested in directly studying topological bulk and surface energy dispersions. So, for instance, I leave to future work a discussion of the fascinating phenomena related to quantized transport or optical response in topological materials.

\section{Theory of invariants}

The simplest topological phases can be understood in a mean-field theory of non-interacting quasiparticles. The system is governed by a Hamiltonian which looks like,
\begin{equation}
H = \sum c^{\dag}_{i \alpha} t_{i \alpha, j \beta} c_{j \beta} = \sum c^{\dag}_{k \alpha} h_{\alpha \beta} (k) c_{k \beta} = \sum \varepsilon_n (k) c^{\dag}_{kn} c_{kn}
\end{equation}
Here $i,j$ refer to lattice sites, $k$ refers to crystal momentum, $\alpha,\beta$ refer collectively to all of the spin, orbital and lattice basis degrees of freedom and $n$ refers to energy bands. Summation over repeated indices is implied. The operators create or annihilate fermionic quasiparticles which either occupy a particular orbital, $c^{\dag}_{i \alpha}$ or an extended Bloch state, $c^{\dag}_{kn}$. In a first-quantized notation, the Bloch states can be written as $c^{\dag}_{kn} | 0 \rangle = | k, n \rangle$. The quantity $h_{\alpha \beta} (k)$ is called the Bloch Hamiltonian.

In the simplest cases, a topological invariant refers to a number that can be calculated using some complicated formula involving the eigenstates of $H$. The invariant typically has three properties: (1) it involves some kind of integral over all the eigenstates, so in this sense it is \textit{global}; (2) it does not change smoothly under small perturbations of $H$ (but can jump from one value to another if the perturbation is strong enough to push the system through a phase transition), so in this sense it is \textit{robust}; (3) it is \textit{quantized} to a set of allowed values, like $\pm1$ or the integers, depending on the details of the specific topological classification. Actually these three properties are similar to the intuitive properties of invariants that we know about from mathematical topology. The ``number of holes in the donut'' is a classic example considered in topology. Counting the number of holes is necessarily a global operation---it involves looking all over the donut to check for holes. Also, deforming the donut smoothly cannot change the number of holes; instead you have to puncture the donut to add a new hole, which is considered a dramatic, non-smooth change. Lastly, the number of holes in the average donut is clearly either $0$ or some positive integer. These similarities between electrons and donuts help explain why the quantities that I'll discuss in condensed matter physics might reasonably be called ``topological'' invariants.

For the cases I discuss here, the key underlying quantity is the Berry connection, $A(k) = i \langle k, n | \nabla_k | k, n \rangle$. The Berry phase $\gamma$ is the integral of $A(k)$ over a closed one-dimensional $k$-space loop,
\beq
\gamma = \oint A(k) \cdot dk
\eeq
For a one-dimensional system---an atomic chain---this integral can be taken over the entire one-dimensional Brillouin zone, in which case it can be written as,
\beq
\gamma = \int_{-\pi}^{\pi} A(k) dk
\eeq
Alternatively, it is meaningful to take the integral over any kind of one-dimensional loop in a two-dimensional or three-dimensional Brillouin zone. Under certain conditions, the Berry phase may be quantized, usually due to some symmetry that requires $\gamma = -\gamma$. Since the Berry phase is meaningful only up to $2 \pi$, this condition means that $\gamma$ is allowed to take on values of $0$ or $\pi$. For a two-band model the Bloch Hamiltonian can always be written in the form $h(k) = d_0(k) + d_i(k) \cdot \sigma_i$ with $i \in {x,y,z}$. We can consider $\hat{\bf{d}} = \bf{d}/|\bf{d}|$. Let $\alpha$ be the solid angle swept out by $\hat{\bf{d}}$ as we traverse a closed path. Then, we have,
\beq
\gamma = \frac{\alpha}{2}
\eeq
This equation provides a way to obtain the Berry phase directly from the Hamiltonian for the case of a two-band model.

Next, the curl of the Berry connection is the Berry curvature,
\beq
\Omega(k) = \nabla_k \times A(k)
\eeq
It can be integrated over a closed two-dimensional manifold to produce a Chern number 
\beq
n = \oint_\mathcal{M} \nabla_k \times A(k) \cdot d \sigma
\eeq

For a gapped band structure $n$ is always quantized. For a two-dimensional system, such as a thin film of crystal, the integral can be taken over the entire two-dimensional Brillouin zone,
\beq
n = \iint_{-\pi,-\pi}^{\pi,\pi} \nabla_k \times A(k) \cdot \hat{z}\ dk_x dk_ y
\eeq
Having defined two topological invariants, let's now look a little closer at some cases of interest.
%A rule of thumb in the case of topological boundary states with symmetry-protected degeneracies is that the difference connectivities that one can draw while respecting the relevant symmetries correspond to the set of possible invariants. For instance, for the case of a $\mathbb{Z}_2$ topological insulator, I can begin with a Kramers degeneracy and ask how I can connect the two outgoing surface states bands to the bulk valence and conduction bands. There are only two distinct possibilities: connect both surface states to the same bulk band (either valence or conduction) or connect one surface states to the bulk valence band and the other surface state to the conduction band. This suggests a $\mathbb{Z}_2$ classification. 
%I've now reviewed a few key properties of the Berry phase, Chern number and. I've also discussed how these invariants appear in each dimension and introduced their appearance in the Su-Schrieffer-Heeger model, the anomalous quantum Hall state, topological insulators, topological crystalline insulators, Weyl point semimetals and Weyl line semimetals.
% Berry phase relation to solid angle on sphere
% Polarization vs. spectral flow
% Explain polarization?

\begin{figure}
\centering
\includegraphics[width=12cm]{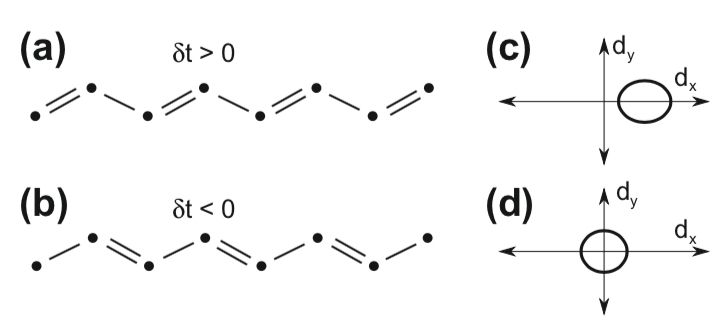}
\caption{\label{SSH_cartoon} The Su-Schrieffer-Heeger (SSH) model describes a one-dimensional atomic chain with a two-atom basis with tow different hopping amplitudes, a strong hopping and a weak hopping. Figure from Ref. \cite{ReviewSSH_Kane}.}
\end{figure}

\section{One dimension: Su-Schrieffer-Heeger model}

The Su-Schrieffer-Heeger (SSH) model exhibits a simple topological phase. It has the drawback that its topological classification is somewhat subtle, but on the other hand the relationship between the bulk invariant and topological boundary modes is arguably less mysterious than in the case of the quantum Hall state. We begin with a one-dimensional atomic chain, with lattice spacing $a$, with two sites per unit cell, as shown in Fig. \ref{SSH_cartoon}. There is one orbital per site and we ignore spin. For the Hamiltonian, we follow the notation of Ref. \cite{ReviewSSH_Kane},
\beq
H = \sum_i (t + \delta t) c_{Ai}^{\dag} c_{Bi} + (t - \delta t) c^{\dag}_{A,i+1} c_{Bi} + h.c.
\eeq
In the crystal momentum basis, we find,
\beq
H = \sum_k c^{\dag}_{\alpha k} h_{\alpha \beta} (k) c_{\beta k}, h_{\alpha \beta} (k) = \bf{d} (k) \cdot \bf{\sigma}
\eeq
where the gap functions are,
\begin{align}
d_x(k) &= (t + \delta t) + (t - \delta t) \cos ka \\
d_y(k) &= (t - \delta t) \sin ka \\ 
d_z(k) &= 0
\end{align}
Since $d_z (k) = 0$, $\hat{\bf{d}}$ is stuck in the $x, y$ plane. As a result, the only solid angles that it can sweep out are $0$ and $2 \pi$. This corresponds to a Berry phase $\gamma = 0$ or $\pi$ and shows that the Berry phase is quantized. We can directly show that this Berry phase invariant corresponds to a protected end mode. We write the polarization as the expectation value of the position of the Wannier function,
\beq
P = e (\langle R | \hat{r} | R \rangle - R)
\eeq
Here, $ | R \rangle $ is the Wannier function associated with lattice site $R$ and $\hat{r}$ is the position operator. By expressing the Wanner function in terms of Bloch states explicitly, we can show,
\beq
P = \frac{ie}{2\pi} \int_{-\pi/a}^{\pi/a} \langle k | \nabla_k | k \rangle dk = \frac{e}{2\pi} \gamma
\eeq
Therefore, up to integer charges, the polarization is zero for the $\gamma = 0$ phase and $e/2$ for the $\gamma = \pi$ phase.
%When discussing polarization in classical continuous media, Griffiths shows that a bulk polarization gives rise to a bulk bound charge density, $\rho_b = -\nabla \cdot \bf{P}$, and a bound surface charge density, $\sigma_b = \bf{P} \cdot \hat{n}$ \footnote{\textit{Introduction to Electrodynamics}, D. J. Griffiths, Eqs. 4.11 \& 4.12.}. This result has an interesting property: for a constant shift, $\bf{P}(\bf{x}) \rightarrow \bf{P}(\bf{x}) + \bf{c}$, $\rho_b$ is unchanged, while $\sigma_b$ changes by $\bf{c} \cdot \hat{n}$. This result is quite intuitive: for instance, if you have two uniform positive and negative charge distributions on top of each other, with a constant relative position shift, then there will be no polarization or bound charge deep in the bulk, but there can be an accumulation of charge on the edges of the system. Adding an extra shift $\bf{c}$ will not change the situation in the bulk, but can change the edge accumulation of charge.

One peculiarity of the SSH model is that the topological invariant flips if we change $\delta t \rightarrow -\delta t$. This can be seen from the gap function $d_x (k)$. If $\delta t > 0$, $d_x(k) > 0\ \forall\ k$, so the area swept out on the Bloch sphere is zero. If $\delta t < 0$, then $d_x(k)$ does change sign as we traverse $k$, sweeping out the equator of the Bloch sphere and giving a Berry phase of $\pi$. At the same time, there is an apparent contradiction: if we simply replace the weak bonds with the strong bonds in Fig. \ref{SSH_cartoon}, it's clear that the physical system remains unchanged---a sign flip of $\delta t$ is the same as a translation of the whole chain. So are the two topological phases truly distinct, or secretly the same? One way to address this ambiguity is to consider the question of the choice of unit cell. In the bulk of the system there is, intuitively, a gauge invariance associated with the choice of unit cell. Any physical quantity, such as the average mass density, should not depend on which unit cell is chosen. However, the introduction of a boundary breaks this gauge invariance and introduces a preferred unit cell. If the unit cell cuts through the strong bond of the atomic chain, a topological end mode arises. If the unit cell cuts through the weak bond, the system is trivial.

We can explore this further by looking at the Hamiltonian. Let us fix $\delta t > 0$. Notice that the stronger hopping is within the unit cell, rather than between unit cells. And, as noted above, the gap functions sweep out zero area on the Bloch sphere and the Berry phase $\gamma = 0$. This is consistent with termination of the atomic chain on a weak bond. In other words, when the chain terminates on a weak bond, the preferred unit cell is associated with $\gamma = 0$.

How can we get $\gamma = \pi$ from this model? Of course, we could simply flip the sign of $\delta t$, but this could be interpreted as gradually tuning the system through a topological phase transition ($\delta t = 0$). Instead, I will flip the invariant by simply redefining the unit cell. I relabel,
\begin{align}
c^{\dag}_{Aj} &\rightarrow c^{\dag}_{Bj} \\
c^{\dag}_{Bj} &\rightarrow c^{\dag}_{A,j+1}
\end{align}
This leads to,
\begin{align}
c^{\dag}_{Ak} &\rightarrow c^{\dag}_{Bk} \\
c^{\dag}_{Bk} &\rightarrow e^{-ika} c^{\dag}_{Ak}
\end{align}
And finally we can see that the gap functions become,
\begin{align}
d_x(k) &= (t + \delta t) \cos ka + (t- \delta t) \\
d_y(k) &= (t + \delta t) \sin ka \\
d_z(k) &= 0
\end{align}

Now indeed $d_x(k)$ switches sign as it winds, so that we encircle the Bloch sphere. This gives a solid angle of $2 \pi$, or $\gamma = \pi$. We can say that this unit cell is preferred if the atomic chain is cut on the strong bond, giving an end mode. So we can view the Berry phase of the SSH model as being gauge dependent and not meaningful. However, with the introduction of a boundary, the system specifies a choice of unit cell and then, using this unit cell, we can calculate the Berry phase and predict the presence or absence of an end mode. In this way, although we can calculate a bulk topological invariant for the SSH model, we need to known information about the boundary in order to know how to calculate the invariant. This is different from the case of the Chern insulator or strong $\mathbb{Z}_2$ invariants, where no ambiguity regarding termination arises. Similarly, if we have a defect in the SSH chain, we can view this defect as the presence of two SSH segments with opposite unit cell conventions which are connected to each other. Since both unit cell conventions are present, exactly one of them has the topological phase and there is a guaranteed boundary mode.

% Chiral symmetry?

%One of the properties of the SSH invariant is that it assumes only nearest neighbor hopping. This can expressed as the presence of a chiral symmetry which is difficult to maintain in

% polyacetylene

\section{Two dimensions: the quantum Hall state}

% Altland Zirnbauer
% Bulk invariant
% Surface states
% Quantum Hall state
% von Klitzing constant
% Observation of edge modes by ARPES?
% Analogy to the magnetic field
% Spectral flow/Laughlin argument
% Time-reversal symmetry

While the SSH model is a fascinating theoretical construct, its one dimensional nature and the requirement of chiral symmetry has limited its direct experimental interest. In fact, historically, it was the discovery of a two-dimensional topological state, the quantum Hall state, which set physicists down the path to topological phases of matter. Specifically, it was observed that for a two-dimensional electron gas under an applied magnetic field, the transverse conductivity is exactly quantized to integer multiples of a universal value, the von Klitzing constant,
\beq
\sigma_{xy} = \frac{n}{R_\textrm{Klitzing}} \textrm{, \ } R_\textrm{Klitzing} = \frac{h}{e^2} \sim 25.81 k \Omega
\eeq
Through unit analysis, it can be seen that the dimensionality is crucial for this effect. Specifically, the units of resistance and resistivity are the same in two dimensions, so that the resistance does not depend on the physical dimensions of the sample.

% and, like the Berry phase, it is closely related to the Berry connection. It can be written as the integral over a closed two-dimensional $k$-space surface,

%\beq
%n = \sum_\textrm{occ. bands} \oint_\mathcal{M} \nabla_k \times A(k) \cdot d \sigma
%\eeq

%The curl of the Berry connection is called the Berry curvature,

%\beq
%\Omega(k) = \nabla_k \times A(k)
%\eeq

%For this quantity to be well-defined, the system must typically have a gap, so that we can integrate over the entire band without any ambiguity regarding which state to include in the integral. 

%In the case of a two-dimensional system, such as a thin film of crystal supported by a substrate, the Chern number can be taken over the entire two-dimensional Brillouin zone,

%\beq
%n = \iint_{-\pi/a, -\pi/a}^{\pi/a, \pi/a} \nabla_k \times A(k) \ dk_x dk_y
%\eeq

%In this case, the manifold of integration is the surface of your average donut, $\mathbb{T}^2$. 

The underlying topological invariant is the Chern number. It can be shown that the Chern number vanishes for a time-reversal symmetric system. Specifically, the Berry curvature is odd under time-reversal symmetry, $\Omega(k) = - \Omega(-k)$, so that in the presence of time-reversal symmetry under integration over the full Brillouin zone, the Chern number vanishes, see Sect. 4.5 in Ref. \cite{Topo_Andrei}. This is simply the statement that the quantum Hall effect requires a magnetic field.

The Chern number is associated with a powerful bulk-boundary correspondence which states that at the boundary between two two-dimensional systems with different Chern numbers, $n_1$ and $n_2$, there must be a net $| n_1 - n_2 |$ number of edge modes crossing the bulk band gap, often referred to as ``chiral modes''. The argument for why such modes arise is not as transparent as in the case of topological polarization states, as in the case of the SSH model. They can be understood by Laughlin's gauge argument, see Sect. 6.1 in Ref. \cite{Topo_Andrei}.

% Fig. \ref{edge_states}

%In contrast to the case of Berry phase, the Chern number does not require any special spatial symmetry to be preserved in order to remain quantized.

% Gapped
% Protected edge modes

\section{Three dimensions: the Weyl point and Dirac line semimetals}

% Weyl bulk invariant
% Fermi arc surface state
% Weyl line invariant, Berry phase, P & T?
% Related properties may arise for other symmetres
% Must be gapped

The Chern number is not only meaningful for two-dimensional systems, but also for two-dimensional slices of a three-dimensional bulk Brillouin zone. Weyl semimetals provide one example of a three-dimensional system where Chern numbers arise. Consider, for instance, the slice defined by a fixed $k_x$. On slices of fixed $k_x$ cutting between two Weyl points, the Chern number may be, for instance, $+1$. If we then sweep the slice along $k_x$ we will find that the Chern number remains $+1$ until we encounter a Weyl point. On passing through the Weyl point, the Chern number increments to $0$ or $+2$. In this way, the momentum parameter $k_x$ can be viewed as a tuning parameter for a two-dimensional system. When we slide the slice through a Weyl point, it is like tuning the two-dimensional system through a gap closing and reopening, incrementing the Chern number invariant. The bulk-boundary correspondence also generalizes in a similar way to three-dimensional crystals. In particular, the Chern number magic works on the one-dimensional edges of the two-dimensional slice. Assembling all the chiral edge modes from all slices together produces topological Fermi arc surface states.

The Weyl points are referred to as chiral charges or, poetically, as ``magnetic monopoles in momentum space''. These names arise from an understanding of Berry curvature as a quantity similar to the magnetic field but defined in momentum space. Specifically, the Berry curvature is the curl of the Berry connection in the same way that the magnetic field is the curl of the electromagnetic vector potential. The Berry curvature further remains unchanged under inversion symmetry, $\Omega (k) = \Omega(-k)$, making it a pseudovector, also like the magnetic field. Crucially, while magnetic monopoles are so far unknown as fundamental particles, monopoles of Berry curvature are known in crystals. These are precisely Weyl points and we have observed them in ARPES in a number of materials \cite{TaAs_Suyang, NbAs_Suyang, TaP_Suyang, NbP_me, MoWTe2_me, MoWTe2_me2, TaIrTe4_me, Review_HasanXuBian}.

\begin{figure}
\centering
\includegraphics[width=14cm]{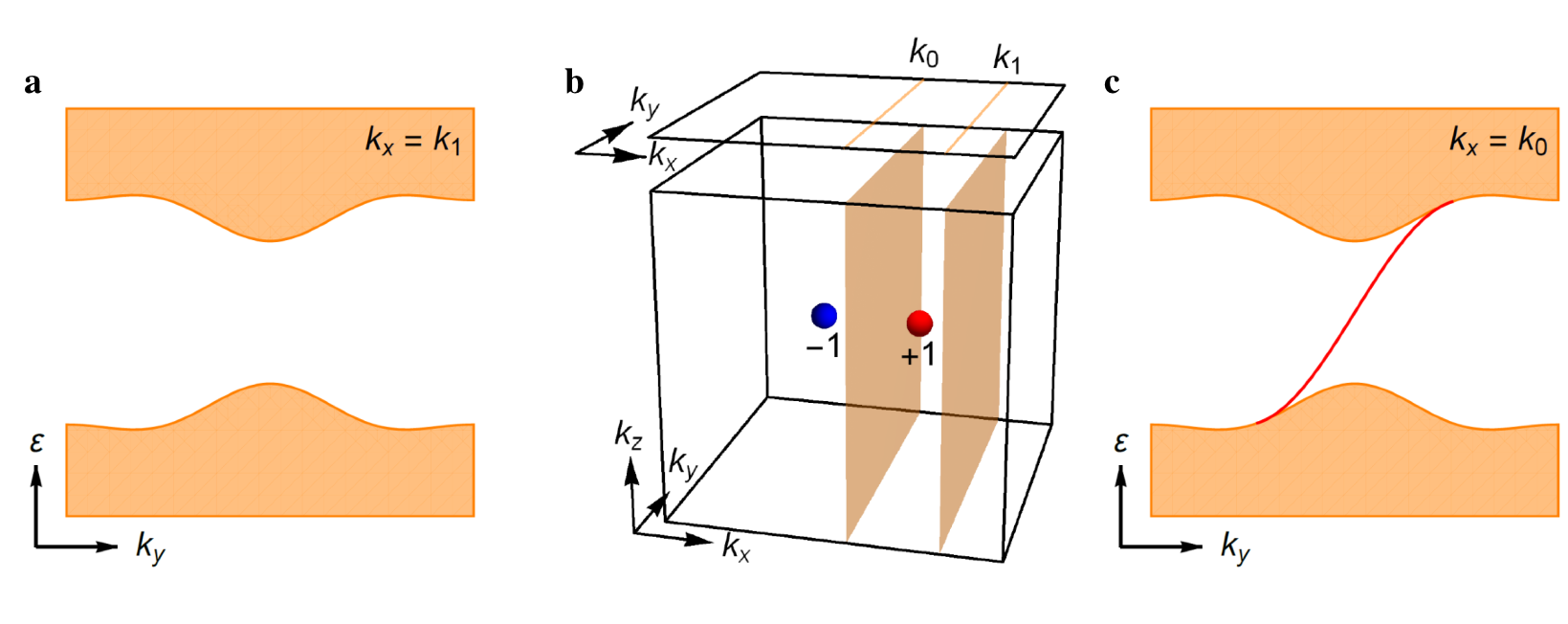}
\caption{\label{Chern_Weyl} Slice with Chern number ({\bf a}) 0 and ({\bf c}) 1, with associated chiral edge mode, taken at different $k_x$ of a Weyl semimetal, as shown in {\bf c}. Figure adapted from \cite{ARCMP_me}.}
\end{figure}

The Berry phase also makes an appearance in three dimensions as the integral over loops in three-dimensional Brillouin zones and can be understood as the topological invariant of a spinless Dirac line \cite{LineNode_ChenFang}. In $P$ and $T$ symmetric systems in three dimensions without spin-orbit coupling, Dirac lines arise generically. $P$ and $T$ also impose the constraint that the Berry phase on an arbitrary closed loop obeys $\gamma  = - \gamma$, so that it is quantized to $0$ or $\pi$ \cite{Ca3P2_Schnyder_2016}. For a closed loop linking a Dirac line, we have $\gamma = \pi$ \cite{LineNode_ChenFang}. $P$ and $T$ symmetries together require that $\Omega(k) = - \Omega(k)$. This would suggest $\Omega(k) = 0$ identically, raising an apparent paradox. To see the paradox, note that by application of Stokes' theorem, a $\pi$ Berry phase on any loop would correspond to non-zero Berry curvature on surfaces bounded by that loop \cite{TINI_Balents}. To resolve this paradox, I argue that the Dirac line is associated with a pathological distribution of Berry curvature which is strictly zero away from the Dirac line and divergent on the Dirac line.

%I can show this with a co-dimension argument. Consider a two-band model without spin. In real life, these bands would be two-fold degenerate (due to spin), so all bands are two-fold degenerate (counting spin) and the Dirac line degenerate would be four-fold degenerate (counting spin). However, suppressing spin, there are two singly-degenerate bands and we would like to see why Dirac lines are generic. $T$ applies the constraint $h^*(k) = h(-k)$. If $P$ takes a two-dimensional representation, then for a two-band model, it will act as, $\sigma_x h(k) \sigma_x = h(-k)$. Together, these two constraints require $\sigma_x h^*(k) \sigma_x = h(-k)$, which requires the $\sigma_z$ gap function to vanish identically. In other words, the Bloch hamiltonian can be written as $h(k) = d_x (k) \sigma_x + d_y (k) \sigma_y$. The generic set of zeros of these gap functions will be some surface in the three-dimensional Brillouin zone and the intersection of the two surface generically form lines. This shows that Dirac lines arise generically in the band structure. 

% Need to check, \sigma_x h^*(k) \sigma_x = h(k) is always imposed by PT?
% Check P and T constraint on Berry phase
% Check gamma = pi for Dirac line

\begin{figure}
\centering
\includegraphics[width=7cm, trim={2in 2.8in 5.8in 2.6in}, clip]{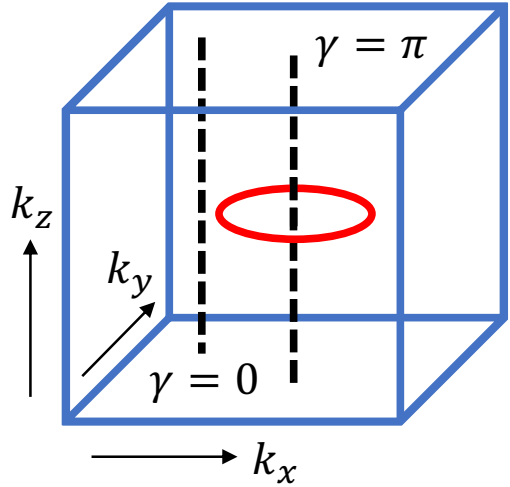}
\caption{\label{Berry_Dirac} $\pi$ Berry phase on a loop through a Dirac line in a three dimensional Brillouin zone.}
\end{figure}

Here it is useful to draw an analogy with graphene, which is a two-dimensional Dirac point semimetal, typically treated without spin-orbit coupling. Because it is two-dimensional rather than three-dimensional, we generically expect Dirac points rather than Dirac lines. Now consider the Berry curvature distribution of the graphene Dirac cone with a $P$ symmetry-breaking mass gap, see Sect. 8.2.3 in Ref. \cite{Topo_Andrei},

\beq
\Omega_{xy} = \frac{m}{2(m^2 + k^2)^{3/2}}
\eeq

At the Dirac point,

\beq
\Omega_{xy} = \frac{m}{2|m|^3} = \frac{\textrm{sgn}(m)}{2|m|^2}
\eeq

For vanishing mass, this quantity diverges. We can see that as we restore inversion symmetry (take $m \rightarrow 0$), the Berry curvature does not quite vanish, but either becomes strictly zero or diverges.

\clearpage
%\cleardoublepage

\ifdefined\phantomsection
  \phantomsection  % makes hyperref recognize this section properly for pdf link
\else
\fi
\addcontentsline{toc}{section}{Bibliography}

{\singlespacing
%\bibliography{../../master_bib}{}
%\bibliographystyle{../../Science}

}

%% file: ch-arpes/arpes.tex
\chapter{Spin \& angle-resolved photoemission spectroscopy}
\label{ch:arpes}

{\singlespacing
\begin{chapquote}{Hayley Kiyoko, \textit{Rich Youth}}
Living, falling in too deep\\
Got no time to meet their needs\\
Oh my God\\
Let me be free
\end{chapquote}}

\lettrine[lines=3]{M}{ost} of the experiments discussed in this thesis were carried out using angle-resolved photoemission spectroscopy (ARPES). ARPES is a ``photon in'', ``electron out'' experiment---incoming light ejects photoelectrons from the surface of the crystal \cite{Hertz, Einstein}. Using what we know about the photon going in as well as what we can measure about the electron coming out, we can learn about the sample. In the simplest treatment, we can apply fundamental considerations of energy and momentum conservation: specifically, in the instrumental apparatus we have the capability to set the energy and momentum of the incoming photon, and we are capable of measuring the energy and momentum of the outgoing photoelectron. Using these known values, perhaps with a few mild assumptions and caveats, we can calculate the energy and momentum of the electron in the crystal and recover the band structure \cite{ARPES_Hufner, ARPES_Damascelli, ARPES_Damascelli2}. In this sense, ARPES is a wonderfully direct technique which allows us to immediately observe the crystal's band structure---one of the fundamental objects of condensed matter physics.

ARPES has evolved into a fantastically sophisticated experiment, particularly for the case of ARPES beamlines at a synchrotron facility. In this case, the light source for the ARPES experiment is a synchrotron particle accelerator, which is often a national-level facility with an annual operating budget of $\$10^7$. The electron analyzer is itself a mini particle accelerator, developed and sold by dedicated commercial companies such as ScientaOmicron and SPECS GmBH. Modern sample manipulators allow arbitrary translations and rotations of the sample over a wide range of distance and angle while maintaining ultrahigh vacuum pressures of $10^{-11}$ Torr, extremely cold temperatures of $\sim 7$ K and ensuring repeatability of sample position to within $\sim$ micrometers. Advanced ancillary equipment is often used to improve the experiment, such as a high-power microscope to give a detailed optical image of the surface under study. Our group co-proposed and developed one such state-of-the-art instrument, Beamline 4 of the Advanced Light Source, including the ARPES and resonant inelastic X-ray scattering (RIXS) endstations as well as the beamline itself, at Lawrence Berkeley Laboratory in CA, USA \cite{BL4_a, BL4_b}.

From the point of view of topological phases of matter, the pay-off of this elaborate experimental setup is that you can demonstrate topological invariants in a powerful and direct way. In particular, for many phases the topological invariants are associated with unique band structure objects, such as Weyl points and Fermi arcs in Weyl semimetals, as discussed in the preceding chapter. Because it directly measures the band structure, ARPES has emerged as perhaps the technique of choice to unambiguously show that a given material is topological. The problem and the technique are further well-matched because ARPES naturally probes the surface of a material, so that topological surface states can be directly studied.

% Should be called photoemission spectro+momentoscopy
% Good to put somewhere here, cleave the sample, etc...

%When you perform an experiment at a synchrotron ARPES endstation, you rise up on the winds of the Zephyr, you rush forth on the streams of Okeanos.

%In this section I will introduce useful concepts in photoemission kinematics and then discuss the key components of the experiment in turn: the light source, the sample and the analyzer.

% Mention analyzer slit here?

\begin{figure}
\centering
\includegraphics[width=12cm]{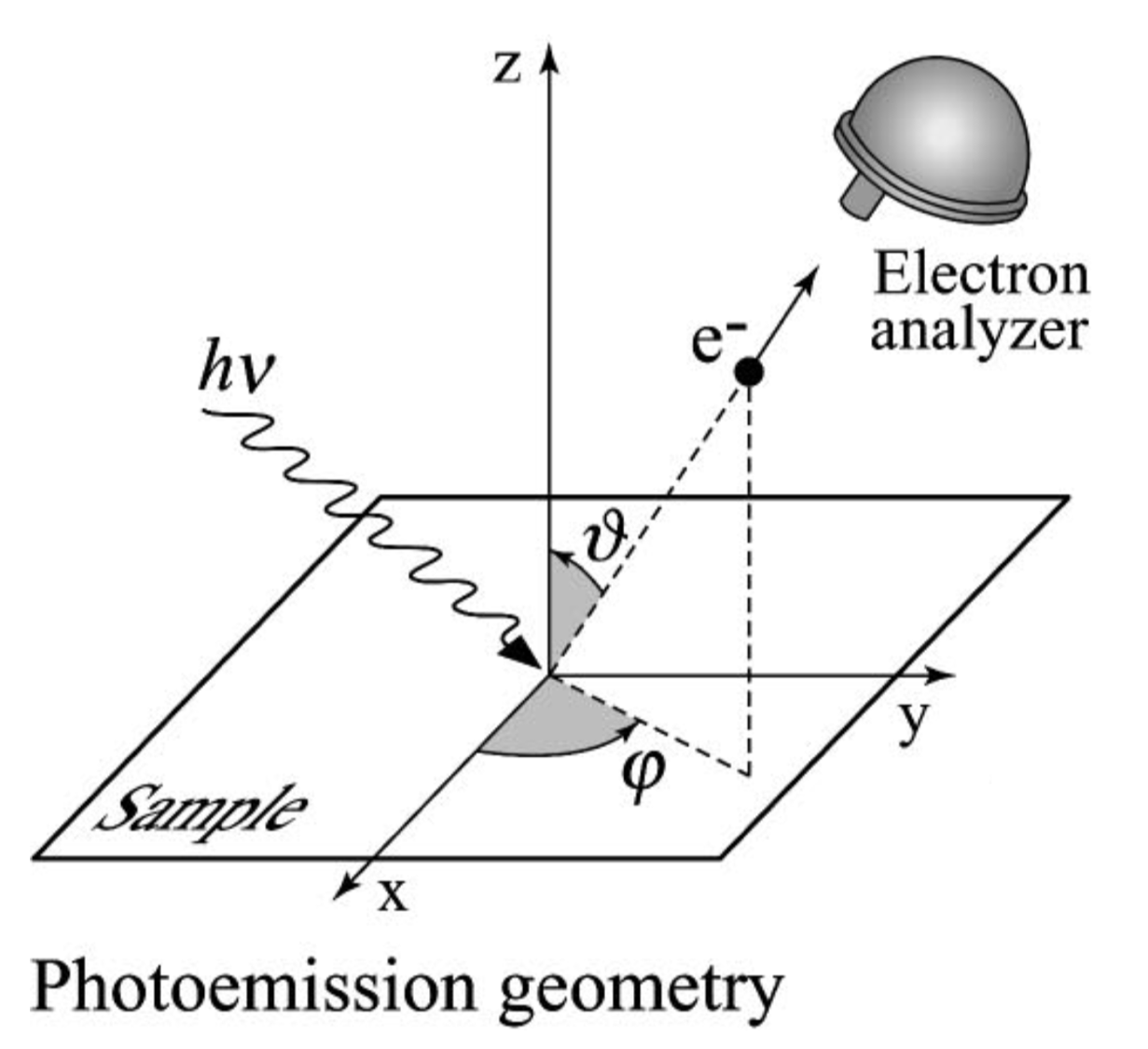}
\caption{\label{arpes_overview} In angle-resolved photoemission spectroscopy (ARPES), incoming photons are absorbed by a sample and produce photoelectrons, which are then measured using an electron analyzer. Figure from Ref. \cite{ARPES_Damascelli}.}
\end{figure}

\section{Geometry \& kinematics}
\label{geokin}

% We focus on kinematics b/c topology mostly uses that

The system under study in ARPES is a half-infinite crystal, a cleaved sample with a surface. Photoelectrons are accepted by the electron analyzer through a narrow entrance slit, parametrized by an angle $\gamma$, which in typical operation may be capable of capturing photoelectrons which fly out in an angular window of $\gamma = 0^{\circ} \pm 15^{\circ}$, with negligible angular acceptance in the direction perpendicular to the slit. This perpendicular direction is parametrized by another angle $\theta$. In a typical case, we are interested in acquiring a Fermi surface by physically rotating the sample (or the analyzer) to access different $\theta$, in a range of perhaps $\theta = 0^{\circ} \pm 30^{\circ}$. We can consider the case where the analyzer is mounted so that the entrance slit is oriented vertically and we consider a cleaved sample directly facing the analyzer. In this case, intuitively, we can see that for $\theta = 0$, the analyzer slit captures photoelectrons with $k_x = 0$ and $k_y$ in a certain range corresponding to the angular window. All photoelectrons with momenta $k_x \neq 0$ or with $k_y$ outside the allowed range will miss the analyzer and will not be observed. By rotating $\theta$ we can begin to capture photoelectrons with $k_x \neq 0$. Apart from having a finite angular acceptance window, the electron analyzer further has a finite energy acceptance window, meaning that the photoelectron can only be analyzed if its kinetic energy falls in a window of something like $\varepsilon_k = \varepsilon_0 \pm 1$ eV, where $\varepsilon_0$ has to be fixed for a given measurement but can be freely changed within a wide range from scan to scan. For photoelectrons which satisfy all of these constraints, the electron analyzer directly measures their kinetic energy and the angle at which they entered the analyzer slit, providing the information,
\beq
\varepsilon_k, \gamma, \theta \textrm{\ \ \ (raw information from the analyzer, in common cases)}
\eeq
The question is then how to convert these highly technical and ARPES-specific quantities to physical information about the band structure of the crystal which produced the photoelectrons,
\beq
E_B, \bf{K} \textrm{\ \ \ (typical desired physical information about the sample)}
\eeq

\begin{figure}
\centering
\includegraphics[width=7cm]{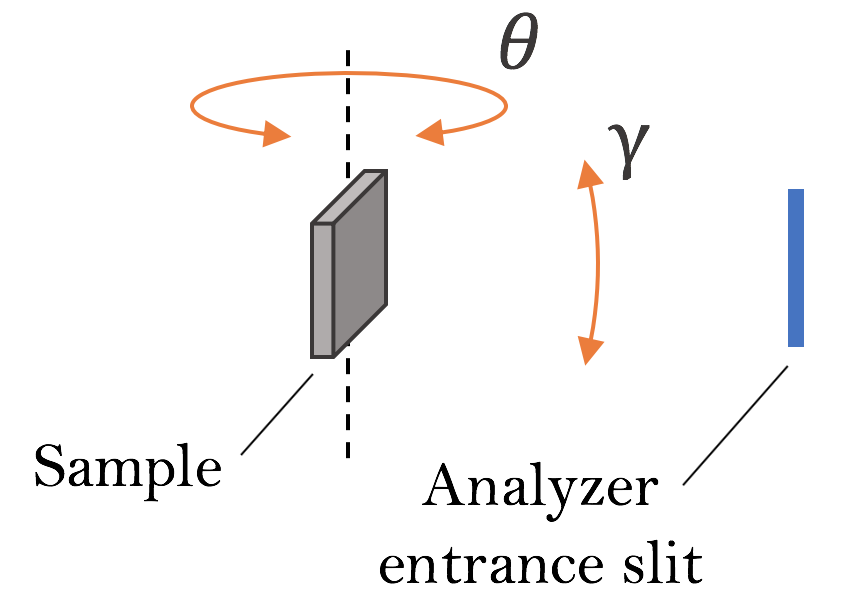}
\caption{\label{angle_conventions} The definitions of the $\gamma$ and $\theta$ angles as used in \ref{geokin}.}
\end{figure}

Here $E_B$ is the binding energy of the electron in the crystal and $\bf{K}$ is the crystal momentum, with magnitude $K$. To study this problem, let's first write down the energy and momentum conservation laws. Call $\bf{k}$ the momentum of the photoelectron in vacuum, with magnitude $k$. The cleaved crystal preserves translation symmetry in the two in-plane dimensions, so the momenta $k_x$ and $k_y$ are conserved. In a repeated zone scheme we can simply write,
\beq
K_x = k_x \textrm{\ \ \ (in-plane momentum conservation, repeated zone scheme)}
\eeq
\beq
K_y = k_y
\eeq
However, really it makes sense to restrict the crystal momentum to the Brillouin zone, so we should write,
\beq
K_x = k_x - n_i b^i_x \textrm{\ \ \ (in-plane momentum conservation, reduced zone scheme)}
\eeq
\beq
K_y = k_y - n_i b^i_x
\eeq
Here $n_i$ refers to a set of three integers which tell us how to get back to the first Brillouin zone. In the $k_z$ direction, normal to the surface, the translation symmetry is broken and momentum is not conserved. We capture this effect by including an energy shift \cite{ARPES_Damascelli2},
\beq
\label{wow}
\frac{K^2}{\alpha^2} = \frac{k^2}{\alpha^2} + V_0
\eeq
%Note that we have made the rather egregious assumption that the dispersion of the electron in the band structure is simply quadratic (really, if we already knew this, we might not even be interested in the ARPES experiment at all).

In practice, $V_0$, the so-called inner potential, is simply a fitting parameter, approximated as a constant, to capture the breaking of translation symmetry along $\hat{z}$. It is positive, $V_0 > 0$ in the above equation---the crystal surface slows down the photoelectron in the $z$ direction. The parameter $\alpha$ is a fundamental constant,
\beq
\alpha = \frac{\sqrt{2m}}{\hbar} \sim 0.5123\ \frac{\textrm{\AA}^{-1}}{\sqrt{\textrm{eV}}}
\eeq
Here $m$ is the mass of the free electron and $\hbar$, the reduced Planck's constant. The choice of units is convenient for converting energies to units of crystal momentum, typically stated in inverse $\textrm{\AA}$ngstr\"oms, $\textrm{\AA}^{-1}$.
Using the in-plane momentum conservation relations, Eq. \ref{wow} reduces to,
\beq
\label{kzeq}
K_z = \sqrt{k_z^2 + \alpha^2 V_0}
\eeq
That's it for momentum (non)-conservation. Next, conservation of energy requires that the kinetic energy of the photoelectron in vacuum be given by,
\beq
\label{kinetic}
\varepsilon_k = \frac{k^2}{\alpha^2} = h \nu - \phi - |E_\textrm{B}| 
\eeq
Here $h \nu$ is the photon energy, written as Planck's constant times the frequency of the photon. In the photoemission community it is typically stated in units of electron volts, eV. The quantity $\phi$ is the work function, roughly $4.5$ eV, which represents the energy difference between the Fermi level of the crystal and vacuum---meaning, the minimum energy needed to produce a photoelectron. It is not a sample-dependent quantity, but it is apparatus-dependent, so you and your friend's ARPES setup will have different work functions, with typical system-to-system variation of the order of $0.1$eV. Finally, the binding energy, $E_\textrm{B}$ is the difference in energy between the Fermi level and the original state occupied by the electron.

% Relationship of hydrogen and work function?
% Why does the work function vary from system to system? What does it represent? Why this value? 

Let's now apply these conservation equations. In general, photoelectrons produced by an incoming beam of light will fly off in all directions. But let us first focus our attention on photoelectrons emitted normal to the surface of the sample. These photoelectrons have $\gamma = 0$ and $\theta = 0$ and, intuitively, $k_x = k_y = 0$. Therefore, all kinetic energy contributes to the $k_z$ momentum component and we have simply,
\beq
k_z = k = \alpha \sqrt{\varepsilon_k} \textrm{\ \ \ (normal emission)}
\eeq
Using Eq. \ref{kzeq}, the momentum of the electron in the band structure is,
\beq
K_z = \alpha \sqrt{\varepsilon_k + V_0} \textrm{\ \ \ (normal emission)} 
\eeq
What about electrons which have $\theta = 0$ but $\gamma \neq 0$? These, intuitively, should have $k_x = 0$ and $k_y \neq 0$ (assuming, again, a vertical analyzer slit). Their momentum is given by rotating the normal emission momentum by an angle $\gamma$ along the analyzer slit. This means,
\beq
\begin{bmatrix}
k_x \\
k_y \\
k_z \\
\end{bmatrix}
=
\begin{bmatrix}
1 & 0 & 0 \\
0 & \cos \gamma & \sin \gamma \\
0 & -\sin \gamma & \cos \gamma \\
\end{bmatrix}
\begin{bmatrix}
0 \\
0 \\
k \\
\end{bmatrix}
=
\begin{bmatrix}
0 \\
k \sin \gamma \\
k \cos \gamma \\
\end{bmatrix}
\eeq
Where, again, $k$ is the magnitude of the momentum of the photoelectron in vacuum. Now, again, applying momentum conservation, we find,
\beq
K_y = k \sin \gamma = \alpha \sin \gamma \sqrt{\varepsilon_k}
\eeq
\beq
K_z = \sqrt{k^2 \cos^2 \gamma + \alpha^2 V_0} = \alpha \sqrt{ \varepsilon_k \cos^2 \gamma + V_0 }
\eeq
\beq
\textrm{(the case where\ } \gamma \neq 0, \ \theta = 0 \textrm{)}
\eeq
This captures the cases of a normal emission measurement where we study all photoelectrons which make it into the electron analyzer slit. However, usually we also want to map the Fermi surface, which requires a non-zero polar angle $\theta$. In this case, we first rotate the sample by $\theta$ and then collect photoelectrons emitted at an angle $\gamma$. Therefore, the vacuum momentum in terms of these natural ARPES angles is,
%\beq
%\begin{bmatrix}
%k_x \\
%k_y \\
%k_z \\
%\end{bmatrix}
%=
%\begin{bmatrix}
%1 & 0 & 0 \\
%0 & \cos\gamma & \sin\gamma \\
%0 & -\sin\gamma & \cos\gamma \\
%\end{bmatrix}
%\begin{bmatrix}
%\cos\theta & 0 & \sin\theta \\
%0 & 1 & 0 \\
%-\sin\theta & 0 & \cos\theta \\
%\end{bmatrix}
%\begin{bmatrix}
%0 \\
%0 \\
%k \\
%\end{bmatrix}
%=
%\begin{bmatrix}
%k \sin\theta \\
%k \cos\theta \sin\gamma \\
%k \cos\theta \cos\gamma \\
%\end{bmatrix}
%\eeq
%The momenta in the crystal are therefore,
%\beq
%K_x = \alpha \sin \theta \sqrt{\varepsilon_k}
%\eeq
%\beq
%K_y = \alpha \cos \theta \sin \gamma \sqrt{\varepsilon_k}
%\eeq
%\beq
%K_z = \alpha \sqrt{\varepsilon_k \cos^2 \theta \cos^2 \gamma + V_0 }
%\eeq
\beq
\begin{bmatrix}
k_x \\
k_y \\
k_z \\
\end{bmatrix}
=
\begin{bmatrix}
\cos\theta & 0 & \sin\theta \\
0 & 1 & 0 \\
-\sin\theta & 0 & \cos\theta \\
\end{bmatrix}
\begin{bmatrix}
1 & 0 & 0 \\
0 & \cos\gamma & \sin\gamma \\
0 & -\sin\gamma & \cos\gamma \\
\end{bmatrix}
\begin{bmatrix}
0 \\
0 \\
k \\
\end{bmatrix}
=
\begin{bmatrix}
k \sin\theta \cos\gamma\\
k \sin\gamma \\
k \cos\theta \cos\gamma \\
\end{bmatrix}
\eeq
The momenta in the crystal are therefore,
\beq
K_x = \alpha \sin \theta \cos \gamma \sqrt{\varepsilon_k}
\eeq
\beq
K_y = \alpha \sin \gamma \sqrt{\varepsilon_k}
\eeq
\beq
K_z = \alpha \sqrt{\varepsilon_k \cos^2 \theta \cos^2 \gamma + V_0 }
\eeq

The binding energy is given by,
\beq
|E_\textrm{B}| = h \nu - \phi - \varepsilon_k
\eeq
Other cases can be worked out analogously depending on the specific geometry of the ARPES setup \cite{Ishida_gap_functions}. These equations express ($E_\textrm{B}, \bf{K}$) in terms of the raw parameters ($\varepsilon_k, \gamma, \theta$). Note that the conversion uses the incident photon energy $h\nu$, which is known, and the work function $\phi$, which is determined by subtracting the measured kinetic energy of the Fermi level from $h\nu$ (the case $E_\textrm{B} = 0$). To determine the kinetic energy of the Fermi level, we typically fit the spectrum to a Fermi edge. There remains the parameter $V_0$, which we typically fix by performing a dependence on photon energy and searching for high-symmetry points. We can judge the accuracy of the inner potential approximation scheme by comparing the extracted value of $V_0$ for several high-symmetry photon energies. In practice when applying these equations we make the approximation $\varepsilon_k \sim h \nu - \phi$ in the equations for $\bf{K}$, since the binding energies of interest are typically only of order $0.1$ eV, which is often  $< 1\%$ of $h \nu - \phi$. We also often approximate the trigonometric functions as $\sin \gamma \sim \gamma$, $\cos \gamma \sim 1$, etc.

\section{Light sources}

The lower limit on the energy of the incoming light for an ARPES experiment is set by the work function $\phi \sim 4.5$ eV. Photons below this energy cannot excite electrons to vacuum---they cannot produce photoelectrons. The upper limit is typically set by the size of the surface Brillouin zone. Since $k$ scales with $\sqrt{h\nu - \phi}$, at higher $h\nu$ a fixed $(\gamma, \theta)$ emission angle corresponds to a larger momentum, so that eventually the surface Brillouin zone shrinks to a point and there is no longer much that is ``angle-resolved'' in the spectra. This limit occurs roughly at $1$ keV. These two fundamental considerations set the frequencies of interest for ARPES. The human eye responds to energies below $\sim 3.3$ eV (380nm, violet), so the lowest-energy light useful for ARPES is ultraviolet\footnote{Actually, this mismatch in energy range is no coincidence: it is optical transitions, not photoelectric effects, that are associated with photoisomerization in the human eye. At frequencies where the photoelectric effect becomes important, the human eye no longer responds.}. The highest-energy useful light falls into the lower region of the X-ray frequencies and so is called soft X-ray light.

The community continues to develop a rich variety of light sources to produce radiation for ARPES experiments. The goal is typically to increase the photon flux while improving the energy resolution, decreasing the size of the beamspot and increasing the available range of photon energies. Typical light sources for ARPES include noble gas plasma lamps, lasers and synchrotron beamlines. Lamps often use helium or xenon atomic emission lines to produce light, especially the He I$\alpha$ line at 21.2 eV, the He II$\alpha$ line at 40.8e eV and the Xe I line at 8.4 eV. They are simple and robust, but often suffer from low flux, a large beamspot and poor monochromaticity (poor energy resolution), although more sophisticated setups can improve their performance. Lasers provide high flux and excellent energy resolution, but it has traditionally been challenging to produce laser light at sufficiently high photon energies for photoemission. Notably, nonlinear crystals have been successfully implemented to up-convert visible laser light to 6 eV or 7 eV while maintaining high flux and energy resolution. Other approaches promise even higher photon energies. For instance, the Artemis facility at the Rutherford Appleton Laboratory in the UK uses high harmonic generation on a gas target to provide light with resolution of $\sim 0.1$ eV, flux of $10^9$ ph/s (photons per second) and a beam spot size of 100$\mu$m$^2$ in the range of photon energies 21 eV to 35 eV as of mid-2018\footnote{See the Artemis website: \href{https://www.clf.stfc.ac.uk/Pages/Artemis.aspx}{https://www.clf.stfc.ac.uk/Pages/Artemis.aspx}}.

Lastly, synchrotron light sources are fantastically sophisticated and complex, but arguably provide the purest and most versatile beams of light known to man. For instance, the state-of-the-art SSRL BL 5-2 in SLAC, CA, USA offers photon energies of 25 eV to 200 eV with beamline resolutions easily better than 5 meV and flux of order $10^{12}$ ph/s with typical spot size $40$ $\mu$m horizontally and $10$ $\mu$m vertically. The ADRESS beamline at the Swiss Light Source in PSI, Villigen, Switzerland produces light in the soft X-ray regime, from 300 eV to 1600 eV with a spot size of $\sim 10 \times 70$ $\mu$m, beamline resolution approaching 30 meV and a phenomenal flux of $10^{13}$ ph/s \cite{ADRESS_Strocov}. This high flux is crucial because photoemission cross-section decays precipitously for soft X-ray photon energies, often requiring measurements with long integration times and sacrifices to resolution.

% , see Ref. \cite{laser_xenon_Harter} for an example
% Add more precise upper bound to hv
% Typical momentum shift due to the momentum of the photon
% Cross-section, and need for soft-x ray high intensity to get signal at those photon energies, universal curve, etc.

\section{The manipulator}

In most ARPES setups, the direction of the photon beam is fixed and the electron analyzer remains fixed during the measurement. As a result, to produce a Fermi surface mapping, it is necessary to rotate the sample, so that we can capture photoelectrons leaving the sample surface in different directions. State-of-the-art manipulators offer six degrees of freedom: translation in $x$, $y$ and $z$ in a range of several millimeters; along with rotation about the axis of the manipulator, typically labelled $\theta$; rotation about an axis perpendicular to the manipulator, $\phi$; and rotation about the plane of the sample surface, called $\omega$ or azimuth, typically all with a range of $\sim 100^{\circ}$. The translation degrees of freedom allow you to search for the highest-quality surface region of the sample. This capability is more and more crucial as micron-scale beam spots become routine. The manipulator also controls the sample temperature, typically through cooling by liquid helium (LHe), supplied at a temperature of $4.2$ K, as well as an onboard heater that allows raising the temperature. The lowest base temperature realistically achievable appears to be $\sim 7$ K on a six-axis manipulator, likely due to its complicated mechanical apparatus which results in the loss of a few degrees temperature. Typically, the thermocouple reading is itself expected to be overly optimistic, because it is mounted slightly away from the sample position. Realistically one only knows that the sample temperature is a few degrees above the recorded thermocouple temperature. Another consideration is that the sample needs to be electrically grounded for ARPES, so that the emitted photoelectrons can be replaced. As a result, strongly insulating samples cannot be measured and the sample must be mounted in a way that maintains electrical contact with the body of the manipulator.

% Sample needs to be grounded
% Temperature error
% Alkali metal deposition
% Gold discussion with example plot

\section{The electron analyzer}

How does the electron analyzer measure the energy and momentum of the incoming photoelectron? The design principle uses a hemispherical analyzer with an applied electric field. When the photoelectron enters the hemisphere, it is deflected by the applied electric field. If the energy is too high, the applied electric field is insufficient to steer the photoelectron through the hemisphere and the photoelectron crashes against the outer shell of the hemisphere. If the energy is too low, the applied electric field causes the photoelectron to crash into the inner shell. However, for a range of allowed energies, the electron makes it through and falls onto a two-dimensional detector, where its position is recorded. The angular degree of freedom also comes along for the ride and different incident angles correspond as well to different points of impact on the two-dimensional detector. In this way, the incident energy and angle ($\varepsilon$,$\gamma$) of the photoelectron is mapped onto the ($x$, $y$) position of a detector. The detector consists of a micro-channel plate (MCP), which uses a cascade effect to amplify the signal associated with a single electron impact, followed by a phosphor screen which converts electron impacts into flashes of light, which are then recorded by an ordinary charge-coupled device (CCD) camera.

The width $w$ of the analyzer slit contributes to the energy resolution. To see this, consider electrons passing through the hemisphere at the pass energy, $P$, the energy for which the electric field and the electron velocity are matched and the electron stays at fixed radius throughout its trajectory. For an infinitely narrow slit, all photoelectrons with $\varepsilon = P$ will land along a line in the center of the MCP. However, for a slit with finite width, these electrons will now land on a strip of finite width. In addition, $w$ contributes to the angular resolution in the direction perpendicular to the slit, the direction of Fermi surface mappings.

\begin{figure}
\centering
\includegraphics[width=15cm]{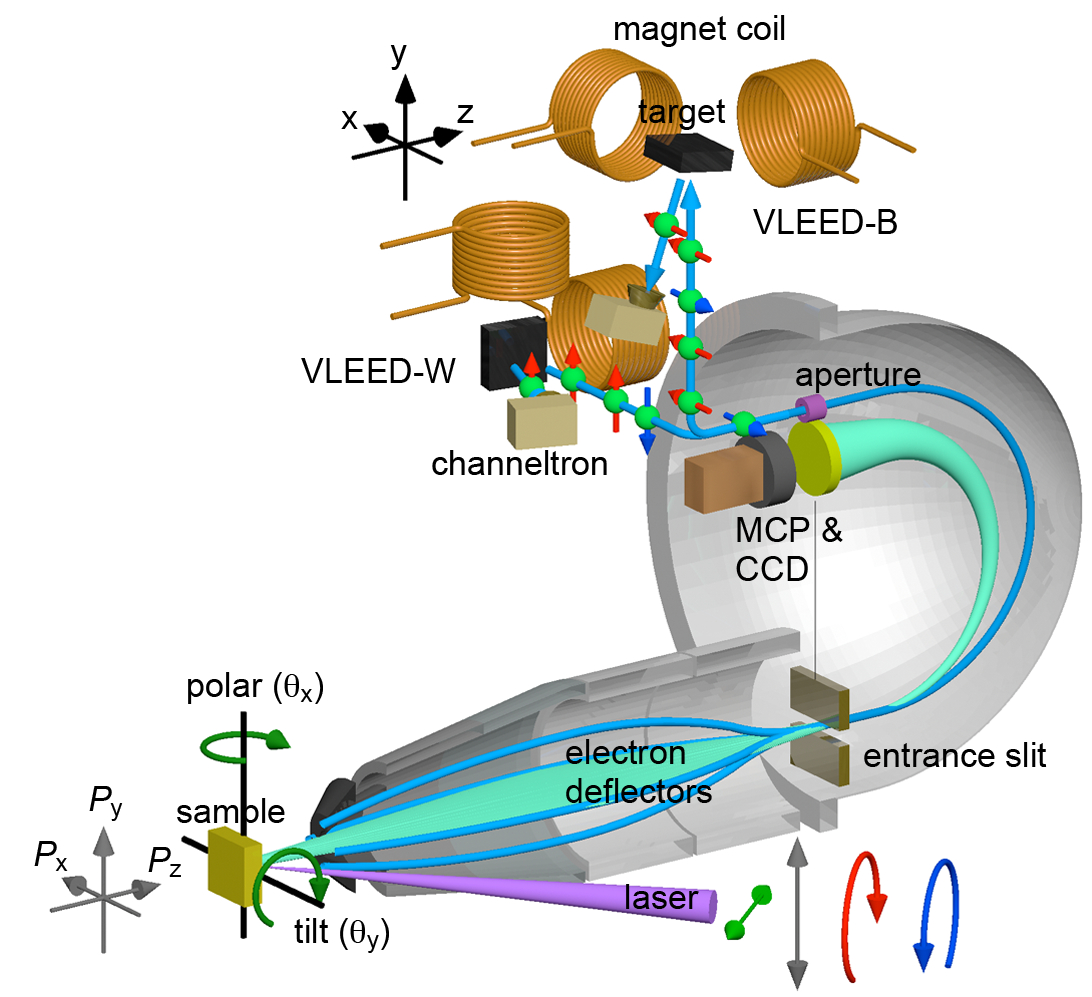}
\caption{\label{electron_analyzer} Key components of an electron analyzer, shown here with a spin resolution capability. Figure from Ref. \cite{Kondo_analyzer}.}
\end{figure}

\begin{figure}
\centering
\includegraphics[width=16cm, trim={3cm, 6cm, 2.5cm, 4.5cm}, clip]{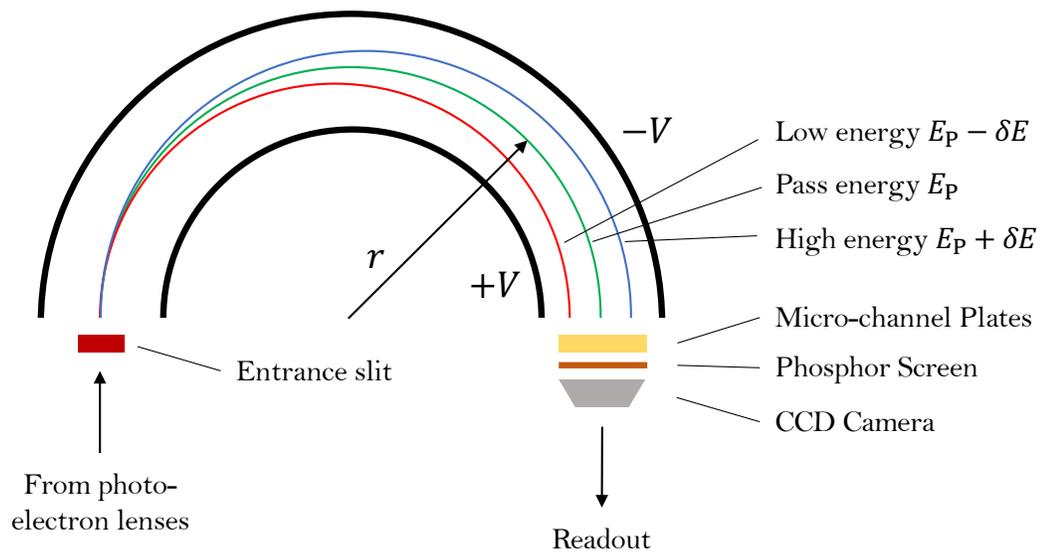}
\caption{\label{electron_analyzer} Operating principle of a hemispherical electron analyzer. Electrons at the pass energy follow a path of constant radius through the analyzer, while those with lower energy approach the inner shell and those with higher energy approach the outer shell. In this way, different energies are converted into different real space positions on the micro-channel plate (MCP) detector.}
\end{figure}

%  From ``DA30-L, Advanced Training'', a presentation by Patrik Karlsson, ScientaOmicron.

%The pass energy $P$ contributes to the energy resolution by setting the energy range corresponding to the width of the MCP. For a higher pass energy, the electrons have less time to disperse in $x$ as they traverse the hemisphere, so there is a larger acceptable energy window.

%The energy resolution can be summarized by,

%\beq
%\Delta \varepsilon = \frac{w P}{2 r}
%\eeq

% Resolution

%\cleardoublepage

\clearpage

\ifdefined\phantomsection
  \phantomsection  % makes hyperref recognize this section properly for pdf link
\else
\fi
\addcontentsline{toc}{section}{Bibliography}

{\singlespacing
%\bibliography{../../master_bib}{}
%\bibliographystyle{../../Science}

}

%% file: ch-TaAs/ch-TaAs.tex
\chapter{Discovery of a Weyl fermion semimetal and topological\\ Fermi arcs in TaAs}
\label{ch:taas}

{\singlespacing
\begin{chapquote}{The Naked and Famous, \textit{What We Want}}
All the chemicals reel in the absence of the noise\\
We are fools in the wake of the physical\\
I just don't know what I want...\\
Just keep it trivial
\end{chapquote}}

%%\documentclass[prl,aps,superscriptaddress,preprint,floatfix,
%%nofootinbib]{revtex4}
%
%\documentclass[aps,prl,preprint,nopacs,superscriptaddress]{revtex4}
%\usepackage{graphicx}
%\usepackage{verbatim}
%\usepackage{mathrsfs}
%\pagestyle{headings}
%
%%\documentclass[pra,aps,showpacs,groupedaddress,preprint]{revtex4}
%
%%\documentclass[pra,aps,showpacs,groupedaddress]{revtex4}
%%\setlength{\oddsidemargin}{.5in}
%%\setlength{\textwidth}{5.5in}
%
%\usepackage{amsmath,amsfonts,amssymb}
%%\usepackage{wrapfig}
%\usepackage{graphicx}
%%\usepackage{bbm}
%%\usepackage{graphics}k
%
%%\setcitestyle{round}
\def\3{2.8in}    %used for figure widths
\def\2{2.5in}
\def\4{3.0in}

\def \beq {\begin{equation}}
\def \eeq {\end{equation}}

\noindent This chapter is based on the article, \textit{Discovery of a Weyl fermion semimetal and topological Fermi arcs} by Su-Yang Xu*, Ilya Belopolski* \textit{et al}., \textit{Science} {\bf 349}, 613 (2015), available at \href{https://science.sciencemag.org/content/349/6248/613}{https://science.sciencemag.org/content/349/6248/613}.\\

%\begin{abstract}
\lettrine[lines=3]{A} Weyl semimetal is a crystal which hosts Weyl fermions as emergent quasiparticles and admits a topological classification that protects Fermi arc surface states on the boundary of a bulk sample. This unusual electronic structure has deep analogies with particle physics, leads to unique transport properties and extends the topological classification beyond insulators. Here, we report the experimental discovery of the first Weyl semimetal, TaAs. Using photoemission spectroscopy, we directly observe Fermi arcs on the surface and Weyl cones in the bulk of single crystals of TaAs. We find that Fermi arcs terminate on the Weyl nodes, consistent with the topological classification. Our work opens the field for the experimental study of Weyl fermions and topological Fermi arcs in condensed matter and materials physics.
%A Weyl semimetal is a crystal which hosts Weyl fermions as emergent quasiparticles and admits a topological classification that protects Fermi arc surface states on the boundary of a bulk sample. This unusual electronic structure has deep analogies with particle physics, may lead to unique properties in transport and extends the classification of topological phases beyond insulators. Here we report the experimental discovery of the first Weyl semimetal, TaAs. Using photoemission spectroscopy, we directly observe Fermi arcs on the surface and Weyl cones in the bulk of single crystals of TaAs. We find that the surface state Fermi arcs terminate at the Weyl nodes in a way consistent with the topological classification. Our work opens up the field for the experimental study of Weyl fermions and topological Fermi arcs in condensed matter and materials physics.
%\end{abstract}
%\date{\today}
%\maketitle

\section{Introduction}

Weyl fermions have long been known in quantum field theory, but have not been observed as a fundamental particle in nature \cite{Weyl, Balents_viewpoint, Wilczek}. Recently, it was understood that a Weyl fermion can arise as an emergent quasiparticle in certain crystals, Weyl semimetals, leading to intense research interest \cite{Weyl, Balents_viewpoint, Wilczek, TI_book_2014, Murakami2007, Wan2011, Thallium, Suyang, Vanderbilt, Hosor, Chiral, Aji2012, Carbotte2013, Nielsen1983, Chiral_Qi, Ojanen, Ashvin2, NMR, Hasan2010, Qi2011, Hsin_TaAs, Dai_TaAs, Hsin_TaAs_PRB}. A Weyl semimetal offers a dramatic example of emergence in physics and enriches the deep correspondence between high-energy physics and condensed matter physics. Despite being a gapless metal, it is further characterized by topological invariants, broadening the classification of topological phases of matter beyond insulators. Specifically, Weyl fermions at zero energy correspond to points of bulk band degeneracy, Weyl points, which are associated with a chiral charge that protects gapless surface states on the boundary of a bulk sample. These surface states take the form of Fermi arcs connecting the projection of bulk Weyl points in the surface Brillouin zone (BZ) \cite{Wan2011}. Such a surface state band structure would violate basic band theory in an isolated two-dimensional system and can only arise on the boundary of a three-dimensional sample, providing a dramatic example of the bulk-boundary correspondence in a topological phase. In contrast to topological insulators where only the surface states are interesting \cite{Hasan2010, Qi2011}, a Weyl semimetal features unusual band structure in the bulk and on the surface. This opens up unparalleled research opportunities, where both bulk and surface sensitive experimental probes can measure the topological nature and detect novel quantum phenomena. Specifically, the Weyl fermions in the bulk provides a condensed matter realization of the chiral anomaly, giving rise to a negative magnetoresistance under parallel electric and magnetic fields, novel optical conductivity, non-local transport and local non-conservation of ordinary current predicted in theories \cite{Chiral, Chiral_Qi, Nielsen1983, NMR}. At the same time, the Fermi arc surface states are predicted to show novel quantum oscillations in magneto-transport, as well as unusual quantum interference effects in tunneling spectroscopy. The realization of Weyl fermions in a condensed matter system, the novel topological classification and the unusual transport phenomena associated with both the bulk Weyl points and Fermi arc surface states have attracted worldwide research interest \cite{TI_book_2014, Thallium, Suyang, Weyl, Balents_viewpoint, Wilczek, Murakami2007, Wan2011, Vanderbilt, Hosor, Chiral, Aji2012, Carbotte2013, Chiral_Qi, Nielsen1983, Ojanen, Ashvin2, Hasan2010, Qi2011, Hsin_TaAs, NMR, Dai_TaAs, Hsin_TaAs_PRB}. However, for many years, a Weyl semimetal has not be found in experiments and has become a much-sought-out treasure of condensed matter physics.

Here we report the experimental realization of the first Weyl semimetal in a single crystalline material tantalum arsenide, TaAs. Utilizing the combination of the vacuum ultraviolet (low-energy) and soft X-ray (SX) angle-resolved photoemission spectroscopy (ARPES), we systematically and differentially study the surface and bulk electronic structure of TaAs. Our ultraviolet (low-energy) ARPES measurements, which are highly surface sensitive, demonstrate the existence of the Fermi arc surface states, consistent with our band calculations presented here. Moreover, our SX-ARPES measurements, which are reasonably bulk sensitive, reveal the three-dimensional linearly dispersive bulk Weyl cones and Weyl nodes. Furthermore, by combining the low-energy and SX-ARPES data, we show that the locations of the projected bulk Weyl nodes correspond to the terminations of the Fermi arcs within our experimental resolution. These systematic measurements demonstrate TaAs as a Weyl semimetal. All previous experimental studies on all candidate materials failed to find the surface states and the bulk Weyl nodes, or anything that resembles any correspondence. It is only in TaAs (this work), both Fermi arcs and Weyl cones, consistent with theory, are observed for the first time. These results represent a breakthrough, which is expected to lead to a flurry of follow-up research activities opening up the field for experimental studies. %Our experimental identification of the first Weyl semimetal opens a new era in the synergy between high-energy and condensed matter physics.

\begin{figure*}
\centering
\includegraphics[width=15cm, trim={1in 0in 1in 0in}, clip]{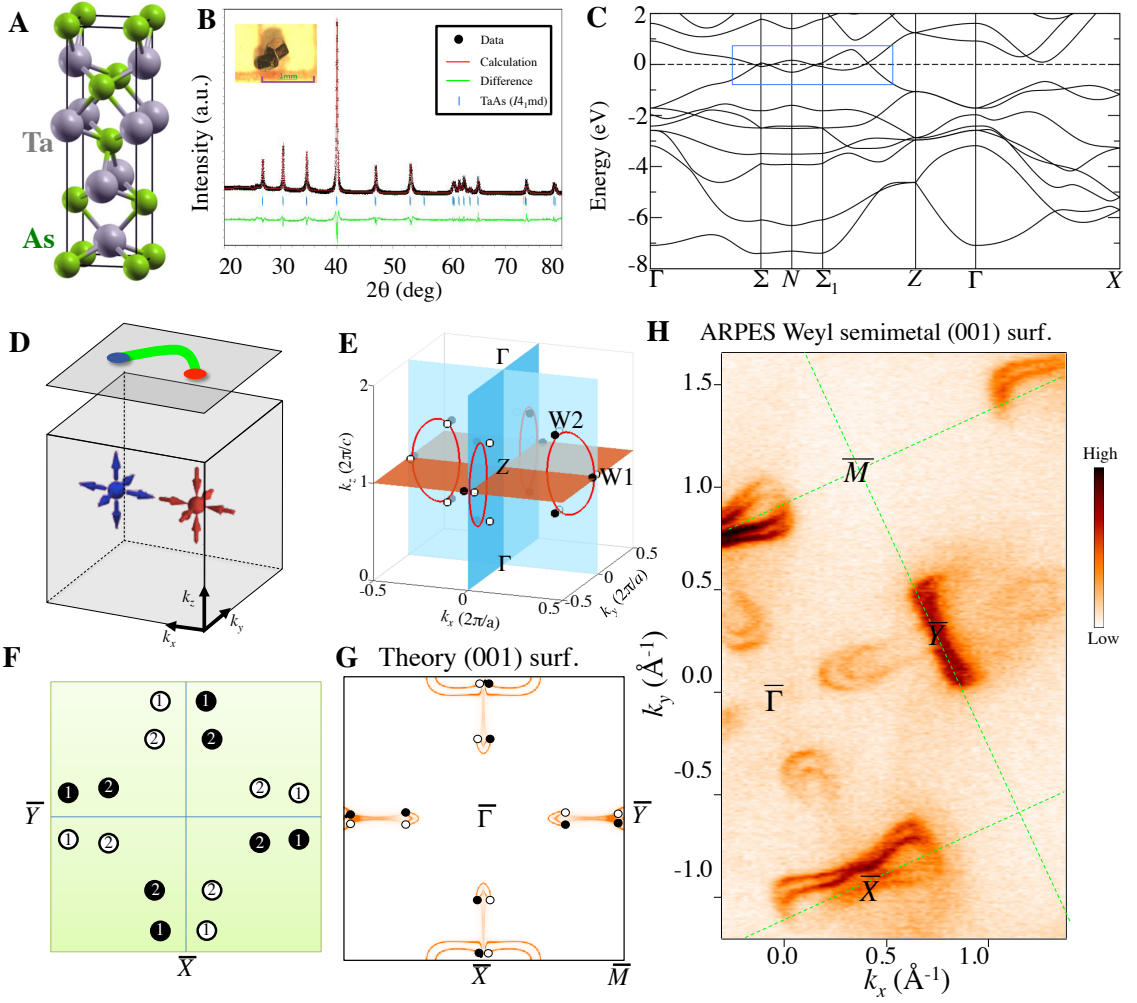}
\caption{\label{TaAsFig1}\textbf{Crystal structure and electronic structure of TaAs.} (\textbf{A}) Body-centred tetragonal structure of TaAs, shown as stacked Ta and As layers. The lattice of TaAs does not have space inversion symmetry. (\textbf{B}) X-ray diffraction measurements on the TaAs samples used in our ARPES experiments. (\textbf{C}) First-principles band structure calculations of TaAs without spin-orbit coupling. The blue box highlights the bulk bands touchings locations in the BZ. (\textbf{D}) Schematic illustration of the simplest Weyl semimetal state that has two single Weyl nodes with the opposite ($\pm1$) chiral charges in the bulk. (\textbf{E}) In the absence of spin-orbit coupling, there are two line nodes on the $k_{x}=0$ mirror plane and two line nodes on the $k_{y}=0$ mirror plane. In the presence of spin-orbit coupling, each line node vaporizes into six Weyl points. The Weyl points are denoted by small circles. Black and white show the opposite chiral charges of the Weyl points. (\textbf{F}) A schematic showing the projected Weyl points and their projected chiral charges. (\textbf{G}) Theoretically calculated band structure of the Fermi surface on the (001) surface of TaAs. (\textbf{H}) The ARPES measured Fermi surface of the (001) cleaving plane of TaAs. The high symmetry points of the surface Brillouin zone are noted.}
\end{figure*}

Tantalum arsenide, TaAs, is a semimetallic material that crystalizes in a body-centered tetragonal lattice system (Fig.~\ref{TaAsFig1}A) \cite{Shuang_TaAs}. The lattice constants are $a=3.437$ \r{A} and $c=11.656$ \r{A}, and the space group is $I4_{1}md$ (\#109, $C_{4v}$), as consistently reported in all previous structural studies \cite{TaAs_Crystal_1, TaAs_Crystal_3, Shuang_TaAs}. The crystal consists of interpenetrating Ta and As sub-lattices, where the two sub-lattices are shifted by $\left( \frac{a}{2},\frac{a}{2},\delta \right) $, $\delta \approx \frac{c}{12}$. It is important to note that the system lacks space inversion symmetry. Our diffraction data (Fig.~\ref{TaAsFig1}B) matches well with the lattice parameters and the space group $I4_{1}md$, confirming the lack of inversion symmetry in our single crystal TaAs samples. Electrical transport measurements on TaAs confirmed its semimetallic transport properties and reported negative magnetoresistance suggesting the anomalies due to Weyl fermions \cite{Shuang_TaAs}. %Furthermore, the intensity ratio between the Ta $4f$ and As $3d$ peaks is close to $1:1$. These measurements further confirm the chemical composition and the high quality of our TaAs samples.reported the ultrahigh electron mobility up to $10^5$ cm$^2/$V$\cdot$s

We discuss the essential aspects of the theoretically calculated bulk band structure \cite{Hsin_TaAs, Dai_TaAs, Hsin_TaAs_PRB} that predicts TaAs as a Weyl semimetal candidate. Without spin-orbit coupling, calculations \cite{Hsin_TaAs, Dai_TaAs, Hsin_TaAs_PRB} show that the conduction and valence bands interpenetrate (dip into) each other to form four 1D line nodes (closed loops) located on the $k_{x}=0$ and $k_{y}=0$ blue planes (Figs.~\ref{TaAsFig1}C,E). Upon the inclusion of spin-orbit coupling, each line node loop is gapped out and shrinks into six Weyl nodes that are away from the $k_{x}=0$ and $k_{y}=0$ mirror planes, shown by the small circles in Fig.~\ref{TaAsFig1}E. In our calculation, in total there are 24 bulk Weyl cones \cite{Hsin_TaAs, Dai_TaAs, Hsin_TaAs_PRB}, all of which are linearly dispersive and are associated with a single chiral charge of $\pm1$ (Fig.~\ref{TaAsFig1}E). We denote the 8 Weyl nodes that are located on the brown plane ($k_z=2\pi/c$) as W1 and the other 16 nodes that are away from this plane as W2. At the (001) surface BZ (Fig.~\ref{TaAsFig1}F), the 8 W1 Weyl nodes are projected near the vicinity of the surface BZ edges, $\bar{X}$ and $\bar{Y}$. More interestingly, for the 16 W2 Weyl nodes, two of the W2 Weyl nodes with the same chiral charge are projected onto the same point on the surface BZ. Therefore, in total there are 8 projected W2 Weyl nodes with a projected chiral charge of $\pm2$, which are located near the midpoints of the $\bar{\Gamma}-\bar{X}$ and the $\bar{\Gamma}-\bar{Y}$ lines. Since the $\pm2$ chiral charge is a projected value, the Weyl cone is still linear \cite{Hsin_TaAs, Hsin_TaAs_PRB}. The number of Fermi arcs terminating on a projected Weyl node must equal its projected chiral charge. Therefore, in TaAs, two Fermi arc surface states must terminate on each projected W2 Weyl node.
 
 \section{Observation of Fermi arcs}
 
We carried out low-energy ARPES measurements to explore surface electronic structure of TaAs. Fig.~\ref{TaAsFig1}H presents an overview of the (001) Fermi surface map. We observe three types of dominant features, namely a crescent-shaped feature in the vicinity of the midpoint of each $\bar{\Gamma}-\bar{X}$ or $\bar{\Gamma}-\bar{Y}$ line, a bowtie-like feature centered at the $\bar{X}$ point, as an extended feature centered at the $\bar{Y}$ point. We find that the Fermi surface and the constant energy contours at shallow binding energies (Fig.~\ref{TaAsFig2}A) violate the $C_4$ symmetry, considering the features at $\bar{X}$ and $\bar{Y}$ points. In the crystal structure of TaAs, where the rotational symmetry is implemented as a screw axis that sends the crystal back into itself after a $C_4$ rotation and a translation by $c/2$ along the rotation axis, such asymmetry is expected in calculation. The crystallinity of (001) surface in fact breaks the rotational symmetry. We now focus on the crescent-shaped features. Their peculiar shape suggests the existence of two arcs since their termination points in $k$-space coincide with the surface projection of the W2 Weyl nodes. Since the crescent feature consists of two non-closed curves, it can either arise from two Fermi arcs or a closed contour, however, the decisive property that clearly distinguishes one case from the other is the way in which the constant energy contour evolves as a function of energy. As shown in Fig.~\ref{TaAsFig2}F, in order for the crescent feature to be Fermi arcs, the two non-closed curves have to move (disperse) to the same direction as one varies the energy (please refer to the Supplementary Materials (SM) for systematic elaborations on this point). We have unbiasedly carried out systematic measurements on the crescent feature. We now provide ARPES data to show that the crescent feature in TaAs indeed exhibits this ``co-propagating'' property. To do so, we single out a crescent feature as shown in Figs.~\ref{TaAsFig2}B,E and show the band dispersions at representative momentum space cuts, Cut I and Cut II, as defined in Fig.~\ref{TaAsFig2}E. The corresponding $E$-$k$ dispersions are shown in Figs.~\ref{TaAsFig2}C,D. The evolution (dispersive ``movement'') of the bands as a function of binding energy can be clearly read from the slope of the bands in the dispersion maps, which is, in turn, drawn in Fig.~\ref{TaAsFig2}E by the white arrows. It can be seen that the evolution of the two non-closed curves are consistent with the co-propagating property. In order to further visualize the evolution of the constant contour throughout $k_x$, $k_y$ space. We take the surface state constant energy contours at two slightly different binding energies, namely $E_{\textrm{B}}=0=E_{\textrm{F}}$ and $E_{\textrm{B}}=20$ meV. Fig.~\ref{TaAsFig2}G shows the difference between these two constant energy contours, namely ${\Delta}I(k_x,k_y) = I(E_{\textrm{B}}=20$ meV$, k_x,k_y)-I(E_{\textrm{B}}=0$ meV$, k_x,k_y)$, where $I$ is the ARPES intensity. By performing this subtraction as a check, the $k$-space regions in Fig.~\ref{TaAsFig2}G that have negative spectral weight (red in color) corresponds to the constant energy contour at $E_{\textrm{B}}=0$ meV, whereas those regions with positive spectral weight (blue in color) corresponds to the contour at $E_{\textrm{B}}=20$ meV. Thus one can visualize the two contours in a single $k_x,k_y$ map. The alternating "red - blue - red - blue" sequence for each crescent feature in Fig.~\ref{TaAsFig2}G again shows the co-propagating property, consistent with the schematic drawing in Fig.~\ref{TaAsFig2}F. Furthermore, we note that there are two crescent features, one located near the $k_x=0$ axis and the other near the $k_y=0$ axis, in Fig.~\ref{TaAsFig2}G. The fact that we observe the co-propagating property for two independent crescent features which are $90^{\circ}$ rotated with respect to each other further shows that this observation is not due to artifacts, such as a $k$ misalignment while performing the subtraction. The above systematic data reveal the existence of Fermi arcs on the (001) surface of TaAs. Just like one can identify a crystal as a topological insulator by observing an odd number of Dirac cone surface states without surface calculations, we emphasize that our systematic data here, alone without referring to any surface band structure calculations, is sufficient to identify TaAs as a Weyl semimetal because of bulk-boundary correspondence in topology.

Theoretically, the co-propagating property of the Fermi arcs is unique to Weyl semimetals because it arises from the nonzero chiral charge of the projected bulk Weyl nodes (see the SM), which in this case is $\pm2$. Therefore, this property distinguishes the crescent Fermi arcs not only from any closed contour but also from the double Fermi arcs in Dirac semimetals \cite{Hasan_Na3Bi, Nagaosa} because the bulk Dirac nodes do not carry any net chiral charges (see the SM). After observing the surface electronic structure containing Fermi arcs in our ARPES data, we are able to slightly tune the free parameters of our surface calculation and obtain a calculated surface Fermi surface that reproduces and explains our ARPES data, as shown in Fig.~\ref{TaAsFig1}G. This serves as an important cross-check that our data and interpretation are self consistent. Specifically, our surface calculation indeed also reveals the crescent Fermi arcs that connect the projected W2 Weyl nodes near the midpoints of each $\bar{\Gamma}-\bar{X}$ or $\bar{\Gamma}-\bar{Y}$ line (Fig.~\ref{TaAsFig1}G). In addition,s our calculation shows the bowtie surface states centered at the $\bar{X}$ point, also consistent with our ARPES data. According to our calculation, these bowtie surface states are in fact Fermi arcs (see the SM) associated with the W1 Weyl nodes near the BZ boundaries. However, our ARPES data cannot resolve the arc character since the W1 Weyl nodes are too close to each other in momentum space compared to the experimental resolution. Additionally, we note that the agreement between the ARPES data and the surface calculation upon the contour at the $\bar{Y}$ point can be further improved by fine-tuning the surface parameters. In order to establish the topology, it is not necessary for the data to have a perfect correspondence with the details of calculation since some changes in the choice of the surface potential allowed by the free parameters do not change the topology of the materials. This is well-known in calculations and experiments on topological insulators \cite{Hasan2010, Qi2011}. In principle, Fermi arcs can coexist with additional closed contours in a Weyl semimetal \cite{Wan2011, Hsin_TaAs, Hsin_TaAs_PRB}, just as Dirac cones can coexist with additional trivial surface states in a topological insulator \cite{Hasan2010, Qi2011}. Particularly, establishing one set of Weyl Fermi arcs is sufficient to prove a Weyl semimetal \cite{Wan2011}. This is achieved by observing the crescent Fermi arcs as we show here by our ARPES data in Fig.~\ref{TaAsFig2}, which is further consistent with our surface calculations.

\section{Demonstration of bulk Weyl cones}

We now present bulk-sensitive SX-ARPES \cite{SXARPES} data, which reveal the existence of bulk Weyl cones and Weyl nodes. This serves as an independent but equally strong proof of the Weyl semimetal state in TaAs. Fig.~\ref{TaAsFig3}B shows the SX-ARPES measured $k_x$-$k_z$ Fermi surface at $k_y=0$ (note that none of the Weyl nodes are located on the $k_y=0$ plane). We emphasize that the clear dispersion along the $k_z$ direction (Fig.~\ref{TaAsFig3}B) firmly shows that our SX-ARPES predominantly images the bulk bands. SX-ARPES allows us the bulk-surface contrast in favor of the bulk band structure, which can be further tested by measuring the band dispersion along the $k_z$ axis in the SX-ARPES setting. This is confirmed by the agreement between the ARPES data (Fig.~\ref{TaAsFig3}B) and the corresponding bulk band calculation (Fig.~\ref{TaAsFig3}A). We now choose an incident photon energy (i.e. a $k_z$ value) that corresponds to the $k$-space location W2 Weyl nodes and map the corresponding $k_x$-$k_y$ Fermi surface. As shown in Fig.~\ref{TaAsFig3}C, the Fermi points that are located away from the $k_x$ or $k_y$ axes are the W2 Weyl nodes. In Fig.~\ref{TaAsFig3}D, we clearly observe two linearly dispersive cones that correspond to the two nearby W2 Weyl nodes along Cut 1. The $k$-space separation between the two W2 Weyl nodes is measured to be $0.08$ $\textrm{\AA}^{-1}$, which is consistent with both the bulk calculation and the separation of the two terminations of the crescent Fermi arcs measured in Fig.~\ref{TaAsFig2}. The linear dispersion along the out-of-plane direction for the W2 Weyl nodes is shown by our data in Fig.~\ref{TaAsFig3}E. Additionally, we also observe the W1 Weyl cones in Figs.~\ref{TaAsFig3}G-I. Notably, our data shows that the energy of the bulk W1 Weyl nodes is lower than that of the bulk W2 Weyl nodes, which agrees well with our calculation shown in Fig.~\ref{TaAsFig3}J and an independent modeling of the bulk transport data on TaAs \cite{Shuang_TaAs}. 

We emphasize that the observation of the Weyl nodes in Fig.~\ref{TaAsFig3} provides another independent demonstration of the Weyl semimetal state in TaAs as Weyl fermions only occur in Weyl semimetals (metals) in spin-orbit coupled bulk crystals. In these materials, point-like band crossings can either be Weyl cones or Dirac cones. Because the observed bulk cones in Figs.~\ref{TaAsFig3}C and D are located neither at Kramers' points nor on a rotational axis, they cannot be identified as bulk Dirac cones and have to be Weyl cones according to all topological theories \cite{Nagaosa, Wan2011}. Therefore, our SX-ARPES data alone, proves the existence of bulk Weyl nodes. Furthermore, the agreement between the SX-ARPES data and our bulk calculation, which only requires the crystal structure and the lattice constants as inputs, provides further cross-check. Thus by revealing the bulk Weyl cones and Weyl nodes, our SX-ARPES data, independent of our low-energy ARPES (surface sensitive) data, demonstrate TaAs as a bulk Weyl semimetal.

\section{Bulk-boundary correspondence between Fermi arcs and Weyl cones}

\begin{figure}
\centering
\includegraphics[width=15.5cm, trim={1cm, 1cm, 1cm, 1cm}, clip]{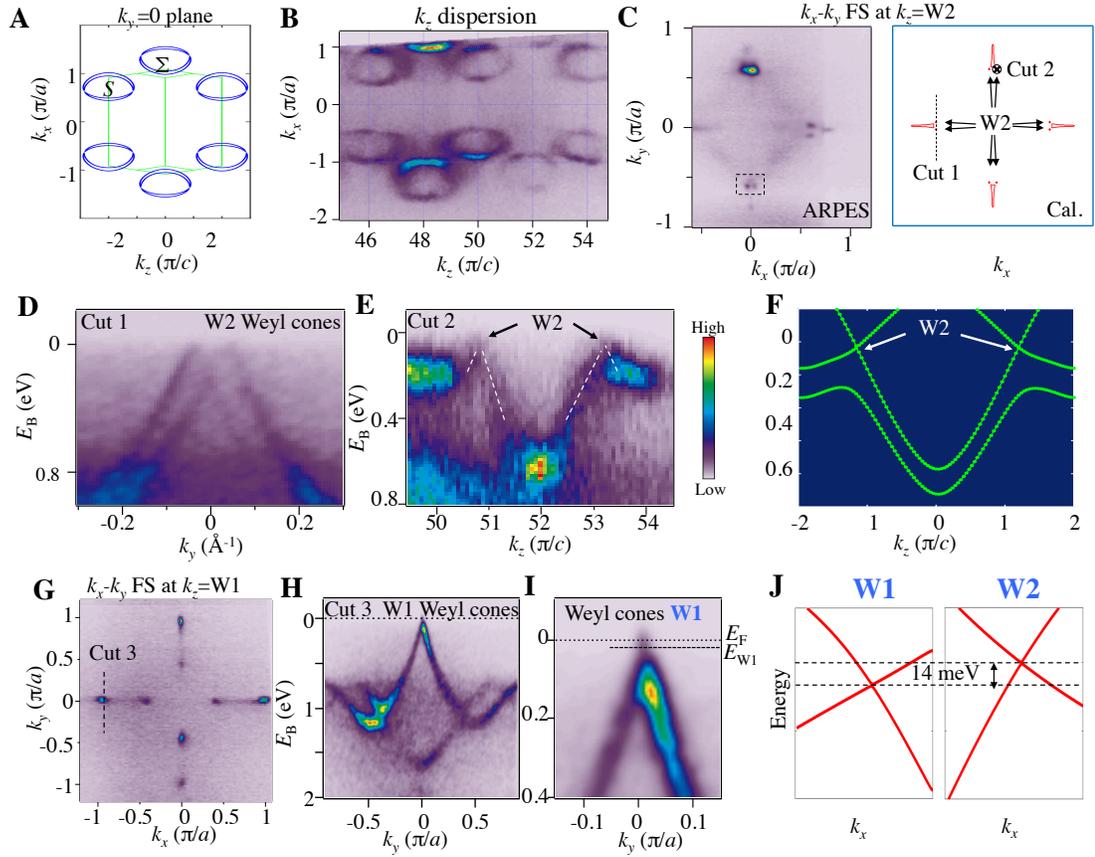}
\caption{\textbf{Observation of bulk Weyl cones and Weyl nodes in TaAs.} (\textbf{A, B}) First-principles calculated and ARPES measured $k_z$-$k_x$ Fermi surface maps at $k_y=0$. (\textbf{C})  ARPES measured and first-principles calculated $k_x$-$k_y$ Fermi surface maps at the $k_z$ value that corresponds to the W2 Weyl nodes. The dotted line defines the $k$-space cut direction for Cut 1, which goes through two nearby W2 Weyl nodes along the $k_y$ direction. The black cross defines Cut 2, which means that the $k_x$, $k_y$ values are fixed at the location of a W2 Weyl node and one varies the $k_z$ value. (\textbf{D}) ARPES $E$-$k_y$ dispersion map along the Cut 1 direction, which clearly shows the two linearly dispersive W2 Weyl cones. (\textbf{E}) ARPES $E$-$k_z$ dispersion map along the Cut 2 direction, showing that the W2 Weyl cone also disperses linearly along the out-of-plane $k_z$ direction. (\textbf{F}) First-principles calculated $E$-$k_z$ dispersion that corresponds to the Cut 2 shown in panel E. (\textbf{G}) ARPES measured $k_x$-$k_y$ Fermi surface maps at the $k_z$ value that corresponds to the W1 Weyl nodes. The dotted line defines the $k$-space cut direction for Cut 3, which goes through the W1 Weyl nodes along the $k_y$ direction. (\textbf{H,I}) ARPES $E$-$k_y$ dispersion map and its zoomed-in version along the Cut 3 direction, revealing the linearly dispersive W1 Weyl cone. (\textbf{J}) First-principles calculation shows a 14 meV energy difference between the W1 and W2 Weyl nodes.}\label{TaAsFig3}
\end{figure}

Finally, we show that the $k$-space locations of the surface Fermi arc terminations match up with the projection of the bulk Weyl nodes on the surface BZ. We superimpose the SX-ARPES measured bulk Fermi surface containing W2 Weyl nodes (Fig.~\ref{TaAsFig3}C) onto the low-energy ARPES Fermi surface containing the surface Fermi arcs (Fig.~\ref{TaAsFig2}A) to-scale. From Fig.~\ref{TaAsFig4}A we see that all the arc terminations and projected Weyl nodes match up with each other within the $k$-space region that is covered in our measurements. To establish this point quantitatively, in Fig.~\ref{TaAsFig4}C, we show the zoomed-in map near the crescent Fermi arc terminations, from which we obtain the $k$-space location of the terminations to be at $\vec{k}_{\textrm{arc}}=(0.04\pm0.01\textrm{\AA}^{-1}, 0.51\pm0.01\textrm{\AA}^{-1})$. Fig.~\ref{TaAsFig4}D shows the zoomed-in map of two nearby W2 Weyl nodes, from which we obtain the $k$-space location of the W2 Weyl nodes to be at $\vec{k}_{\textrm{W2}}=(0.04\pm0.015\textrm{\AA}^{-1}, 0.53\pm0.015\textrm{\AA}^{-1})$. In our bulk calculation, the $k$-space location of the W2 Weyl nodes is found to be at $(0.035\textrm{\AA}^{-1}, 0.518\textrm{\AA}^{-1})$. Since the SX-ARPES bulk data and the low-energy ARPES surface data are completely independent measurements using two different beamlines, the fact that they match up provides another independent proof of the topological nature (the surface-bulk correspondence) of the Weyl semimetal state in TaAs. In Figs. S5-7 in the SM, we further show that the bulk Weyl cones can also be observed in our low-energy ARPES data, although their spectral weight is much lower than the surface state intensities that dominate the data. In summary, we have shown the existence of Fermi arc surface states, the bulk Weyl cones, and demonstrated that the Fermi arcs terminate at the projection points of Weyl nodes based on our systematic and thorough ARPES data within our experimental resolution, which in the degree of rigor is comparable to the experimental resolution and analysis of previous ARPES studies that proved the first 3D topological insulators \cite{Hasan2010, Qi2011, Review_HasanXuBian_2015}. Our first demonstration of the Weyl semimetal state in TaAs paves the way for the realization of many fascinating topological quantum phenomena predicted in the Weyl semimetal state in the coming years.

\clearpage
%\addtocounter{figure}{-1}
%\begin{figure*}[t!]
%\caption{(\textbf{F}) A schematic showing the projected Weyl points and their projected chiral charges. (\textbf{G}) A first-principles band structure calculation of the Fermi surface on the (001) surface of TaAs. (\textbf{H}) The ARPES measured Fermi surface of the (001) cleaving plane of TaAs. The high symmetry points of the surface Brillouin zone (BZ) are noted.}
%\end{figure*}

\clearpage
\begin{figure}
\centering
\includegraphics[width=15cm, trim={1cm, 1cm, 1cm, 1cm}, clip]{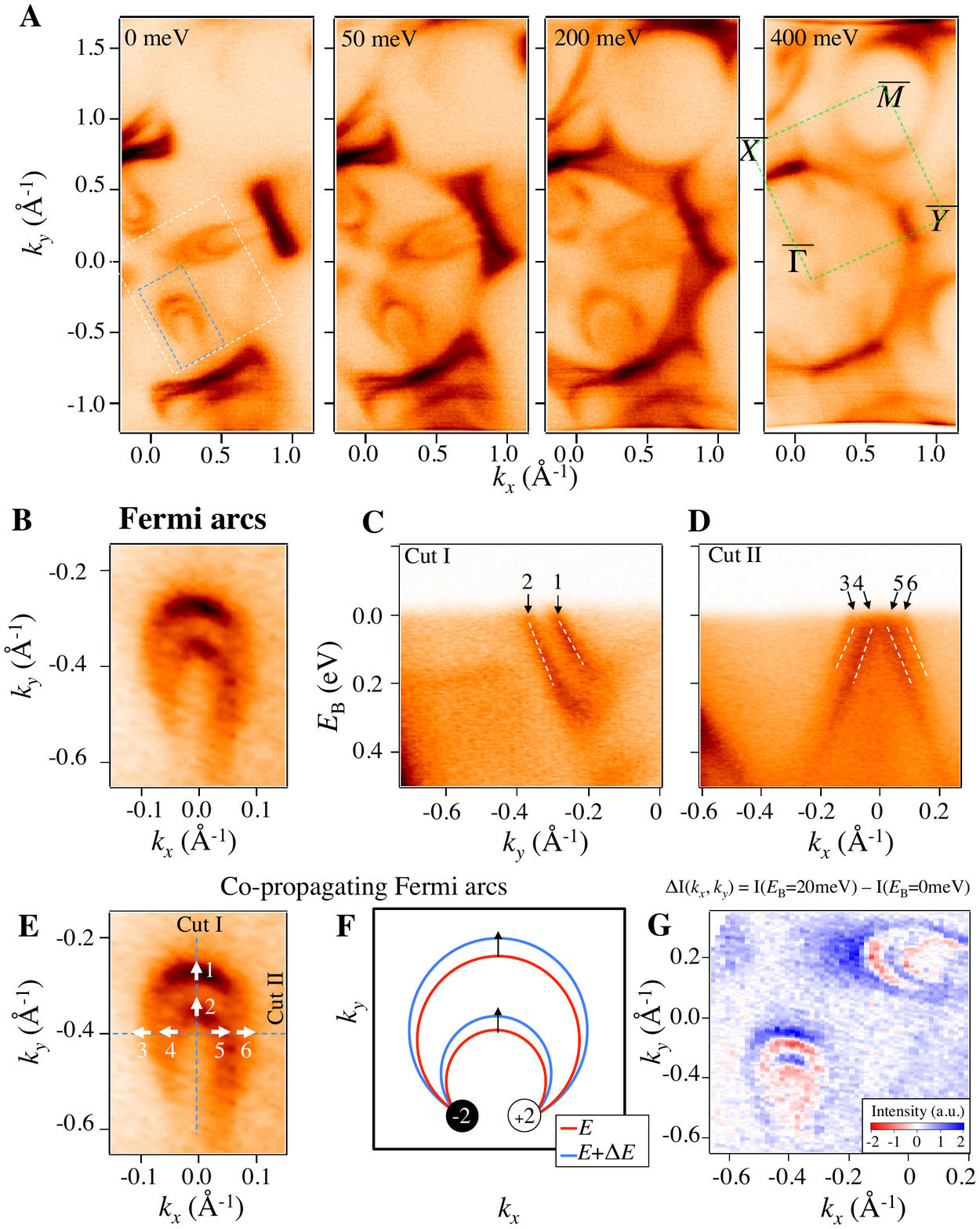}
\end{figure}

\begin{figure*}[t!]
\caption{\label{TaAsFig2}\textbf{Observation of Fermi arc surface states on the (001) surface of TaAs.} (\textbf{A}) ARPES Fermi surface map and constant binding energy contours measured using incident photon energy of 90 eV. (\textbf{B}) High-resolution ARPES Fermi surface map of the crescent Fermi arcs. The $k$-space range of this map is defined by the blue box in panel A. (\textbf{C,D}) Energy dispersion maps along Cuts I and II. (\textbf{E}) Same Fermi surface map as in panel B. The dotted lines defines the $k$-space direction for Cuts I and II. The numbers 1-6 note the Fermi crossings that are located on Cuts I and II. The white arrows show the evolution of the constant energy contours as one varies the binding energy, which is obtained from the dispersion maps in panels \textbf{C} and \textbf{D}. (\textbf{F}) A schematic showing the evolution of the Fermi arcs as a function of energy, which clearly distinguish between two Fermi arcs and a closed contour. (\textbf{G}) The difference between the constant energy contours at the binding energy $E_{\textrm{B}}=20$ meV and the binding energy $E_{\textrm{B}}=0$ meV, from which one can visualize the evolution of the constant energy contours through $k_x$-$k_y$ space. The range of this map is shown by the white dotted box in panel A.}
\end{figure*}

\clearpage

\clearpage
\begin{figure}
\centering
\includegraphics[width=13.5cm]{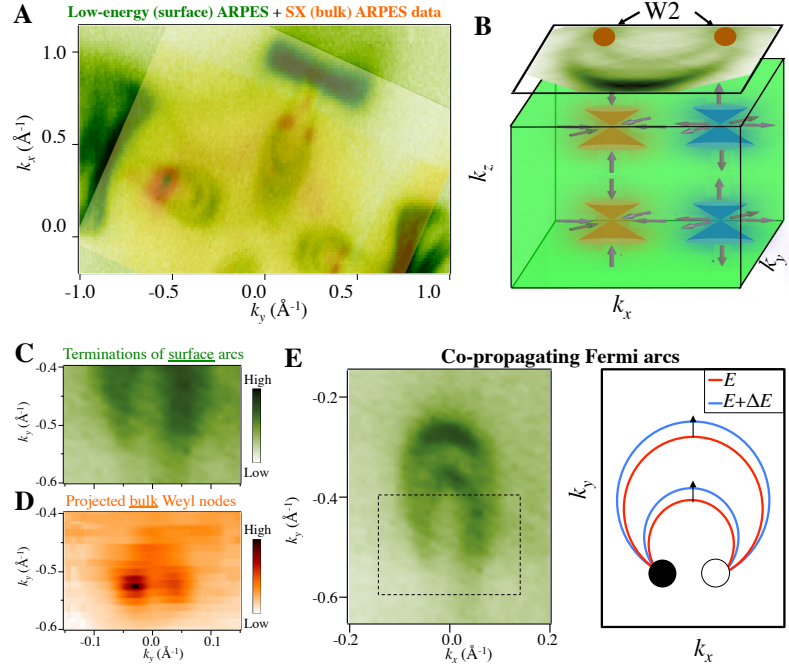}
\caption{\label{TaAsFig4} \textbf{Surface-bulk correspondence and the topologically nontrivial nature in TaAs.} (\textbf{A}) Low photon energy ARPES Fermi surface map ($h\nu=90$ eV) from Fig.~\ref{TaAsFig2}A, with the soft X-ray ARPES map ($h\nu=650$ eV) from Fig.~\ref{TaAsFig3}C overlaid on top of it to-scale, showing that the locations of the projected bulk Weyl nodes correspond to the terminations of the surface Fermi arcs. (\textbf{B}) The bottom shows a rectangular tube in the bulk BZ that encloses four W2 Weyl nodes. These four W2 Weyl nodes project onto two points at the (001) surface BZ with projected chiral charges of $\pm2$, shown by the brown circles. The top surface shows the ARPES measured crescent surface Fermi arcs that connect these two projected Weyl nodes. (\textbf{C}) Surface state Fermi surface map at the $k$-space region corresponding to the terminations of the crescent Fermi arcs. The $k$-space region is defined by the black dotted box in panel E. (\textbf{D}) Bulk Fermi surface map at the $k$-space region corresponding to the W2 Weyl nodes. The $k$-space region is defined by the black dotted box in Fig.~\ref{TaAsFig3}C. (\textbf{E}) ARPES and schematic of the crescent-shaped co-propagating Fermi arcs.}
\end{figure}

\section{Materials and methods}

\subsection{Sample growth and ARPES measurement techniques}

High quality single crystals of TaAs were grown by the standard chemical vapor transport method reported in Ref. \cite{TaAs_Crystal_1}. High resolution vacuum ultraviolet angle-resolved photoemission spectroscopy (ARPES) measurements were performed at Beamlines 4.0.3, 10.0.1 and 12.0.1 of the Advanced Light Source (ALS) at the Lawrence Berkeley National Laboratory (LBNL) in Berkeley, California, USA, Beamline 5-4 of the Stanford Synchrotron Radiation Light source (SSRL) at the Stanford Linear Accelerator Center (SLAC) in Palo Alto, California, USA and Beamline I05 of the Diamond Light Source (DLS) in Didcot, UK. The energy and momentum resolution was better than 30 meV and 1\% of the surface Brillouin zone. Samples were cleaved in situ under a vacuum condition better than $5 \times 10^{-11}$ Torr at all beamlines. The SX-ARPES measurements were performed at the Adress Beamline at the Swiss Light Source in the Paul Scherrer Institut (PSI) in Villigen, Switzerland. The experimental geometry of SX-ARPES has been described in \cite{SXARPES}. The present results were acquired with circular polarization of incident X-rays. The sample was cooled down to 12 K to quench the electron-phonon interaction effects reducing the $k$-resolved spectral fraction. In our photon energy range from 300 to 1000 eV, the combined (beamline and analyzer) experimental energy resolution varied between 40 and 140 meV. The angular resolution of the ARPES analyzer was $0.07^{\circ}$.

\subsection{First-principles calculation methods}

First-principles calculations were performed by the OPENMX code based on norm-conserving pseudopotentials generated with multi-reference energies and optimized pseudoatomic basis functions within the framework of the generalized gradient approximation (GGA) of density functional theory (DFT) \cite{Perdew}. Spin-orbit coupling was incorporated through $j$-dependent pseudo-potentials. For each Ta atom, three, two, two, and one optimized radial functions were allocated for the $s$, $p$, $d$, and $f$ orbitals ($s^3p^2d^2f^1$), respectively, with a cutoff radius of 7 Bohr. For each As atom, $s^3p^3d^3f^2$ was adopted with a cutoff radius of 9 Bohr. A regular mesh of 1000 Ry in real space was used for the numerical integrations and for the solution of the Poisson equation. A $k$ point mesh of $17 \times 17 \times 5$ for the conventional unit cell was used and experimental lattice parameters \cite{TaAs_Crystal_1} were adopted in the calculations. Symmetry-respecting Wannier functions for the As $p$ and Ta $d$ orbitals were constructed without performing the procedure for maximizing localization and a real-space tight-binding Hamiltonian was obtained \cite{Weng}. This Wannier function based tight-binding model was used to obtain the surface states by constructing a slab with 80-atomic-layer thickness with Ta on the top and As on the bottom.

\clearpage

%\cleardoublepage
\ifdefined\phantomsection
  \phantomsection  % makes hyperref recognize this section properly for pdf link
\else
\fi
\addcontentsline{toc}{section}{Bibliography}

{\singlespacing

}

%\end{document}

%% file: ch-heterostructure/heterostructure.tex
\chapter{Emergent topological multilayer superlattice}
\label{ch:hetero}

{\singlespacing
\begin{chapquote}{Les BB Brunes, \textit{Nico Teen Love}}
Je fais un `hic' et je me marre\\ 
En pensant \`a ce con d'Icare\\
Et nos ailes br\^ul\'ees 
\end{chapquote}}

\newcommand{\topo}{Bi$_2$Se$_3$}
\newcommand{\triv}{In$_x$Bi$_{2-x}$Se$_3$}
\newcommand{\samA}{4QL/4QL 20\%}
\newcommand{\samB}{4QL/2QL 25\%}
\newcommand{\samC}{4QL/2QL 15\%}
\newcommand{\samD}{4QL/1QL 20\%}
\newcommand{\samE}{4QL/1QL 10\%}
\newcommand{\samF}{4QL/1QL 15\%}
\newcommand{\samG}{3QL/3QL 20\%}
\newcommand{\samH}{10QL/10QL 20\%}

\noindent This chapter is based on the article, \textit{A novel artificial condensed matter lattice and a new platform for one-dimensional topological phases} by Ilya Belopolski \textit{et al}., \textit{Sci. Adv.} {\bf 3}, e1501692 (2017), available at \href{http://advances.sciencemag.org/content/3/3/e1501692}{http://advances.sciencemag.org/content/3/3/\\e1501692}.\\

%\begin{abstract}
\lettrine[lines=3]{E}{ngineered} lattices in condensed matter physics, such as cold atom optical lattices or photonic crystals, can have fundamentally different properties from naturally-occurring electronic crystals. Here, we report a novel type of artificial quantum matter lattice. Our lattice is a multilayer heterostructure built from alternating thin films of topological and trivial insulators. Each interface within the heterostructure hosts a set of topologically-protected interface states, and by making the layers sufficiently thin, we demonstrate for the first time a hybridization of interface states across layers. In this way, our heterostructure forms an emergent atomic chain, where the interfaces act as lattice sites and the interface states act as atomic orbitals, as seen from our measurements by angle-resolved photoemission spectroscopy (ARPES). By changing the composition of the heterostructure, we can directly control hopping between lattice sites. We realize a topological and a trivial phase in our superlattice band structure. We argue that the superlattice may be characterized in a significant way by a one-dimensional topological invariant, closely related to the invariant of the Su-Schrieffer-Heeger model. Our topological insulator heterostructure demonstrates a novel experimental platform where we can engineer band structures by directly controlling how electrons hop between lattice sites.
%\end{abstract}

%\date{\today}
%\maketitle

\section{Introduction}

While crystals found in nature offer a great richness of phenomena, recent progress in physics has also been driven by the study of engineered systems, where key material parameters or Hamiltonian matrix elements can be directly controlled in experiment to give rise to novel emergent properties. For example, the Haldane model for a Chern insulator was recently realized in an optical lattice of ultracold atoms \cite{ColdAtoms} and a Weyl semimetal was observed in a photonic crystal with double-gyroid lattice structure \cite{PhotonicWeyl}. Each advance required engineering a specific ultracold atom optical lattice or photonic crystal. In both cases, these engineered lattices host an emergent band structure, analogous to the band structure formed by electrons in a crystal lattice, but with distinct properties and highly-tuned parameters that gave rise to a novel phase of matter.

Here, we demonstrate an emergent band structure in a novel type of condensed matter lattice based on topological insulators \cite{Mele, HasanKane}. Specifically, we stack together layers of a topological and trivial insulator to create a one-dimensional topological insulator heterostructure. Each interface in the heterostructure hosts a set of topologically-protected interface states which hybridize with each other across the layers, giving rise to a superlattice band structure where the interfaces play the role of lattice sites, the topological surface states play the role of atomic orbitals and the heterostructure acts as a one-dimensional atomic chain. Such a superlattice band structure is emergent in the sense that it arises only when many layers are stacked together, in the same way that an ordinary band structure arises as an emergent property of a crystal lattice. A similar heterostructure has recently been proposed as a simple theoretical model for a Dirac or Weyl semimetal \cite{BurkovBalents}. Here, we experimentally realize and directly observe the first superlattice band structure of this type. We use molecular beam epitaxy (MBE) to build the topological insulator heterostructure and we use angle-resolved photoemission spectroscopy (ARPES) to study its band structure. By adjusting the pattern of layers in the heterostructure, we realize both a topological and a trivial phase, demonstrating that we can tune our system through a topological phase transition. Our work may lead to the realization of novel three-dimensional and one-dimensional symmetry-protected topological phases. At the same time, a topological insulator heterostructure provides a highly-tunable emergent band structure in a true electron system, relevant for transport experiments and device applications.

\begin{figure*}
\centering
\includegraphics[width=15.5cm]{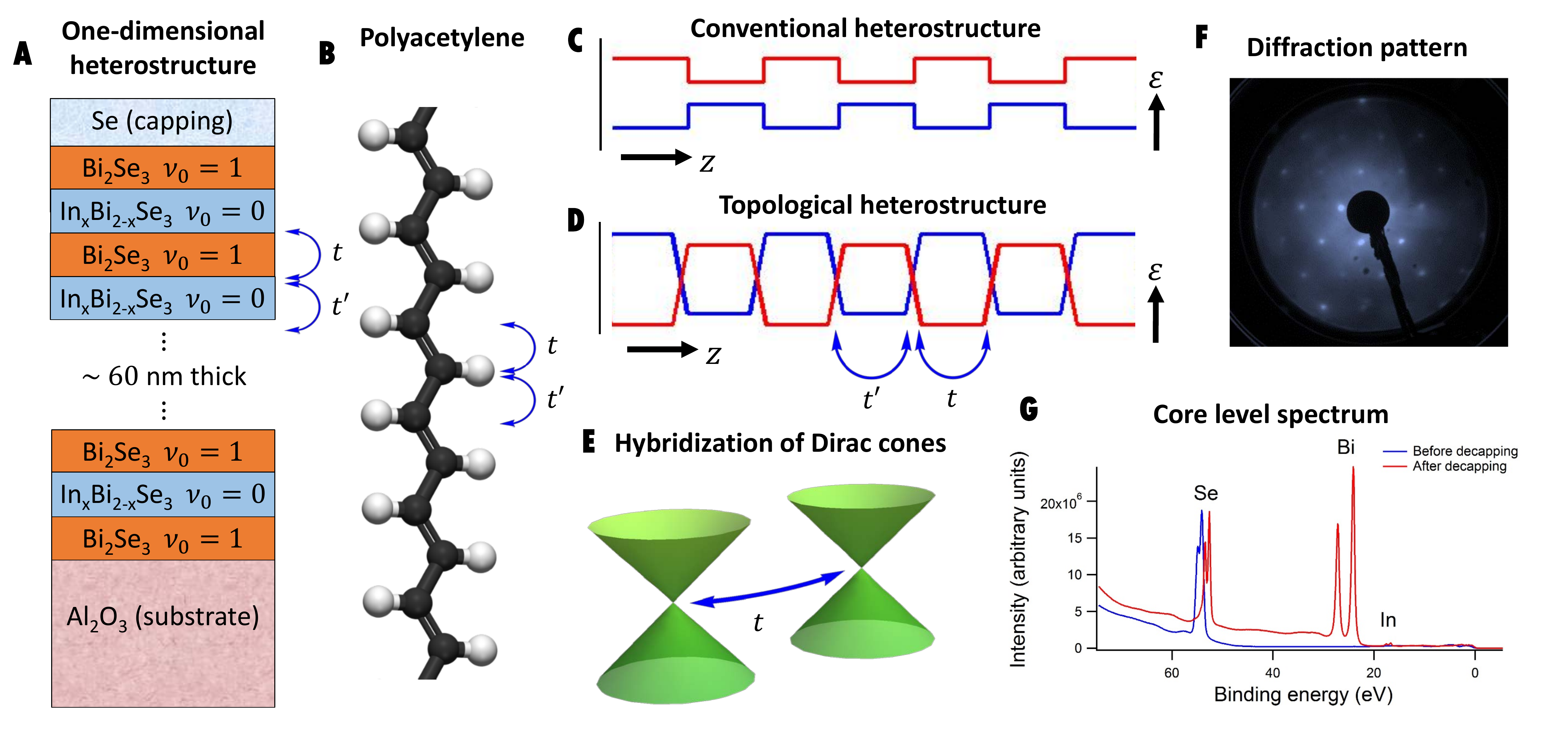}
\caption{\label{HFig1}\textbf{Overview of the topological insulator heterostructure.} (\textbf{A}) The heterostructure consists of a stack of alternating $\mathbb{Z}_2$ topological insulator layers and trivial insulator layers. We use \topo\ as the topologically non-trivial layer and \triv\ as the topologically trivial layer. The $\mathbb{Z}_2$ invariant is denoted by $\nu_0 = 1,0$. An amorphous Se capping layer protects the sample in atmosphere and is removed by heating the sample $\textit{in situ}$. (\textbf{B}) Our system realizes an emergent version of a polyacetylene chain, well-known as a toy model in the study of one-dimensional topological phases. The model has two carbon atoms per unit cell, with one orbital each and hopping amplitudes $t$ and $t'$ associated with the double and single carbon bonds. In the topological insulator heterostructure, the topological and trivial layers play the role of the double and single carbon bonds. (\textbf{C}) A conventional semiconductor heterostructure consists of an alternating pattern of materials with different band gaps. (\textbf{D}) In a topological insulator heterostructure, the band gaps in adjacent layers are inverted, giving rise to topologically-protected Dirac cone interface states between layers. If the layers are thin, then adjacent Dirac cones may hybridize. This hybridization can be described by a hopping amplitude $t$ across the topological layer and $t'$ across the trivial layer. (\textbf{E}) Illustration of the Dirac cone surface states at each interface in the heterostructure, assuming no hybridization. With a hybridization $t$, the topological interface states form a superlattice band structure. (\textbf{F}) A standard low-energy electron diffraction pattern and (\textbf{G}) core level photoemission spectrum show that the samples are of high quality and the Se capping layer was successfully removed by \textit{in situ} heating, exposing a clean sample surface in vacuum.}
\end{figure*}

We first provide a more detailed introduction to the topological insulator heterostructure and argue that our system forms a novel type of condensed matter lattice. A topological insulator heterostructure is built up of alternating layers of topologically trivial and non-trivial insulators. In this work, we use \topo\ as the topological insulator and \triv\ as the trivial insulator, see Fig. \ref{HFig1} A. We note that heterostructures of \topo\ and In$_2$Se$_3$ have already been successfully synthesized \cite{Nikesh, HKU, transport}. Bulk \topo\ has a non-trivial $\mathbb{Z}_2$ invariant, $\nu_0 = 1$ \cite{MatthewBiSe, ZhangBiSe}. Under doping by In, bulk \topo\ undergoes a topological phase transition to a topologically trivial phase with $\mathbb{Z}_2$ invariant, $\nu_0 = 0$ \cite{BrahlekTPT}. In all samples considered here, all \triv\ layers are well into the trivial phase. As a result, the topological invariant flips back and forth from layer to layer in the heterostructure, so that each interface hosts a set of topological interface states \cite{Mele, HasanKane, QiZhang, BAB, HasanMoore, HasanSusu}. Next, we note that these interface states have a finite penetration depth into the bulk of the crystal, so that for sufficiently thin layers, the topological interface states hybridize with each other across the layers and will be subject to an energy level repulsion \cite{MadhabQL, XueQikunQL}. This hybridization can be captured by a hopping amplitude $t$ across the topological layer and $t'$ across the trivial layer. In this way, the topological insulator heterostructure can be viewed as an analog of a polyacetylene chain, shown in Fig. \ref{HFig1} B, where the topological insulator corresponds to the carbon double bond and the trivial insulator corresponds to the carbon single bond (or vice versa). We note that our heterostructure is similar to a conventional semiconductor heterostructure in that the band gap varies in $z$, Fig. \ref{HFig1} C \cite{YuCardona, EsakiTsu}. However, we see that new phenomena arise in a topological insulator heterostructure because the band gap inverts from layer to layer, Fig. \ref{HFig1} D. Specifically, the hybridization of topological interface states, illustrated schematically in Fig. \ref{HFig1} E, gives rise to a superlattice dispersion. In this way, the topological insulator heterostructure is a novel type of condensed matter lattice, where the interfaces correspond to the atomic sites and the topological insulator interface states correspond to the atomic orbitals. Because we can precisely control the thickness of the topological and trivial layers, this superlattice band structure is highly tunable. At the same time, it remains a true electron system, relevant for transport experiments and device applications. Further, if the Fermi level can be placed in the bulk band gap, then the underlying bulk bands of each heterostructure layer become irrelevant, and the transport properties are determined only by the superlattice band structure. This emergent band structure provides a new platform for realizing novel phases in three dimensions, such as a magnetic Weyl semimetal \cite{BurkovBalents}. Further, we note that the energy level repulsion occurs between the electron states at the Dirac points, at $k_{||} = 0$. As a result, the system can be well-described as an atomic chain with a two-site basis and one orbital per site, leading to a two-band single-particle Hamiltonian. By implementing a chiral symmetry in the layer pattern of the heterostructure, it may be possible to realize a phase with strictly one-dimensional topological invariant protected by chiral symmetry, although further theoretical study may be necessary \cite{SSHOriginalPaper, SSHK, PeriodicTable}. We note that prior to measurement by ARPES, we measure standard core level spectra and diffraction patterns to confirm the high quality of our samples, see Fig. \ref{HFig1} F,G.

\section{Superlattice dispersion in \topo/\triv\ heterostructures}

We next demonstrate that we have observed an emergent superlattice dispersion in our topological insulator heterostructure. We present a systematic study of four different \topo/\triv\ superlattices, which we write as \samA, \samB, \samC\ and \samD, in Fig. \ref{HFig2} A-D. In this notation, the first parameter refers to the thickness of the \topo\ layer, the second parameter refers to the thickness of the \triv\ layer and the percentage refers to the In doping $x$ of the \triv\ layer. In Fig. \ref{HFig2} E-H, we present ARPES spectra of the four heterostructures along a cut through the $\bar{\Gamma}$ point and in Fig. \ref{HFig2} I-L the same ARPES spectra with additional hand-drawn lines to mark the bands observed in the data. In Fig. \ref{HFig2} M-P, we further present the energy distribution curves (EDCs) of photoemission intensity as a function binding energy at $k_{||} = 0$. These curves correspond to a vertical line on the image plot, indicated by the green arrows in Fig. \ref{HFig2} I-L. The gapless surface state, labeled (1) in \samC\ and \samD, is entirely inconsistent with a \topo\ film 4QL thick. It can only be explained by considering hybridization across the \triv\ layer, demonstrating an emergent superlattice band structure arising from hopping of Dirac cone interface states within the heterostructure. In addition, a gap opens from \samC\ to \samB\ \textit{without any observable bulk band inversion} and \textit {without time-reversal symmetry breaking}. This is an apparent contradiction with the basic theory of $\mathbb{Z}_2$ topological insulators. Again, this result demonstrates a superlattice dispersion. As we increase $t/t'$, the top two lattice sites show larger hybridization and a gap opens in the Dirac cone on the last site. Because ARPES is only sensitive to the top surface of the heterostructure, we cannot observe the inversion of the superlattice bands, but only the increased coupling to the top surface. Our observation of (1) a gapless surface state on a 4QL film of \topo\ and (2) a gap opening in a \topo\ surface state without apparent band inversion each demonstrate that we have observed a superlattice dispersion in our system. In this way, we have shown a completely novel type of electronic band structure, which arises from a lattice of topological insulator Dirac cones which are allowed to hybridize.

\section{One-dimensional atomic-chain picture of the emergent superlattice}

We provide a one-dimensional picture of the topological and trivial phases of the superlattice. We consider again \samD, shown in Fig. \ref{HFig3} A-D. Above, we argued that \samD\ has a gapless Dirac cone because $t < t'$. We note that the different hopping amplitude arises from the large bulk band gap and large thickness of the \topo\ layer relative to the \triv\ layer, as illustrated in Fig. \ref{HFig3} E. Alternatively, we can consider how orbitals pair up and gap out in real space. We see that in the topological phase, there are two end modes left without a pairing partner, see Fig. \ref{HFig3} F. We contrast the topological phase with the trivial phase observed in \samA, shown in Fig. \ref{HFig3} G-J. In this case, $t > t'$, because the trivial phase has larger band gap than the topological phase, illustrated in Fig. \ref{HFig3} K. Alternatively, the real space pairing leaves no lattice site without a pairing partner, see Fig. \ref{HFig3} L. We see that our observation of a topological and trivial phase in a topological insulator heterostructure can be understood in terms of an emergent one-dimensional atomic chain where the termination of the chain is either on a strong or weak bond.

We present ARPES spectra of other compositions in the topological and trivial phase to provide a systematic check of our results. First, we compare \samE\ and \samF\ with \samD. The unit cells of these lattices are illustrated in Fig. \ref{HFig4} A-C. The ARPES spectra are shown in Fig. \ref{HFig4} D-F. We also show an EDC at $k_{||} = 0$ in Fig. \ref{HFig4} G-I, as indicated by the green arrows in Fig. \ref{HFig4} D-F. We find that \samE\ and \samF\ also host a gapless surface state and are also topological. This is expected because we have further decreased the In concentration in the trivial layer, increasing the hybridization $t$ and pushing the sample further into the topological phase. Next, we compare \samG\ with \samA. The unit cells are illustrated in Fig. \ref{HFig4} J-L. The ARPES spectra are shown in Fig. \ref{HFig4} M-O and an EDC at $k_{||} = 0$ is shown in Fig. \ref{HFig4} P-R. We find that \samG\ is also in the trivial phase, with a gap in the surface states larger than in \samD. This is consistent with earlier ARPES studies of single thin films of \topo\ on a topologically trivial substrate \cite{MadhabQL, XueQikunQL}. For \samG\ we further observe the same bands at two different photon energies, providing a check that the gapped surface state is not an artifact of low photoemission cross-section at a special photon energy. Our results on \samE, \samF\ and \samG\ provide a systematic check of our results.

\section{Discussion}

Our ARPES spectra show that we have realized the first emergent band structure in a lattice of topological interface states. We have further demonstrated that we can tune this band structure through a topological phase transition. We summarize our results by plotting the compositions of Fig. \ref{HFig2} on a phase diagram as a function of $t'/t$, shown in Fig. \ref{Fig5} A. Our results can also be understood in terms of a one-dimensional atomic chain terminated on either a strong or weak bond. It is natural to ask whether any rigorous analogy exists between our topological insulator heterostructure and a one-dimensional topological phase \cite{PeriodicTable}. An interesting property of our heterostructure is that it naturally gives rise to an approximate chiral symmetry along the stacking direction. In particular, because the wavefunction of a topological insulator surface state decays rapidly into the bulk, we expect that the only relevant hybridization in our heterostructure takes place between adjacent interface states. Further, along the stacking direction, the lattice is bipartite. As a result of nearest-neighbor hopping on a bipartite lattice gives rise to an emergent chiral symmetry along the stacking direction. We note that chiral symmetry is required for a number of topological phases in one dimension, including the BDI, CII, DIII and CI classes, as well as the well-known Su-Schrieffer-Heeger model \cite{PeriodicTable, SSHOriginalPaper, SSHK}. We suggest that future theoretical work could determine whether a one-dimensional topological phase might be realized in our system. Specifically, we might consider topological invariants on the one-dimensional band structure along $k_z$ at fixed $k_{||} = 0$. We also note that we may introduce a second nearest-neighbor hopping by using even thinner layers, see Fig. \ref{Fig5} B. This may break the chiral symmetry, removing the one-dimensional topological invariant and shifting the surface state out of the superlattice bulk band gap. There may further arise an unusual behavior in the surface states if the heterostructure is also characterized by a non-trivial three-dimensional $\mathbb{Z}_2$ invariant. Another application of our topological insulator heterostructure is to break inversion symmetry using a unit cell consisting of four layers of different thickness, see Fig. \ref{Fig5} C. This may lead to a superlattice band structure of spinful bands. More complicated topological insulator heterostructures may allow us to engineer time reversal broken one-dimensional topological phases or large degeneracies implemented by fine-tuning hopping amplitudes. Our systematic ARPES measurements demonstrate the first chain of topological insulator surface states. In this way, we provide not only an entirely novel type of condensed matter lattice but a new platform for engineering band structures in true electron systems by directly controlling how electrons hop between lattice sites.

\section{Materials and methods}

High quality \topo/\triv\ heterostructures were grown on $10 \times 10 \times 0.5$ mm$^3$ Al$_2$O$_3$ (0001) substrates using a custom-built SVTA-MOS-V-2 MBE system with base pressure of $2 \times 10^{-10}$ Torr. Substrates were cleaned \textit{ex situ} by 5 minute exposure to UV-generated ozone and \textit{in situ} by heating to $800^\circ$C in an oxygen pressure of $1 \times 10^{-6}$ Torr for 10 minutes. Elemental Bi, In, and Se sources, 99.999\% pure, were thermally evaporated using effusion cells equipped with individual shutters. In all samples, the first layer of the superlattice was \topo, 3QL of which was grown at $135^\circ$C and annealed to $265^\circ$C for the rest of the film growth \cite{MBE}. For \triv\ growth, Bi and In were co-evaporated by opening both shutters simultaneously, while the Se shutter was kept open at all times during the growth. To accurately determine the concentration of In in the \triv\ layers, source fluxes were calibrated \textit{in situ} by a quartz crystal microbalance and \textit{ex situ} by Rutherford back scattering, which together provide a measure of the In concentration accurate to within ~1\%. All heterostructures were in total 59QL or 60QL in thickness, depending on the unit cell. In this way, all heterostructures consisted of $\sim 10$ unit cells. All samples were capped at room temperature \textit{in situ} by a $\sim 100$ nm thick protective Se layer to prevent surface contamination in atmosphere. Prior to capping by Se, the high quality of the heterostructure was checked by reflective high-energy electron diffraction (RHEED), see Fig. S1.

Angle-resolved photoemission spectroscopy (ARPES) measurements were carried out at several synchrotron lightsources, in particular at ANTARES, the Synchrotron Soleil, Saint-Aubin, France; ADRESS, the Swiss Light Source (SLS), Villigen, Switzerland; the HRPES endstation of SIS, SLS; I05, Diamond Light Source, Oxfordshire, UK; and CASSIOPEE, Soleil. Samples were clamped onto the sample holder either using an Mo or Ta clamp screwed onto the sample base or thin strips of Ta foil spot-welded onto the sample base. In addition to securely mounting the sample, the clamp also provided electrical grounding. The Se capping layer was removed by heating the sample in a vacuum preparation chamber at temperatures between $200^{\circ}$C and $300^{\circ}$C, at pressures better than $10^{-9}$ Torr, for $\sim 1$hr. Following decapping, the quality of the exposed sample surface was checked by low-energy electron diffraction (LEED) at a typical electron beam energy of 100V. The presence of sharp Bragg peaks and their six-fold rotation symmetry, as shown in main text Fig. 1H, shows the high quality of the exposed sample surface. ARPES measurements were carried out at pressures better than $10^{-10}$ Torr at incident photon energies between 15 eV and 320 eV. We take a short Fermi surface mapping near the center of the Brillouin zone for each sample, at relevant photon energies, to identify the rotation angles corresponding to $\bar{\Gamma}$. This allows us to take $E$-$k$ cuts through the $\bar{\Gamma}$ point to within a rotation angle of $\pm 0.15^{\circ}$, minimizing the error in our measurement of the band gap due to sample misalignment.

\clearpage

\clearpage
\begin{figure}
\centering
\includegraphics[width=15cm, trim={0.9cm, 0cm, 1cm, 0cm}, clip]{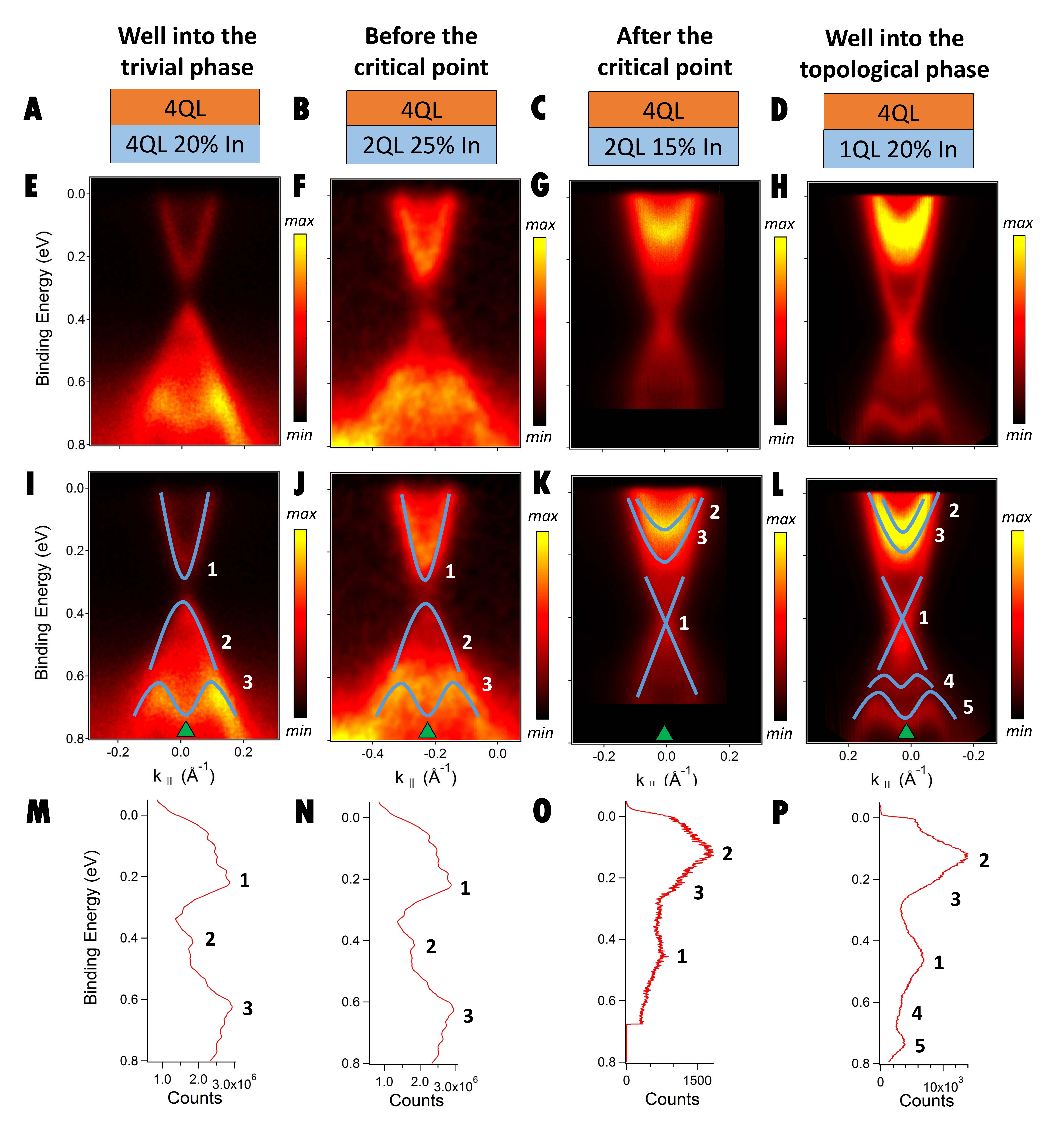}
\end{figure}

\clearpage
\begin{figure}
\caption{\label{HFig2}\textbf{Observation of an emergent superlattice band structure in trivial and topological phases.} (\textbf{A-D}) The unit cells of the heterostructures studied, with different thickness and In-doping of the \triv\ layer. (\textbf{E-H}) ARPES spectra of heterostructures. We see that (\textbf{E}) and (\textbf{F}) showed gapped surface states while (\textbf{G}) and (\textbf{H}) show gapless surface states. Note that in all samples measured, the top layer of the superlattice, which is the only layer directly measured by ARPES, is 4QL of \topo. Nonetheless, the spectra differ strikingly. (\textbf{I-L}) The same spectra as in (\textbf{E-H}), but with additional hand-drawn lines showing the key features of the spectra. In the \samA\ and \samB\ samples, we observe a gapped topological surface state, (1) and (2), along with a valence band quanutm well state (3). By contrast, in \samC\ and \samD, we observe a gapless Dirac cone surface state, (1), two conduction band quantum well states, (2) and (3) and, in \samD, two valence quantum well states, (4) and (5). We emphasize that the gapless Dirac cone is observed even though the top \topo\ layer is only 4QL thick. (\textbf{M-P}) Energy distribution curves through the $\bar{\Gamma}$ point of each spectrum in (\textbf{E-H}). The peaks corresponding to the bulk quantum well states and surface states are numbered based on the correspondance with the full spectra, (\textbf{E-H}).}
\end{figure}

\clearpage
\begin{figure}
\centering
\includegraphics[width=15cm, trim={1.1cm, 1cm, 1cm, 1cm}, clip]{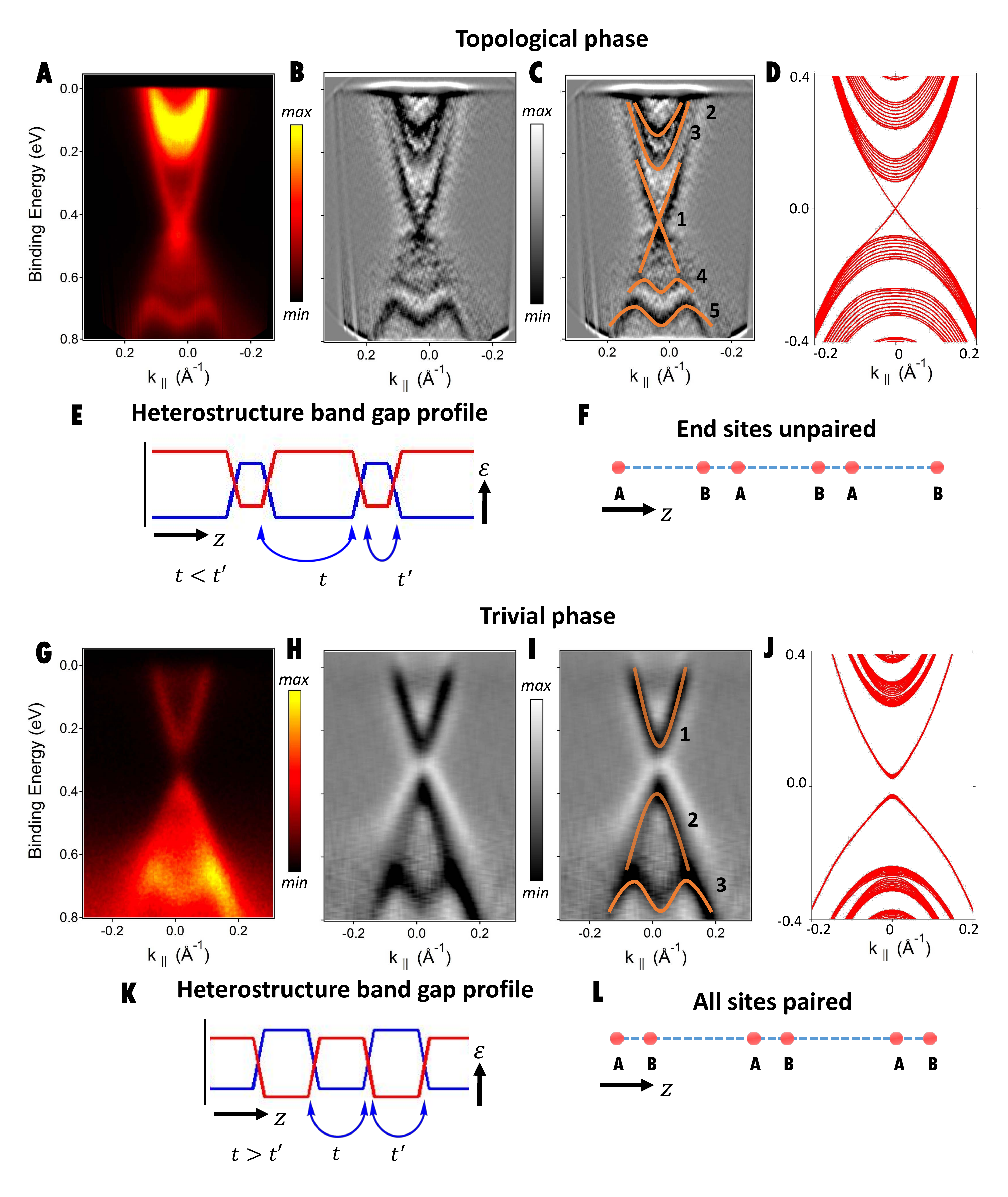}
\end{figure}

\clearpage
\begin{figure}
\caption{\label{HFig3}\textbf{Realization of the Su-Schrieffer-Heeger model.} (\textbf{A}) ARPES spectrum of \samD, with (\textbf{B}) a second-derivative map of the same spectrum and (\textbf{C}) the same second-derivative map with additional hand-drawn lines highlighting the gapless surface state and the two quantum well states in the conduction and valence bands. (\textbf{D}) A tight-binding calculation demonstrating a topological phase qualitatively consistent with our experimental result. (\textbf{E}) A cartoon of the band gap profile of the \samD\ heterostructure. The trivial layer is much thinner than the topological layer, so $t < t'$ and we are in the SSH topological phase. (\textbf{F}) The SSH topological phase can be understood as a phase where the surface states pair up with their nearest neighbors and gap out but where the pairing takes place in such a way that there is an unpaired lattice site at the end of the atomic chain. (\textbf{G}) ARPES spectrum of \samA, with (\textbf{H}) a second-derivative map of the same spectrum and (\textbf{I}) the same second-derivative map with additional hand-drawn lines highlighting the gapped surface state and the quantum well state in the valence band. (\textbf{J}) A tight-binding calculation demonstrating a trivial phase qualitatively consistent with our experimental result. (\textbf{K}) A cartoon of the band gap profile of the \samA\ heterostructure. The trivial layer is as thick as the topological layer, with a larger band gap due to high In-doping, so $t > t'$ and we are in the SSH trivial phase. (\textbf{L}) The SSH trivial phase can be understood as a phase where all lattice sites in the atomic chain have a pairing partner.}
\end{figure}

\clearpage
\begin{figure}
\centering
\includegraphics[width=9cm, trim={1cm, 0cm, 1cm, 0cm}, clip]{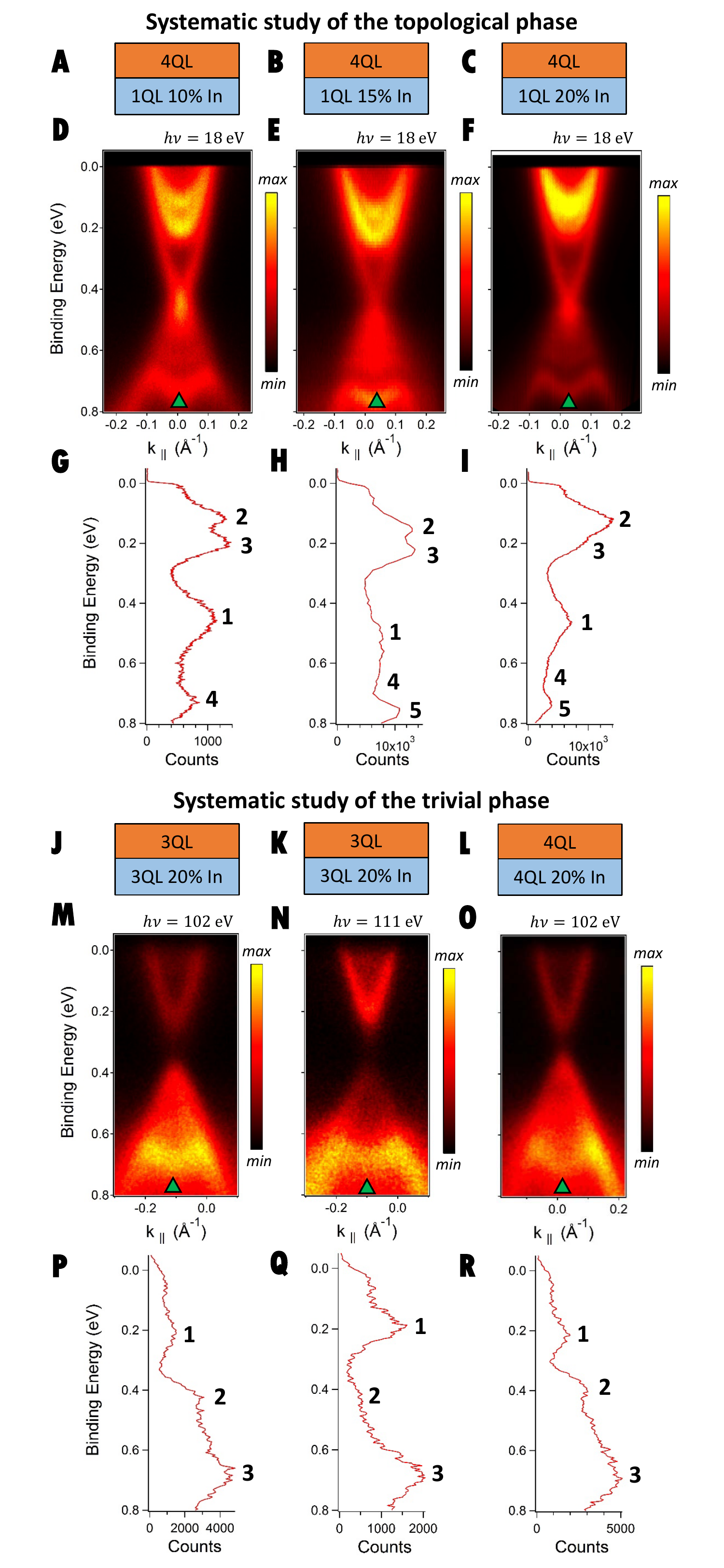}
\end{figure}

\clearpage
\begin{figure}
\caption{\label{HFig4}\textbf{Systematic study of the topological and trivial phases.} (\textbf{A-F}) ARPES spectra of 4QL/1QL samples at $h \nu = 18$ eV with varying concentrations of In, showing a robust topological phase. (\textbf{G-I}) Energy distribution curves through the $\bar{\Gamma}$ point of each spectrum in (\textbf{D-F}). (\textbf{J-O}) ARPES spectra of 3QL/3QL 20\% and 4QL/4QL 20\% samples at $h \nu = 102$ eV and a 3QL/3QL 20\% sample at $h \nu = 111$ eV, showing a robust trivial phase. (\textbf{P-R}) Energy distribution curves through the $\bar{\Gamma}$ point of each spectrum in (\textbf{M-O}).}
\end{figure}

\clearpage
\begin{figure}[h!]
\centering
\includegraphics[width=10cm]{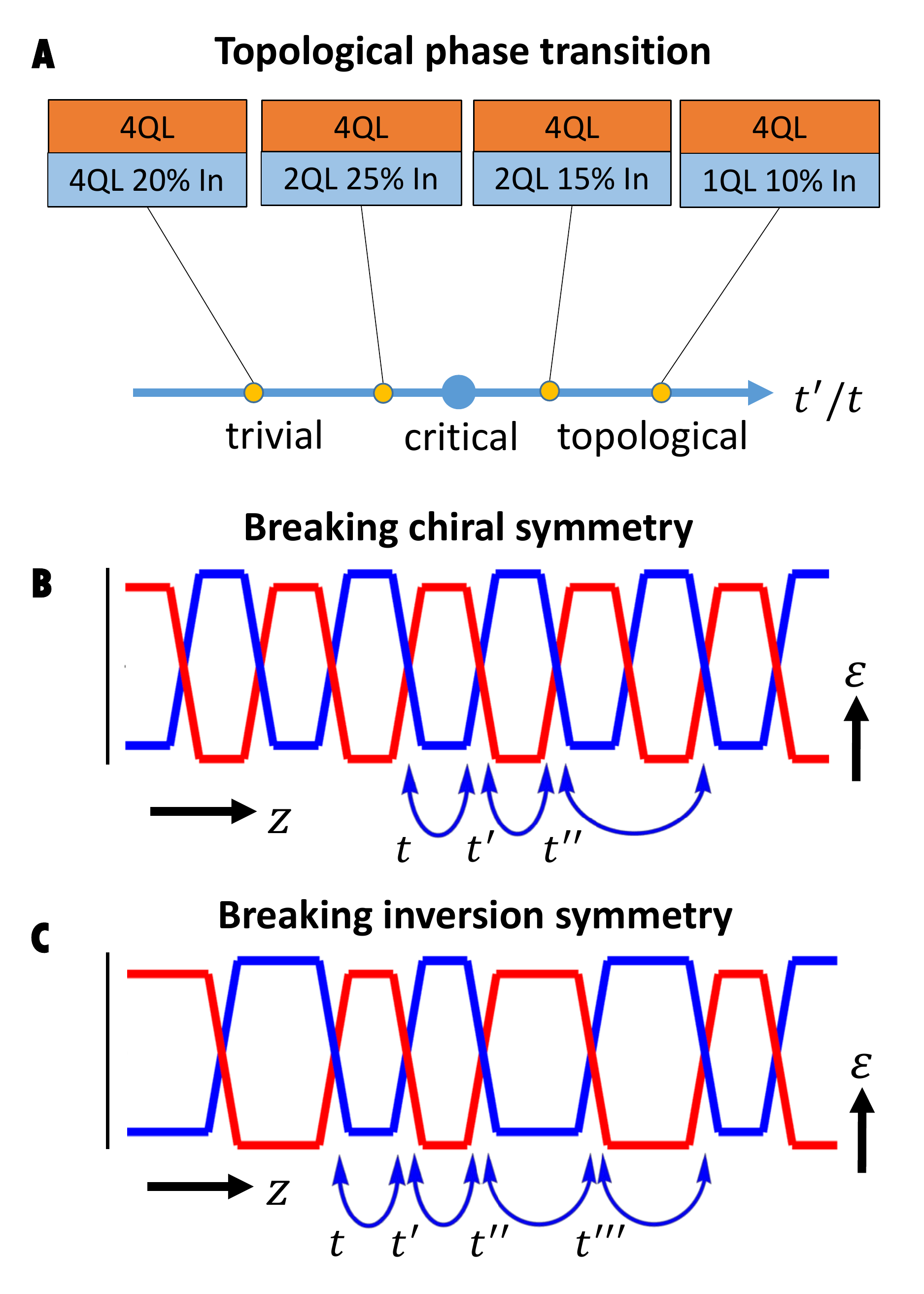}
\caption{\label{Fig5}\textbf{Phase diagram of a tunable emergent superlattice band structure.} (\textbf{A}) By varying the hopping $t'$ across the trivial layer we realize a trivial phase and topological phase in our superlattice band structure. (\textbf{B}) In the heterostructures demonstrated here, we expect that only nearest-neighbor hopping is relevant, giving rise to an emergent chiral symmetry along the stacking direction. By using thinner layers we propose to introduce a next nearest-neighbor hopping, breaking this chiral symmetry. This may change the topological classification of our system. (\textbf{C}) By doubling the unit cell, we can break inversion symmetry. This will give rise to a spinful superlattice dispersion.}
\end{figure}

%\clearpage

%\begin{center}
%\textbf{\large Supplementary Materials}
%\end{center}

%\setcounter{equation}{0}
%\setcounter{figure}{0}
%\setcounter{table}{0}
%%\setcounter{page}{1}
%\makeatletter
%\renewcommand{\theequation}{S\arabic{equation}}
%\renewcommand{\thefigure}{S\arabic{figure}}
%\renewcommand{\thetable}{S\arabic{table}}

\subsection{Estimate of Indium diffusion in the heterostructure}

\begin{figure}[h!]
\centering
\includegraphics[width=15cm]{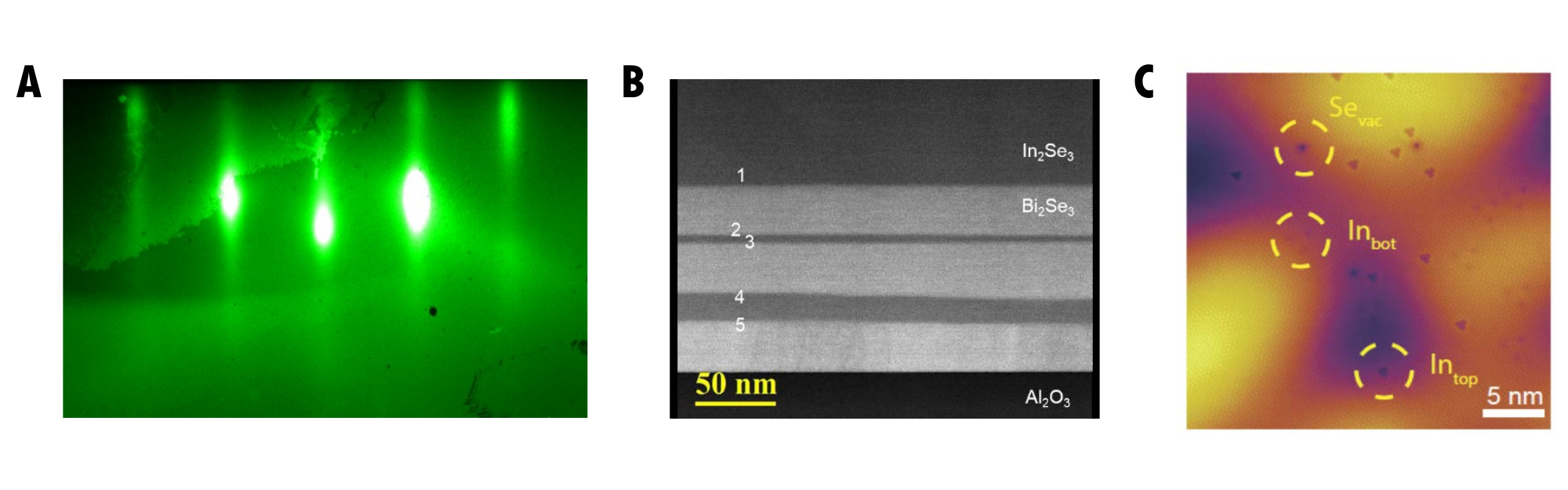}
\caption{\label{FigS4}\textbf{Characterization of In diffusion in the heterostructures.} (\textbf{A}) Reflection high-energy electron diffraction (RHEED) image showing an atomically flat single crystalline growth, as indicated by the sharp streaks and Kikuchi lines. This confirms the high quality of our heterostructures. (\textbf{B}) High-angle annular dark-field scanning transmission electron microscopy (HAADF-STEM) image of a \topo/In$_2$Se$_3$ heterostructure, showing sharp interfaces between the layers. The sample has layer pattern 100 In$_2$Se$_3$ / 30 \topo\ / 5 In$_2$Se$_3$30 / 30 \topo\ / 20 In$_2$Se$_3$ / 30 \topo, demonstrating at the same time the sharp interfaces in our heterostructure as well as our ability to precisely control the layer thickness in MBE. (\textbf{C}) Indium atoms observed on \topo\ by scanning tunneling microscopy (STM) in a 5QL/20QL 50\% heterostructure with one unit cell. In$_\textrm{top}$ and In$_\textrm{bot}$ represent In atoms diffused to the top and bottom Bi layer within the topmost QL of \topo, while Se$_\textrm{vac}$ represent Se vacancies. (Reprinted from Ref. [24], $\copyright$ 2015 American Chemical Society).}
\end{figure}

It is important to check that diffusion of In within the sample does not remove the layered \topo/\triv\ pattern of the heterostructure. In this section, we characterize In diffusion in our heterostructures. In subsequent sections, we also provide a number of checks in ARPES which confirm indirectly that the heterostructure consists of clean interfaces. We show a high-angle annular dark-field scanning transmission electron microscopy (HAADF-STEM) image, Fig. \ref{FigS4} B, of a thin film with alternating layers of \topo\ and In$_2$Se$_3$ of different thicknesses [9]. Sharp interfaces between the In$_2$Se$_3$ and Bi$_2$Se$_3$ indicate no substantial diffusion between the layers. We also measure a heterostructure consisting of a single \topo/\triv\ unit cell, with the \topo\ on top, see Fig. \ref{FigS4} C. We use scanning tunneling microscopy (STM) to count the number of In atoms that diffuse to the top QL of Bi$_2$Se$_3$, see Ref. 24 for details of the STM measurement. For both 30QL/20QL 50\% and 5QL/20QL 50\% films grown at $275^{\circ}$C, we find $\sim 0.2 \%$ In diffusion to the top QL of \topo\ [24]. This indicates that there is minimal In diffusion at $x = 50\%$ and also that there is no significant variation of In diffusion with the thickness of the \topo\ layer. To gain insight into the effect of growth temperature and doping $x$, we further studied 3QL/20QL 50\% grown at $300^{\circ}$C and 30QL/20QL 100\% grown at $275^{\circ}$C. We found $\sim 2\%$ and $\sim 1.3\%$ In diffusion to the top QL of the \topo\ layer, respectively (data not shown). Clearly (and perhaps not unexpectedly) In diffusion is suppressed with lower growth temperature and lower $x$. Since the heterostructures studied here are grown at $265^{\circ}$C with $x \leq 25\%$, In diffusion should be considerably suppressed even compared to the $\sim 0.2 \%$ diffusion observed for 30QL/20QL 50\% and 5QL/20QL 50\% grown at $275^{\circ}$C. Finally, we note that the topological phase transition for \triv\ occurs at $x \sim 4\%$, so we can conclude that In diffusion in our heterostructures is easily low enough to ensure that the \topo\ and \triv\ layers remain topological and trivial, respectively. Our direct characterization of In diffusion in our samples shows that In diffusion does not remove the layered pattern of the heterostructure.

\subsection{Observation of a topological phase transition in numerics}

\begin{figure}
\centering
\includegraphics[width=15cm]{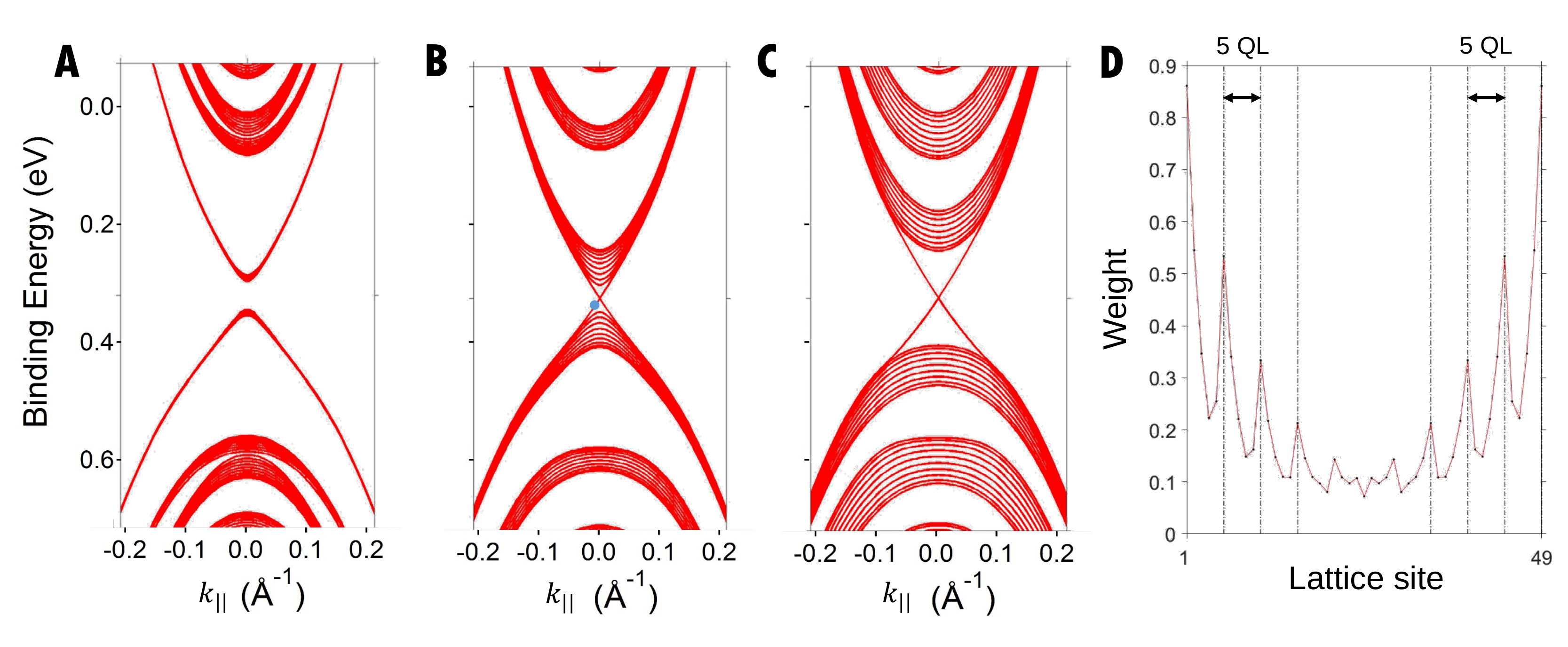}
\caption{\label{FigS1}\textbf{Topological phase transition in numerics.} Dispersion of a tight-binding model taking into account hopping between Se $p$ orbitals in a topological insulator heterostructure. Changing the hopping amplitudes shows (\textbf{A}) a trivial insulator with gapped superlattice band, (\textbf{B}) a phase near the critical point, with a superlattice band at the critical point and (\textbf{C}) a topological insulator with a gapped superlattice band and gapless surface state. (\textbf{D}) Wavefunction of a superlattice state, showing that the state is spatially localized at the interfaces of the heterostructure. The plotted state is marked by the blue circle in (\textbf{B}). Note that in this model we do not \textit{a priori} assume any topological surface or interface states.}
\end{figure}

We perform a tight-binding calculation of the band structure of our topological insulator heterostructure. We consider only hopping between Se $p$ orbitals within the Se planes and between the Se planes of Bi$_2$Se$_3$, with hopping $t$ within the topological layer, $t'$ within the trivial layer and $t''$ at the interface between layers. By tuning the hopping parameters, we observe a band inversion between the two superlattice bands, associated with a topological phase transition between a topological insulator and a trivial insulator, shown in Fig. \ref{FigS1}. We also plot the wavefunction of a specific state in real space, showing that the state is localized only at the heterostructure interfaces.

\subsection{Fine dependence on Indium doping}

\begin{figure}
\centering
\includegraphics[width=14cm]{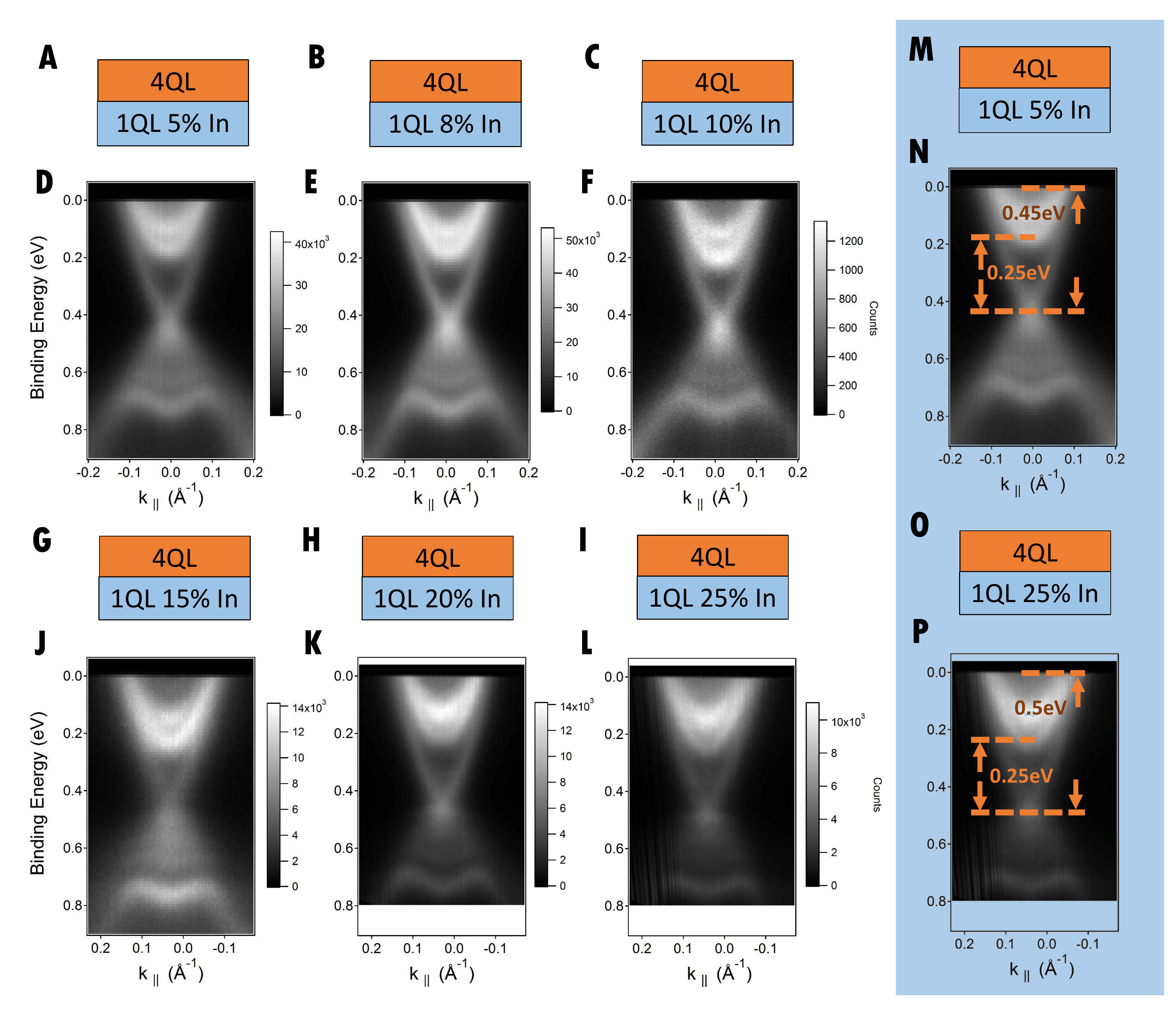}
\caption{\label{FigS2}\textbf{Change in Fermi level with In doping.} Systematic study of heterostructures with compositions shown in (\textbf{A}-\textbf{C}) and (\textbf{G}-\textbf{I}). All samples have 4QL topological layers and 1QL trivial layers, but have varying In doping in the trivial layer. (\textbf{D}-\textbf{F}, \textbf{J}-\textbf{L}) ARPES spectra, showing an $E$-$k$ cut through $\bar{\Gamma}$. We see that the bulk valence and conduction band quantum well states do not change as a function of In doping. In particular, the bulk band gap does not decrease despite a 20\% increase in In doping in the adjacent trivial layer, showing that there is no observable In diffusion to the topmost topological layer. At the same time, the Fermi level rises further into the bulk conduction band. This $n$ doping clearly increases with In doping and we speculate that it is due to the large band gap of the trivial layer under large In doping. Blue panel: we show again the first and last composition in this series, with 5\% In (\textbf{M}) and 25\% In (\textbf{O}). (\textbf{N}, \textbf{P}) The distance between the conduction band minimum and the Dirac point is $\sim 0.25$ eV for both compositions, but the Fermi level moves upward, so the Dirac point is $\sim 0.45$ eV below the Fermi level for 5\% In and $\sim 0.5$ eV below the Fermi level for 25\% In. We see that while the bulk band gap of the topmost topological layer does not begin to close, the In doping in the trivial layer is different in the different samples. This shows that the heterostructure consists of sharp interfaces.}
\end{figure}

We provide an additional, independent check that there is no In diffusion at least near the top of our heterostructure. We present $E$-$k$ cuts through $\bar{\Gamma}$ of 4QL/1QL heterostructures with In doping 5\%, 8\%, 10\%, 15\%, 20\% and 25\% at an incident photon energy of $h\nu  = 18$ eV, shown in Fig.  \ref{FigS2}. We consider the separation between the inner bulk conduction quantum well state and the Dirac point and we observe that this energy difference is $\sim 0.2$ eV independent of the In doping. We make similar observations about the other bulk conduction quantum well state as well as the bulk valence quantum well states. We note that bulk Bi$_2$Se$_3$ is very sensitive to In doping, with a topological phase transition at only $\sim 4\%$ In. However, the trivial In doping in these heterostructures changes by 20\% without any gap closing in the quantum well states of the top layer. This shows that In does not diffuse into the topmost topological layer from the topmost trivial layer. At the same time, we observe that the Fermi level rises with increasing In doping. This rise in the Fermi level is again inconsistent with In diffusion into the \topo\ layer because the bulk conduction band recedes above the Fermi level with increasing $x$ in \triv\ [12]. We suggest that the rise in the Fermi level with In doping may result from the large band gap of the topologically trivial layers. Specifically, the topological and trivial layers have different carrier density, leading to a band-bending effect that may correspond to an effective $n$ doping on the top \topo\ layer. Regardless of the mechanism of $n$ doping, we note that this systematic shift of the Fermi level with In doping confirms that the In concentration indeed increases in the trivial layers. The dependence of the bulk gap and Fermi level on In doping shows that the heterostructure consists of high-quality interfaces.

\subsection{Comparison with a single thin film of B\lowercase{i}$_2$S\lowercase{e}$_3$}

\begin{figure}
\centering
\includegraphics[width=13cm]{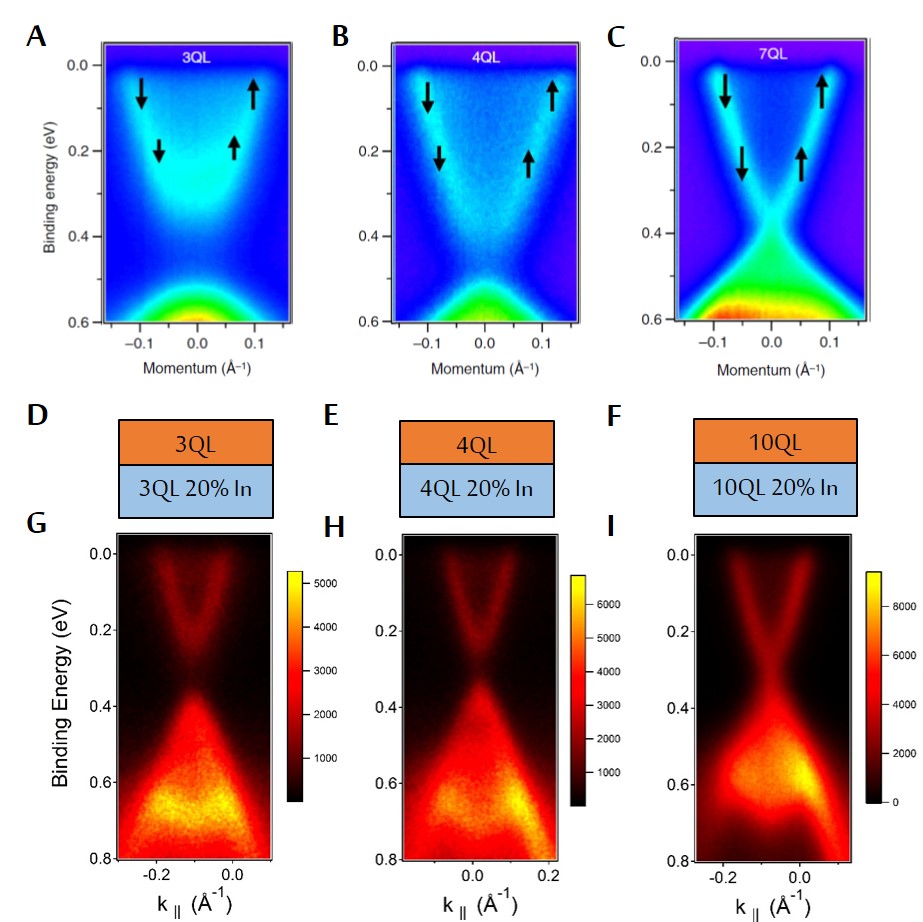}
\caption{\label{FigS3}\textbf{Dimerized-limit heterostructure and a single thin film of Bi$_2$Se$_3$.} (\textbf{A}-\textbf{C}) Gapped topological surface states in a single thin film of Bi$_2$Se$_3$, of thickness 3QL, 4QL and 7QL, respectively, adapted from [14]. (\textbf{D}-\textbf{F}) Unit cells of heterostructures \samG, \samA\ and \samH. (\textbf{G}-\textbf{I}) ARPES spectra, showing an $E$-$k$ cut through $\bar{\Gamma}$. The band gap is $\sim 0.2$ eV in the single 3QL thin film, but $\sim 0.15$ eV in \samG. The band gap is $\sim 0.1$ eV in both the single 4QL thin film and \samA. The surface state is gapless for the single 7QL thin film and \samH. The difference in the band gap at 3QL cannot be attributed to In diffusion because In diffusion should shrink the effective thickness of the topological layer and increase the size of the gap. The smaller gap must therefore be due to superlattice dispersion. Because the gap in the heterostructure is no larger than the gap in the single thin film, it is clear that there is no In diffusion from the topmost trivial layer into the topmost topological layer and the heterostructure consists of sharp interfaces.}
\end{figure}

We present yet another, independent check that the heterostructure is well-defined by considering dimerized-limit heterostructures. We repeat the $E$-$k$ cuts through $\bar{\Gamma}$ of \samG, \samA\ and \samH, discussed in the main text, and we show earlier results on single thin films of Bi$_2$Se$_3$, from [17]. We see that the size of the gap in the topological surface states is $\sim 0.2$ eV for a single 3QL film of Bi$_2$Se$_3$, but $\sim 0.15$ eV for \samG, as shown in Fig. \ref{FigS3}. Further, the gap is $\sim 0.1$ eV for both \samA\ and a single thin film 4QL thick. Lastly, the gap for a single thin film vanishes above 7QL, and we observe a gapless surface state in \samH. If there were In diffusion from the topmost trivial layer to the topmost topological layer, then the effective thickness of the topological layer should decrease, causing the surface state gap to increase. Since the gap in the heterostructure is no larger than the gap in the corresponding single thin film, there is no In diffusion into the topmost topological layer. Incidentally, the $\sim 0.15$ eV gap in \samG\ therefore shows an emergent superlattice band structure away from the dimerized limit, with small but observable hopping across the trivial layer. Our comparison of topological insulator heterostructures with single thin films of Bi$_2$Se$_3$ again shows that the heterostructure consists of high-quality interfaces.

\subsection{Detailed analysis of bulk quantum well states}

\begin{figure}[h!]
\centering
\includegraphics[width=15cm]{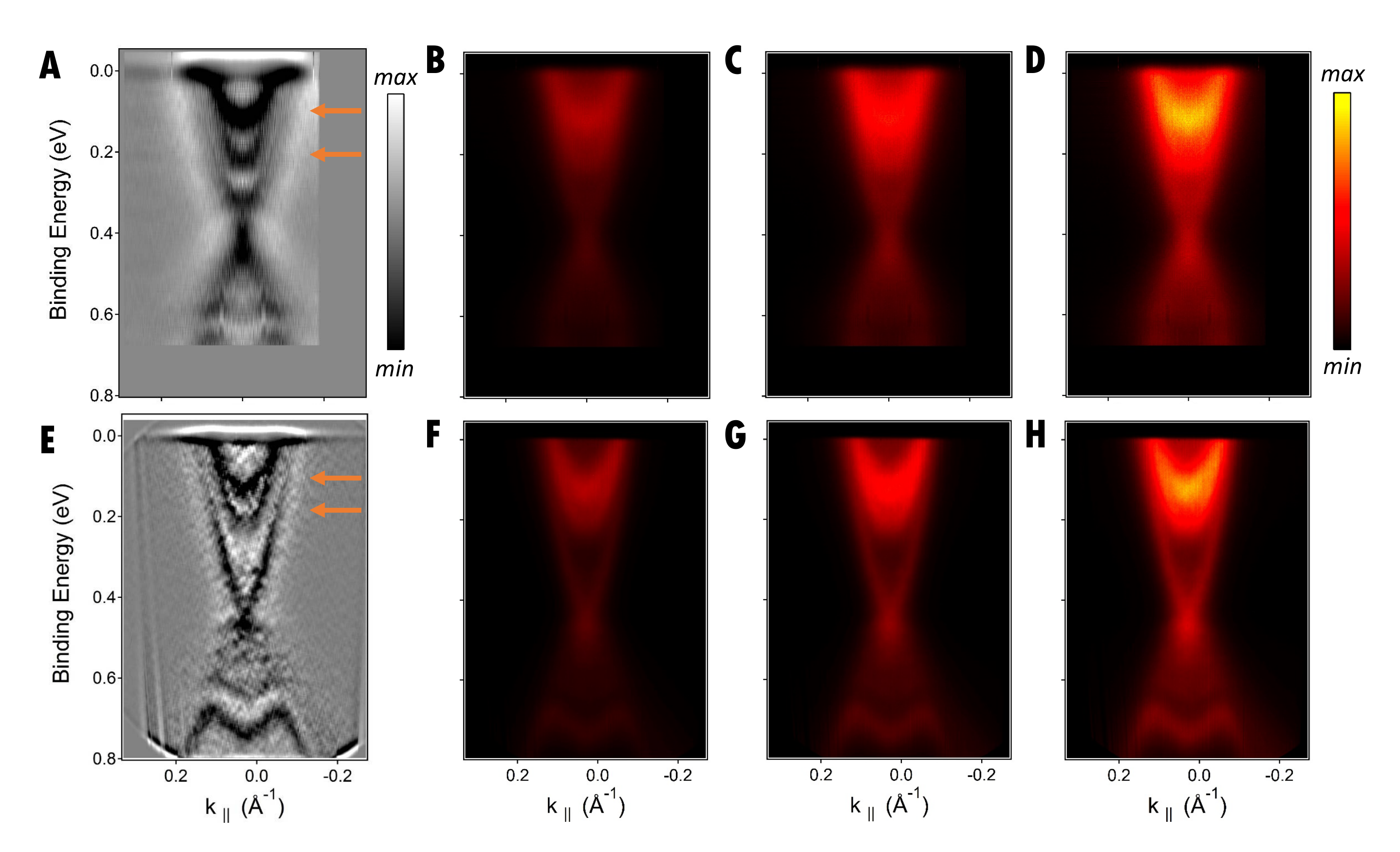}
\caption{\label{FigS5}\textbf{Bulk quantum well states of \samC\ and \samD.} (\textbf{A}) A second-derivative map of the ARPES spectrum of \samC, shown in Fig. 2 G of the main text. We clearly see two quantum well states in the bulk conduction band, marked by the orange arrows. (\textbf{B}-\textbf{D}) The same cut, with different color scales, to make the two quantum well states visible in the raw data. (\textbf{E}) The same as main text Fig. 3 B, repeated here for completeness. Again, we clearly see two quantum well states in the bulk conduction band. (\textbf{F}-\textbf{H}) The same cut, with different color scales, to make the two quantum well states visible in the raw data.}
\end{figure}

To clearly demonstrate the pair of bulk quantum well states in \samC\ and \samD, we present additional analysis of the spectra that were shown in Figs. 2 G, H of the main text. In Fig. \ref{FigS5} A, we show a second-derivative map for \samC, to complement the second-derivative map shown for \samD\ in main text Fig. 3 B. We also show the raw data in several color scales, to make the features more visible. We clearly observe two bulk quantum well states in both spectra, as well as a gapless topological surface state and valence band quantum well states.

\clearpage
%\cleardoublepage
\ifdefined\phantomsection
  \phantomsection  % makes hyperref recognize this section properly for pdf link
\else
\fi
\addcontentsline{toc}{section}{Bibliography}

{\singlespacing
}

%% file: ch-weylcriteria/ch-weylcriteria.tex
\chapter{Experimental criteria for detecting topological Fermi arc surface states in Weyl semimetals}
\label{ch:weylcriteria}

{\singlespacing
\begin{chapquote}{Electric Youth, \textit{The Best Thing}}
They say you have no name,\\
They say you show no signs,\\
And who you are by day,\\
Isn't the same in the night
\end{chapquote}}

\renewcommand{\beq}{\begin{equation*}}
\renewcommand{\eeq}{\end{equation*}}

\noindent This chapter is based on the article, \textit{Criteria for directly detecting topological Fermi arcs in Weyl semimetals} by Ilya Belopolski \textit{et al}., \textit{Phys. Rev. Lett.} {\bf 116}, 066802 (2016), available at \href{https://journals.aps.org/prl/abstract/10.1103/PhysRevLett.116.066802}{https://journals.aps.org/prl/abstract/10.1103/PhysRevLett.116.\\066802}.\\

%\begin{abstract}
\lettrine[lines=3]{T}{he} recent discovery of the first Weyl semimetal in TaAs provides the first observation of a Weyl fermion in nature and demonstrates a novel type of anomalous surface state, the Fermi arc. Like topological insulators, the bulk topological invariants of a Weyl semimetal are uniquely fixed by the surface states of a bulk sample. Here, we present a set of distinct conditions, accessible by angle-resolved photoemission spectroscopy (ARPES), each of which demonstrates topological Fermi arcs in a surface state band structure, with minimal reliance on calculation. We apply these results to TaAs and NbP. For the first time, we rigorously demonstrate a non-zero Chern number in TaAs by counting chiral edge modes on a closed loop. We further show that it is unreasonable to directly observe Fermi arcs in NbP by ARPES within available experimental resolution and spectral linewidth. Our results are general and apply to any new material to demonstrate a Weyl semimetal.
%\end{abstract}
%
%\date{\today}
%\maketitle

\section{Introduction}

\begin{figure}
\centering
\includegraphics[width=13cm, trim={30 45 80 80}, clip]{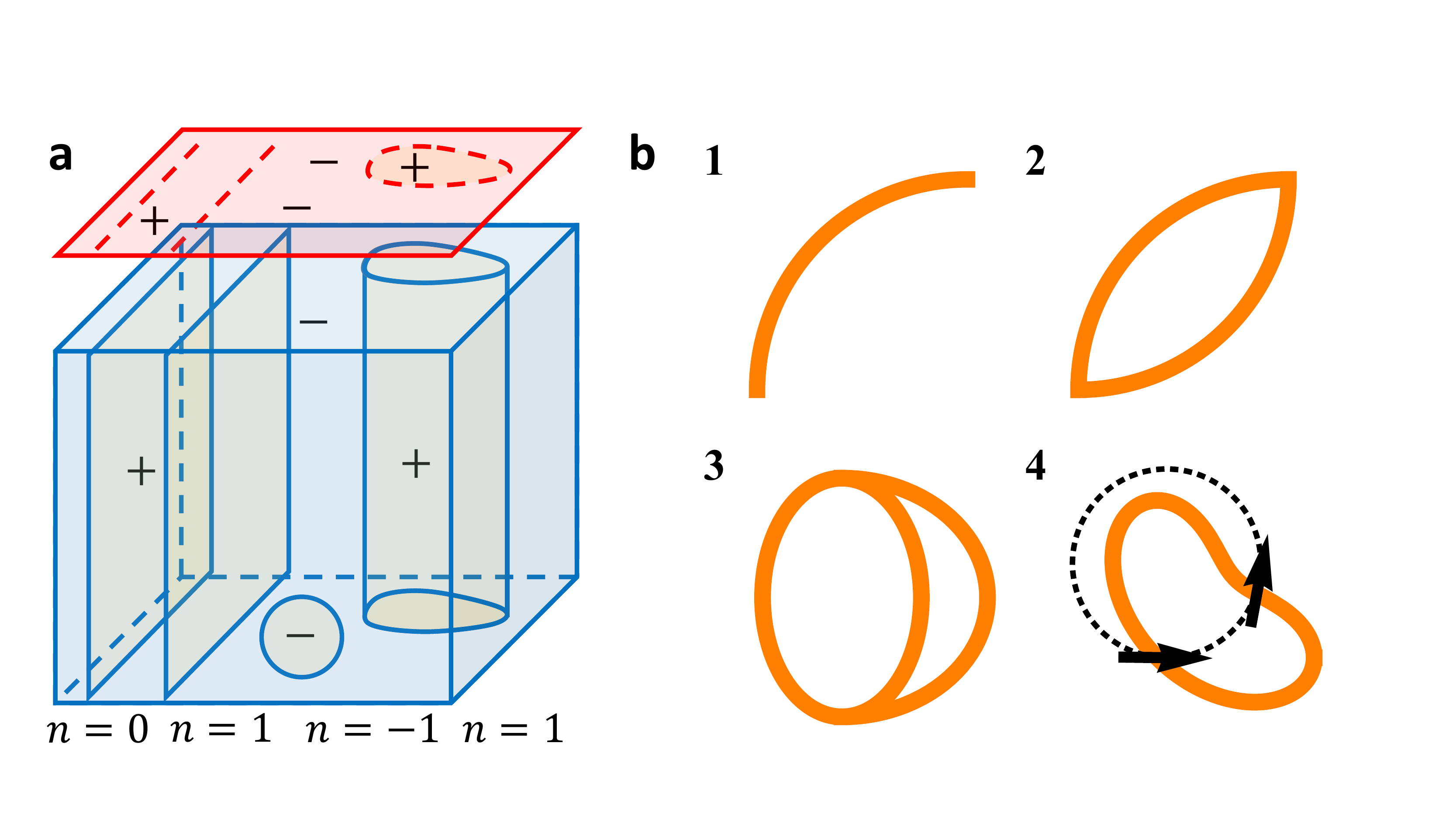}
\caption{\label{critFig1}\textbf{Four criteria for Fermi arcs.} (a) A bulk (blue) and surface (red) Brillouin zone with four Weyl points ($\pm$) and various 2D manifolds with Chern number $n$. Sweeping a plane through a Weyl point changes $n$. Also, any closed loop on the surface hosts chiral edge modes protected by the enclosed bulk chiral charge. For instance, a closed loop enclosing a $+$ Weyl point will host one right-moving chiral edge mode \cite{Hosur, Pyrochlore, Vish}. (b) The criteria: (1) A disjoint contour. (2) A closed contour with a kink. (3) No kinks within experimental resolution, but an odd set of closed contours. (4) An even number of contours without kinks, but net non-zero chiral edge modes.}
\end{figure}

A Weyl semimetal is a crystal which hosts Weyl fermions as emergent quasiparticles \cite{Weyl,Herring,Abrikosov,Nielsen,Volovik,Murakami,Pyrochlore,Multilayer,Hosur,Vish}. Although Weyl fermions are well-studied in quantum field theory, they have not been observed as a fundamental particle in nature. The recent experimental observation of Weyl fermions as electron quasiparticles in TaAs offers a beautiful example of emergence in science \cite{TaAsThyUs, TaAsThyThem, TaAsUs, LingLu, TaAsThem, TaAsChen, TaAsNodesDing, NbAs, TaPUs, TaPThem, HaoNbP}. Weyl semimetals also give rise to a topological classification closely related to the Chern number of the integer quantum Hall effect \cite{Pyrochlore, Vish, Hosur, Bernevig}. In the bulk band structure of a three-dimensional sample, Weyl fermions correspond to points of accidental degeneracy, Weyl points, between two bands \cite{Murakami, Multilayer}. The Chern number on a two-dimensional slice of the Brillouin zone passing in between Weyl points can be non-zero, as illustrated in Fig. \ref{critFig1}(a) \cite{Pyrochlore, Vish, Hosur}. Exactly as in the quantum Hall effect, the Chern number in a Weyl semimetal protects topological boundary modes. However, the Chern number changes when the slice is swept through a Weyl point, so the chiral edge modes associated with each slice terminate in momentum space at the locations of Weyl points, giving rise to Fermi arc surface states. This bulk-boundary correspondence makes it possible to demonstrate that a material is a Weyl semimetal by measuring Fermi arc surface states alone. As a result, to show novel Weyl semimetals, it is crucial to understand the possible signatures of Fermi arcs in a surface state band structure.

\begin{figure*}
\centering
\includegraphics[width=15cm, trim={20 20 10 30}, clip]{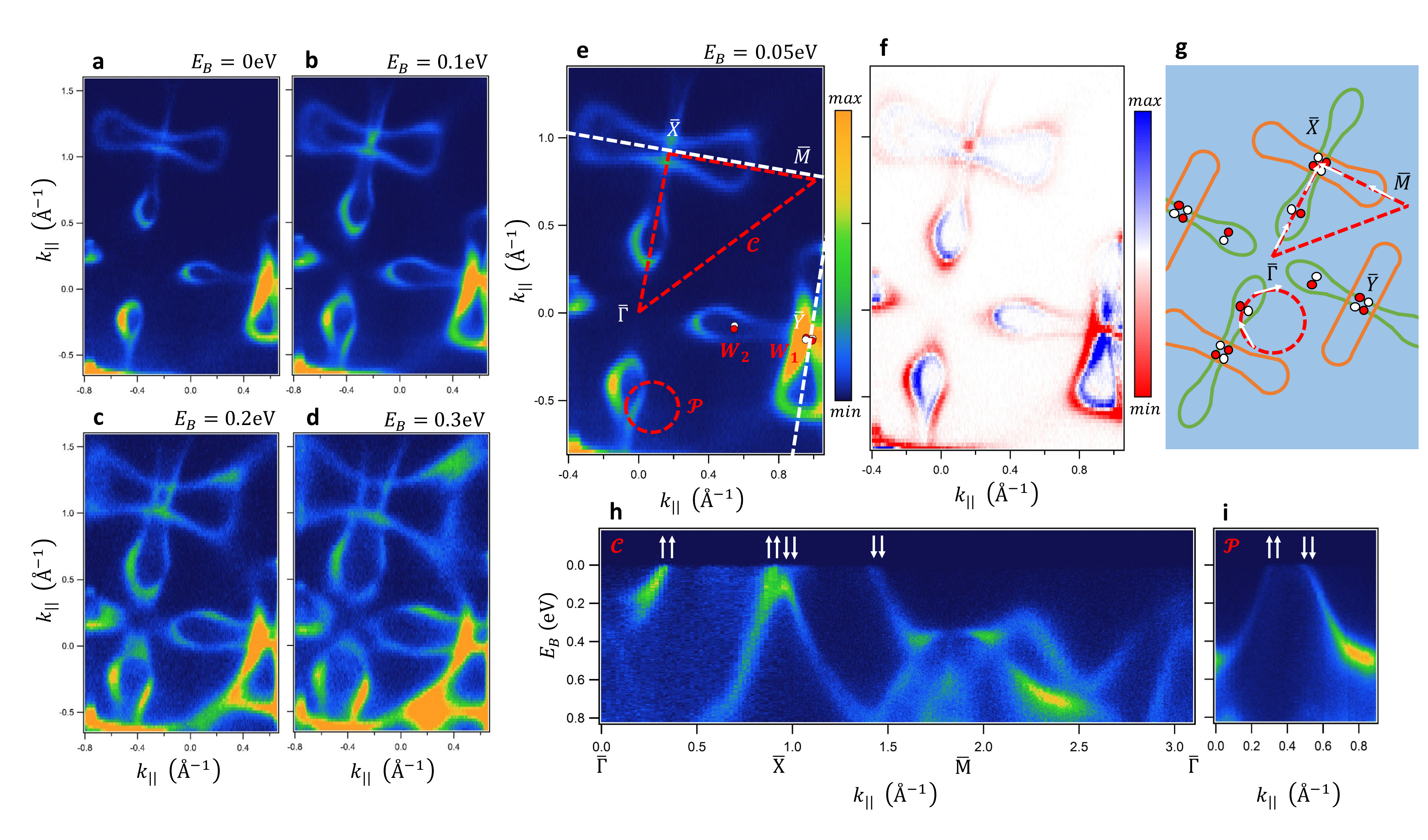}
\caption{\label{critFig2}\textbf{Surface states of NbP by ARPES.} (a)-(d) Constant energy cut by vacuum ultraviolet APRES, at incident photon energy $h\nu = 30$eV, on the (001) surface of NbP. (e) Same as (a)-(d), at $E_B = 0.05$ eV, with Weyl points marked and the two paths $\mathcal{C}$ and $\mathcal{P}$ on which we measure Chern numbers. (f) The difference of ARPES spectra at $E_B = 0.05$ eV and $E_B = 0.1$ eV, illustrating the direction of the Fermi velocity all around the lollipop and peanut pockets. The blue contour is always inside the red contour, indicating that the sign of the Fermi velocity is the same going around each contour. This result excludes the possibility that the lollipop actually consists of Fermi arcs attached to the $W_2$. (g) Cartoon summarizing the band structure. (h) Band structure along $\mathcal{C}$, with chiralities of edge modes marked by the arrows. We associate one arrow to each spinful crossing, even where we cannot observe spin splitting due to the weak SOC of NbP. There are the same number of arrows going up as down, so the Chern number is zero. (i) Same as (h) but along $\mathcal{P}$. Again, the Chern number is zero.}
\end{figure*}

\begin{figure*}
\centering
\includegraphics[width=15cm, trim={120 150 60 90}, clip]{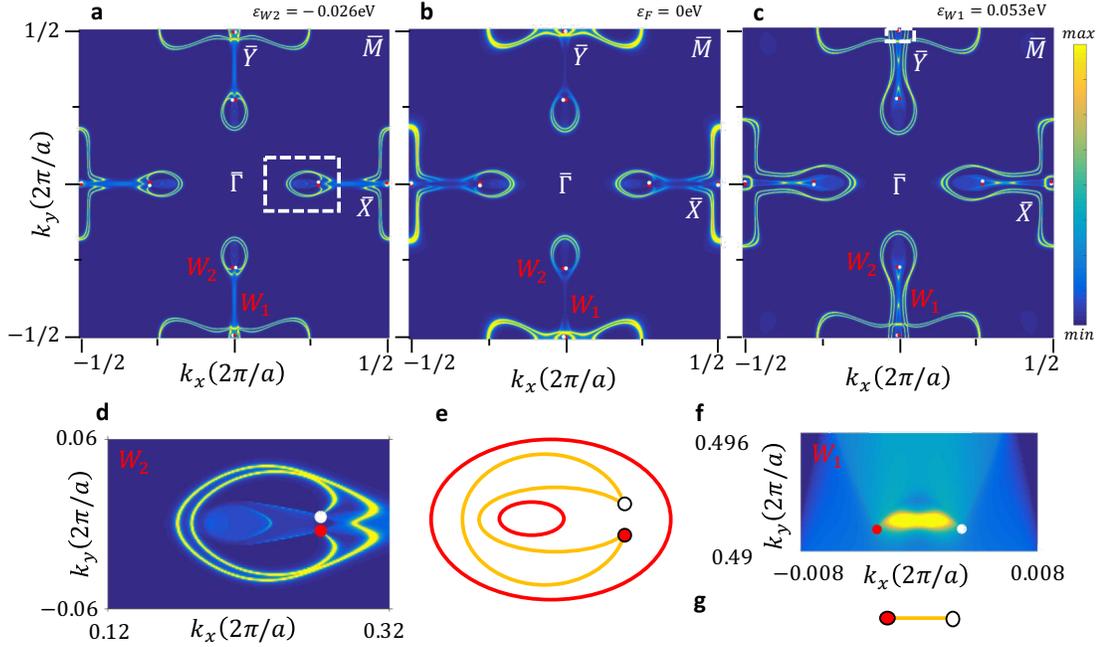}
\caption{\label{critFig3}\textbf{Numerical calculation of Fermi arcs in NbP.} First-principles band structure calculation of the (001) surface states of NbP at (a) the energy of the $W_2$, above the Fermi level, (b) the Fermi level and (c) the energy of the $W_1$, below the Fermi level. In (a) we see (1) a small set of surface states near the $W_2$ and (2) larger surface states near $\bar{X}$ and $\bar{Y}$. Near the energy of (b) there is a Lifshitz transition between the surface states at (1) and (2), giving rise to lollipop and peanut-shaped pockets. At (c) we see that the lollipop and peanut pockets enlarge, so they are hole-like. We also observe short Fermi arcs connecting the $W_1$. The numerical calculation shows excellent overall agreement with our ARPES spectra, suggesting that NbP is a Weyl semimetal. (d) Zoom-in of the surface states around the $W_2$, indicated by the white box in (a). We find two Fermi arcs and two trivial closed contours, illustrated in (e). (f) Zoom-in of the surface states around $W_1$, indicated by the white box at the top of (c). We find one Fermi arc, illustrated in (g).}
\end{figure*}

\begin{figure*}
\centering
\includegraphics[width=15.5cm, trim={0.9in 1.5in 0.5in 1.3in}, clip]{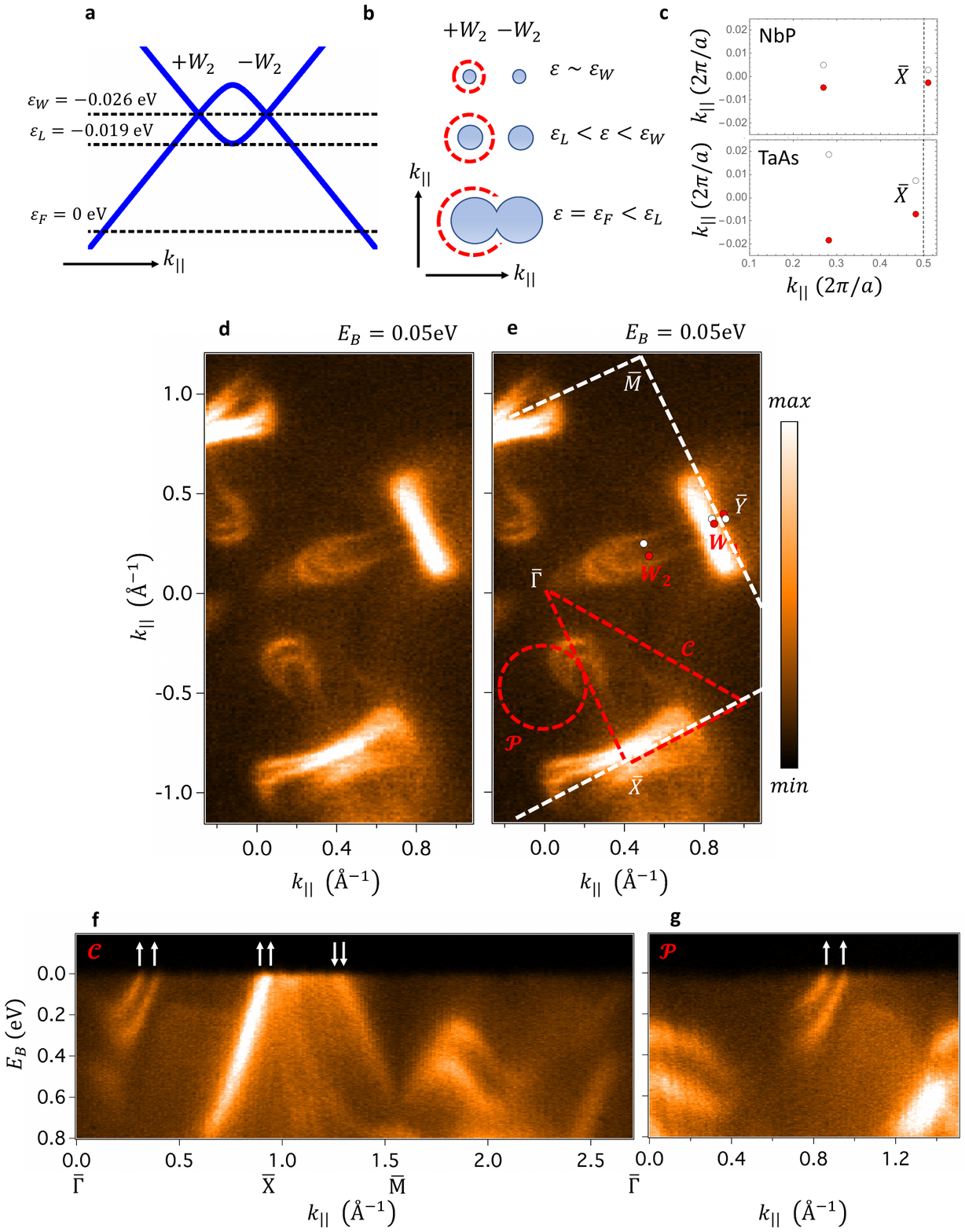}
\end{figure*}

\begin{figure*}
\caption{\label{critFig4}\textbf{Comparison of NbP and TaAs.} (a) Energies of the $W_2$, their Lifshitz transition and the Fermi level from the numerical calculation. One consequence of the small spin-splitting is that the Lifshitz transition between the $W_2$ is only $\sim 0.007$ eV below the energy of the $W_2$ and $\sim 0.019$ eV above the Fermi level. (b) Because $\varepsilon_F < \varepsilon_L$, it makes no sense to calculate the Chern number on $\mathcal{C}$ and $\mathcal{P}$, because the system on that cut is not an insulator, see the interrupted dotted red line in the last row of (b). (c) Plot of the positions of the Weyl point projections in NbP and TaAs arising from one Dirac line (see $\S 1$, SI). We see that the separation of the Weyl points is $\sim 4$ times larger in TaAs due to the large SOC. (d) Fermi surface by vacuum ultraviolet APRES, at incident photon energy $h\nu = 90$eV, on the (001) surface of TaAs. (e) Same as (d), but with Weyl points marked and two paths $\mathcal{C}$ and $\mathcal{P}$. (f) Band structure along $\mathcal{C}$, with chiralities of edge modes marked by the arrows. The net Chern number appears to be $+2$, inconsistent with the calculation. This can be explained by considering the small separation between the $W_1$. (g) Same as (f) but along $\mathcal{P}$. The path encloses only the well-spaced Weyl points and we find a Chern number $+2$, consistent with the calculation.}
\end{figure*}

Here, we present four distinct signatures of Fermi arcs, all in principle experimentally accessible in an ARPES measurement of a surface state band structure. Each signature alone, observed in any set of surface state bands, is sufficient to show that a material is a Weyl semimetal. Then, we apply these criteria to TaAs and NbP. We find that we can rigorously demonstrate a Weyl semimetal in TaAs by directly measuring a non-zero Chern number in ARPES. For NbP, we find that it is unreasonable to directly observe Fermi arcs within experimental resolution because the spin-orbit coupling is too weak.

\section{The four criteria}

Any surface state band structure which satisfies any one of the following four conditions contains topological Fermi arcs and is necessarily a Weyl semimetal:

\begin{enumerate}
\item \label{dis} \textit{Disjoint arc}: Any surface state constant-energy contour with an open curve is a Fermi arc and demonstrates a Weyl semimetal.
\item \label{kink} \textit{Kink off a rotation axis}: A Weyl point is characterized by chiral charge $n$, equal to the Chern number on a small spherical manifold enclosing the Weyl point in the bulk Brillouin zone, illustrated by the small sphere in Fig. \ref{critFig1}(a) \cite{Vish, Hosur, Pyrochlore}. For a Weyl point of chiral charge $|n| > 1$ or if multiple Weyl points project onto the same point of the surface Brillouin zone, there may not be a disjoint constant-energy contour because multiple arcs will emanate from the same Weyl point projection. However, the arcs will generically attach to the Weyl point at different slopes, giving rise to a kink in the constant-energy contour. Moreover, such a kink can only arise from the attachment of two Fermi arcs. A kink on the projection of a rotation axis may arise in a topological Dirac semimetal \cite{Na3Bi}. Off a rotation axis, such a kink guarantees a Weyl semimetal.
\item \label{odd} \textit{Odd number of curves}: For projected chiral charge $|n| > 1$, the constant-energy contours may consist entirely of closed curves and the kink may be below experimental resolution, so the constant-energy contour appears everywhere smooth. However, if $|n|$ is odd, the constant-energy contour will consist of an odd number of curves, so at least one curve must be a Fermi arc.
\item \label{Chern} \textit{Non-zero Chern number}: Consider any closed loop in the surface Brillouin zone where the bulk band structure is everywhere gapped and, at some energy, add up the signs of the Fermi velocities of all surface states along this loop, with $+1$ for right movers and $-1$ for left movers. The sum is the projected chiral charge enclosed in the curve, corresponding to a Chern number on a bulk \cite{Pyrochlore, Vish, Hosur}. A non-zero sum on at least one loop shows a Weyl semimetal, provided the loop is chosen to be contractible on the torus of the surface Brillouin zone.
\end{enumerate}

\noindent Note that while (\ref{dis}), (\ref{kink}) and (\ref{odd}) describe properties of a constant-energy slice of the Fermi surface, the counting argument (\ref{Chern}) requires a measurement of the dispersion. We note also that criterion (\ref{Chern}) allows us to determine all bulk topological invariants and Weyl points of a material by studying only its surface states.

In the rest of this Letter, we apply these criteria to TaAs and NbP. We find that criterion (\ref{Chern}) shows a Weyl semimetal in TaAs, but that all criteria fail for NbP because the spin-orbit coupling is too weak and the Fermi level is too low. We also present a calculation of NbP and show that it may be possible to demonstrate a Fermi arc in NbP by ARPES by observing a kink, criterion (\ref{kink}), but only if the Fermi level can be raised $>$ 20 meV. Lastly, we point out that the counting argument (\ref{Chern}) as recently applied to TaAs \cite{TaAsThem, TaAsChen} and NbP \cite{NbPThem} is invalid because certain Weyl points are too close together.

\section{Application to NbP}

We show that the surface state band structure of NbP as measured by ARPES does not satisfy any of the Fermi arc criteria. We note that vacuum ultraviolet ARPES is sensitive to the surface states of NbP, see Supplementary Information (SI). The surface states consist of lollipop and peanut-shaped pockets and we find that both are hole-like, see Figs. \ref{critFig2}(a)-(d). \textit{Criterion (1)}. Both the lollipop and peanut pockets are closed, so we observe no single disconnected arc. \textit{Criterion (2)}. We see no evidence of a kink in the constant-energy contours, so we do not observe a pair of arcs connecting to the same $W_2$ in a discontinuous way. \textit{Criterion (3)}. We observe an even number of contours everywhere, so no arc. \textit{Criterion (4)}. We can study the Fermi velocity using a difference map of two ARPES spectra, at $E_B = 0.05$ eV and $E_B = 0.1$ eV, shown in Fig. \ref{critFig2}(f). We see that the Fermi velocities have the same sign all the way around both the lollipop and the peanut pockets. If the lollipop consisted of Fermi arcs, one arc should evolve in a hole-like way, while the other arc should evolve in an electron-like way, so the different regions of the lollipop pocket should have opposite Fermi velocities in this sense. Because all points on both pockets have the same Fermi velocity, within the resolution of our ARPES measurements these pockets are trivial, hole-like surface states, and we observe no topological Fermi arcs. We can also consider a closed path through the surface Brillouin zone and count chiralities of edge modes along the path, illustrated in Figs. \ref{critFig2}(e,g). We check a triangular path, $\mathcal{C}$, along $\bar{\Gamma}-\bar{X}-\bar{M}-\bar{\Gamma}$, which encloses net chiral charge $+1$, Fig. \ref{critFig2}(h), and a small circular path, $\mathcal{P}$, which encloses net chiral charge $-2$, Fig. \ref{critFig2}(i). For each path, we label each spinless crossing with an up or down arrow to indicate the sign of the Fermi velocity. We find that going around either $\mathcal{C}$ and $\mathcal{P}$ we have net zero chirality, showing zero Chern number on the associated bulk manifold. The surface states of NbP fail all criteria for Fermi arcs. We cannot show that NbP is a Weyl semimetal using only the surface state band structure from ARPES.

Next, we compare our experimental results to numerical calculations on NbP and show that it is challenging to observe Fermi arcs in our spectra because of the low spin-orbit coupling. We present a calculation of the (001) surface states in NbP for the P termination, at the binding energy of $W_2$, $\varepsilon_{W2} = -0.026$ eV, at the Fermi level and at the binding energy of $W_1$, $\varepsilon_{W1} = 0.053$ eV, see Figs. \ref{critFig3}(a)-(c). We also plot the Weyl point projections, obtained from a bulk band structure calculation \cite{FourCompounds}. We observe surface states (1) near the mid-point of the $\bar{\Gamma}-\bar{X}$ and $\bar{\Gamma}-\bar{Y}$ lines and (2) near $\bar{Y}$ and $\bar{X}$. The surface states (1) form two Fermi arcs and two closed contours at $\varepsilon_{W2}$, see Fig. \ref{critFig3}(d),(e). These states undergo a Lifshitz transition near $\varepsilon_F$ with the surface states (2), forming a large hole-like pocket below the Fermi level. The surface states (2) also form a large hole-like pocket. They contain within them, near the $\bar{X}$ and $\bar{Y}$ points, a short Fermi arc connecting each pair of $W_1$, see Fig. \ref{critFig3}(f),(g). We note the excellent agreement with our ARPES spectra, where we also see lollipop and peanut contours which evolve into trivial, closed, hole-like pockets below $\varepsilon_F$. We also find in our calculation that the separation of Weyl points and the surface state spin-splitting is small. This result is consistent with our ARPES spectra, which do not show spin-splitting in the surface states near the Fermi level.

The small spin-splitting observed in our numerical calculations underlines the difficulty in observing topological Fermi arc surface states in NbP. The separation of the Weyl points is $< 0.01 \textrm{\AA}^{-1}$ for the $W_1$ and $< 0.02 \textrm{\AA}^{-1}$ for the $W_2$, both well below the typical linewidth of our ARPES spectra, $\sim 0.05 \textrm{\AA}^{-1}$. For this reason, we cannot resolve the momentum space region between the $W_1$ or the $W_2$ to determine if there is an arc. We emphasize that we cannot surmount this difficulty by considering Fermi level crossings on $\mathcal{P}$ or $\mathcal{C}$, as shown in Fig. \ref{critFig2}(e). It is obvious that if we cannot resolve the two Weyl points in a Fermi surface mapping, then we also cannot resolve a Fermi arc connecting them in a cut passing through the Weyl points. In this way, on $\mathcal{P}$ we cannot verify the arc connecting the $W_1$ and on $\mathcal{P}$ and $\mathcal{C}$ we cannot verify the empty region between the $W_2$. As an additional complication, it is difficult to use $\mathcal{P}$ or $\mathcal{C}$ because the Fermi level is below the Lifshitz transition for the $W_2$ in NbP, see Fig. \ref{critFig4}a. This invalidates use of criterion (4) because for $\varepsilon_F < \varepsilon_L$, there is no accessible binding energy where the bulk band structure is gapped along an entire loop passing in between a pair of $W_2$, illustrated in Fig. \ref{critFig4}b by the broken dotted red line for $\varepsilon = \varepsilon_F$. However, if we could access $\varepsilon > \varepsilon_L$, then it may be possible to use criterion (2) to demonstrate a Fermi arc in NbP without resolving the $W_2$ and counting chiralities. In particular, while the Fermi arc may appear to form a closed contour due to the small separation of $W_2$, it could have a kink at the location of the $W_2$. Unlike criterion (4), applying criterion (2) to (001) NbP would not depend on resolving the $W_2$ or the spin-splitting of the surface states. Improvements in the quality of NbP crystals or the cleaved surface could also allow the Fermi arcs to be resolved by reducing the spectral linewidth. Lastly, we point out that our results invalidate recent claims that $\mathcal{C}$ \cite{NbPThem} or any direct measurement of Fermi arcs \cite{NbPAndo, NbPChen} can be used to demonstrate a Weyl semimetal in NbP, due to the large linewidth of available ARPES spectra.

\section{Application to TaAs}

We apply criterion (4) to demonstrate Fermi arcs in TaAs. The larger spin-splitting in TaAs increases the separation of Weyl points as compared to NbP, see Fig. \ref{critFig4}c. However, we emphasize that only the $W_2$ are well within experimental resolution. The separation of the $W_1$ in TaAs is comparable to the separation of the $W_2$ in NbP, which as noted above, cannot be resolved. In contrast to NbP, the (001) surface states of TaAs near $\bar{\Gamma}$ consist of crescent pockets with clear spin-splitting, see Fig. \ref{critFig4}(d). We also observe bowtie surface states near $\bar{X}$ and $\bar{Y}$. We apply criterion (4) to search for Fermi arcs in TaAs on paths $\mathcal{C}$ and $\mathcal{P}$, shown in Fig. \ref{critFig4}(e). We mark the Fermi velocity of each crossing with an arrow, but now with one arrow per spinful crossing. Along $\mathcal{C}$, see Fig. \ref{critFig4}(f), we see two well-resolved crossings not far from $\bar{\Gamma}$. However, the spinful crossings from the bowtie pocket are difficult to resolve. Based on the constant-energy contour, Fig. \ref{critFig4}(d), we may interpret the bowtie pocket as consisting of two slightly-separated spinful contours, so we associate two arrows with each of the remaining two crossings along $\mathcal{C}$. We find that the enclosed Chern number is $+2$, while it is $+3$ according to a numerical calculation of the band structure \cite{FourCompounds}. Again, this inconsistency results from the small separation of the $W_1$. As a result, we cannot resolve the additional crossing from the Fermi arc connecting the $W_1$ near $\bar{X}$, which would give a Chern number of $+3$. We can avoid the bowtie pocket by using $\mathcal{P}$, see Fig. \ref{critFig4}g. Here, we see only the two well-separated states near $\bar{\Gamma}$. We find two edge modes of the same chirality, unambiguously showing a Chern number $+2$ on the associated bulk manifold, satisfying criterion (4) for a Weyl semimetal. In this way, we demonstrate that TaAs is a Weyl semimetal by studying only the surface states as measured by ARPES. We note that our results invalidate earlier measurements on $\mathcal{C}$ used to demonstrate a Weyl semimetal in TaAs \cite{TaAsThem, TaAsChen}. Here, we have shown that the small separation between the $W_1$ in TaAs makes it impossible to calculate the Chern number on $\mathcal{C}$. In the same way, our results on NbP invalidate recent works claiming a Weyl semimetal in NbP using $\mathcal{C}$ \cite{NbPThem} or by directly observing Fermi arcs \cite{NbPAndo, NbPChen}. We emphasize that to show a Weyl semimetal it is not enough to present an overall agreement between ARPES and numerics. In the case of TaAs and NbP, there are many trivial surface states, so an overall agreement is not entirely relevant for the topological invariants or the Weyl semimetal state. In addition, an overall agreement is precarious in cases where the system is near a critical point or where there may be several closely-related crystal structures. Rather, to show a Weyl semimetal, it is sufficient to pinpoint a topological Fermi arc in an ARPES spectrum of surface states. Here, we have presented a set of general and distinct criteria, applicable to any material, any one of which pinpoints a topological Fermi arc. By presenting criteria for Fermi arcs, we provide a useful reference for demonstrating novel Weyl semimetals.

\section{Materials and methods}

Single crystal TaAs and NbP samples were grown by chemical vapor transport methods. Angle-resolved photoemission spectroscopy (ARPES) measurements were performed using a Scienta R4000 at Beamline 5-4 of the Stanford Synchrotron Radiation Lightsource, SLAC and Beamline 4 of the Advanced Light Source, LBNL in CA, USA. The angular resolution was better than 0.2$^{\circ}$ and the energy resolution better than 20meV. Samples were cleaved $\textit{in situ}$ and measured under vacuum better than $10^{-10}$ Torr at temperatures $<$ 30K. First principles electronic structure calculations of NbP were carried out with the OpenMX code \cite{OpenMX}, based on the generalized gradient approximation (GGA) \cite{Perdew}. Spin-orbit coupling was incorporated through $j$-dependent pseudo-potentials. For each Nb atom three, three, three and one optimized radial functions were allocated for the s, p, d, and f orbitals ($s3p3d3f1$), respectively. For each P atom, $s3p3d2f1$ were adopted. A $k$-point mesh of $17 \times 17 \times 5$ for the conventional unit cell was used and experimental lattice parameters were adopted in the calculations \cite{Crystal1, Crystal2, Crystal3}. We use Nb $s$ and $d$ orbitals and P $p$ orbitals to construct Wannier functions without performing the procedure for maximizing localization \cite{Wannier1, Wannier2}.  We calculated the surface spectral weight of a semi-infinite (001) slab using an iterative Green's function method based on the Wannier function basis set.

%\vspace{0.2cm}

%{\bf Acknowledgments.} Work at Princeton University and Princeton-led synchrotron-based ARPES measurements are supported by the Emergent Phenomena in Quantum Systems Initiative of the Gordon and Betty Moore Foundation under grant GBMF4547 (M.Z.H.). Single-crystal growth is supported by the National Basic Research Program of China under grants 2013CB921901 and 2014CB239302 and characterization is supported by the U.S. Department of Energy, Office of Science, Basic Energy Sciences under grant DE-FG-02-05ER46200. H.L. acknowledges the Singapore National Research Foundation under award NRF-NRFF2013-03. The work at Northeastern University is supported by the U.S. Department of Energy, Office of Science, Basic Energy Sciences under grant DE-FG02-07ER46352, and benefited from theory support at the Advanced Light Source and the allocation of supercomputer time at the National Energy Research Scientific Computing Center under U.S. DOE grant DE-AC02-05CH11231. I.B. acknowledges the support of the U.S. National Science Foundation GRFP. M.N. is supported by the Los Alamos National Laboratory LDRD program and start-up funds from the University of Central Florida. We thank Makoto Hashimoto and Donghui Lu for technical assistance with ARPES measurements at SSRL Beamline 5-4, SLAC, Menlo Park, CA, USA. We also thank Nicholas Plumb and Ming Shi for technical assistance with ARPES measurements at the HRPES endstation of the SIS beamline, Swiss Light Source, Villigen, Switzerland. I.B. thanks Titus Neupert for useful discussions.

In this supplementary information section, we first present an overview of the crystal structure and electronic band structure of NbP. Then, we argue that vacuum ultraviolet ARPES is sensitive to the surface states of (001) NbP. Lastly, we discuss spin-splitting and phenomena arising from the different orbital contributions of the surface states.

\subsection{Crystal, electronic structure of the Weyl semimetal NbP}

%\bigskip
%{\bf $\S 1.$ Overview of the crystal, electronic structure of the Weyl semimetal NbP}
%\bigskip

Niobium phosphide (NbP) crystallizes in a body-centered tetragonal Bravais lattice, in point group $C_{4v}$ ($4mm$), space group $I4_1md$ (109), isostructural to TaAs, TaP and NbAs \cite{Crystal1, Crystal2, Crystal3}. The crystal structure can be understood as a stack of alternating Nb and P square lattice layers, see Fig. \ref{critFigS1}a. Each layer is shifted with respect to the one below it by half an in-plane lattice constant, $a/2$, in either the $\hat{x}$ or $\hat{y}$ direction. The crystal structure can also be understood as arising from intertwined helices of Nb and P atoms which are copied in-plane to form square lattices, with one Nb (or P) atom at every $\pi/2$ rad along the helix. The conventional unit cell consisting of one period of the helices is shown in Fig. \ref{critFigS1}b. This helical structure is related to the non-symmorphic $C_4$ symmetry, where a $C_4$ rotation followed by a translation by $c/4$ is required to take the crystal back into itself. We note that NbP has no inversion symmetry, so that all bands are generically singly-degenerate. This is a crucial requirement for NbP to be a Weyl semimetal. We show a photograph of the sample taken through an optical microscope, suggesting that it is a single crystal and of high quality, in Fig. \ref{critFigS1}c. A scanning tunneling microscopy (STM) topography of the sample shows a square lattice surface, demonstrating that NbP cleaves on the (001) plane, see Fig. \ref{critFigS1}d. The lack of defects further suggests the high quality of the single crystals. From the ionic model, we expect that the conduction and valence bands in NbP arise from Nb $4d$ and P $3p$ orbitals, respectively. However, an \textit{ab initio} bulk band structure calculation along high-symmetry lines shows that NbP does not have a full gap but is instead a semimetal, see $\Sigma-\Gamma$, $Z-\Sigma'$, $\Sigma'-N$ in Fig. \ref{critFigS1}e, with the bulk Brillouin zone in Fig. \ref{critFigS1}f. In the absence of spin-orbit coupling, the band structure near the Fermi level consists of four Dirac lines, shown in purple in Fig. \ref{critFigS1}g, see also \cite{FourCompounds}. These Dirac lines are protected by two vertical mirror planes, shown in blue. After spin-orbit coupling is included, each Dirac line vaporizes into six Weyl points shifted slightly off the mirror plane, marked by the dots in Fig. \ref{critFigS1}g. Two Weyl points are on the $k_z = 0$ plane, shown in red, and we call these Weyl points $W_1$. The other four, we call $W_2$. We note that on the (001) surface, two $W_2$ of the same chirality project onto the same point of the surface Brillouin zone, giving rise to a projected Weyl point of chiral charge $\pm 2$. The $W_1$ give projections of chiral charge $\pm 1$, see Fig. \ref{critFigS1}h and again \cite{FourCompounds}. We also point out that the chiralities of the $W_1$ are flipped between NbP and TaAs, compare main text Figs. 2g and 4e. This is because the Dirac line is larger in NbP and crosses over the edge of the first Brillouin zone, into the second Brillouin zone. As a result, some of the Weyl points arising from the Dirac line end up in the second Brillouin zone, while Weyl points from the Dirac line of the second Brillouin zone show up in the first Brillouin zone. This flips the chiralities of the $W_1$ between NbP and TaAs. This can also be seen from main text Fig. 4c. See also Fig. 1H of Ref. \cite{TaPUs}.

\subsection{The surface states of (001) NbP by vacuum ultraviolet ARPES}

%\bigskip
%{\bf $\S 2.$ The surface states of (001) NbP by vacuum ultraviolet ARPES}
%\bigskip

Here, we show that we observe surface states but not bulk states in vacuum ultraviolet ARPES on the (001) surface of NbP. In our ARPES spectra, we observe a Fermi surface consisting of lollipop-shaped pockets along the $\bar{\Gamma} - \bar{X}$ and $\bar{\Gamma} - \bar{Y}$ lines and peanut-shaped pockets on the $\bar{M} - \bar{X}$ and $\bar{M} - \bar{Y}$ lines, see Fig. 2(a)-(d) in main text. The spectra are consistent with spectra of surface states of other compounds in the same family, suggesting that these pockets are surface states rather than bulk states. Because $C_4$ symmetry is implemented as a screw axis in NbP, the (001) surface breaks $C_4$ symmetry and the surface state dispersion is not $C_4$ symmetric. Our data suggest that the peanut pockets at $\bar{X}$ and $\bar{Y}$ differ slightly, showing a $C_4$ breaking that demonstrates surface states. We note, however, that this effect is much weaker than in TaAs or TaP \cite{TaAsUs, TaPUs}. This result could be explained by reduced coupling between the square lattice layers, restoring the $C_4$ symmetry of each individual layer. In particular, we note that the lattice constants of NbP are comparable to those of TaAs, while the atomic orbitals are smaller due to the lower atomic number. We expect this effect to be particularly important for surface states derived from the $p_x$, $p_y$, $d_{xy}$ and $d_{x^2-y^2}$ orbitals and indeed we observe no $C_4$ breaking at all for the lollipop pockets, which arise from the in-plane orbitals \cite{FourCompounds}. We conclude that we observe the surface states of NbP in our ARPES spectra. 

\subsection{Spin-splitting, orbital contributions of the surface states of NbP}

%\bigskip
%{\bf $\S 3.$ Spin-splitting and orbital contributions of the surface states of NbP}
%\bigskip

Here, we point out several other features of the (001) surface states of NbP. First, we note that we can observe a spin splitting in the lollipop pocket at $E_B \sim 0.2$ eV, shown in Fig. \ref{critFigS4}a and repeated with guides to the eye in Fig. \ref{critFigS4}b. Next, we observe in our ARPES spectra that the lollipop and peanut pockets do not hybridize anywhere in the surface Brillouin zone. We present a set of dispersions near the intersection of the lollipop and peanut pockets and we find no avoided crossings, see Fig. \ref{critFigS4}c for the locations of the cuts, shown in Fig. \ref{critFigS4}e. As mentioned above, we attribute this effect to the different orbital character of the two pockets. In particular, the lollipop pocket arises mostly from in-plane $p$ and $d$ orbitals and the peanut pocket arises mostly from out-of-plane $p$ and $d$ orbitals. The suppressed hybridization may be related to the $C_2$ symmetry of the (001) surface. In particular, we note that the in-plane and out-of-plane orbitals transform under different representations of $C_2$. The contributions from different, unhybridized orbitals to the surface states in NbP may give rise to novel phenomena. For example, we propose that quasiparticle interference between the lollipop and peanut pockets will be suppressed in an STM experiment on (001) NbP. Also, the rich surface state structure may explain why the bulk band structure of NbP is invisible to vacuum ultraviolet ARPES. Specifically, because the surface states take full advantages of all available orbitals, there are no orbitals left near the surface to participate in the bulk band structure. Other phenomena may also arise from the rich surface state structure in NbP. Finally, in Fig. \ref{critFigS4}d we present a calculation of the Fermi arcs near the $W_2$ in TaAs, equivalent to the calculation shown in main text Fig. 3d and in agreement with the Fermi arcs we observe by ARPES in TaAs.

\clearpage
\begin{figure*}
\centering
\includegraphics[width=15cm, trim={70 210 70 90}, clip]{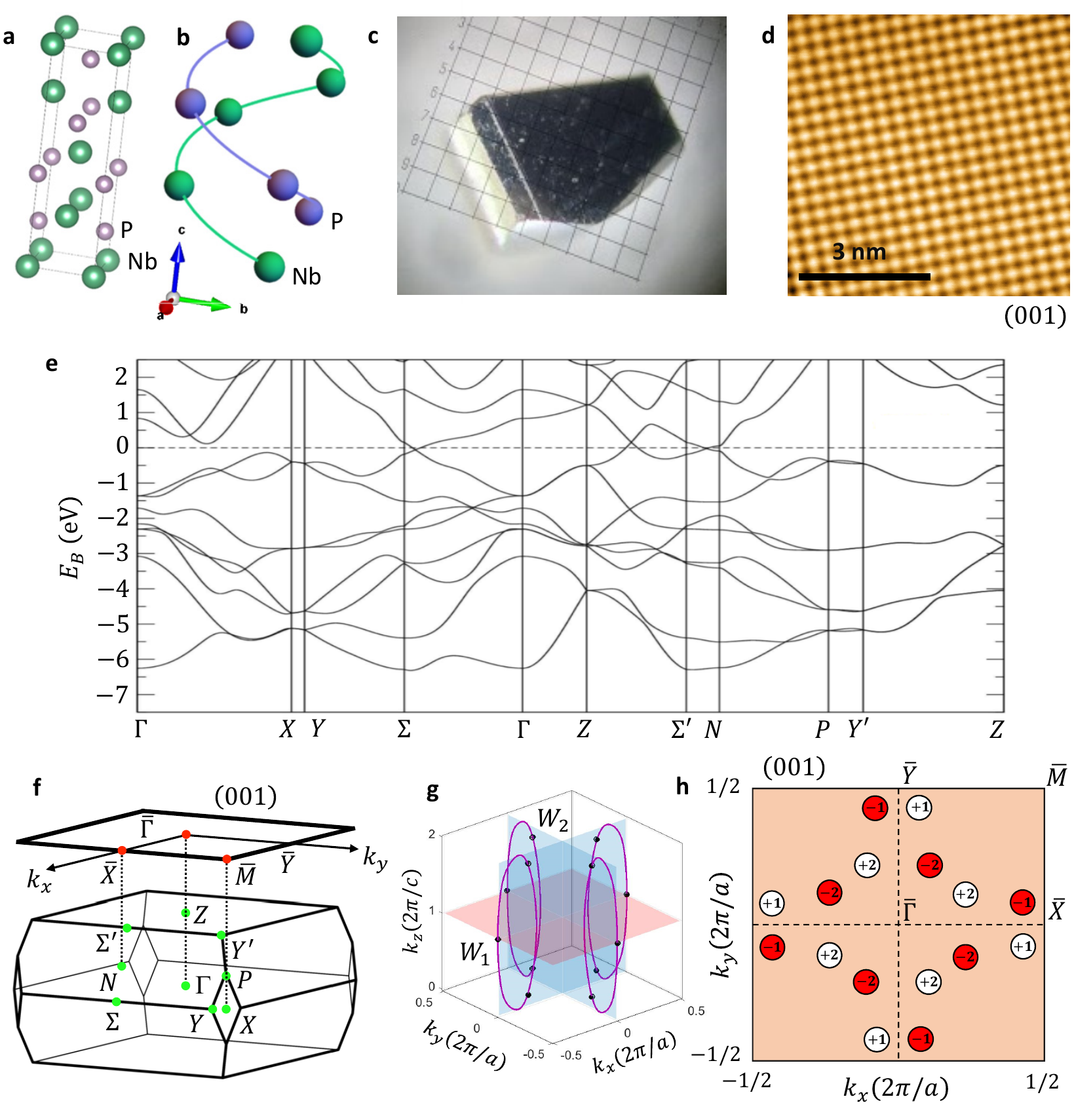}
\end{figure*}

\clearpage
\begin{figure*}
\caption{\label{critFigS1}\textbf{Overview of the Weyl semimetal candidate NbP.} (a) The crystal structure of NbP, which can be understood as a stack of square lattices of Nb and P, with a stacking pattern which involves an in-plane shift of each layer relative to the one below it. (b) The crystal structure can also be understood as a pair of intertwined helices of Nb and P atoms which are copied in-plane to form square lattice layers. The axis of the helix is the $\hat{z}$ direction of the conventional unit cell, the center of the helix is $1/4$ of the way along the diagonal of one plaquette of a square lattice layer, and the radius of the helix is $a/2\sqrt{2}$. (c) Photograph of the sample taken through an optical microscope, showing a beautiful crystal. (d) An STM topography of the (001) surface of NbP, showing the high quality of the sample surface, with no defects within a 6.2 nm $\times$ 6.2 nm window. The image was taken at bias voltage $-0.3$ eV and temperature $4.6$ K. (e) \textit{Ab initio} bulk band structure calculation of NbP, using GGA exchange correlation functionals, without spin-orbit coupling (SOC), showing that NbP is a semimetal with band inversions along $\Sigma-\Gamma$, $Z-\Sigma'$ and $\Sigma'-N$. (f) The bulk Brillouin zone and (001) surface Brillouin zone of NbP, with high-symmetry points labeled. (g) Without SOC, NbP has four Dirac lines protected by two mirror planes (blue). With SOC, each Dirac line vaporizes into six Weyl points (indicated by the dots), two on the $k_z = 0$ plane (red, labelled as $k_z = 1$ and equivalent to $k_z = 0$), called $W_1$, and four away from $k_z = 0$, called $W_2$. Note that the Weyl points have very small separation in momentum space, so that each dot corresponds to two Weyl points, one of each chirality. (h) Illustration of the Weyl point projections in the (001) surface Brillouin zone. Two $W_2$ of the same chiral charge project onto the same point in the surface Brillouin zone, giving Weyl point projections of chiral charge $\pm 2$. The separation is not to scale, but the splitting between pairs of $W_2$ is in fact larger than the splitting between pairs of $W_1$, see also main text Fig. 4(c).}
\end{figure*}

\begin{figure}
\centering
\includegraphics[width=15cm, trim={70 430 120 130}, clip]{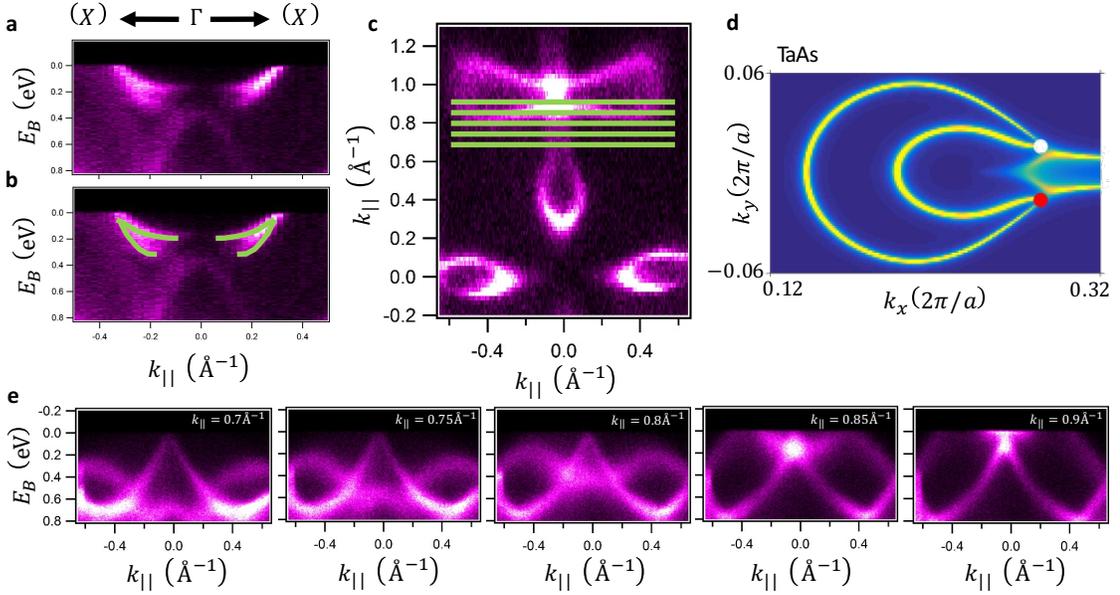}
\caption{\label{critFigS4}\textbf{Spin-splitting of the (001) surface states in NbP.} (a) Surface states by ARPES along $\bar{X}-\bar{\Gamma}-\bar{X}$, showing a spin splitting below the Fermi level. (b) Same as (a) but with guides to the eye to mark the spin splitting. (c). Fermi surface by APRES at $E_B = 0.1$ eV, marking the cuts shown in (e). (d) Calculation from \textit{ab initio} of the Fermi arcs near $W_2$ in TaAs, equivalent to main text Fig. 3d for NbP but with much larger spin-splitting between the surface states due to the large spin-orbit coupling. (e) Dispersion of the lollipop and peanut pockets, showing that the two pockets move through each other without any observable hybridization. This absence of an avoided crossing may be related to the different orbital character of the two pockets. Specifically, the lollipop pocket arises from in-plane orbitals and the peanut pocket from out-of-plane orbitals.}

\end{figure}
\clearpage

%\cleardoublepage
\ifdefined\phantomsection
  \phantomsection  % makes hyperref recognize this section properly for pdf link
\else
\fi
\addcontentsline{toc}{section}{Bibliography}

{\singlespacing

}

%\end{document}

%% file: ch-mwt1/ch-mwt1.tex
%\documentclass[aps,prl,reprint,twocolumn,nopacs,superscriptaddress]{revtex4}
%\usepackage{amsmath}
%\usepackage{amssymb}
%\usepackage{graphicx}
%\usepackage{hyperref}
%\pagestyle{headings}
%\newcommand{\beq}{\begin{equation*}}
%\newcommand{\eeq}{\end{equation*}}
\newcommand{\comp}{Mo$_x$W$_{1-x}$Te$_2$}
\newcommand{\half}{Mo$_{0.45}$W$_{0.55}$Te$_2$}
\newcommand{\ten}{Mo$_{0.07}$W$_{0.93}$Te$_2$}
\newcommand{\twenty}{Mo$_{0.2}$W$_{0.8}$Te$_2$}
\newcommand{\forty}{Mo$_{0.4}$W$_{0.6}$Te$_2$}

%\begin{document}

\chapter{Fermi arc electronic structure and Chern numbers in \comp}
\label{ch:mwt1}

{\singlespacing
\begin{chapquote}{Superbus, \textit{Addictions}}
J'affectionne les tourments,\\
Et j'apprends\\
\end{chapquote}}

\noindent This chapter is based on the article, \textit{Fermi arc electronic structure and Chern numbers in the type-II Weyl semimetal candidate} \comp\ by Ilya Belopolski {\it et al}., {\it Phys. Rev. B} {\bf 94}, 085127 (2016), available at \href{https://journals.aps.org/prb/abstract/10.1103/PhysRevB.94.085127}{https://journals.aps.org/prb/abstract/10.1103/ PhysRevB.94.085127}.\\

%\begin{abstract}
\lettrine[lines=3]{I}{t} has recently been proposed that electronic band structures in crystals can give rise to a previously overlooked type of Weyl fermion, which violates Lorentz invariance and, consequently, is forbidden in particle physics. It was further predicted that \comp\ may realize such a Type II Weyl fermion. Here, we first show theoretically that it is crucial to access the band structure above the Fermi level, $\varepsilon_F$, to show a Weyl semimetal in \comp. Then, we study \comp\ by pump-probe ARPES and we directly access the band structure $> 0.2$ eV above $\varepsilon_F$ in experiment. By comparing our results with \textit{ab initio} calculations, we conclude that we directly observe the surface state containing the topological Fermi arc. We propose that a future study of \comp\ by pump-probe ARPES may directly pinpoint the Fermi arc. Our work sets the stage for the experimental discovery of the first Type II Weyl semimetal in \comp.

\section{Introduction}

\begin{figure}
\centering
\includegraphics[width=12cm, trim={0 0 0 0}, clip]{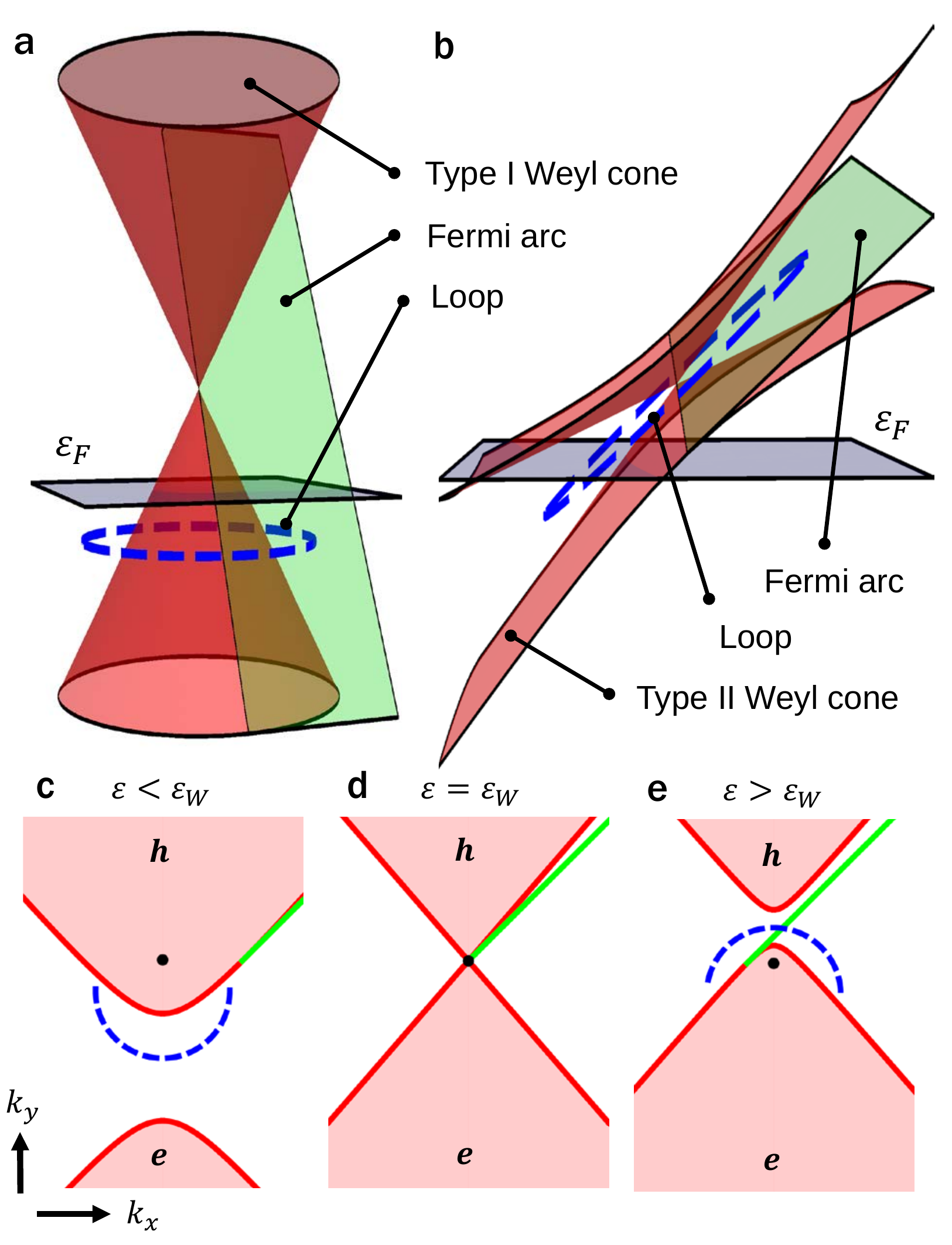}
\caption{\label{mwt1Fig1}\textbf{Chern numbers in Type II Weyl semimetals.} A loop (dashed blue line) to show a nonzero Chern number in (a) a Type I Weyl cone and (b) a Type II Weyl cone. Suppose the Weyl point lies above $\varepsilon_F$. In the simplest case, it is clear that for a Type I Weyl cone, we can show a nonzero Chern number by counting crossings of surface states on a closed loop which lies entirely below the Fermi level. This is not true for a Type II Weyl semimetal. (c-e) Constant-energy cuts of (b). We can attempt to draw a loop below $\varepsilon_F$, as in (c), but we find that the loop runs into the bulk hole pocket. We can close the loop by going to $\varepsilon > \varepsilon_W$, as in (e). However, then the loop must extend above $\varepsilon_F$.}
\end{figure}

\begin{figure*}
\centering
\includegraphics[width=15.4cm, trim={0 0 0 0}, clip]{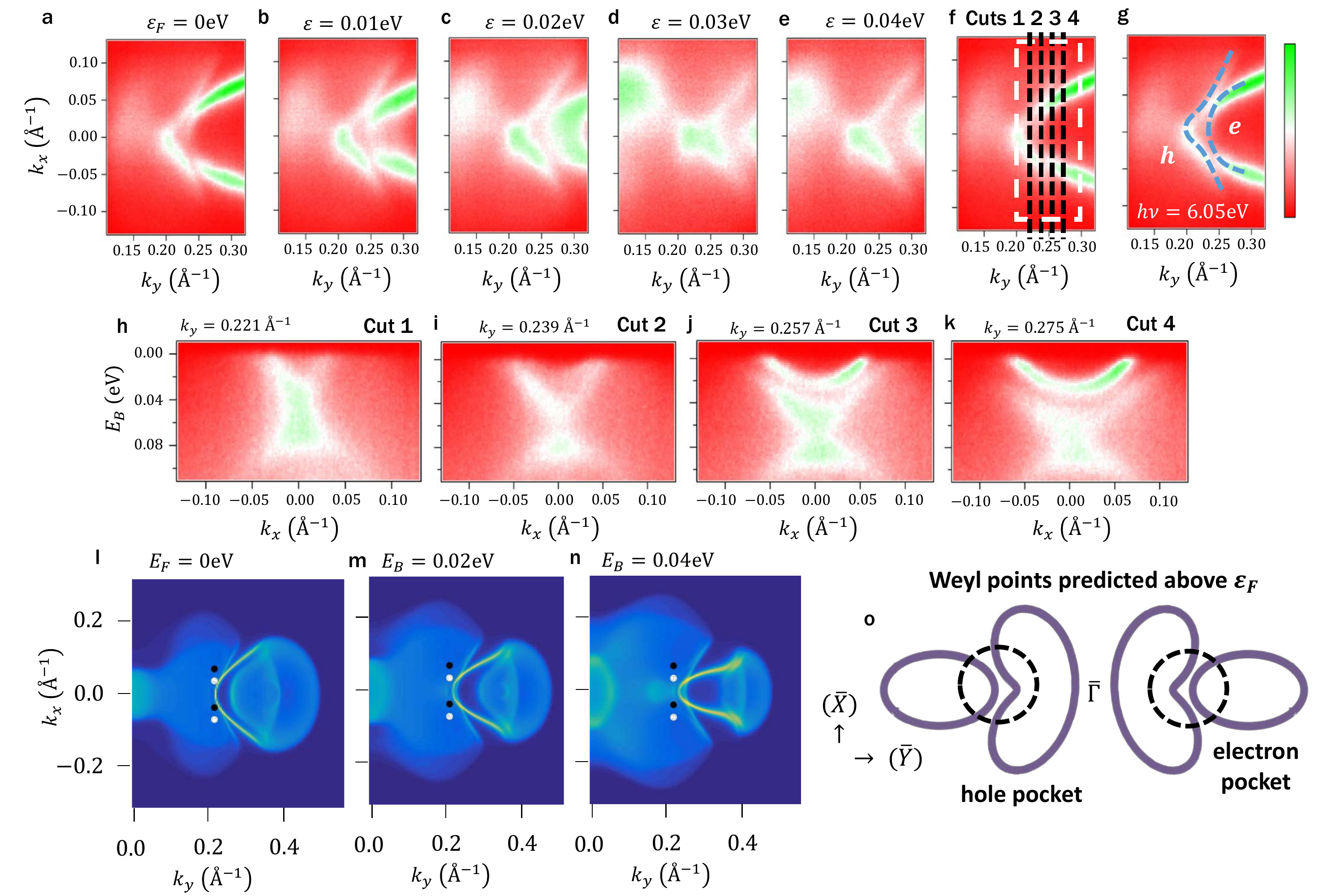}
\caption{\label{mwt1Fig2}\textbf{\half\ below the Fermi level.} (a-g) Conventional ARPES spectra of the constant-energy contour. We observe a palmier-shaped hole pocket and an almond-shaped electron pocket. The two pockets approach each other and we directly observe a beautiful avoided crossing near $\varepsilon_F$ where they hybridize, in (a). This hybridization is expected to give rise to Weyl points above $\varepsilon_F$. (h-k) ARPES measured $E_{\textrm{B}}-k_y$ dispersion maps along the cuts shown in (f). We expect Weyl points or Fermi arcs above the Fermi level at certain $k_y$ where the pockets approach. (l-n) Constant-energy contours for \forty\ from \textit{ab initio} calculations. The black and white dots indicate the Weyl points, above $\varepsilon_F$. Note the excellent overall agreement with the ARPES spectra. The offset on the $k_y$ scale on the ARPES spectra is set by comparison with calculation. (o) Cartoon of the palmier and almond at the Fermi level. Based on calculation, we expect Weyl points above $\varepsilon_F$ where the pockets intersect.}
\end{figure*}

\begin{figure*}
\centering
\includegraphics[width=15.5cm, trim={20 10 10 10}, clip]{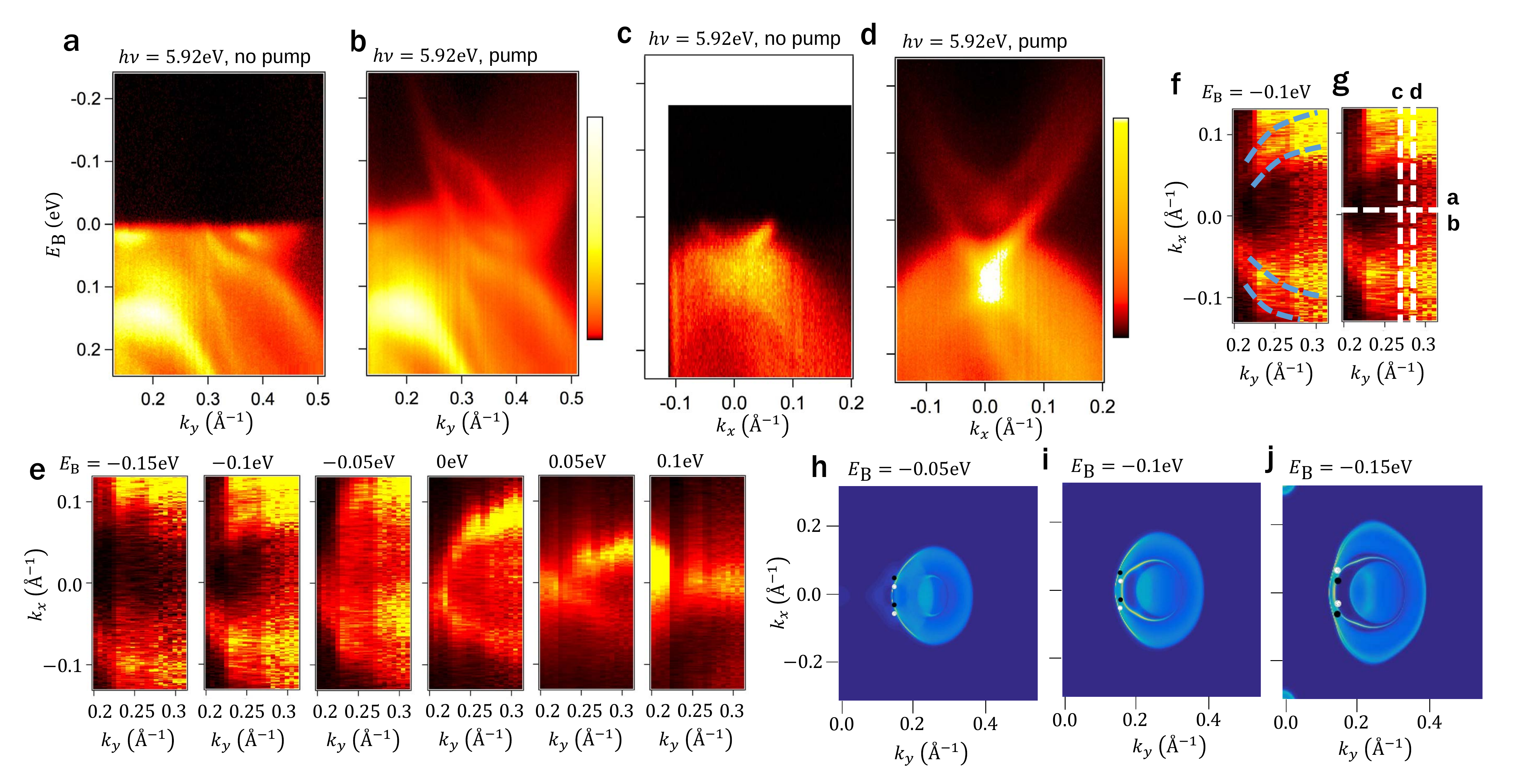}
\caption{\label{mwt1Fig3}\textbf{\half\ above the Fermi level.} (a,b) $E_{\textrm{B}}-k_y$ dispersion maps of \half\ along the $\bar{\Gamma}-\bar{Y}$ direction at $k_{x} = 0$ with and without the pump laser. The sample responds beautifully to the pump laser, allowing us access to the band structure $ > 0.2$eV above $\varepsilon_{F}$, well above the predicted energies of the Weyl points. (c,d) Dispersion maps of \half\ with and without pump on a cut parallel to $\bar{\Gamma}$-$\bar{X}$ at (c) $k_{y}\sim 0.29 \textrm{\AA}$ and (d) $k_{y}\sim 0.26 \textrm{\AA}$. (e-g) The evolution of the almond pocket in energy. We observe that the almond pocket evolves into two nested contours, seen most clearly at $E_{B}=-0.1$eV. (h-j) Calculation of the Fermi surface for \forty\ above $\varepsilon_F$. We see two nested electron pockets, consistent with the measured Fermi surface at $E_B = -0.1$eV.}
\end{figure*}

Weyl fermions have been known since the early twentieth century as chiral particles associated with solutions to the Dirac equation at zero mass \cite{Weyl, Peskin}. In particle physics, imposing Lorentz invariance uniquely fixes the dispersion for a Weyl fermion. However, effective field theories in condensed matter physics are not required to obey Lorentz invariance, leaving a freedom in the Weyl fermion dispersion. Recently, it was discovered that this freedom allows a new type of Weyl fermion to arise in a crystalline band structure, distinct from the Weyl fermion relevant to particle physics \cite{AndreiNature, Grushin, Bergholtz, Trescher, Beenakker, Zyuzin, Isobe}. This Type II Weyl fermion strongly violates Lorentz invariance and has a dispersion characterized by a Weyl cone which is tilted over on its side. It was further predicted that a Type II Weyl semimetal arises in WTe$_2$ \cite{AndreiNature}. Concurrently, MoTe$_2$ and \comp\ were predicted to be Weyl semimetals \cite{TayRong, Binghai, Zhijun} and, more recently, several additional Type II Weyl semimetal candidates have been proposed \cite{Ta3S2, Koepernik, Autes}. All theoretical studies found that all Weyl points in the \comp\ series are above the Fermi level, $\varepsilon_F$. While angle-resolved photoemission spectroscopy (ARPES) would be the technique of choice to directly demonstrate a Type II Weyl semimetal in \comp, conventional ARPES can only study occupied electron states, below $\varepsilon_F$, making it challenging to access the Weyl semimetal state in \comp. Nonetheless, several ARPES works attempt to access the Weyl semimetal state in MoTe$_2$ and WTe$_2$ by studying the band structure above $\varepsilon_F$ in the tail of the Fermi-Dirac distribution \cite{Adam1, Adam2, Xinjiang2}, while other works have tried to demonstrate a Type II Weyl semimetal in MoTe$_2$ and WTe$_2$ in ARPES by studying only the band structure below $\varepsilon_F$ \cite{Shuyun, Chen, Xinjiang, HongDing, Baumberger}. We note also a recent experimental study of Type II Weyl fermions in an unrelated compound \cite{LaAlGe, RAlX}.

Here, we first argue that it's crucial to study the band structure above $\varepsilon_F$ to show a Weyl semimetal in \comp, even if the Fermi arcs fall partly below the Fermi level. Next, we experimentally demonstrate that we can access states sufficiently far above $\varepsilon_F$ using a state-of-the-art photoemission technique known as pump-probe ARPES. We find excellent agreement between our pump-probe ARPES data and \textit{ab initio} calculations, suggesting that \comp\ is a Type II Weyl semimetal. We suggest that a future pump-probe ARPES study may directly pinpoint the topological Fermi arc. In this way, our results set the theoretical and experimental groundwork for demonstrating the first Type II Weyl semimetal in \comp. Our work also opens the way to studying the unoccupied band structure and time-relaxation dynamics of transition metal dichalcogenides by pump-probe ARPES.

%Using this technique, we provide what is likely the strongest evidence accessible with available experimental techniques for a Weyl semimetal in \comp. Specifically,  However, within our experimental resolution, the Fermi arc connects smoothly to adjacent trivial surface states. As a result, we find that we \textit{cannot} directly show a Weyl semimetal from the experimental data alone.

%Our work sets the stage for experimentally demonstrating the first Type II Weyl semimetal in \comp. We also show that we can study the unoccupied band structure and time-relaxation dynamics of related transition metal dichalcogenides using pump-probe ARPES.

\section{Limitations on measuring the Chern number for Type II Weyl cones}

Can we show that a material is a Weyl semimetal if the Weyl points are above the Fermi level, $\varepsilon_W > \varepsilon_F$? For the simple case of a well-separated Type I Weyl point of chiral charge $\pm 1$, it is easy to see that this is true. Specifically, although we cannot see the Weyl point itself, the Fermi arc extends below the Fermi level, see Fig. \ref{mwt1Fig1}(a). Therefore, we can consider a closed loop in the surface Brillouin zone which encloses the Weyl point. By counting the number of surface state crossings on this loop we can demonstrate a nonzero Chern number \cite{NbPme}. In our example, we expect one crossing along the loop. By contrast, this approach fails for a well-separated Type II Weyl point above $\varepsilon_F$. In particular, recall that when counting Chern numbers, the loop we choose must stay always in the bulk band gap. As a result, for the Type II case we cannot choose the same loop as in the Type I case since the loop would run into the bulk hole pocket. We might instead choose a loop which is slanted in energy, see Fig. \ref{mwt1Fig1}(b), but such a loop would necessarily extend above $\varepsilon_F$. Alternatively, we can consider different constant energy cuts of the Type II Weyl cone. In Figs. \ref{mwt1Fig1}(c-e), we show constant-energy cuts of the Type II Weyl cone and Fermi arc. We see that we cannot choose a closed loop around the Weyl point by looking only at one energy, because the loop runs into a bulk pocket. However, we can build up a closed loop from segments at energies above and below the Weyl point, as in Figs. \ref{mwt1Fig1}(c,e). But again, we must necessarily include a segment on a cut at $\varepsilon > \varepsilon_W$. We find that for a Type II Weyl semimetal, if the Weyl points are above the Fermi level, we must study the unoccupied band structure.

Next, we argue that in the specific case of \comp\ we must access the unoccupied band structure to show a Weyl semimetal. In the Supplemental Material, we present a detailed discussion of the band structure of \comp\  \cite{SM}. Here, we only note a key result, consistent among all \textit{ab initio} calculations of \comp, that all Weyl points are Type II and are above the Fermi level \cite{AndreiNature, TayRong, Zhijun, Binghai}. These facts are essentially sufficient to require that we access the unoccupied band structure. However, it is useful to provide a few more details. Suppose that the Fermi level of \comp\ roughly corresponds to the case of Fig. \ref{mwt1Fig1}(c). To count a Chern number using only the band structure below the Fermi level, we need to find a path enclosing a nonzero chiral charge while avoiding the bulk hole and electron pockets. We can try to trace a path around the entire hole pocket. However, as we will see in Fig. \ref{mwt1Fig2}, the Weyl point projections all fall in one large hole pocket at $\varepsilon_F$. As a result, tracing around the entire hole pocket encloses zero chiral charge, see also an excellent related discussion in Ref. \cite{Trivedi}. Therefore, demonstrating a Weyl semimetal in \comp\ requires accessing the unoccupied band structure.

We briefly introduce the occupied band structure of \comp. We present ARPES spectra of \half\ below the Fermi level, see Figs. \ref{mwt1Fig2}(a-k). We observe a palmier-shaped hole pocket and an almond-shaped electron pocket which chase each other as we scan in binding energy. We find excellent agreement between our ARPES results and \textit{ab initio} calculation, see Fig. \ref{mwt1Fig2}(l-n). Based on calculation, at two energies above $\varepsilon_F$, the pockets catch up to each other and intersect, forming two sets of Weyl points $W_1$ and $W_2$, see Fig. \ref{mwt1Fig2}(o) and also the Supplemental Material \cite{SM}.

\section{Accessing Weyl cones \& Fermi arcs by pump-probe ARPES}

Next, we show that we can directly access the relevant unoccupied states in \comp\ with pump-probe ARPES. In our experiment, we use a $1.48$eV pump laser to excite electrons into low-lying states above the Fermi level and a $5.92$eV probe laser to perform photoemission \cite{IshidaMethods}. We first study \half\ along $\bar{\Gamma}-\bar{Y}$ at fixed $k_x = 0 \textrm{\AA}^{-1}$, see Figs. \ref{mwt1Fig3}(a,b). The sample responds beautifully to the pump laser and we observe a dramatic evolution of the bands up to energies $> 0.2$eV above $\varepsilon_F$. We find, similarly that we can directly access the unoccupied band structure on a cut at fixed $k_y \sim k_W$, see Figs. \ref{mwt1Fig3}(c,d). Further, by plotting constant-energy cuts, we can directly observe that the almond pocket continues to grow above $\varepsilon_F$, while the palmier pocket recedes, consistent with calculation, see Figs. \ref{mwt1Fig3}(e-j). We note that all available calculations of \comp\ place the Weyl points $< 0.1$eV above the Fermi level. In addition, the Weyl point projections are all predicted to lie within $0.25 \textrm{\AA}^{-1}$ of the $\bar{\Gamma}$ point \cite{AndreiNature, TayRong, Binghai, Zhijun}. We see that our pump-probe measurement easily accesses the relevant region of reciprocal space to show a Weyl semimetal in \comp\ for all $x$.

\section{Demonstrating a Weyl semimetal}% in \half}

\begin{figure*}
\centering
\includegraphics[width=15.5cm, trim={0 0 0 0}, clip]{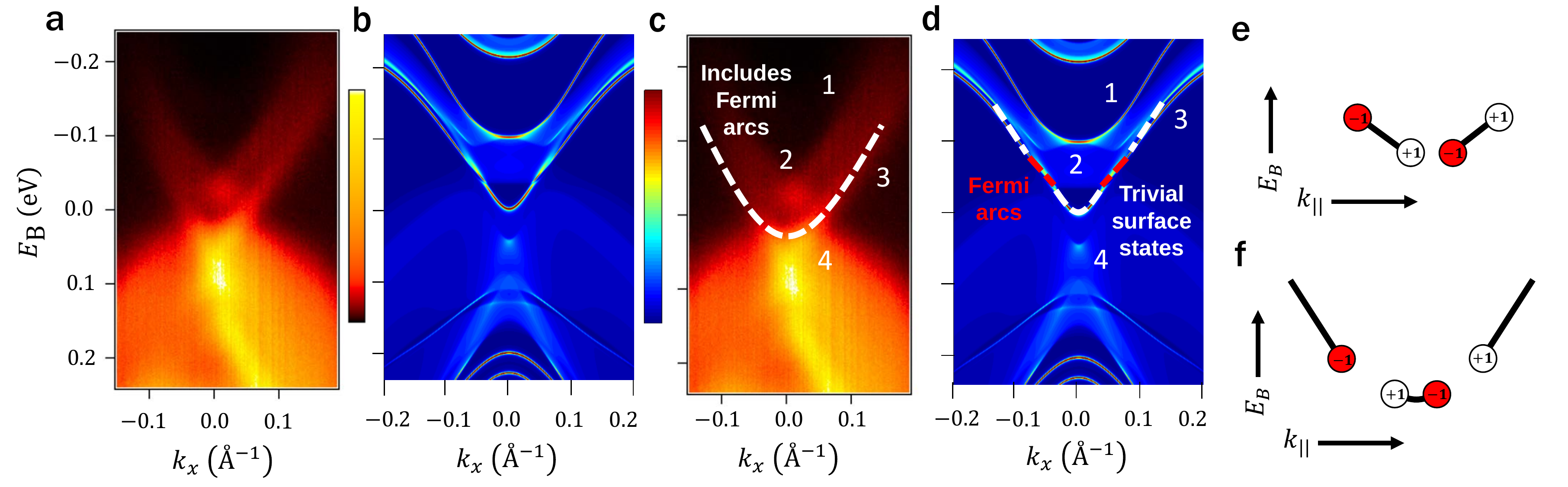}
\caption{\label{mwt1Fig4}\textbf{Signatures of Fermi arcs in \comp.} (a) Pump-probe ARPES spectrum at $k_y = 0.225 \textrm{\AA}^{-1}$. (b) \textit{Ab initio} calculation at $k_y = k_\textrm{W1} = 0.215 \textrm{\AA}^{-1}$, showing excellent overall agreement with the data. (c,d) Same as (a,b) but with key features in the data marked. (e,f) There are two possible scenarios for the connectivity of the arcs. The scenario in (e) is favored by our \textit{ab initio} calculations. The surface state electron pocket (3) contains both the topological Fermi arcs and adjacent trivial surface states. While the Fermi arc is formally disconnected from the trivial surface states, in practice the trival surface state merges into the bulk very close to the Weyl points, so that we do not observe a disjoint arc. Nonetheless, future measurements by pump-probe ARPES may allow us to observe a kink or ``ripple'' where the topological Fermi arc and trivial surface states meet, demonstrating that \comp\ is a Type II Weyl semimetal.}
\end{figure*}

We next present evidence for a Weyl semimetal in \half. The agreement between our pump-probe ARPES spectra and \textit{ab initio} calculation strongly suggests that \half\ is a Weyl semimetal. We consider the spectrum at fixed $k_y \sim k_W$, see Figs. \ref{mwt1Fig4}(a-d). We find an upper electron pocket (1) with a short additional surface state (2), a lower electron pocket (3) and the approach between hole and electron bands (4), all in excellent agreement with the calculation. We also note the excellent agreement in the constant energy contours both above and below $\varepsilon_F$, as discussed above. In addition to this overall agreement between experiment and theory, we might ask if there is any direct signature of a Weyl semimetal that we can pinpoint from the experimental data alone \cite{NbPme}. First, we consider a measurement of the Chern number, following the prescription discussed above for Type II Weyl cones. Formally, we can use the loop in Fig. \ref{mwt1Fig1}(b) around either $W_1$ or $W_2$ to measure a Chern number of $\pm 1$ \cite{NbPme}. Specifically, the topological Fermi arc will contribute one crossing, while the trivial surface state will contribute either zero crossings or two crossings of opposite Fermi velocity, with net contribution zero. However, in our experiment, the finite resolution prevents us from carrying out this counting. In particular, the linewidth of the surface state is comparable to its energy dispersion, so we cannot determine the sign of the Fermi velocity.

Next, we note that since the chiral charges of all Weyl point projections are $\pm 1$, we expect a disjoint arc connecting pairs of Weyl points, in one of two possible configurations, Figs. \ref{mwt1Fig4}(e,f). From our calculation, we expect case Fig. \ref{mwt1Fig4}(e). However, we observe no such disjoint arc in (3) in our spectrum. From calculation, we see that this is perhaps reasonable because the Fermi arc is adjacent to trivial surface states, see the dotted lines in Figs. \ref{mwt1Fig4}(c,d). Indeed, we can understand the Fermi arcs in \comp\ as arising from the large electron-like surface state (3) of Figs. \ref{mwt1Fig4}(c,d), which we can imagine as being present whether or not there are Weyl points. Then, we can tune the system through a topological phase transition. Before the transition, the surface state is entirely trivial. After the transition, the Weyl point projections sit on (3) and ``snip out'' a topological Fermi arc from the large surface state. Formally, the Fermi arc terminates strictly on the Weyl points, while the remaining trivial surface states merge into the bulk in some generic way near the Weyl points. However, within any reasonable resolution, the topological and trivial surface states appear to connect at the Weyl points. As a result, we see no disjoint Fermi arc.

We might then ask if we can observe a kink, since the Fermi arc and the trivial surface state will generically meet at some angle. We note that we observe a kink in calculation, but not in the ARPES spectra presented here. We propose that a more complete pump-probe ARPES study of \comp\ may show such a kink. In particular, a full $k_y$ dependence may catch a kink or ``ripple'' in the surface state, signaling a topological Fermi arc. If there is a difference in how well localized the Fermi arc is on the surface of the sample, compared to the trivial surface state, then a difference in the photoemission cross section may make one or the other feature brighter, allowing us to directly detect the arc. A composition dependence may further show a systematic evolution of the kink, which would also prove an arc. Such an analysis is beyond the scope of this work. Here, we have shown theoretically that accessing the unoccupied band structure is crucial to show a Weyl semimetal in \comp\ and, in addition, we have directly accessed the unoccupied band structure in experiment and observed the surface state containing the topological Fermi arc. These results set the stage for directly demonstrating that \comp\ is a Type II Weyl semimetal.

\section{Additional systematics}

\subsection{Overview of the band structure of \comp}
%\bigskip

\begin{figure*}[h]
\centering
\includegraphics[width=15cm]{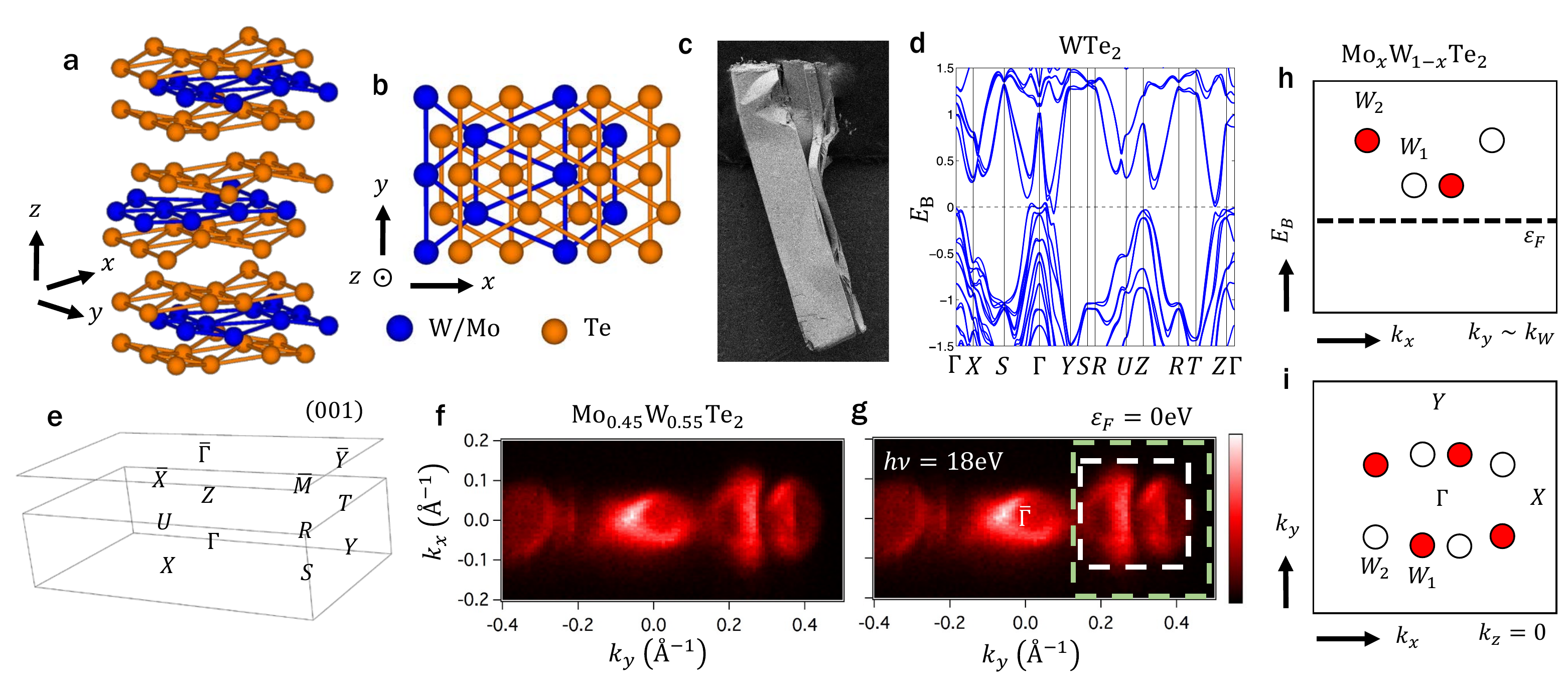}
\caption{\label{mwt1SFig1} \textbf{Overview of \comp}. (a) Side view of the \comp\ crystal structure. \comp\ crystallizes in an orthorhombic structure with space group $Pmn2_1$ ($\#31$). The W/Mo atoms are shown in blue and the Te atoms in orange. (b) Top view of \comp, showing only one monolayer, consisting of a layer of W/Mo atoms sandwiched between two layers of Te atoms. The structure lacks inversion symmetry. (c) A beautiful scanning electron microscope (SEM) image of a $\sim{0.4 \times 2}$ mm rectangular shaped \half\ bulk crystal clearly demonstrates a layered structure. (d) The electronic bulk band structure of WTe$_2$. There is a band inversion near $\Gamma$ which gives rise to electron and hole pocket along the $\Gamma$-$Y$ direction. (e) The bulk and (001) surface Brillouin zone (BZ) of \comp\ with high symmetry points labeled. (f,g) Fermi surface by vacuum ultraviolet APRES on the (001) surface of \half\ at $\varepsilon_{F}=0$eV with an incident photon energy of $18$ eV. We observe a palmier pocket and an almond pocket on either side of $\bar{\Gamma}$ along the $\bar{\Gamma}$-$\bar{Y}$ direction. The pockets correspond well to the electron and hole pockets we find in our \textit{ab initio} calculations, as presented in main text Fig. 2(l-n). We indicate the Fermi surface region presented in main text Figs. 2(a-e) (white dotted line) and the region presented below (green dotted line). We expect to see Weyl points and Fermi arcs above the Fermi level in the region where the electron and hole pockets intersect. (h,i) Distribution of Weyl points from \textit{ab initio} calculation in \comp. All Weyl points lie above the Fermi level, they are all Type II, they lie on $k_z = 0$ and roughly on the same $k_y \sim \pm k_\textrm{W}$. It is clear that to directly observe the Weyl points or calculate Chern numbers in \comp, it is necessary to access the unoccupied band structure.}
\end{figure*}

We briefly present an overview of the system under study. \comp\ crystallizes in an orthorhombic Bravais lattice, space group $Pmn2_1$ ($\#31$), lattice constants $a = 6.282\textrm{\AA}$, $b = 3.496\textrm{\AA}$ and $c = 14.07\textrm{\AA}$ \cite{MoTe2WTe2}. The atomic structure is layered, with single layers of W/Mo sandwiched in between Te bilayers, see Figs. \ref{mwt1SFig1}(a,b). The structure has no inversion symmetry, a crucial condition for a Weyl semimetal. Shown in Fig. \ref{mwt1SFig1}(c) is a scanning electron microscope (SEM) image of a typical \half\ sample showing a layered crystal structure. Energy-dispersive spectroscopy (EDS) measurements confirm that the crystals have composition \half, see Methods: \half\ growth, below. We display the bulk band structure of WTe$_2$ along high-symmetry lines in Fig. \ref{mwt1SFig1}(d), with the bulk and (001) surface Brillouin zone (BZ) of \comp\ shown in Fig. \ref{mwt1SFig1}(e). The bulk band structure of WTe$_2$ is gapped throughout the Brillouin zone except near the $\Gamma$ point where the bulk valence and conduction bands approach each other, forming an electron and hole pocket near the Fermi level. Because the crystal structure breaks inversion symmetry, we expect Weyl points to arise generically where the bands hybridize. Although Weyl points have been found in WTe$_2$ in calculation, it is now understood that WTe$_2$ is very near the critical point for a transition to a trivial phase and the Weyl semimetal state is not expected to be robust \cite{TayRong}. However, it has been shown in calculation that an Mo doping causes the bands to further invert, increasing the separation of the Weyl points \cite{TayRong}. This calculation result motivates our study of \comp. Next, we provide an overview of the (001) Fermi surface of \half\ under vacuum ultraviolet ARPES, see Fig. \ref{mwt1SFig1}(f,g). We observe the palmier and almond pockets on either side of $\bar{\Gamma}$ along $\bar{\Gamma}-\bar{Y}$. Our results are in excellent overall agreement with \textit{ab initio} calculations of \forty, as shown in Figs. 2(l-n) of the main text. One important discrepancy is the distance between the features and $\bar{\Gamma}$, which is underestimated by $\sim 0.1\textrm{\AA}^{-1}$ in calculation. We speculate that this might be an artifact of the ARPES Fermi surface mapping procedure. Specifically, if the sample tends to form a curved surface \textit{in situ} after cleaving, there may be a drift in the measured slice of momentum space. In Fig. \ref{mwt1SFig1}(g) we also mark the Fermi surface region we measure in Fig. \ref{mwt1SFig2}(a-e) (white dotted line) and the Fermi surface region presented below in Fig. \ref{mwt1SFig2} (green dotted line). From \textit{ab initio} calculation, we find that \comp\ has 8 Weyl points, all above the Fermi level, all on the $k_z = 0$ plane of the Brillouin zone and all roughly located at $k_y \sim \pm k_{\textrm{W}}$, as shown schematically in Fig. \ref{mwt1SFig1}(h,i). We can understand the Weyl points as arising from valence and conduction bands which approach each other more or less tangentially, forming pairs of Type II Weyl points. 

%\bigskip
\subsection{Systematic ARPES data on \half}
%\bigskip

\begin{figure*}[h]
\centering
\includegraphics[width=15cm]{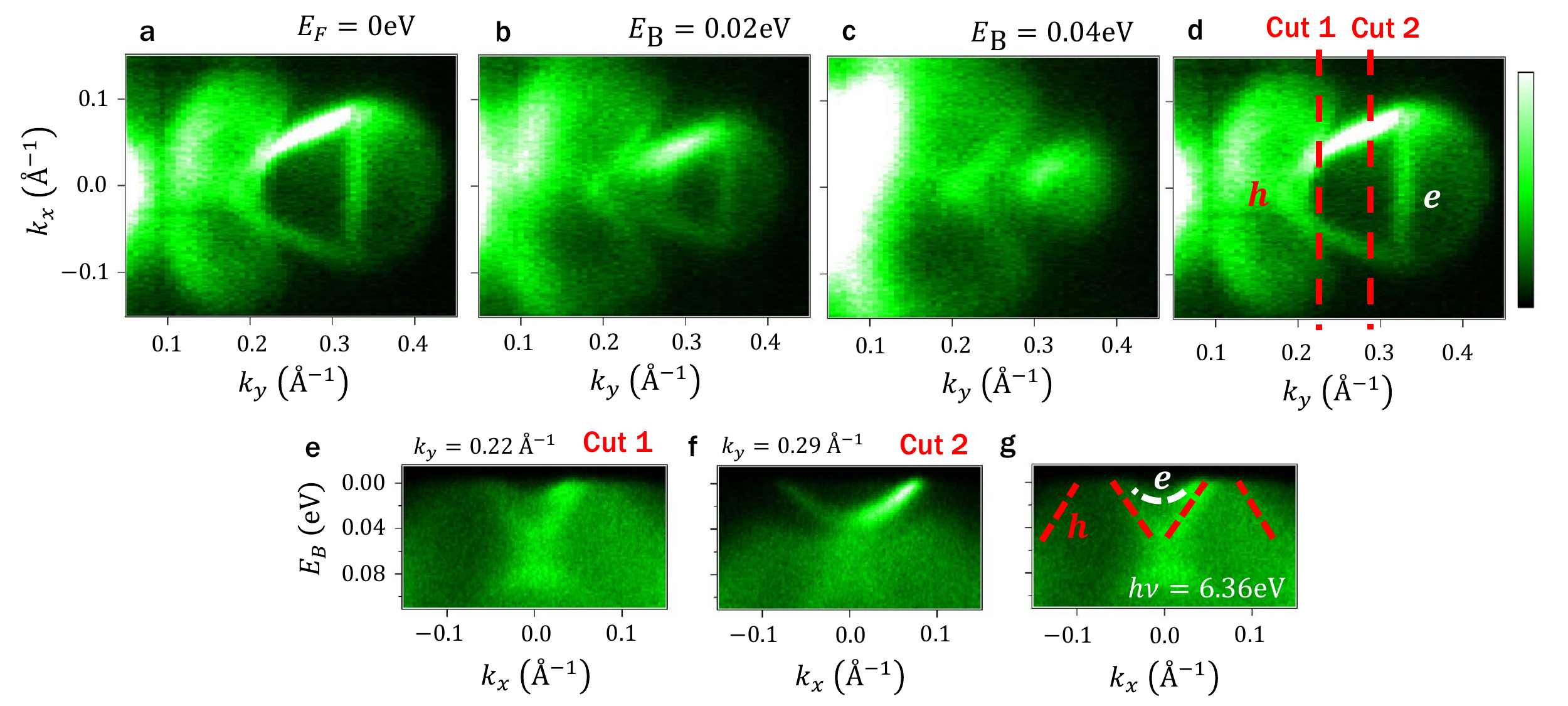}
\caption{\label{mwt1SFig2} \textbf{Occupied band structure of \comp}. (a-c) Fermi surface at different binding energies, clearing showing a hole-like palmier pocket and an electron-like almond pocket. (d) Same as (a), with the hole and electron pockets labelled, and with the cuts shown below marked. (e,f) Cuts near $k_y \sim k_\textrm{W}$ showing the electron and hole bands approach close to one another. We expect the electron and hole pockets to intersect above the Fermi level at $k_y \sim k_\textrm{W}$, forming Weyl points and Fermi arcs, see main text. (g) Same as (e), but with the electron and hole pockets marked. The hole pocket is partly concave, allowing it to hug the electron pocket.}
\end{figure*}

We present an additional Fermi surface mapping at low photon energy to complement the mapping presented in main text Fig. 2(a-e). Specifically, in Figs. \ref{mwt1SFig2}(a-c), we present a mapping at $h\nu = 6.36$ eV, which highlights the entire palmier and almond, unlike the mapping at $h\nu = 6.05$ eV, which provides beautiful contrast where the two pockets approach each other. We clearly see that the palmier is hole-like, while the almond is electron-like, supporting our discussion in the main text. We label the electron and hole pockets and two $E_\textrm{B}-k_x$ cuts in Fig. \ref{mwt1SFig2}(d). In Figs. \ref{mwt1SFig2}(e,f) we see that the electron and hole pockets closely approach each other, as seen also at $h\nu = 6.05$ eV and discussed in the main text. In Fig. \ref{mwt1SFig2}(g) we mark the electron and hole pockets, emphasizing that the hole pocket has a concave shape so that it nests the electron pocket.

%\bigskip
\subsection{Time-delay dependence in pump-probe ARPES}
%\bigskip

\begin{figure}
\centering
\includegraphics[width=15cm, trim={0 470 240 20}, clip]{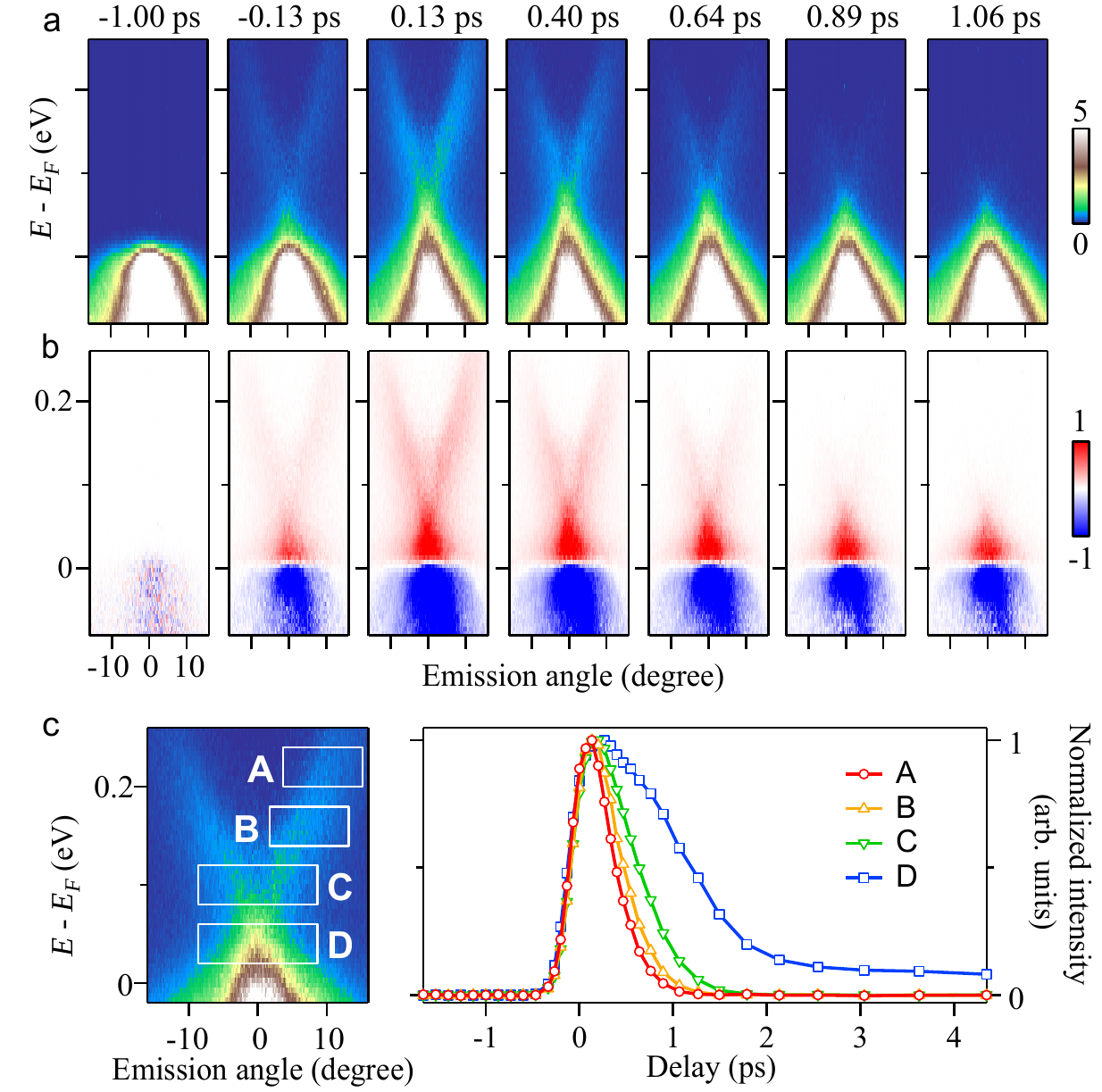}
\caption{\label{mwt1SFig3}\textbf{Time-delay dependence on \half.}. (a) Pump-probe ARPES spectra at fixed $k_y \sim 0.2 \textrm{\AA}^{-1}$ for $\Delta t$ from $-1.00$ ps to $+1.06$ ps. We note that all pump-probe ARPES data presented in the main text are taken at $\Delta t = 0.13$ ps. (b) Difference between each spectrum in (a) with a reference spectrum before $\Delta t = 0$. (c) Change in the photoemission intensity with $\Delta t$ in four momentum-energy windows, as indicated by the boxes. The peak in intensity above the Fermi level occurs just after $\Delta t = 0$, followed by a rapid decay.}
\end{figure}

It is important to check that the states observed above $E_F$ in our pump-probe ARPES spectra correspond to the unoccupied bands of \half, rather than artifacts of the photoemission process itself. To show this, we study the dependence of the pump-probe ARPES spectrum on the delay time $\Delta t$ between the pump and probe pulses, see Fig. \ref{mwt1SFig3}. If the unoccupied states were induced by the pump pulse, then we would expect that the energy of the state would change with time as the system relaxes. By constrast, in our data we see that the unoccupied states gradually empty out with time, but show no shift in energy. This strongly suggests that our pump-probe ARPES spectra above $E_F$ can be interpreted as a direct image of the unoccupied bands of \half. We further note that the excellent agreement with \textit{ab initio} calculation further confirms that we observe the unoccupied bands of \half, rather than a pump-induced state. Lastly, we consider the possibility that our spectra capture unoccupied states produced by an inverse process, where the probe populates intermediate states photoemitted by the pump. To rule out this process, we note that for $\Delta t < 0$, where we might expect a signature of the inverse process, we do not observe populated states above the Fermi level, see again Fig. \ref{mwt1SFig3}. These results confirm that our pump-probe ARPES spectra directly measure the unoccupied band structure of \half.

\section{Materials and methods}

%\bigskip
\subsection{ARPES measurements}
%\bigskip

Synchrotron-based ARPES measurements were performed at the CASSIOPEE beamline at Soleil in Saint-Aubin, France and the I05 beamline at the Diamond Light Source (DLS) in Didcot, United Kindom. ARPES measurements were also carried out using a home-built laser-based ARPES setup at the Ames Laboratory in Ames, Iowa, United States. All measurements were conducted under ultra-high vacuum and at temperatures $\leq 10$K. The angular and energy resolution of the synchrotron-based ARPES measurements was better than $0.2^{\circ}$ and $20$ meV, respectively. Photon energies from 15 eV to 100 eV were used. The angular and energy resolution of the home-built laser-based ARPES measurements was better than $0.1^{\circ}$ and $5$ meV, respectively, with photon energies from 5.77 eV to 6.67 eV \cite{KaminskiMethods}.

The pump-probe ARPES apparatus consisted of a hemispherical analyzer and a mode-locked Ti:Sapphire laser system that delivered $1.48$ eV pump and $5.92$ eV probe pulses at a $250$kHz repetition \cite{IshidaMethods}. For all pump-probe ARPES data presented here, the delay between the pump and probe pulses was $0.13$ ps. The time and energy resolution was $300$ fs and $27$ meV, respectively. The spot diameter of the pump and probe beams at the sample position was $250$ and $85 \mu$m, respectively. Samples were cleaved in the spectrometer at $<5 \times 10^{-11}$ Torr, and measurements were conducted at $\sim8$ K.

%\bigskip
\subsection{\half\ growth}
%\bigskip

High quality ribbon-like single crystals of Mo$_x$W$_{1-x}$Te$_2$ were grown by chemical vapor transport (CVT) with iodine (I) as the agent. Before growing the crystals, the quartz tubes were thoroughly cleaned using thermal and ultrasonic cleaning treatments to avoid contamination. Stoichiometric amounts of W (99.9\% powder, Sigma-Aldrich), Mo (99.95\% powder, Sigma-Aldrich) and Te (99.95\%, Sigma-Aldrich) were mixed with iodine, were sealed in a $20$ cm long quartz tube under vacuum $\sim 10^{-6}$ Torr, and then placed in a three-zone furnace. The reaction zone dwelled at $850$ $^{\circ}$C for 40 hours with the growth zone at $900$ $^{\circ}$C and then heated to $1070$ $^{\circ}$C for seven days with the growth zone at $950$ $^{\circ}$C. Lastly, the furnace was allowed to cool naturally down to room temperatures. The Mo$_x$W$_{1-x}$Te$_2$ single crystals were collected from the growth zone. Excess iodine adhering to the single crystals was removed by using acetone or ethanol. Nominal compositions of the Mo$_x$W$_{1-x}$Te$_2$ crystals were $x = 0.0, 0.1, 0.15, 0.2, 0.3$. An energy dispersive spectroscopy (EDS) measurement was carried out to precisely determine the composition of the samples. Samples were first surveyed by an FEI Quanta 200FEG environmental scanning electron microscope (SEM). The chemical compositions of the samples were characterized by an Oxford X-Max energy dispersive spectrometer that was attached to the SEM. All the samples were loaded at once in the SEM chamber to ensure a uniform characterization condition. Both the SEM imaging and EDS characterization were carried out at an electron acceleration voltage of $10$ kV with a beam current of $0.5$ nA. For each sample, three different spatial positions were randomly picked to check the uniformity. The measured compositions were: MoTe$_2$ for $x = 0$, Mo$_{0.42}$W$_{0.58}$Te$_2$ for $x= 0.1$,  Mo$_{0.46}$W$_{0.54}$Te$_2$ for $x = 0.15$, Mo$_{0.44}$W$_{0.56}$Te$_2$ for $x = 0.2$ and Mo$_{0.43}$W$_{0.57}$Te$_2$ for $x = 0.3$. For $x \neq 0$, we considered the samples to all have the approximate composition Mo$_{0.45}$W$_{0.55}$Te$_2$.

%\bigskip
\subsection{Single-crystal XRD}
%\bigskip

A black, bar-shaped crystal with approximate dimensions $0.11 \times 0.09 \times 0.20$ mm on the tip of a glass fiber was selected for data collection. The single crystal diffraction data were collected on a SuperNova X-ray diffraction system from Agilent Technologies equipped with a graphite-monochromated MoK$\alpha$ radiation ($\lambda = 0.71073 \AA$) at room temperature. The data were corrected for Lorentz factors, polarization, air absorption, and absorption due to variations in the path length through the detector faceplate. Absorption correction based on a multi-scan technique was also applied. Absorption corrections were performed by the SADABS program \cite{XRD1,XRD2}. The space group was determined to be $Pmn2_1$ ($\#31$) based on systematic absences, E-value statistics, and subsequent successful refinements of the crystal structure. The structure was solved by the direct method and refined by full-matrix least-squares fitting on $F^2$ by SHELX-97. \cite{XRD3} There are two metal sites (M1 and M2) which are all octahedral coordination, and four sites for tellurium atoms. The M1 and M2 sites with mixed occupancy by W and Mo were refined and the formula was W$_{0.69}$Mo$_{0.31}$Te$_2$ where the occupancy for W and Mo in M1 and M2 sites were $0.679(4)$, $0.321(4)$ and $0.705(4)$, $0.295(4)$, whose charge was neutral. All atoms were refined anisotropically ($R_1 = 0.0469$, $wR_2 = 0.1006$). See Table \ref{tab1} for crystallographic data and structural refinements.

\begin{table}[]
\centering
\begin{tabular}{| c | c |} 
\hline
Formula & \ \half\ \ \\ \hline
fw & 411.97 \\ \hline
crystal system & orthorhombic \\ \hline
crystal color & black \\ \hline
space group & $Pmn2_1$ ($\#31$) \\ \hline
$a$ $(\textrm{\AA})$ & 3.4834(6) \\ \hline
$b$ $(\textrm{\AA})$ & 6.3042(10) \\ \hline
$c$ $(\textrm{\AA})$ & 13.9152(18) \\ \hline
$\alpha = \beta = \gamma$ (deg.) & 90 \\ \hline
$V$ $(\textrm{\AA}^3)$ & 293.79(4) \\ \hline
$Z$ & 4 \\ \hline
$D_c$ $(\textrm{gcm}^{-3})$ & 8.95418 \\ \hline
GOOF on $F^2$ & 1.002 \\ \hline
Flack	 & 0.00(4) \\ \hline
$R_1$, $wR_2 (I > 2\sigma(I))$ & 0.0402, 0.0954 \\ \hline
$R_1$, $wR_2 (\textrm{all data})$ & 0.0469, 0.1006 \\ \hline
largest diff peak/hole, $e\textrm{\AA}^{-3}$ & 2.65 / -2.95 \\ \hline
\end{tabular}
\caption{Crystal data and structure refinements for \half.}
\label{tab1}
\end{table}

%\bigskip
\subsection{\textit{Ab initio} calculations}
%\bigskip

We computed the electronic structures by using the projector augmented wave method \cite{PAW1,PAW2} as implemented in the VASP \cite{TransitionMetals, PlaneWaves1, PlaneWaves2, GGA} package within the generalized gradient approximation (GGA) schemes \cite{GGA}. For WTe$_2$, the experimental lattice constants used were from \cite{MoTe2WTe2}. A $8 \times 16 \times 4$ Monkhorst Pack $k$-point mesh was used in the computations. The lattice constants and atomic positions of MoTe$_2$ were fully optimized in a self-consistent calculation for an orthorhombic crystal structure until the force became less than $0.001$ eV$/\textrm{\AA}$. Spin-orbit coupling was included in our calculations. To calculate the bulk and surface electronic structures, we constructed a first-principles tight-binding model Hamiltonian for both WTe$_2$ and MoTe$_2$, where the tight-binding model matrix elements were calculated by projecting onto the Wannier orbitals \cite{MLWF1,MLWF2,Wannier90}, which used the VASP2WANNIER90 interface \cite{MLWF3}.  We used the $s$- and $d$-orbitals for W(Mo) and the $p$-orbitals for Te to construct Wannier functions, without performing the procedure for maximizing localization. The electronic structure of Mo$_x$W$_{1-x}$Te$_2$ samples was calculated by a linear interpolation of the tight-binding model matrix elements of WTe$_2$ and MoTe$_2$. The surface states were calculated by the surface Green's function technique \cite{Green}, which computed the spectral weight near the surface of a semi-infinite system.

\clearpage

%\cleardoublepage
\ifdefined\phantomsection
  \phantomsection  % makes hyperref recognize this section properly for pdf link
\else
\fi
\addcontentsline{toc}{section}{Bibliography}

{\singlespacing
}

%\end{document}

%% file: ch-mwt2/ch-mwt2.tex
%\documentclass[aps,prl,preprint,nopacs,superscriptaddress]{revtex4}
%\usepackage{amsmath}
%\usepackage{amssymb}
%\usepackage{graphicx}
%\usepackage{hyperref}
%\pagestyle{headings}
\renewcommand{\beq}{\begin{equation*}}
\renewcommand{\eeq}{\end{equation*}}
\newcommand{\mowte}{Mo$_x$W$_{1-x}$Te$_2$}
\newcommand{\tf}{Mo$_{0.25}$W$_{0.75}$Te$_2$}
%
%\begin{document}

\chapter{Discovery of a new type of Weyl semimetal state in \mowte}
\label{ch:mwt2}

{\singlespacing
\begin{chapquote}{Hailee Steinfeld, \textit{Rock Bottom}}
What are we fighting for?\\
Seems like we do it just for fun\\
\end{chapquote}}

\noindent This chapter is based on the article, {\it Discovery of a new type of topological Weyl fermion semimetal state in} \mowte\ by Ilya Belopolski {\it et al}., {\it Nat. Commun.} {\bf7}, 13643 (2016), available at \href{https://www.nature.com/articles/ncomms13643}{https://www.nature.com/articles/ncomms13643}.\\

\lettrine[lines=3]{T}{he} recent discovery of a Weyl semimetal in TaAs offers the first Weyl fermion observed in nature and dramatically broadens the classification of topological phases. However, in TaAs it has proven challenging to study the rich transport phenomena arising from emergent Weyl fermions. The series \mowte\ are inversion-breaking, layered, tunable semimetals already under study as a promising platform for new electronics and recently proposed to host Type II, or strongly Lorentz-violating, Weyl fermions. Here we report the discovery of a Weyl semimetal in \mowte\ at $x = 25\%$. We use pump-probe angle-resolved photoemission spectroscopy (pump-probe ARPES) to directly observe a topological Fermi arc above the Fermi level, demonstrating a Weyl semimetal. The excellent agreement with calculation suggests that \mowte\ is the first Type II Weyl semimetal. We also find that certain Weyl points are at the Fermi level, making \mowte\ a promising platform for transport and optics experiments on Weyl semimetals.
%\end{abstract}

%\date{\today}
%\maketitle

\section{Introduction}

The recent discovery of the first Weyl semimetal in TaAs has opened a new direction of research in condensed matter physics \cite{TPT, TaAsUs, LingLu, TaAsThem, TaAsThyUs, TaAsThyThem}. Weyl semimetals are fascinating because they give rise to Weyl fermions as emergent electronic quasiparticles, have an unusual topological classification closely related to the integer quantum Hall effect, and host topological Fermi arc surface states \cite{Weyl, Peskin, Herring, Abrikosov, Nielsen, Volovik, Murakami, Multilayer, Pyrochlore, Vish}. These properties give rise to many unusual transport phenomena, including negative longitudinal magnetoresistance from the chiral anomaly, an anomalous Hall effect, the chiral magnetic effect, non-local transport and novel quantum oscillations \cite{Param, Hosur, Potter}. Although many recent works have studied transport properties in TaAs \cite{ZhangC, HuangXiao, NbPTransport}, transport experiments are challenging because TaAs and its isoelectronic cousins have a three-dimensional crystal structure with irrelevant metallic bands and many Weyl points. As a result, there is a need to discover new Weyl semimetals better suited for transport and optics experiments and eventual device applications.

% Toxic
% Tunable
% Type II

% Discovery of Weyl opens a new chapte
% Interesting because of all of this this and this
% Also has fascinating transport properties
% Transport hard in TaAs
% Need for new materials

Recently, the \mowte\ series has been proposed as a new Weyl semimetal \cite{TR, AndreiNature, Zhijun, Binghai}. Unlike TaAs, \mowte\ has a layered crystal structure and is rather widely available as large, high-quality single crystals. Indeed, MoTe$_2$, WTe$_2$ and other transition metal dichalcogenides are already under intense study as a platform for novel electronics \cite{ReviewTMDC, LiangFu, LaserPattern, Heinz, Valley}. Moreover, \mowte\ offers the possiblity to realize a tunable Weyl semimetal, which may be important for transport measurements and applications. Recently, it was also discovered theoretically that WTe$_2$ hosts a novel type of strongly Lorentz-violating Weyl fermion, or Type II Weyl fermion, long ignored in quantum field theory \cite{AndreiNature, Grushin, Bergholtz, Trescher, Beenakker, Ta3S2, Koepernik, Autes, Suyang}. This offers a fascinating opportunity to realize in a crystal an emergent particle forbidden as a fundamental particle in particle physics. There are, moreover, unique transport signatures associated with strongly Lorentz-violating Weyl fermions \cite{AndreiNature, Grushin, Bergholtz, Trescher, Zyuzin, Isobe}. For all these reasons, there is considerable interest in demonstrating that \mowte\ is a Weyl semimetal. However, it is important to note that \textit{ab initio} calculations predict that the Weyl points in \mowte\ are above the Fermi level \cite{TR, Zhijun, Binghai}. This makes it challenging to access the Weyl semimetal state with conventional angle-resolved photoemission spectroscopy (ARPES). Recently, we have demonstrated that we can access the unoccupied band structure of \mowte\ by pump-probe ARPES to the energy range necessary to study the Weyl points and Fermi arcs \cite{myothermowte}. At the same time, despite the promise of \mowte\ for transport, if the Weyl points are far from the Fermi level, then the novel phenomena associated with the emergent Weyl fermions and violation of Lorentz invariance will not be relevant to the material's transport properties.

% Recently, the \mowte family of materials has been proposed as a new class of Weyl semimetals.
% Layered, easy to grow
% Already under study for applications
% In addition, have a novel type of Weyl fermion, strongly Lorentz-violating

Here we report the discovery of a Weyl semimetal in \mowte\ at doping $x = 25\%$. We use pump-probe ARPES to study the band structure above the Fermi level and we directly observe two kinks in a surface state band. We interpret the kinks as corresponding to the end points of a topological Fermi arc surface state. We apply the bulk-boundary correspondance and argue that since the surface state band structure includes a topological Fermi arc, \mowte\ is a Weyl semimetal \cite{NbPme}. The end points of the Fermi arc also allow us to fix the energy and momentum locations of the Weyl points. We find excellent agreement with our \textit{ab initio} calculation. However, crucially, we find that certain Weyl points have lower binding energy than expected from calculation and, in fact, are located very close to the Fermi level. This unexpected result suggests that our \tf\ samples may be useful to study the unusual transport phenomena of Weyl semimetals and, in particular, those particularly exotic phenomena arising from strongly Lorentz-violating Weyl fermions. Our work also sets the stage for the first tunable Weyl semimetal. Our discovery of a Weyl semimetal in \mowte\ provides the first Weyl semimetal outside the TaAs family, as well as a Weyl semimetal which may be tunable and easily accessible in transport studies. Taken together with calculation, our experimental results further show that we have realized the first Weyl semimetal with Type II, or strongly Lorentz-violating, emergent Weyl fermions.
 
% Here, we report the discovery of a Weyl semimetal in \mowte\ at $x = 25\%$, the first Weyl semimetal outside the TaAs class.
% We directly observe a surface state kink marking the positions of Weyl points
% We find the energy separation is slight larger than in calculation
% Unexpectedly, we observe that the Weyl points are very close to the Fermi level, useful for transport
% Sets the stage for the first tunable Weyl semimetal
% Our discovery of a Weyl semimetal in \mowte\ provides a Weyl semimetal in a tunable, layered compound which may serve as an accessible platform for transport studies of Weyl semimetals and the exotic properties of strongly Lorentz-violating Weyl fermions.

%\section{Results}

\section{Overview of the crystal and electronic structure}

We first provide a brief background of \mowte\ and study the band structure below the Fermi level. WTe$_2$ crystallizes in an orthorhombic Bravais lattice, space group $Pmn2_1$ ($\#31$), lattice constants $a = 6.282\textrm{\AA}$, $b = 3.496\textrm{\AA}$, and $c = 14.07\textrm{\AA}$, as shown in Fig. \ref{mwt2Fig1}a \cite{MoTe2WTe2}. Crucially, the crystal has no inversion symmetry, a requirement for a Weyl semimetal \cite{Murakami}. The crystals we study are flat, shiny, layered and beautiful, see Fig. \ref{mwt2Fig1}b. The natural cleaving plane is (001), with surface and bulk Brillouin zones as shown in Fig. \ref{mwt2Fig1}c. We first consider the overall band structure of WTe$_2$. There are two bands, one electron and one hole pocket, near the Fermi level, both very near the $\Gamma$ point of the bulk Brillouin zone, along the $\Gamma-Y$ line. Although the bands approach each other and Weyl points might be expected to arise where the bands cross, it is now understood that WTe$_2$ is in fact very close to a phase transition between a Weyl semimetal phase and a trivial phase, so that the electronic structure of WTe$_2$ is too fragile to make it a compelling candidate for a Weyl semimetal \cite{TR}. Next, we interpolate between \textit{ab initio} Wannier function-based tight-binding models for WTe$_2$ and MoTe$_2$ to study \mowte\ at arbitrary $x$ \cite{TR}. For a wide range of $x$, we find a robust Weyl semimetal phase \cite{TR}. In Fig. \ref{mwt2Fig1}e,f, we show where the Weyl points sit in the Brillouin zone. They are all located close to $\Gamma$ in the $k_z = 0$ momentum plane. There are two sets of Weyl points, $W_1$ at binding energies $E_\textrm{B} = - 0.045$ eV and $W_2$ at $E_\textrm{B} = - 0.066$ eV, all above the Fermi level $E_\textrm{F}$. In addition, the Weyl points are almost aligned at the same $k_y = \pm k_\textrm{W}$, although this positioning is not known to be in any way symmetry-protected. We also note that the Weyl cones are all tilted over, corresponding to strongly Lorentz-violating or Type II Weyl fermions, see Fig. \ref{mwt2Fig1}g \cite{AndreiNature}. Next, we study a Fermi surface of \mowte\ at $x = 45\%$ using incident light with photon energy $h \nu = 5.92$ eV, shown in Fig. \ref{mwt2Fig1}h. We observe two pockets, a palmier-shaped pocket closer to the $\bar{\Gamma}$ point of the surface Brillouin zone and an almond-shaped pocket sitting next to the palmier pocket, further from $\bar{\Gamma}$. The palmier pocket is a hole pocket, while the almond pocket is an electron pocket \cite{myothermowte}. We note that we see an excellent agreement between our results and an \textit{ab initio} calculation of \mowte\ for $x = 40\%$, shown in Fig. \ref{mwt2Fig1}i. At the same time, we point out that the electron pocket of the Weyl points is nearly absent in this ARPES spectrum, possibly due to low photoemission cross section at the photon energy used \cite{TR, myothermowte}. However, as we will see below, we do observe this electron pocket clearly in our pump-probe ARPES measurements, carried out at a slightly different photon energy, $h \nu = 5.92$ eV. Based on our calculations and preliminary ARPES results, we expect that the Weyl points sit above the Fermi level, where the palmier and almond pockets approach each other. We also present an $E_\textrm{B}$-$k_x$ spectrum in Fig. \ref{mwt2Fig1}j where we see how the plamier and almond pockets nest into each other. We expect the two pockets to chase each other as they disperse above $E_\textrm{F}$, giving rise to Weyl points, see Fig. \ref{mwt2Fig1}k.

% For Fig. 1 text:
% Cite our earlier paper somewhere in Fig. 1 maybe...
% Weyl points near the same ky

\section{Unoccupied band structure of \mowte}

Next, we show that pump-probe ARPES at probe photon energy $h\nu = 5.92$ eV gives us access to the bulk and surface bands participating in the Weyl semimetal state in \tf, both below and above $E_\textrm{F}$. In Fig. \ref{mwt2Fig2}a-c, we present three successive ARPES spectra of \tf\ at fixed $k_y$ near the predicted position of the Weyl points. We observe a beautiful sharp band near $E_\textrm{F}$, whose sharpness suggests that it's a surface band, and broad continua above and below the Fermi level, whose broad character suggests that they are bulk valence and conduction bands. In Fig. \ref{mwt2Fig2}d-f, we show the same cuts, with guides to the eye to mark the bulk valence and conduction band continua. We also find that we can track the evolution of the bulk valence and conduction bands clearly in our data with $k_y$. Specifically, we see that both the bulk valence and conduction bands disperse toward negative binding energies as we sweep $k_y$ closer to $\bar{\Gamma}$. At the same time, we note that the bulk valence band near $\bar{\Gamma}$ is only visible near $k_x \sim 0$ and drops sharply in photoemission cross-section away from $k_x \sim 0$. In Fig. \ref{mwt2Fig2}g,h we present a comparison of our ARPES data with an \textit{ab initio} calculation of \tf\ \cite{TR}. We also mark the location of the three successive spectra on a Fermi surface in Fig. \ref{mwt2Fig2}i. We include as well the approximate locations of the Weyl points, as expected from calculation. We find excellent correspondence between both bulk and surface states. We add that there is an additional surface state detaching from the bulk conduction band well above the Fermi level and that we can also directly observe this additional surface state both in our ARPES spectra and in calculation. Our pump-probe ARPES results clearly show both the bulk and surface band structure of \tf, both below and above $E_\textrm{F}$, and with an excellent correspondence with calculation.

\section{Observation of a topological Fermi arc above the Fermi level}

Now we show that we observe signatures of a Fermi arc in \tf. We consider the cut shown in Fig. \ref{mwt2Fig3}a, repeated from Fig. \ref{mwt2Fig2}b, and we study the surface state. We observe two kinks in each branch, at $E_\textrm{B} \sim -0.005$ eV and $E_\textrm{B} \sim - 0.05$ eV. This kink is a smoking-gun signature of a Weyl point \cite{NbPme}. We claim that each kink corresponds to a Weyl point and that the surface state passing through them includes a topological Fermi arc. To show these kinks more clearly, in Fig. \ref{mwt2Fig3}b, we show a second derivative plot of the spectrum in Fig. \ref{mwt2Fig3}a. In Fig. \ref{mwt2Fig3}c we also present a cartoon of the kink in our data, with the positions of the $W_1$ and $W_2$ Weyl points marked. Again, note that although the $W_1$ and $W_2$ are not located strictly at the same $k_y$, we expect the $k_y$ separation to be on the order of $10^{-4} \textrm{\AA}^{-1}$ from calculation, so that we can consider them to lie at the same $k_y$ within experimental resolution. We emphasize that from our pump-probe ARPES spectrum, we can directly read off that the energy separation of the Weyl points is $\sim 0.05$ eV and that the $W_1$ are located at $\sim -0.005$ eV. We also present a quantitative analysis of our data, showing a kink. To do this, we fit the surface state momentum distribution curves (MDCs) to a Lorentzian distribution and we plot the train of peaks corresponding to the surface state band. We note that we simultaneously fit the topological surface state, the bulk valence and conduction states and the trivial surface state above the conduction band. In Fig. \ref{mwt2Fig3}d we plot the resulting band dispersions in white and observe an excellent fit to our spectrum. Next, we define a kink as a failure of the train of Lorentzian maxima to fit to a quadratic band. In particular, over a small energy and momentum window, any band should be well-characterized by a quadratic fit, so the failure of such a fit in a narrow energy window implies a kink. After fitting the topological surface state to a quadratic polynomial we find two mismatched regions, marked in Fig. \ref{mwt2Fig3}e, demonstrating two kinks. For comparison, we plot the energy positions of the $W_1$ and $W_2$ as read off directly from Fig. \ref{mwt2Fig3}a. We find an excellent agreement between the qualitative and quantitative analysis, although we note that the fit claims that the $W_2$ kink is lower in energy. To illustrate the success of the Lorentzian fit, in Figs. \ref{mwt2Fig3}f,g we present two representative MDCs at energies indicated by the green arrows. We see that the Lorentzian distributions provide a good fit and take into account all bands observed in our spectra. The raw data, the second derivative plots and the Lorentzian fitting all show two kinks, providing a strong signature of Fermi arcs.

To show that we have observed a topological Fermi arc, we compare our experimental observation of two surface state kinks with a numerical calculation of \tf. In Fig. \ref{mwt2Fig4}a,b, we mark the energies of the Weyl points as well as the band minimum of the surface state in our ARPES spectrum and in calculation. We see that the energy difference between the Weyl points is $\sim 0.02$ eV in calculation but $\sim 0.05$ eV in experiment. Moreover, the band minimum $E_\textrm{min}$ is at $\sim E_\textrm{F}$ in calculation, but at $E_\textrm{B} \sim 0.06$ eV in experiment. The difference in $E_\textrm{min}$ suggests either that our sample is electron-doped or that the $k_y$ position of the Weyl points differs in experiment and theory. Next, crucially, we observe that, in disagreement with calculation, the $W_1$ are located only $\sim 0.005$ eV above $E_\textrm{F}$. This suggests that the Weyl points and Fermi arcs in our \tf\ samples may be accessible in transport. This result is particularly relevent because MoTe$_2$, WTe$_2$ and other transition metal dichalcogenides are already under study as platforms for novel electronics \cite{ReviewTMDC, LiangFu, LaserPattern, Heinz, Valley}. Since the Weyl points of \mowte\ may be at the Fermi level, it is possible that transport measurements may detect a signature of the strongly Lorentz-violating Weyl fermions or other unusual transport phenomena associated with Weyl semimetals in \mowte. We summarize our results in Fig. \ref{mwt2Fig4}c. We directly observe, above the Fermi level, a surface state with two kinks (shown in red). By comparing our results with \textit{ab initio} calculation, we confirm that the kinks correspond to Weyl points. Furthermore, the excellent agreement of our experimental results with calculation shows that we have realized the first Type II Weyl semimetal.

\section{Limits on directly observing Type II Weyl cones}

So far we have studied the surface states of \mowte\ and we have argued that \mowte\ is a Weyl semimetal because we observe a topological Fermi arc surface state. However, topological Fermi arcs cannot strictly distinguish between bulk Weyl cones that are of Type I or Type II. While the excellent agreement with calculation suggests that \mowte\ is a Type II Weyl semimetal, we might ask if we can directly observe a Type II Weyl cone in \mowte\ by ARPES. This corresponds to observing the two branches of the bulk Weyl cone, as indicated by the purple dotted circles in Fig. \ref{mwt2Fig4}c. We reiterate that one crucial obstacle in observing a Type II Weyl cone is that all the recent calculations on WTe$_2$, \mowte\ and MoTe$_2$ predict that all Weyl points are above the Fermi level \cite{TR,AndreiNature,Zhijun,Binghai}. As we have seen, using pump-probe ARPES, we are able to measure the unoccupied band structure and show a Fermi arc. However, in our pump-probe ARPES measurements, we find that the photoemission cross-section of the bulk bands is too weak near the Weyl points. At the same time, our calculations suggest that for a reasonable quasiparticle lifetime and spectral linewidth, the broadening of bands will make it difficult to resolve the two branches of the Weyl cone. We conclude that it is challenging to directly access the Type II Weyl cones in \mowte.

\section{Considerations regarding trivial surface states}

One obvious concern with our experimental result is that we observe two kinks in the surface state, but we expect a disjoint segment based on topological theory. In particular, all calculations show that all Weyl points in \mowte\ have chiral charge $\pm 1$ \cite{TR, AndreiNature, Zhijun, Binghai}. However, our observation of a kink suggests that there are two Fermi arcs connecting to the same Weyl point, which requires a chiral charge of $\pm 2$. To resolve this contradiction, we study the calculation of the surface state near the Weyl points, shown in Fig. \ref{mwt2Fig4}g. We observe, as expected, a Fermi arc (red arrow) connecting the Weyl points. However, at the same time, we see that trivial surface states (yellow arrows) from above and below the band crossing merge with the bulk bands in the vicinity of the Weyl points. As a result, there is no disjoint arc but rather a large, broadband surface state with a ripple arising from the Weyl points. We can imagine that this broadband surface state exists even in the trivial phase. Then, when the bulk bands cross and give rise to Weyl points, a Fermi arc is pulled out from this broadband surface state. At the same time, the remainder of the broadband surface state survives as a trivial surface state. In this way, the Fermi arc is not disjoint but shows up as a ripple. We observe precisely this ripple in our ARPES spectra of \tf.

As a further check of our analysis, we perform a Lorentzian fit of an ARPES spectrum at $k_y$ shifted away from the Weyl points, shown in Fig. \ref{mwt2Fig4}d, the same cut as Fig. \ref{mwt2Fig2}c. We show the Lorentzian fit in Fig. \ref{mwt2Fig4}e and a quadratic fit to the train of peaks in Fig. \ref{mwt2Fig2}f. In sharp contrast to the result for $k_y \sim k_\textrm{W}$, there is no ripple in the spectrum and the quadratic provides an excellent fit. This result is again consistent with our expectation that we should observe a ripple only at $k_y$ near the Weyl points. Our results also set the stage for the realization of the first tunable Weyl semimetal in \mowte. As we vary the composition, we expect to tune the relative separation of the Weyl points and $E_\textrm{F}$. In Fig. \ref{mwt2Fig4}h-k, we present a series of calculations of \mowte\ for $x = 10\%$, $25\%$, $40\%$ and $100\%$. We see that the separation of the Weyl points increases with $x$ and that the $W_1$ approach $E_\textrm{F}$ for larger $x$. We propose that a systematic composition dependence can demonstrate the first tunable Weyl semimetal in \mowte. 

\section{Discussion}

We have demonstrated a Weyl semimetal in \mowte\ by directly observing kinks and a Fermi arc in the surface state band structure. Taken together with calculation, our experimental data show that we have realized the first Type II Weyl semimetal, with strongly Lorentz-violating Weyl fermions. We point out that in contrast to concurrent works on the Weyl semimetal state in MoTe$_2$ \cite{Adam, Shuyun, Baumberger, Adam2, Baumberger2, Chen, Xinjiang2, Xinjiang, HongDing}, we directly access the unoccupied band structure of \mowte\ and directly observe a Weyl semimetal with minimal reliance on calculation. In particular, our observation of a surface state kink at a generic point in the surface Brillouin zone requires that the system be a Weyl semimetal \cite{NbPme}. The excellent agreement with calculation serves as an additional, independent check of our experimental results. We also reiterate that unlike MoTe$_2$, \mowte\ opens the way to the realization of the first tunable Weyl semimetal. Lastly, we note that MoTe$_2$ is complicated because it is near a critical point for a topological phase transition. Indeed, one recent theoretical work \cite{Zhijun} shows that MoTe$_2$ has four Weyl points, while another \cite{Binghai} finds eight Weyl points. Indeed, this is similar to the case of WTe$_2$, which is near the critical point for a transition between eight Weyl points and zero Weyl points. By contrast, \mowte\ sits well within the eight Weyl point phase for most $x$, as confirmed explicity here and in Ref. \cite{TR}. The stability of the topological phase of \mowte\ simplifies the interpretation of our data. By directly demonstrating a Weyl semimetal in \mowte, we provide not only the first Weyl semimetal beyond the TaAs family, but the first Type II Weyl semimetal, as well as a Weyl semimetal which may be tunable and which may be more accessible for transport and optics studies of the fascinating phenomena arising from emergent Weyl fermions in a crystal.

\clearpage
\begin{figure*}
\centering
\includegraphics[width=11cm, trim={1.7in 0.4in 1.7in 0.4in}, clip]{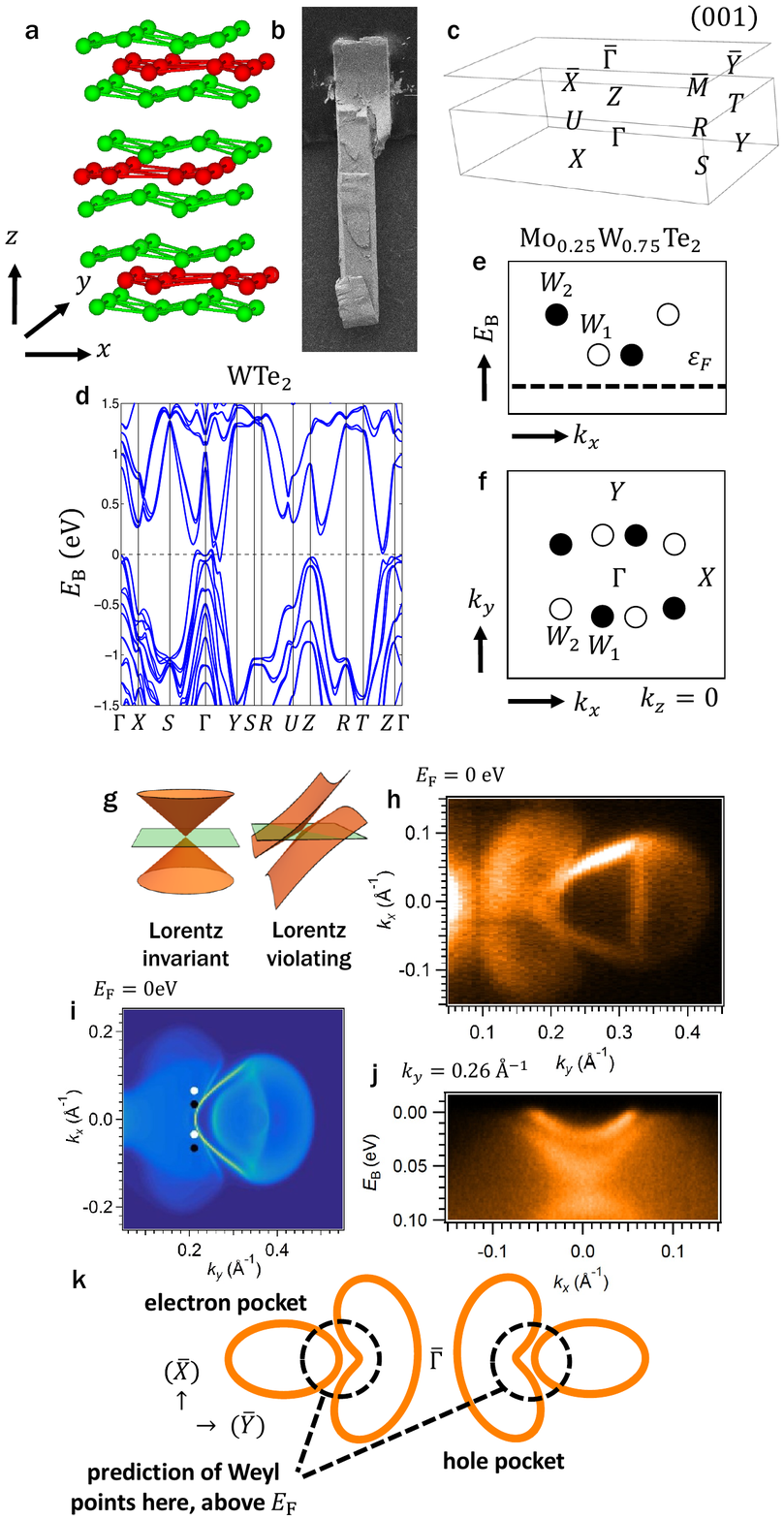}
\end{figure*}

%  It is orthorhombic, space group $Pmn2_1$ ($\#31$), showing the Te layers (green) and W/Mo layers (red).

\clearpage
\begin{figure}
\caption{\label{mwt2Fig1} \textbf{Overview of \mowte.} {\bf a,} The crystal structure of the system is layered, with each monolayer consisting of two Te layers (green) and one W/Mo layer (red). {\bf b,} A wonderful scanning electron microscope (SEM) image of a typical single crystal of \mowte, $x = 45\%$. The layered structure is visible in the small corrugations and breaks in the layers. {\bf c,} Bulk and (001) surface Brillouin zone, with high-symmetry points marked. {\bf d,} Bulk band structure of WTe$_2$ along high-symmetry lines. There are two relevant bands near the Fermi level, an electron band and a hole band, both near the $\Gamma$ point and along the $\Gamma-Y$ line, which approach each other near the Fermi level. {\bf e, f,} Upon doping by Mo, \mowte\ enters a robust Weyl semimetal phase \cite{TR}. Schematic of the positions of the Weyl points in the bulk Brillouin zone. The opposite chiralities are indicated by black and white circles. Crucially, all Weyl points are above the Fermi level. {\bf g,} The Weyl cones in \mowte\ are unusual in that they are all tilted over, associated with strongly Lorentz-violating or Type II Weyl fermions, prohibited in particle physics \cite{AndreiNature}. {\bf h,} Fermi surface of \mowte\ at $x = 45\%$ measured by ARPES at $h \nu = 6.36$ eV, showing a hole-like palmier pocket and an electron-like almond pocket \cite{myothermowte}. {\bf i,} There is an excellent correspondence between our ARPES data and our calculation. Note that the $k_y$ axis on the Fermi surface from ARPES is set by comparison with calculation. {\bf j,} An $E_\textrm{B}$-$k_x$ cut showing the palmier and almond pockets below the Fermi level. {\bf k,} In summary, the Fermi surface of \mowte\ consists of a palmier hole pocket and an almond electron pocket near the $\bar{\Gamma}$ point. The two pockets chase each other as they disperse, eventually intersecting above $E_\textrm{F}$ to give Weyl points.}
\end{figure}

\clearpage
\begin{figure}
\centering
\includegraphics[width=13cm]{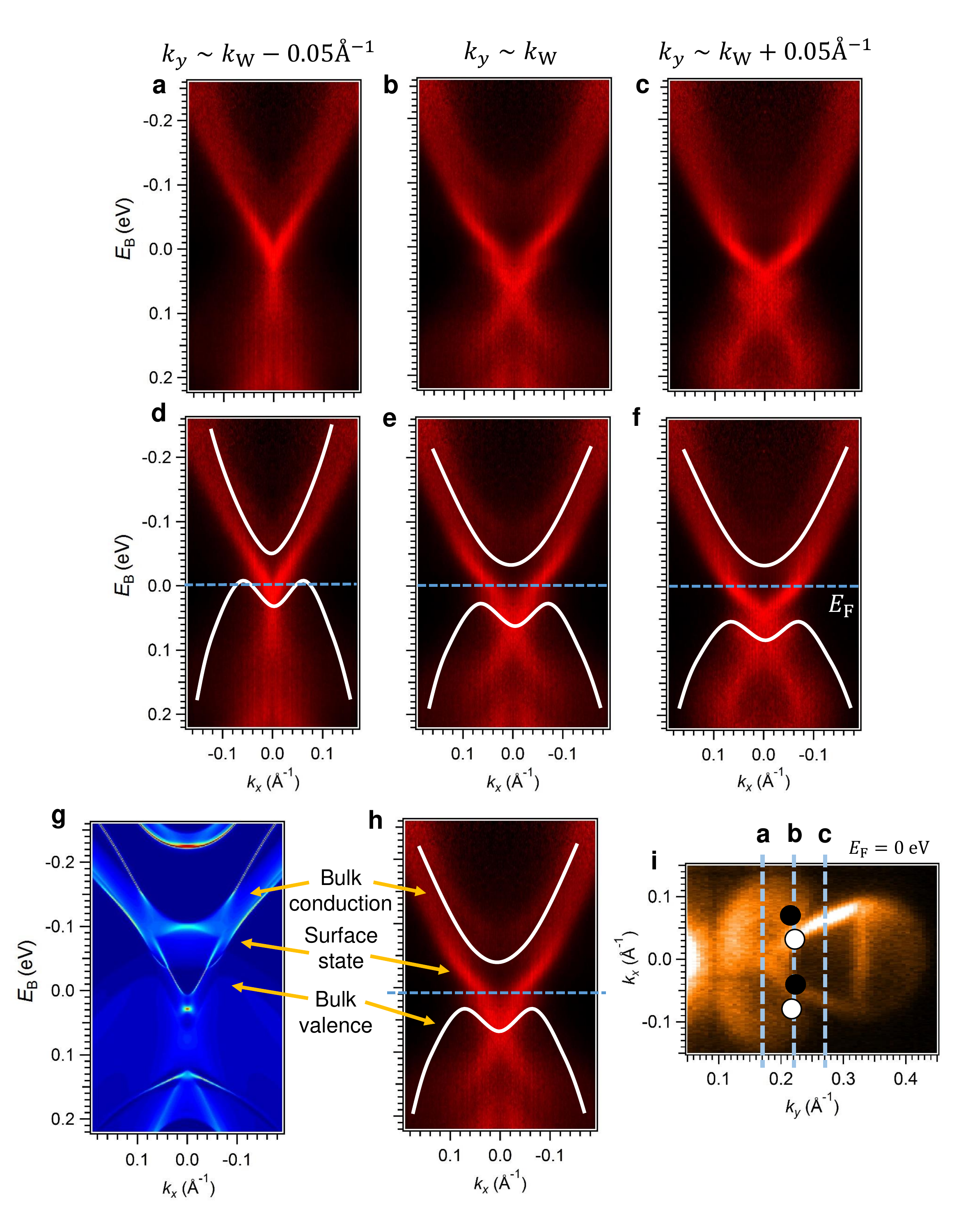}
\end{figure}

\clearpage
\begin{figure}
\caption{\label{mwt2Fig2} \textbf{Dispersion of the unoccupied bulk and surface states of \tf.} {\bf a-c,} Three successive ARPES spectra for \tf\ at fixed $k_y$ near the expected position of the Weyl points, $k_{\textrm{Weyl}}$, using pump-probe ARPES at probe $h\nu = 5.92$ eV. A strong pump response allows us to probe the unoccupied states $\sim 0.3$ eV above $E_F$, which is well above the expected $E_{\textrm{W1}}$ and $E_{\textrm{W2}}$. {\bf d-f,} Same as (a-c), but with the bulk valence and conduction band continuum marked with guides to the eye. We see that we observe all bulk and surface states participating in the Weyl semimetal state. As expected, both the bulk valence and conduction bands move towards more negative binding energies as $k_y$ moves towards $\bar{\Gamma}$. {\bf g, h,} Comparison of our calculations with experimental results for $k_y \sim k_{\textrm{Weyl}}$. As can be seen from (h), our spectra clearly display all bulk and surface bands of \tf\ relevant for the Weyl semimetal state, both below and above $E_F$, and with an excellent agreement with the corresponding calculation in panel g. {\bf i,} The locations of the cuts in (a-c).}
\end{figure}

\clearpage
\begin{figure}
\centering
\includegraphics[width=14cm,trim={1in 1in 1.8in 1in},clip]{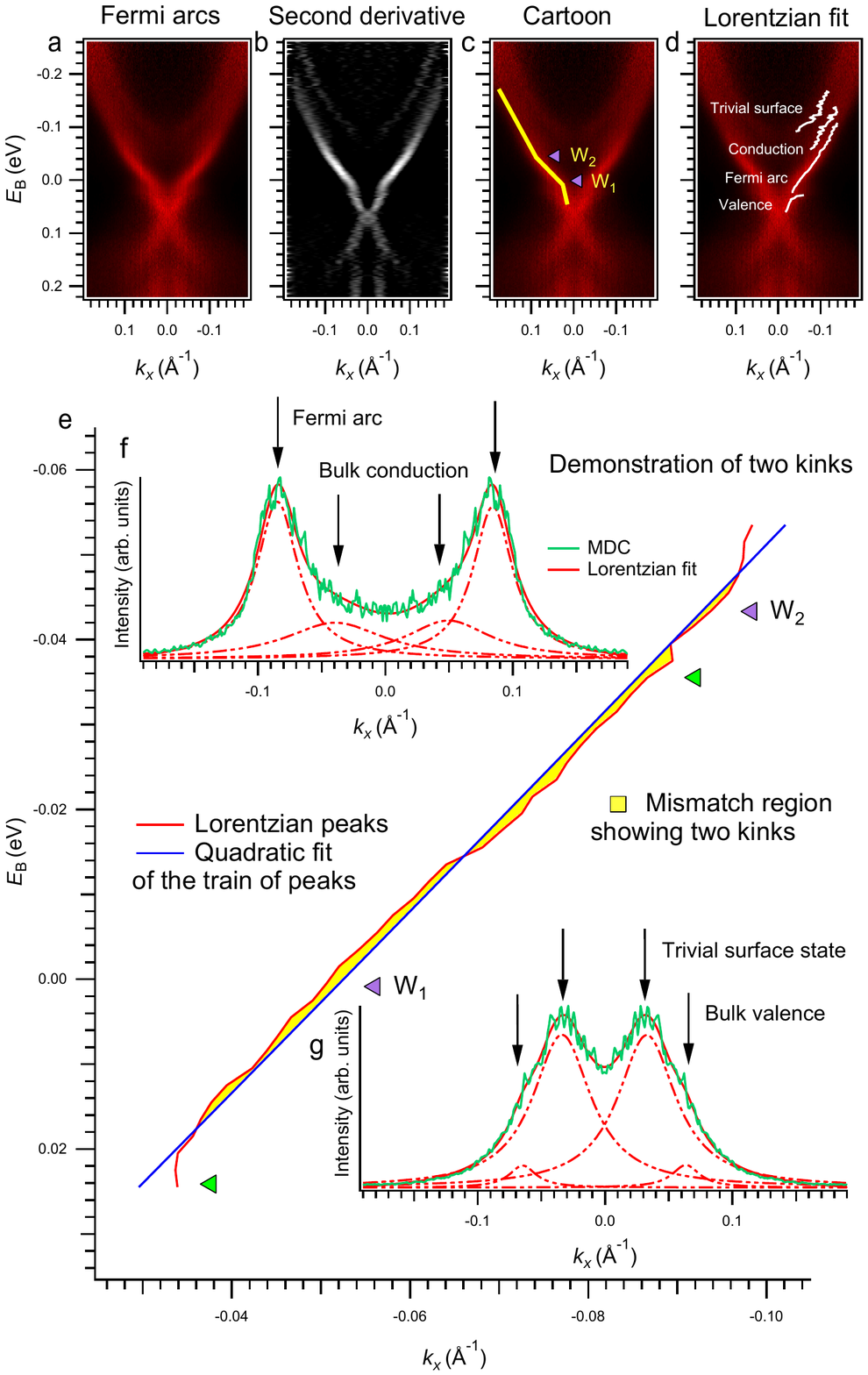}
\end{figure}

\clearpage
\begin{figure}
\caption{\label{mwt2Fig3} \textbf{Direct experimental observation of Fermi arcs in \tf.} {\bf a,} To establish the presence of Fermi arcs in \tf\ we focus on spectrum shown in Fig. \ref{mwt2Fig2}b, with $k_y \sim k_{\textrm{Weyl}}$. We observe two kinks in the surface state, at $E_B \sim -0.005$ eV and $E_B \sim -0.05$ eV. {\bf b,} The kinks are easier to see in a second-derivative plot of panel a. {\bf c,} Same as panel a, but with a guide to the eye showing the kinks. The Weyl points are at the locations of the kinks. The surface state with the kinks contains a topological Fermi arc. {\bf d,} To further confirm the presence of a kink, we fit Lorentzian distributions to our data. We capture all four bands in the vicinity of the kinks: the bulk conduction and valence states, the topological surface state and an additional trivial surface state merging into the conduction band at more negative $E_B$. We define a kink as a failure of a quadratic fit to a band. We argue that for a small energy and momentum window, any band should be well-characterized by a quadratic fit and that the failure of such a fit shows a kink. {\bf e,} By matching the train of Lorentzian peaks of the topological surface state (red) to a quadratic fit (blue) we find two mismatched regions (shaded in yellow), showing two kinks. The purple arrows show the location of the Weyl points, taken from panel c, and are consistent with the kinks we observe by fitting. {\bf f, g,} Two characteristic MDCs at energies indicated by the green arrows in panel e. We see that the Lorentzian distributions provide a good fit and capture all bands observed in our spectra.}
\end{figure}

\clearpage
\begin{figure}
\centering
\includegraphics[width=15cm,trim={1in 1in 1.3in 1in},clip]{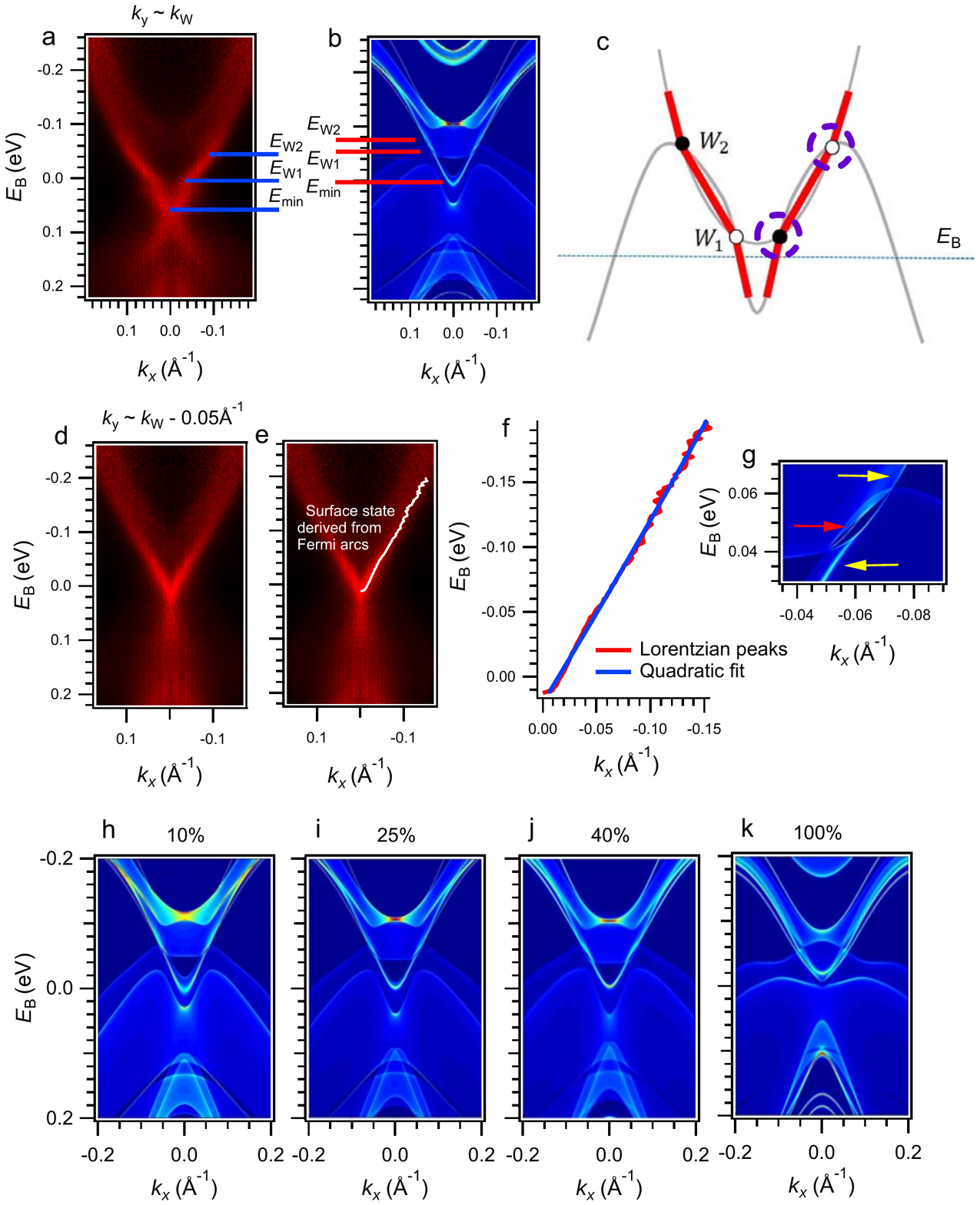}
\end{figure}

\clearpage
\begin{figure}
\caption{\label{mwt2Fig4} \textbf{Demonstration of a Weyl semimetal in \mowte.} {\bf a,} The same spectrum as Fig. \ref{mwt2Fig3}a but with the energies $\varepsilon_{W1}$, $\varepsilon_{W2}$, $\varepsilon_{\textrm{min}}$ marked. {\bf b,} The same energies marked in an \textit{ab initio} calculation of \tf. We note that this cut is not taken at fixed $k_y \sim k_W$. Instead, we cut along the exact line defined by $W_1$ and $W_2$ in the surface Brillouin zone. Since $k_y^{\textrm{W1}}$ is exceedingly close to $k_y^{\textrm{W2}}$, this cut essentially corresponds to our experimental data. The Weyl points are $\sim 0.05$ eV separated in energy in our data, compared to $\sim 0.02$ eV in calculation. In addition, crucially, the $W_1$ are lower in energy than we expect from calculation and in fact are located only $0.005$ eV above $E_F$. {\bf c,} A cartoon of our interpretation of our experimental results. We observe the surface state (red) with a kink at the locations of the Weyl points (black and white circles). Each surface state consists of a Fermi arc (middle red segment) and two trivial surface states which merge with bulk bands near the location of the Weyl points. We observe certain portions of the bulk bands (gray), but not the bulk Weyl cones, see the Supplementary Information. {\bf d,} The same spectrum as Fig. \ref{mwt2Fig2}c, at $k_y$ shifted toward $\bar{\Gamma}$. {\bf e, f,} A Lorentzian fit of the surface state and a quadratic fit to the train of peaks, showing no evidence of a kink. This is precisely what we expect from a cut away from the Weyl points. {\bf g,} A close-up of the band inversion, showing a Fermi arc (red arrow) which connects the Weyl points and trivial surface states (yellow arrows) from above and below which merge with the bulk bands in the vicinity of the Weyl points. {\bf h-k,} Composition dependence of \mowte\ from first principles, showing that the separation of the Weyl points increases with $x$. Our observation of a Weyl semimetal in \tf\ sets the stage for the first tunable Weyl semimetal in \mowte.}
\end{figure}

\section{Materials and methods}

\subsection{Pump-probe ARPES}

Pump-probe ARPES measurements were carried out using a hemispherical Scienta R4000 analyzer and a mode-locked Ti:Sapphire laser system that delivered $1.48$ eV pump and $5.92$ eV probe pulses at a repetition rate of $250$ kHz \cite{IshidaMethods}. The time and energy resolution were $300$ fs and $15$ meV, respectively. The spot diameters of the pump and probe lasers at the sample were $250\ \mu$m and $85\ \mu$m, respectively. Measurements were carried out at pressure $<5 \times 10^{-11}$ Torr and temperature $\sim8$ K.

\subsection{Sample growth}

Single crystals of \mowte\ were grown using a chemical vapor transport (CVT) technique with iodine as the transport agent. Stoichiometric Mo, W and Te powders were ground together and loaded into a quartz tube with a small amount of I. The tube was sealed under vacuum and placed in a two-zone furnace. The hot zone was maintained at $1050^{\circ}$C for 2 weeks and the cold zone was maintained at $950^{\circ}$C.The dopant distribution is not uniform particularly near the crystal surface. The composition of the selected sample was determined by an energy dispersive spectroscopy (EDS) measurement with a scanning electron microscope (SEM). 

\subsection{\textit{Ab initio} calculations}

The \textit{ab initio} calculations were based on the generalized gradient approximation (GGA) \cite{GGA} used the full-potential projected augmented wave method \cite{PAW1,PAW2} as implemented in the VASP package \cite{PlaneWaves1}. Experimental lattice constants were used for both WTe$_2$ \cite{Bonding} and MoTe$_2$. A $15 \times 11 \times 7$ Monkhorst-Pack $k$-point mesh was used in the computations. The spin-orbit coupling effects were included in calculations. To calculate the bulk and surface electronic structures, we constructed first-principles tight-binding model Hamilton by projecting onto the Wannier orbitals \cite{MLWF1,MLWF2,Wannier90}, which use the VASP2WANNIER90 interface \cite{MLWF3}. We used W $d$ orbitals, Mo $d$ orbitals, and Te $p$ orbitals to construct Wannier functions and without perform the procedure for maximizing localization. The electronic structure of the \mowte\ samples with finite doping was calculated by a linear interpolation of tight-binding model matrix elements of WTe$_2$ and MoTe$_2$. The surface states were calculated from the surface Green's function of the semi-infinite system \cite{Green2, Green}.

\clearpage
%\cleardoublepage
\ifdefined\phantomsection
  \phantomsection  % makes hyperref recognize this section properly for pdf link
\else
\fi
\addcontentsline{toc}{section}{Bibliography}

{\singlespacing
}

%\end{document}

%% file: ch-TaIrTe4/ch-TaIrTe4.tex
\chapter{Weyl semimetal with four Weyl points}
\label{ch:tairte4}

{\singlespacing
\begin{chapquote}{C\oe{}ur de Pirate, \textit{Crier Tout Bas}}
Et quand tu chantais\\
plus fort dans ton silence,\\
Je voyais les larmes couler\\
toujours \`a contresens
\end{chapquote}}

\renewcommand{\beq}{\begin{equation*}}
\renewcommand{\eeq}{\end{equation*}}
\newcommand{\mwt}{Mo$_x$W$_{1-x}$Te$_2$}
\newcommand{\tai}{TaIrTe$_4$}
\newcommand{\inv}{$\mathcal{I}$}
\newcommand{\tr}{$\mathcal{T}$}
\newcommand{\eb}{$E_\textrm{B}$}
\newcommand{\ef}{$E_\textrm{F}$}

\noindent This chapter is based on the article, \textit{Signatures of a time-reversal symmetric Weyl semimetal with only four Weyl points} by Ilya Belopolski \textit{et al}., \textit{Nat. Commun.} {\bf 8}, 942 (2017), available at \href{https://www.nature.com/articles/s41467-017-00938-1}{https://www.nature.com/articles/s41467-017-00938-1}.\\

%\begin{abstract}
\lettrine[lines=3]{T}{hrough} intense research on Weyl semimetals during the past few years, we have come to appreciate that typical Weyl semimetals host many Weyl points. Nonetheless, the minimum nonzero number of Weyl points allowed in a time-reversal invariant Weyl semimetal is four. Realizing such a system is of fundamental interest and may simplify transport experiments. Recently, it was predicted that \tai\ realizes a minimal Weyl semimetal. However, the Weyl points and Fermi arcs live entirely above the Fermi level, making them inaccessible to conventional angle-resolved photoemission spectroscopy (ARPES). Here we use pump-probe ARPES to directly access the band structure above the Fermi level in \tai. We observe signatures of Weyl points and topological Fermi arcs. Combined with \textit{ab initio} calculation, our results show that \tai\ is a Weyl semimetal with the minimum number of four Weyl points. Our work provides a simpler platform for accessing exotic transport phenomena arising in Weyl semimetals.

\section{Introduction}

A Weyl semimetal is a crystal which hosts emergent Weyl fermions as electronic quasiparticles. In an electronic band structure, these Weyl fermions correspond to accidental degeneracies, or Weyl points, between two bands \cite{Weyl, Peskin, Abrikosov, Nielsen, Volovik}. It is well-understood that Weyl points can only arise if a material breaks either spatial inversion symmetry, \inv, or time-reversal symmetry, \tr\ \cite{Murakami, Pyrochlore, ARCMP, Hosur}. At the same time, in a Weyl semimetal, symmetries of the system tend to produce copies of Weyl points in the Brillouin zone. As a result, typical Weyl semimetals host a proliferation of Weyl points. For instance, the first Weyl semimetals observed in experiment, TaAs and its isoelectronic cousins, have an \inv\ breaking crystal structure which gives rise to a band structure hosting $24$ Weyl points distributed throughout the bulk Brillouin zone \cite{TaAsUs, TaAsThyUs, TaAsThem, TaAsThyThem, NbAs, TaPUs, HaoNbP, TaAsChen}. However, most of these Weyl points can be related to one another by the remaining symmetries of TaAs, namely two mirror symmetries, $C_4$ rotation symmetry and \tr. In the \mwt\ series, which has recently been under intensive theoretical and experimental study as a Weyl semimetal with strongly Lorentz-violating, or Type II, Weyl fermions, mirror symmetry and \tr\ relate subsets of the eight Weyl points \cite{AndreiNature, TayRong, Binghai, Zhijun, myothermowte, Adam1, Shuyun, Baumberger}. As another example, according to calculation, the Weyl semimetal candidate SrSi$_2$ hosts no fewer than 108 Weyl points, copied in sets of 18 by three $C_4$ rotation symmetries \cite{SrSi2}. However, as we review below, it is well-known that the minimal nonzero number of Weyl points allowed is 4 for a \tr\ invariant Weyl semimetal. Realizing such a minimal Weyl semimetal is not only of fundamental interest, but is also practically important, because a system with fewer Weyl points may exhibit simpler properties in transport and be more suitable for device applications.

Recently, \tai\ was predicted to be a Weyl semimetal with only four Weyl points, the minimum allowed for a \tr\ invariant Weyl semimetal \cite{Koepernik}. It was further noted that the Weyl points are associated with strongly Lorentz-violating, or Type II, Weyl fermions, providing only the second example of a Type II Weyl semimetal after the \mwt\ series \cite{AndreiNature}. Moreover, the Weyl points are well-separated in momentum space, with substantially larger topological Fermi arcs as a fraction of the size of the surface Brillouin zone than other known Weyl semimetals. Lastly, \tai\ has a layered crystal structure, which may make it easier to carry out transport experiments and develop device applications. All of these desirable properties have motivated considerable research interest in \tai. At the same time, one crucial challenge is that the Weyl points and topological Fermi arcs are predicted to live entirely above the Fermi level in \tai, so that they are inaccessible to conventional angle-resolved photoemission spectroscopy (ARPES). 

Here we observe direct signatures of Weyl points and topological Fermi arcs in \tai, realizing the first minimal \tr\ invariant Weyl semimetal. We first briefly reiterate a well-known theoretical argument that the minimum number of Weyl points for a \tr\ invariant Weyl semimetal is four. Then, we present \textit{ab initio} calculations showing a nearly ideal configuration of Weyl points and Fermi arcs in \tai. Next, we use pump-probe ARPES to directly access the band structure of \tai\ above the Fermi level in experiment. We report the observation of signatures of Weyl points and topological Fermi arcs. Combined with \textit{ab initio} calculations, our results demonstrate that \tai\ has four Weyl points. We conclude that \tai\ can be viewed as the a minimal Weyl semimetal, with the simplest configuration of Weyl points allowed in a \tr\ invariant crystal.

% Why don't mention the \tr breaking case

\section{Minimum number of Weyl points under time-reversal symmetry}

\begin{figure}
\centering
\includegraphics[width=15cm, trim={1in 0.3in 1in 0.3in}, clip]{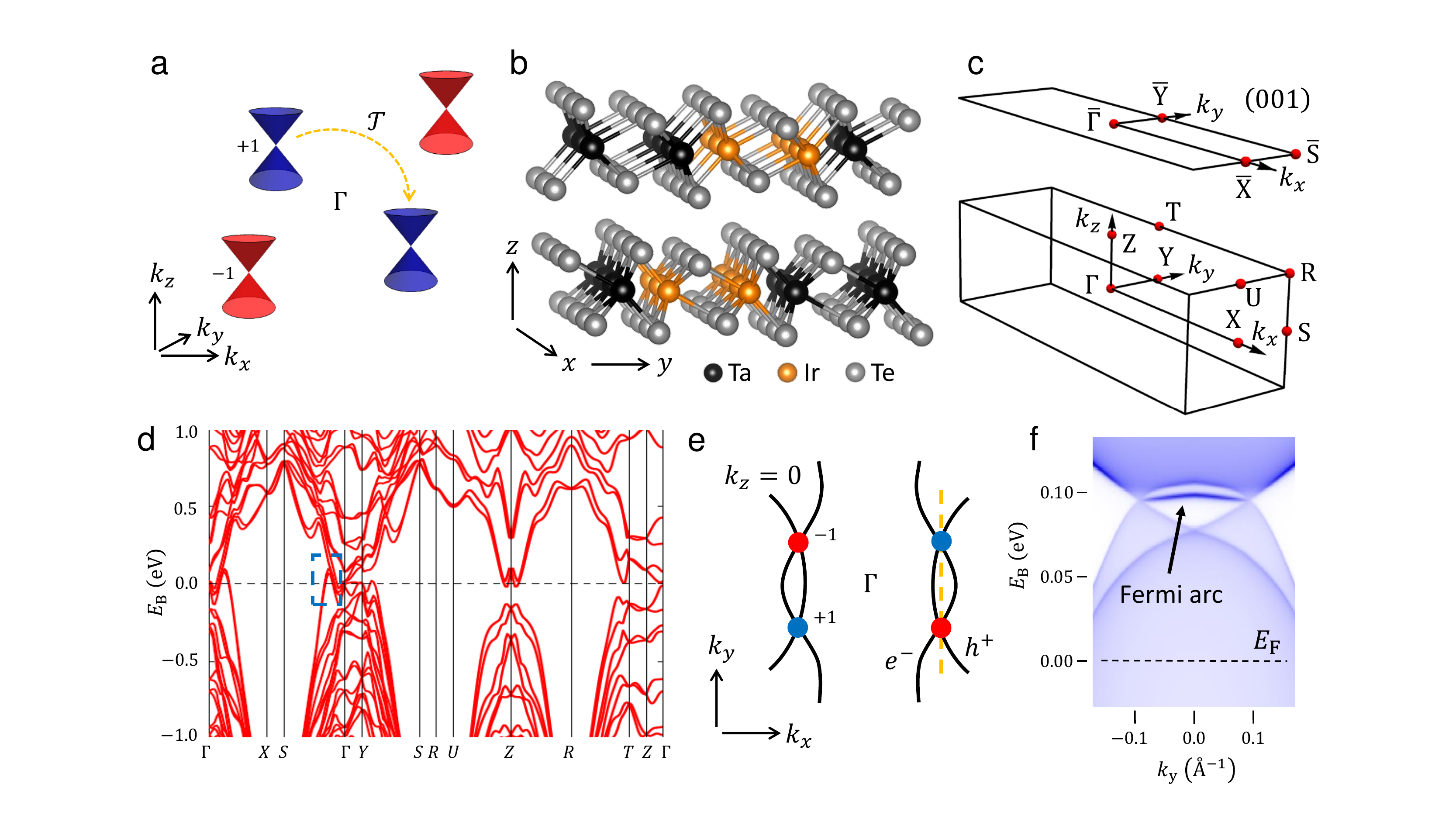}
%\end{figure}
%\clearpage
%
%\begin{figure*}
\caption{\label{titFig1}\textbf{Constraints on Weyl points in $\mathcal{T}$ symmetric systems.} (a) Illustration of the minimal number of Weyl points in a $\mathcal{T}$ invariant Weyl semimetal. The blue and red circles and cones represent Weyl points and Weyl cones with $\pm{1}$ chiral charge at generic $k$-points. In a $\mathcal{T}$ invariant Weyl semimetal the minimal number of Weyl points is four because $\mathcal{T}$ symmetry sends a Weyl point of a given chiral charge at $k$ to a Weyl point of the same chiral charge at $-k$ (orange arrow). To preserve net zero chiral charge, four Weyl points are needed. (b) The crystal structure of TaIrTe$_4$ is layered, in space group 31, which breaks inversion symmetry. (c) The bulk Brillouin zone (BZ) and (001) surface BZ of \tai\ with high-symmetry points marked in red. (d) The electronic band structure of TaIrTe$_4$ along high-symmetry lines. There is a band crossing in the region near $\Gamma$, with Weyl points off $\Gamma-\textrm{S}$ (blue box). (e) Cartoon illustration of the constant-energy contour at $E_{\textrm{B}}=E_{\textrm{W}}$ and $k_z = 0$, with bulk electron and hole pockets which intersect to form Weyl points. A detailed calculation shows that there are in total four Type II Weyl points (blue and red circles) \cite{Koepernik}. (f) Energy-dispersion calculation along a pair of Weyl points in the $k_{y}$ direction, marked by the orange line in (e). The Weyl points and Fermi arcs live at $\sim 0.1$ eV above $E_{\textrm{F}}$, requiring the use of pump-probe ARPES to directly access the unoccupied band structure to demonstrate a Weyl semimetal.}
\end{figure}

We first reiterate well-known arguments that four is the minimum number of Weyl points allowed in a \tr\ invariant Weyl semimetal. A Weyl point is associated with a chiral charge, directly related to the chirality of the associated emergent Weyl fermion. It can be shown that for any given band the sum of all chiral charges in the Brillouin zone is zero. Further, under $\mathcal{T}$ a Weyl point of a given chiral charge at $k$ is mapped to another Weyl point of the same chiral charge at $-k$. This operation of $\mathcal{T}$ on a chiral charge is illustrated in Fig. \ref{titFig1}a on the blue Weyl points with $+1$ chiral charge (the same arrow applies for the red Weyl points but is not drawn explicitly). Now, if an $\mathcal{I}$ breaking Weyl semimetal has no additional symmetries which produce copies of Weyl points, then the minimum number of Weyl points is fixed by \tr\ symmetry and the requirement that total chiral charge vanish. In the simplest case, \tr\ will produce two copies of Weyl points of chiral charge $+1$, as shown in Fig. \ref{titFig1}a. To balance these out the system must have two chiral charges of $-1$, also related by \tr. In this way, the minimum number of Weyl points in a \tr\ invariant Weyl semimetal is four. This simple scenario is realized in TaIrTe$_4$. The crystal structure of TaIrTe$_4$ is described by space group 31 ($Pmn2_1$), lattice constants $a=3.77$ \AA, $b=12.421$ \AA, and $c=13.184$ \AA, with layered crystal structure, see Fig. \ref{titFig1}b. We note that \tai\ takes the same space group as Mo$_x$W$_{1-x}$Te$_2$, has a unit cell doubled along $b$. To study how the Weyl points emerge in TaIrTe$_4$ we present the electronic band structure along various high-symmetry directions, see Brillouin zone and \textit{ab initio} calculation in Fig. \ref{titFig1}c, d. Enclosed by the rectangular box along $\Gamma-\textrm{S}$ is a crossing region between the bulk conduction and valence bands that gives rise to Weyl points. A more detailed calculation shows that the Weyl points have tilted over, or Type II, Weyl cones and that they live above the Fermi level, \ef\ at $k_z = 0$ \cite{Koepernik}. A cartoon schematic of the resulting constant energy contour at the energy of the Weyl points, $E_\textrm{B} = E_\textrm{W}$, is shown in Fig. \ref{titFig1}e. The electron and hole pockets form Type II Weyl cones where they touch (red and blue marks). In this way, TaIrTe$_4$ has four Weyl points, the minimal number allowed in an $\mathcal{I}$ breaking Weyl semimetal. The overall electronic structure of \tai\ near \ef\ is similar to \mwt, but we note that the role of the electron and hole pockets is reversed in \tai\ relative to \mwt. Furthermore, \mwt\ has eight Weyl points, so it is not minimal, and \tai\ also hosts larger Fermi arc surface states than \mwt. To study the expected Fermi arcs in TaIrTe$_4$, we present an energy-dispersion cut along a pair of projected Weyl points along $k_{y}$, Fig.\ref{titFig1}f. We clearly observe a large single Fermi arc surface state at $\sim 0.1$ eV above the Fermi level that is $\sim 0.25$ \AA$^{-1}$ long and connecting a pair of $\pm{1}$ chiral charged Weyl points along $k_y$. In this way, TaIrTe$_4$ provides a minimal Weyl semimetal with large Fermi arcs. Lastly, we note that the Weyl points and Fermi arcs live well above the Fermi level, making them inaccessible by conventional ARPES.

\section{Unoccupied band structure of TaIrTe$_4$ by pump-probe ARPES}

\begin{figure*}
\centering
\includegraphics[width=15cm, trim={1in 4.5in 1in 1in}, clip]{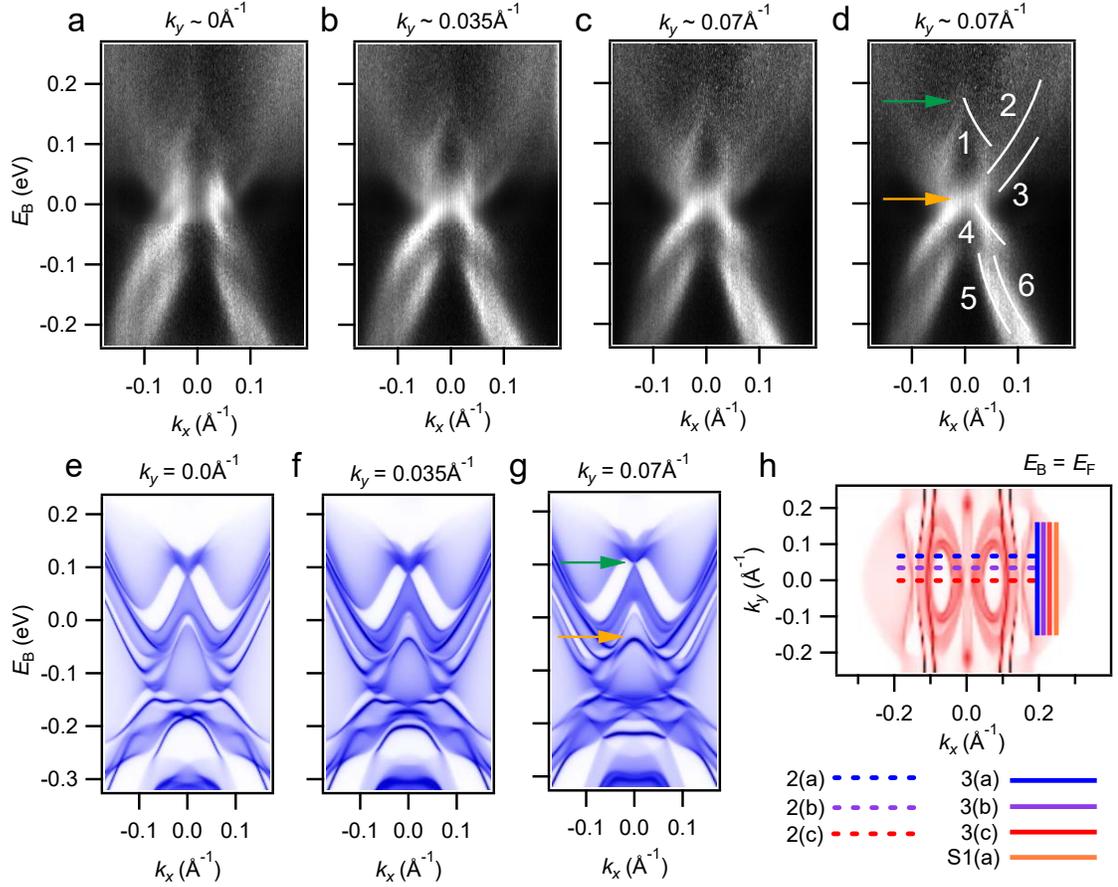}
%\end{figure*}
%
%\begin{figure*}
\caption{\label{titFig2}\textbf{Unoccupied electronic structure of \tai.} (a-c) Pump-probe ARPES dispersion maps of \tai, showing dispersion above \ef\ at fixed $k_y$ near $\bar{\Gamma}$. (d) Same as (c) but with key features marked. (e-f) \textit{Ab initio} calculation of \tai. The data is captured well by calculation, but the sample appears to be hole doped by $\sim 50$ meV, comparing the green and orange arrows in (d) and (g). (h) Calculation of the nominal Fermi surface, showing weak dispersion along $k_y$ near $\bar{\Gamma}$, consistent with the data. All cuts in Fig. ~\ref{titFig2}, Fig. ~\ref{titFig3} and Fig. ~\ref{titSFig1} are marked (blue lines).}
\end{figure*}

Next, we use pump-probe ARPES to directly access the unoccupied band structure of \tai\ up to $E_\textrm{B} > 0.2$ eV and we find excellent agreement with calculation. In our experiment, we use a $1.48$eV pump laser pulse to excite electrons into low-lying states above the Fermi level, followed by a $5.92$eV probe laser pulse to perform photoemission \cite{IshidaMethods}. We study $E_\textrm{B}$-$k_x$ cuts near $\bar{\Gamma}$, Fig. \ref{titFig2}a-c, with key features marked by guides to the eye in Fig. \ref{titFig2}d. Above the Fermi level, we see a crossing-like feature near $E_\textrm{B} \sim 0.15$ eV, labelled 1, and two electron-like bands, 2 and 3, extending out above \ef. Below the Fermi level, we observe a general hole-like structure consisting of three bands, labelled 4-6. As we shift $k_y$ off $\bar{\Gamma}$, we find little change in the spectrum, suggesting that the band structure is rather flat along $k_y$ near $\bar{\Gamma}$. However, we can observe that band 4 moves downward in energy and becomes more intense with increasing $k_y$. We find an excellent match between our ARPES data and \textit{ab initio} calculation, Fig. \ref{titFig2}e-g. Specifically, we identify the same 
crossing-like feature (green arrow) and top of band 4 (orange arrow). We can also track band 4 in $k_y$ in calculation and we find that the band moves down and becomes brighter as $k_y$ increases, in excellent agreement with the data. The electron-like structure of bands 2 and 3 and the hole-like structure of bands 5 and 6 are also both captured well by the calculation. Crucially, however, we notice a shift in energy between experiment and theory, showing that the sample is hole-doped by $\sim 0.05$ eV. Lastly, we plot a constant energy $k_x$-$k_y$ cut at $E_\textrm{B} = E_\textrm{F}$, where we see again that there is little dispersion along $k_y$ near $\bar{\Gamma}$, Fig. \ref{titFig2}h. We also indicate the locations of the $E_\textrm{B}$-$k_x$ cuts of Fig. ~\ref{titFig2} and the $E_\textrm{B}$-$k_y$ cuts of Fig. ~\ref{titFig3}, to be discussed below (blue lines). Our pump-probe ARPES measurements allow us to directly measure the electronic structure above \ef\ in \tai\ and we find excellent match with calculation.

\section{Evidence for a Weyl semimetal in TaIrTe$_4$}

Now we demonstrate that \tai\ is a Weyl semimetal by directly studying the unoccupied band structure to pinpoint Weyl cones and topological Fermi arcs. We study \eb-$k_y$ cuts where, based on calculation, we fix $k_x$ near the locations of the Weyl points, $k_x = k_\textrm{W} = 0.2 \textrm{\AA}^{-1}$, see Fig. \ref{titFig3}a-c, with key features marked by guides to the eye in Fig. \ref{titFig3}d. We observe two cone features, labelled 1 and 2, connected by a weak, rather flat arc feature, labelled 3. We find that the cones are most pronounced at $k_x \sim k_\textrm{W}$, but fade for larger $k_x$. Next, we pinpoint the Fermi arc as a small peak directly on the energy distribution curve (EDC) passing through $k_y = 0$, see the blue curve in Fig. \ref{titFig3}e, where the dotted black line is a fit to the surrounding features. We further track the arc candidate for $k_x$ moving away from $k_\textrm{W}$ and we find that the arc disperses slightly upwards, by about $\sim 10$ meV, see also Supplementary Fig. 1. This dispersion is consistent with a topological Fermi arc, which should connect the Weyl points and sweep upward with increasing $k_x$ \cite{NbPme}. We further pinpoint the upper Weyl cone on a momentum distribution curve (MDC) of the $k_x \sim k_\textrm{W}$ cut, Fig. \ref{titFig3}f. We find an excellent fit of the Weyl cone peaks to Lorentzians. Using this analysis, we can quantitatively track the dispersions of the Weyl cones and Fermi arc on the $k_x \sim k_\textrm{W}$ cut, Fig. \ref{titFig3}g and Supplementary Fig. 2. We note that for the upper Weyl cone we track the bands by Lorentzian fits on the MDC. However, for the Fermi arc and lower Weyl cone, the relatively flat dispersion requires us to track the bands in the EDCs. The EDC peak is challenging to fit, in part because the population distribution is strongly dependent on binding energy for a pump-probe ARPES spectrum. As a result, we track the Fermi arc and lower Weyl cone through a naive quadratic fit of the band peaks, again see Fig. \ref{titFig3}g. We find that the peak trains are nearly linear, see also Supplementary Fig. 2. Based on our pump-probe ARPES spectra, we propose that \tai\ hosts two pairs of Weyl points of chiral charge $\pm 1$ at $k_x = \pm k_\textrm{W}$, connected by Fermi arcs. This particular structure of two Weyl cones connected by a Fermi arc is arguably the simplest possible, Fig. \ref{titFig3}h. We compare our results to calculation, Fig. \ref{titFig3}i-k. We can easily match the Weyl cones, the Fermi arc and an upper electron-like band, labelled 5 in Fig. \ref{titFig3}d. However, we note that from calculation we expect bands 1 and 4 to attach to form a single band, while in our data they appear to be disconnected. We suggest that this discrepancy may arise because photoemission from part of the band is suppressed by low cross-section at the photon energy used in our measurement. In addition, we do not observe good agreement with the lower feature labelled 6 in our calculation, suggesting that this intensity may arise as an artifact of our measurement. At the same time, we consistently observe the broad featureless intensity below the Fermi level in both theory and experiment. Crucially, again we find a mismatch in the Fermi level. In particular, the Weyl points are expected at $E_\textrm{B} \sim 0.1$ eV, but we find the Weyl points at $E_\textrm{B} \sim 0.07$ eV. We note that this sample was grown in a different batch than the sample of Fig. \ref{titFig2} and a comparison with calculation suggests that the second sample is electron doped by $\sim 30$ meV, in contrast to a $\sim 50$ meV hole doping in the first sample. We propose that the difference in doping of the two samples may arise because they were grown under slightly varying conditions. Lastly, we note that the $k_y$ position of the Weyl points shows excellent agreement in theory and experiment. In summary, we observe an arc which (1) terminates at the locations of two Weyl points; (2) appears where expected in momentum space, based on calculation; and (3) disperses upward with $k_x$, as expected from calculation. The cones (1) are gapless at a specific $k_x \sim k_\textrm{W}$; (2) fade for larger $k_x$; (3) appear where expected, based on calculation; (4) are connected by the arc; (5) show up in pairs only on $k_x \sim k_\textrm{W}$, so that there are four in the entire Brillouin zone. This provides strong evidence that \tai\ is a minimal Weyl semimetal with four Weyl points.

\section{Discussion}

We compare \tai\ with other Weyl semimetals and consider our results in the context of general topological theory. Weyl semimetals known to date in experiment host a greater number of Weyl points than \tai. In particular, the well-explored TaAs family of Weyl semimetals hosts 24 Weyl points and \mwt\ hosts eight Weyl points \cite{ARCMP, PS}. We plot the configuration of Weyl points for TaAs, \mwt, and \tai, where red and blue circles denote Weyl points of opposite chirality, Fig. \ref{titFig4}a-c. It is also interesting to note that the length of the Fermi arc in \tai\ is much longer as a fraction of the Brillouin zone than that of TaAs or \mwt, which can be seen clearly in the projections of the Weyl points on the (001) surface of all three systems, Fig. \ref{titFig4}d-f. We see that our discovery of a Weyl semimetal in \tai\ provides the first example of a minimal \inv\ breaking, \tr\ invariant Weyl semimetal. One immediate application of our results is that \tai\ in pump-probe ARPES may provide a platform to observe the time dynamics of carrier relaxation in a Weyl semimetal. More broadly, our results suggest that \tai\ holds promise as a simpler material platform for studying novel properties of Weyl semimetals in transport and applying them in devices. 

\section{Materials and methods}

%Preliminary ARPES measurements were carried out using a home-built laser-based ARPES setup at the Ames Laboratory in Ames, Iowa, United States. Measurements were conducted under ultra-high vacuum and at temperatures $\leq 10$K. The angular and energy resolution was better than $0.1^{\circ}$ and $5$ meV, respectively, with photon energies from 5.77 eV to 6.67 eV \cite{KaminskiMethods}.

\subsection{Pump-probe angle-resolved photoemission spectroscopy}

Pump-probe ARPES measurements were carried out using a hemispherical electron analyzer and a mode-locked Ti:Sapphire laser system that delivered $1.48$ eV pump and $5.92$ eV probe pulses at a repetition rate of $250$ kHz \cite{IshidaMethods}. The system is state-of-the start, with a demonstrated energy resolution of $10.5$ meV, the highest among any existing femtosecond pump-probe setup to date \cite{IshidaSciRepCuprates}. The time and energy resolution used in the present measurements were $300$ fs and $15$ meV, respectively. The spot diameters of the pump and probe lasers at the sample were $250\ \mu$m and $85\ \mu$m, respectively. The delay time between the pump and probe pulses was $\sim 106$ fs. Measurements were carried out at pressures $<5 \times 10^{-11}$ Torr and temperatures $\sim8$ K.

\subsection{Single crystal growth and characterization}

For growth of \tai\ single crystals, all the used elements were stored in an argon-filled glovebox with moisture and oxygen levels less than $0.1$ ppm and all manipulations were carried out in the glovebox. \tai\ single crystals were synthesized by solid state reaction with the help of Te flux. Ta powder (99.99\%), Ir powder (99.999\%) and a Te lump (99.999\%) with an atomic ratio of Ta/Ir/Te = 1:1:12, purchased from Sigma-Aldrich (Singapore), were loaded in a quartz tube and then flame-sealed under a vacuum of $10^{-6}$ Torr. The quartz tube was placed in a tube furnace, slowly heated up to 1000$^\circ$C and held for 100 h, then allowed to cool to 600$^\circ$C at a rate of 0.8$^\circ$C/h, and finally allowed to cool down to room temperature. The shiny, needle-shaped TaIrTe$_4$ single crystals, see Supplementary Fig. 3a, were obtained from the product and displayed a layered structure, confirmed by the optical micrograph, Supplementary Fig. 3b, and SEM images, Supplementary Fig. 3c. The EDX spectrum displays an atomic ratio Ta:Ir:Te of 1.00:1.13(3):3.89(6), consistent with the composition of \tai, Supplementary Fig. 3d.

\subsection{\textit{Ab initio} band structure calculations}

We computed electronic structures using the projector augmented wave method \cite{PAW1,PAW2} as implemented in the VASP \cite{TransitionMetals, PlaneWaves1, PlaneWaves2} package within the generalized gradient approximation (GGA) \cite{GGA} schemes. Experimental lattice constants were used \cite{TaIrTe4Structure}. A 15 $\times$ 7 $\times$ 7 Monkhorst-Pack $k$-point mesh was used in the computations with a cutoff energy of 400 eV. The spin-orbit coupling (SOC) effects were included self-consistently. To calculate the bulk and surface electronic structures, we constructed first-principles tight-binding model Hamilton, where the tight-binding model matrix elements are calculated by projecting onto the Wannier orbitals \cite{MLWF1, MLWF2, Wannier90}, which use the VASP2WANNIER90 interface \cite{MLWF3}. We used Ta $d$, Ir $d$, and Te $p$ orbitals to construct Wannier functions and without perform the procedure for maximizing localization.

\clearpage

\clearpage

\begin{figure*}
\centering
\includegraphics[width=15cm, trim={1.1in 1in 1in 1in}, clip]{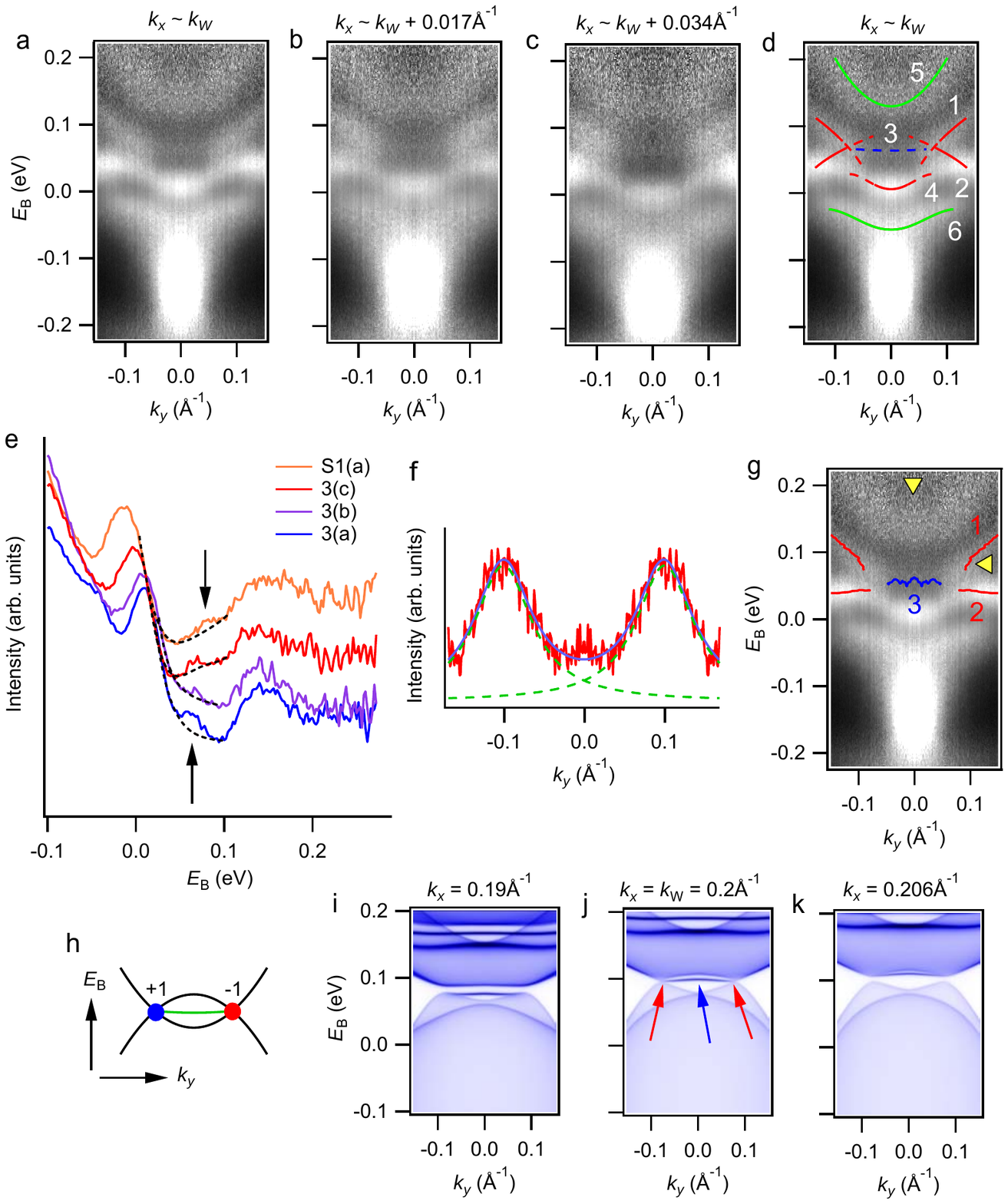}
\end{figure*}
\clearpage

\begin{figure*}
\caption{\label{titFig3}\textbf{Weyl points and Fermi arcs above the Fermi level in \tai.} (a-c) Pump probe ARPES spectra of \tai, showing dispersion above \ef\ at fixed $k_x$ expected to be near the Weyl points. (d) Same spectrum as (a) but with key features marked. The Weyl cone candidates are labelled 1 and 2, the Fermi arc candidate is labelled 3. (e) Energy distribution curves (EDCs) through the Fermi arc at $k_x \sim k_\textrm{W}$, $k_\textrm{W} + 0.017\AA^{-1}$, $k_\textrm{W} + 0.034\AA^{-1}$ and $k_\textrm{W} + 0.045\AA^{-1}$. The dotted black lines are fits to the surrounding features, to emphasize the Fermi arc peak, marked by the black arrows. We observe signatures of the upward dispersion of the Fermi arc with increasing $k_x$, consistent with \textit{ab initio} calculations and basic topological theory. (f) An MDC with two large peaks corresponding to the upper Weyl cones. The dotted green lines show an excellent fit of the peaks to Lorentzian functions. (g) Same spectrum as (a), but with key features marked by a quantitative fits to EDCs and MDCs. The yellow arrows correspond to the location of the EDCs in (e) and the MDCs in (f). (h) Cartoon of the cones and arc observed in the data, showing what is perhaps the simplest configuration of Weyl points and Fermi arcs that can exist in any Weyl semimetal. (i-k) \textit{Ab initio} calculation of \tai\ showing the Weyl points (red arrows) and Fermi arcs (blue arrow). The excellent agreement with calculation demonstrates that we have observed a Weyl semimetal in \tai.}
\end{figure*}
\clearpage

\begin{figure*}
\centering
\includegraphics[width=15cm, trim={60 370 75 60}, clip]{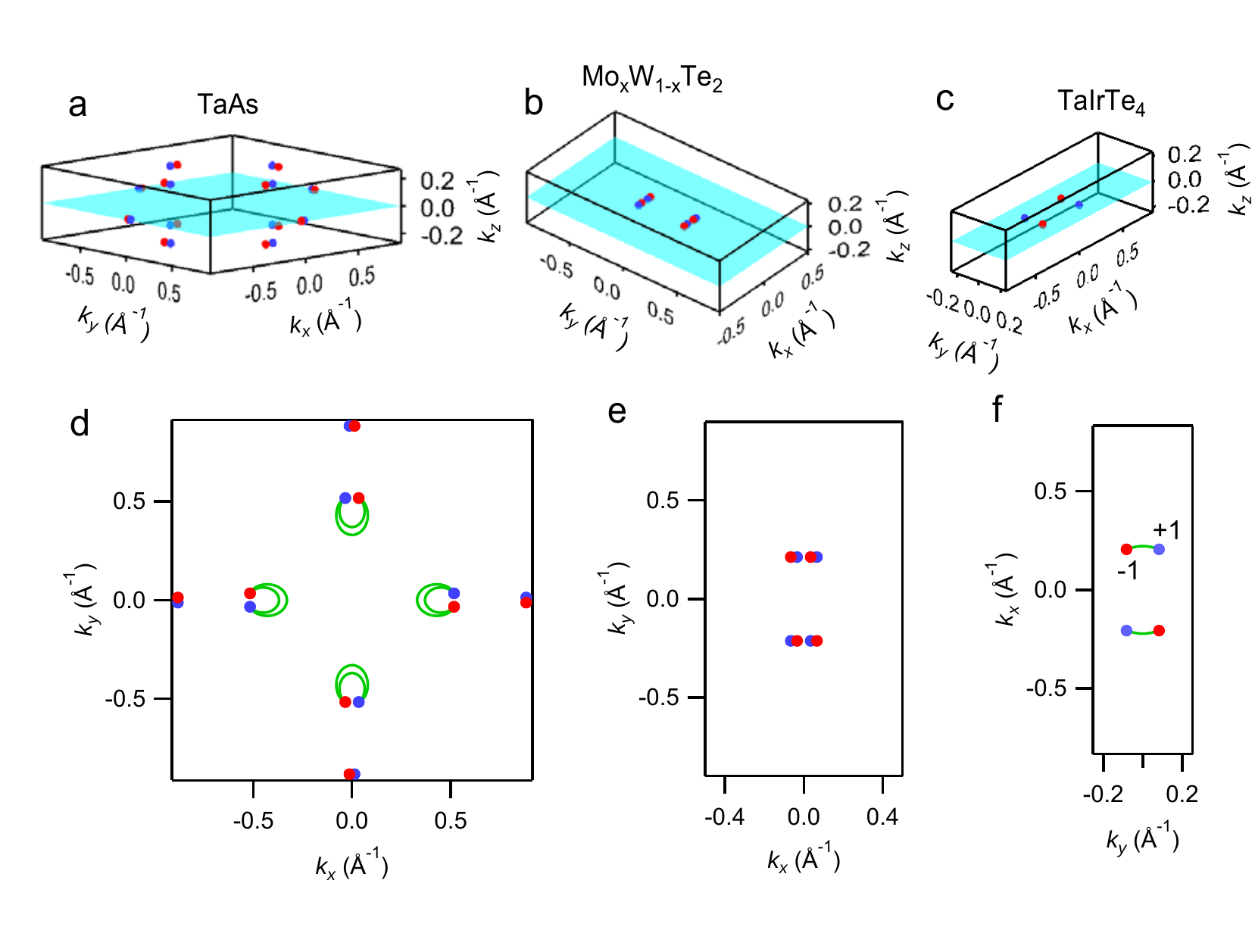}
%\end{figure*}
%\clearpage
%
%\begin{figure*}
\caption{\label{titFig4}\textbf{Comparison of Weyl point configurations.} Weyl points, plotted in red and blue for opposite chiralities, for (a) TaAs, with 24 Weyl points, (b) \mwt, with eight Weyl points and (c) \tai, with the minimal number, only four Weyl points, making \tai\ a minimal \tr\ invariant Weyl semimetal. The $k_z = 0$ plane is marked in cyan. (d-f) The projection of the Weyl points on the (001) surface, with a cartoon of the topological Fermi arcs. The black frame marks the first Brillouin zone. The length of the Fermi arcs in \tai\ is longer as a fraction of the Brillouin zone as compared to TaAs and \mwt.}
\end{figure*}
\clearpage

%\documentclass[aps,prl,preprint,nopacs,superscriptaddress]{revtex4}
%\usepackage{amsmath}
%\usepackage{amssymb}
%\usepackage{graphicx}
%\usepackage{hyperref}
%\pagestyle{headings}
%\newcommand{\beq}{\begin{equation*}}
%\newcommand{\eeq}{\end{equation*}}
%\newcommand{\mowte}{Mo$_x$W$_{1-x}$Te$_2$}
%\newcommand{\tf}{Mo$_{0.25}$W$_{0.75}$Te$_2$}
%\newcommand{\tai}{TaIrTe$_4$}
%\renewcommand{\thefigure}{\arabic{figure}}

%\makeatletter
%\renewcommand{\fnum@figure}{Supplementary Figure \thefigure}
%\makeatother

%\begin{document}

\begin{figure*}
\centering
\includegraphics[trim={5.5cm, 18cm, 5.5cm, 2cm},clip, width=14cm]{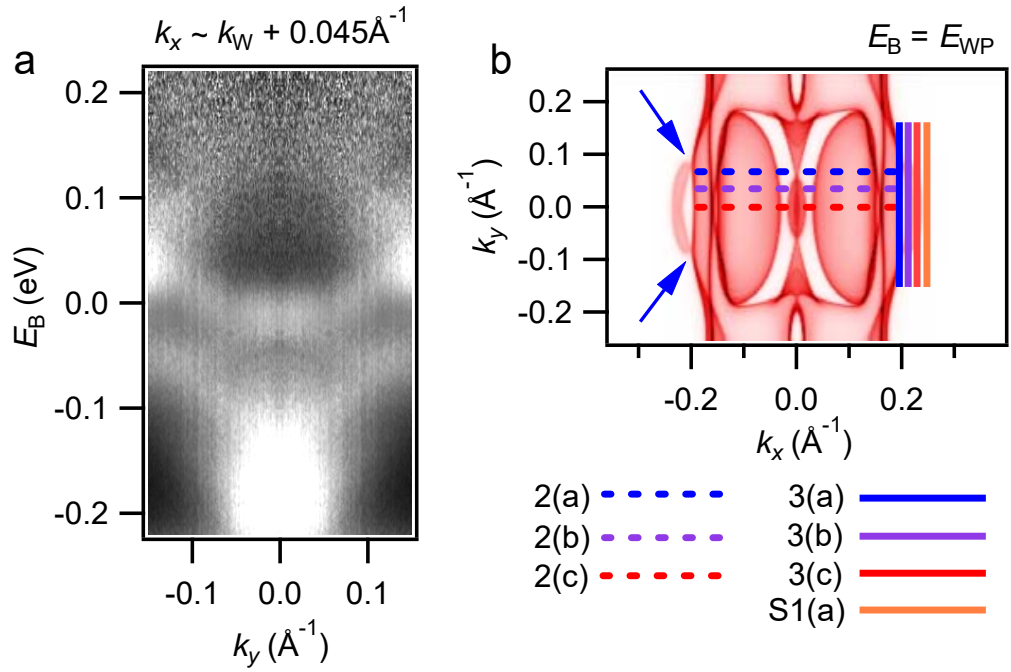}
\caption{\label{titSFig1} \textbf{Systematics above $E_\textrm{\normalfont{F}}$.} (a) Cut along $k_y$ at $k_x \sim k_\textrm{W} + 0.045\textrm{\AA}^{-1}$, showing the Fermi arc receding into the upper Weyl cone, consistent with \textit{ab initio} calculation and basic band theory, see also main text Fig. 3e. (b) Calculated Fermi surface above the Fermi level, at $E = E_\textrm{W}$, with Weyl points marked by the arrows and the locations of the measured cuts marked by the solid and dashed lines.}
\end{figure*}

\clearpage

\begin{figure*}
\centering
\includegraphics[trim={3cm, 8cm, 3cm, 3cm},clip, width=15cm]{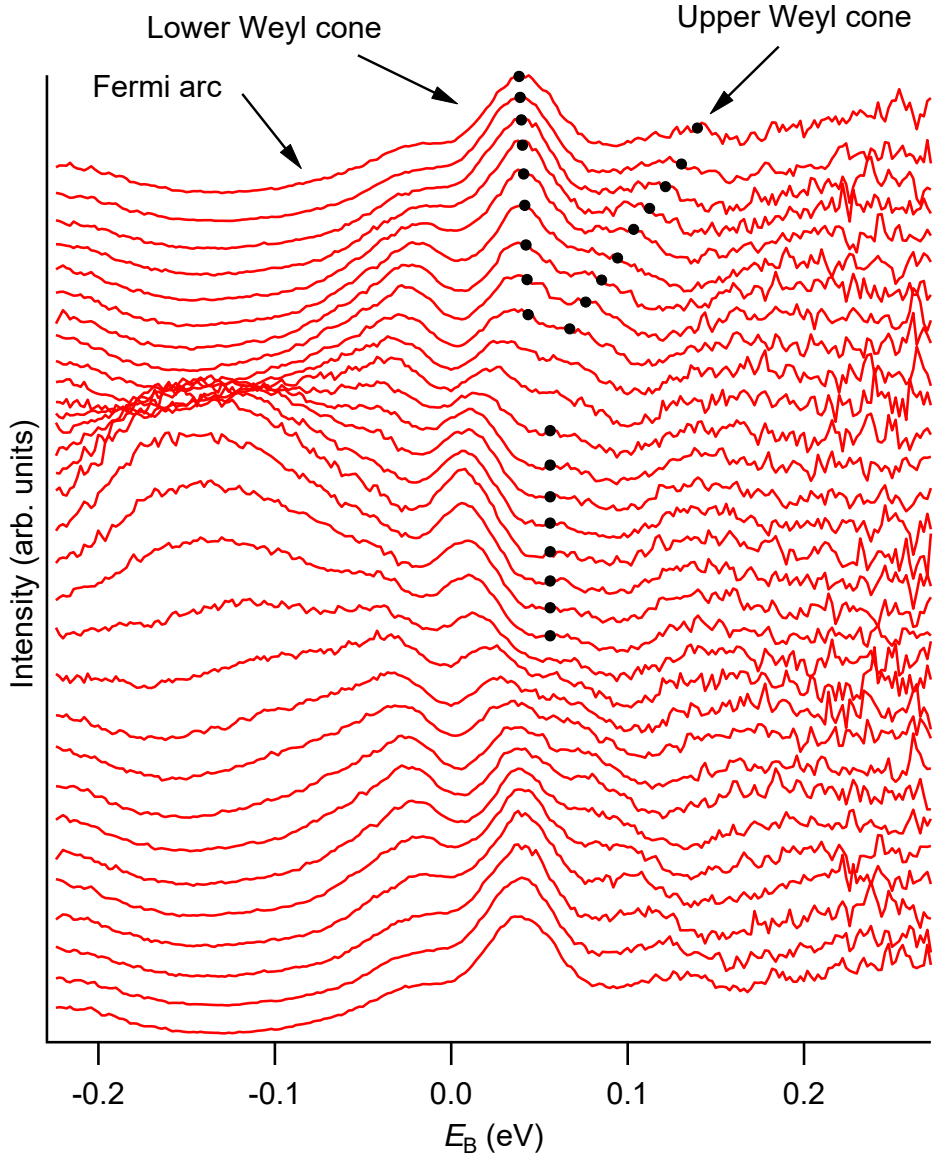}
\caption{\label{titSFig2} \textbf{Systematic EDC stack at $k_x$ approximately equal to $k_\textrm{\normalfont{W}}$.} A systematic stack of energy distribution curves (EDCs) of the $E_\textrm{B}$-$k_y$ cuts at $k_x \sim k_\textrm{W}$, Fig. 3a in the main text. The three trains of black dots indicate linear fits of three sets of peaks obtained by fitting EDCs and momentum distribution curves (MDCs). For the upper Weyl cone we plotted fits of the MDCs by Lorentzians, as illustrated in Fig. 3f of the main text. For the lower Weyl cone and the Fermi arc, we plotted fits of the EDCs by naive quadratics in the region of the peak. We can clearly follow the train of peaks. Near the band crossing it becomes difficult to perform the fit because there are several bands lying close together, so we omit this region from our fitting.}
\end{figure*}

\clearpage

\begin{figure*}
\centering
\includegraphics[width=15cm]{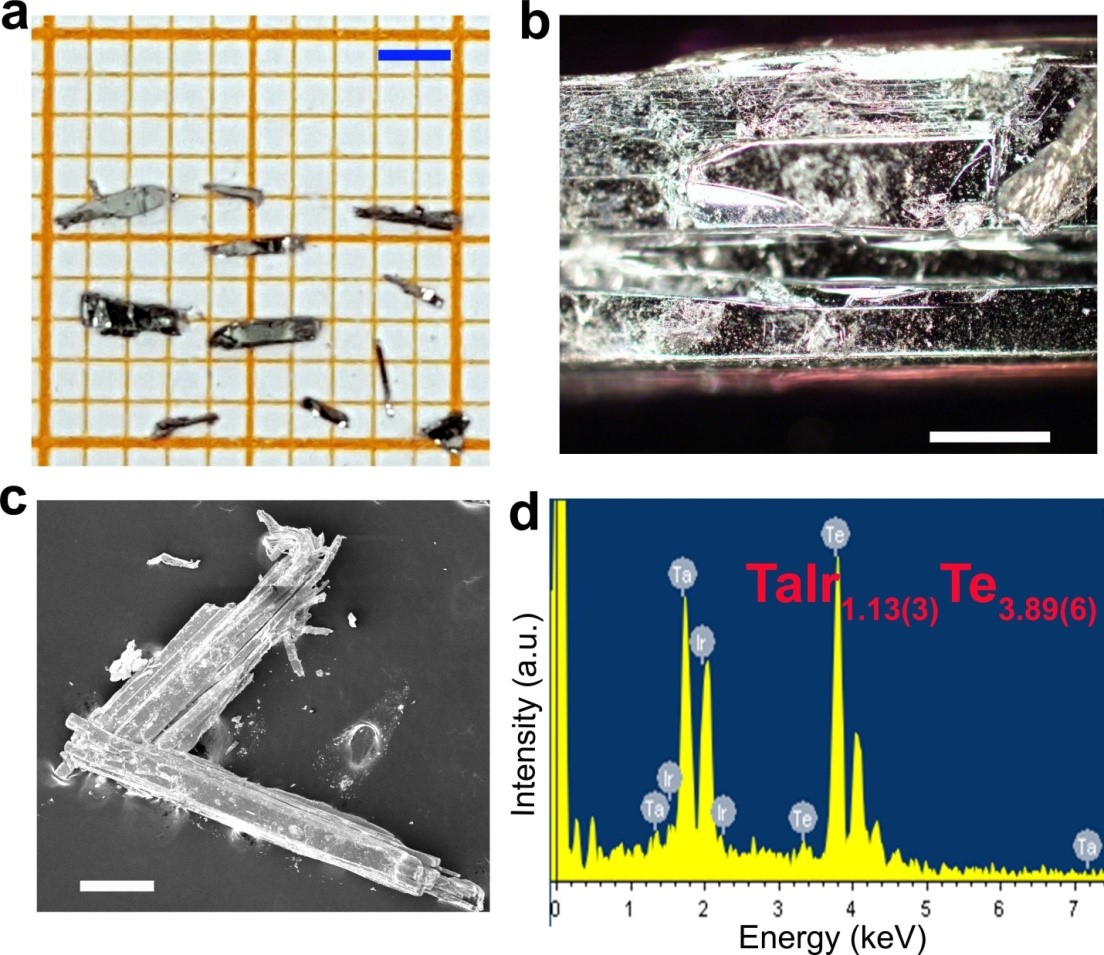}
\caption{\label{Fig2} \textbf{Characterization of \tai\ single crystals.} (a) Photo of as-grown \tai\ single crystals. (b) Optical micrograph of the surface of a \tai\ single crystal. (c) SEM image and (d) EDX spectrum of \tai\ single crystals, with the elemental ratio determined from the EDX spectrum. The scale bars are: (a) 2 mm, (b) 100 $\mu$m and (c) 200 $\mu$m.}
\end{figure*}

\clearpage

%\cleardoublepage
\ifdefined\phantomsection
  \phantomsection  % makes hyperref recognize this section properly for pdf link
\else
\fi
\addcontentsline{toc}{section}{Bibliography}

{\singlespacing

}

%\end{document}

%\end{document}

%% file: ch-chiral/ch-chiral.tex
\chapter{Topological chiral crystals with helicoid-arc quantum states}
\label{ch:chiral}

{\singlespacing
\begin{chapquote}{Belle and Sebastian, \textit{The Boy with the Arab Strap}}
Colour my life\\
With the chaos of trouble\\
\ 
\end{chapquote}}

%What did you learn from your time
%in the solitary cell of your mind?

%\documentclass[prl,twocolumn, aps,superscriptaddress,floatfix,nofootinbib]{revtex4}

%\documentclass[aps,prb,preprint,nopacs,superscriptaddress,]{revtex4}

%\usepackage{graphicx}
%\usepackage{verbatim}
%\usepackage{mathrsfs}
%\pagestyle{headings}
%\usepackage{gensymb}
%\usepackage{color}
%\usepackage{ulem}
%\usepackage{times}
%\usepackage[super]{natbib}

%\usepackage[demo]{graphicx}
%\usepackage{comment}
%\excludecomment{figure}
%\documentclass[pra,aps,showpacs,groupedaddress,preprint]{revtex4}
%\documentclass[pra,aps,showpacs,groupedaddress]{revtex4}
%\setlength{\oddsidemargin}{.5in}
%\setlength{\textwidth}{5.5in}
%\usepackage{etoolbox}
%\pretocmd{\abstractname}{\newpage}{}{}

%\usepackage{amsmath,amsfonts,amssymb}
%\usepackage{wrapfig}
%\usepackage{graphicx}
%\usepackage{bbm}
%\usepackage{graphics}k

%\newcommand{\beginsupplement}{%
%       \setcounter{table}{0}
%        \renewcommand{\thetable}{S\arabic{table}}%
%        \setcounter{figure}{0}
%        \renewcommand{\thefigure}{S\arabic{figure}}%
%     }

\newcommand{\bs}[1]{{\boldsymbol{#1}}}
\newcommand{\red}[1]{{\textcolor{red}{#1}}}
\newcommand{\blue}[1]{{\textcolor{blue}{#1}}}
\newcommand{\magenta}[1]{{\textcolor{magenta}{#1}}}
\newcommand{\green}[1]{{\textcolor[rgb]{0,0.5,0}{#1}}}

\renewcommand{\ef}{$E_\textrm{F}$}
\newcommand{\degree}{^\circ}

\newcommand{\pana}{a}
\newcommand{\panb}{b}
\newcommand{\panc}{c}
\newcommand{\pand}{d}
\newcommand{\pane}{e}
\newcommand{\panf}{f}
\newcommand{\pang}{g}
\newcommand{\panh}{h}
\newcommand{\pani}{i}
\newcommand{\panj}{j}

%\setcitestyle{round}
\def\3{2.5in}    %used for figure widths
\def\2{2.5in}
\def\4{3.0in}\def \beq {\begin{equation}}
\def \eeq {\end{equation}}
%\pagestyle{headings}

%\begin{document}

%\title{Topological chiral crystals with helicoid Fermi arc states\\(2018-08-10995)}

%\maketitle

\noindent This chapter is based on the article, \textit{Topological chiral crystals with helicoid-arc quantum states} by Daniel S. Sanchez*, Ilya Belopolski*, Tyler Cochran* \textit{et al}., \textit{Nature} {\bf 567}, 500 (2019), available at \href{https://www.nature.com/articles/s41586-019-1037-2}{https://www.nature.com/articles/s41586-019-1037-2}.\\

\lettrine[lines=3]{T}{he} unusual behaviour of electrons in materials lays the foundation for modern electronic and information technology \cite{natnews, kmoore, revHK, revQZ, rev4, rev6, rev7}. Materials with topological electronic properties have been proposed as the next frontier, but much remains to be understood. Here we report the experimental observation of a topological chiral crystal in cobalt silicide, CoSi, belonging to the RhSi family. We show that this material hosts a novel phase of matter which exhibits nearly ideal topological surface properties that emerge as a direct consequence of the crystals' structural chirality. In particular, CoSi materials realise the simplest possible topological conductor. We demonstrate that the electrons on the surface of this crystal show a highly unusual helicoid structure that spirals around two high-symmetry momenta. Such helicoid arcs experimentally show projected topological charges of $\pm{2}$, which arise from the unconventional chiral fermions in the bulk. Remarkably, these topological conductors exhibit Fermi arcs which are of length $\pi$, stretching across the entire Brillouin zone and orders of magnitude larger than those found in conventional Weyl semimetals. Our results demonstrate an electronically topological chiral crystal with helicoid Fermi arc surface states. The exotic chiral fermion physics realised in these materials can be used to detect a quantised photogalvanic current or the chiral magnetic effect in future devices.

\section{Introduction}

The discovery of topological insulators has inspired the search for a wide variety of topological conductors \cite{natnews, kmoore, revHK, revQZ, rev4, rev6, rev7}. One example of a topological conductor is the Weyl semimetal (WSM), featuring emergent Weyl fermions as low-energy excitations of the crystal. These Weyl fermions are associated with momentum-space topological chiral charges that live at two-fold degenerate band crossings \cite{natnews, kmoore, rev4, rev6, rev7, NielsenNinomiya1, UniverseinHe, Topo.gapless.phase, Wan, Weyl-Multilayer, Chiral-anomaly}. In general, such emergent topological chiral fermions may appear in a variety of types including not only two-fold degenerate Weyl fermions \cite{rev4, rev7, rev6, NielsenNinomiya1, UniverseinHe, Topo.gapless.phase, Wan, Weyl-Multilayer, Chiral-anomaly, HgCr2Se4, typeIIWeyl}, but also unconventional higher-fold fermions \cite{unconventionalWeyl, RhSi, CoSi}. Recently, a few non-centrosymmetric crystals were identified where a band inversion gives rise to a WSM state \cite{TaAs1, TaAs2, ARPES-TaAs1, ARPES-TaAs2, rev4, rev7, rev6}. However, these materials suffer from several drawbacks: a large number of Weyl fermions, Weyl fermions close to each other in momentum space, too many trivial electronic states near the Fermi level, and small Fermi arcs which are less topologically robust. In order to thoroughly explore and utilise the robust and unusual quantum phenomena induced by chiral fermions, novel topological conductors with near-ideal electronic properties are needed \cite{natnews, kmoore}.

A different approach toward searching for ideal topological conductors is to examine crystalline symmetries, which can also lead to topological band crossings \cite{revHK, revQZ, rev6}. For instance, it has been shown that non-symmorphic symmetries can guarantee the existence of band crossings for certain electron fillings \cite{unconventionalWeyl, filling_constraint1, Ben1}. As another example, we might consider structurally chiral crystals, defined as having a space group that has no inversion symmetry, no mirror symmetries and no roto-inversion symmetries. Structurally chiral crystals are expected to host a variety of topological band crossings which are guaranteed to be pinned to time-reversal invariant momenta (TRIMs) \cite{KramersWeyl, Manes}. Moreover, structurally chiral topological crystals naturally give rise to a quantised circular photogalvanic current, the chiral magnetic effect and other novel transport and optics effects forbidden in known topological conductors, such as TaAs \cite{rev6, TaAs1, TaAs2, ARPES-TaAs1, ARPES-TaAs2}.

Incorporating these paradigms into a broader search, we have studied various candidate nonmagnetic and magnetic conductors, such as Ge$_3$Sn, Mn$_3$Sn, the LuPtSb family, GaMnCo$_2$, HgCr$_2$Se$_4$, Co$_3$Sn$_2$S$_2$, Fe$_3$Sn$_2$ and the CoSi family, with advanced spectroscopic techniques. Many of the materials exhibit either large co-existing trivial bulk Fermi surfaces or surface reconstruction masking topological states. And as often is the case with surface-sensitive techniques, the experimentally realised surface potential associated with a cleaved crystal may or may not allow these unusual electronic states to be observed. Thus, despite the new search paradigms, the discovery of topological materials that are suitable for spectroscopic experiments has remained a significant challenge. Of all the materials we explored, we observed that the $X$Si ($X=$ Co, Rh) family of chiral crystals comes close to the experimental realization of the sought-after ideal topological conductor. Here we report high-resolution angle-resolved photoemission spectroscopy (ARPES) measurements in combination with state-of-the-art \textit{ab initio} calculations to demonstrate novel topological chiral crystals in CoSi and RhSi. These chiral crystals approach ideal topological conductors because of their large Fermi arcs and because they host the minimum non-zero number of chiral fermions---topological properties which we experimentally visualise for the first time.

%Figure1
The $X$Si ($X=$ Co, Rh) family of materials crystallises in a structurally chiral cubic lattice, space group $P2_{1}3$, No. 198 (Fig.~\ref{Fig1}\pana). We confirmed the chiral crystal structure of our CoSi samples by single crystal X-ray diffraction (XRD; Fig.~\ref{Fig1}\panb; Extended Data Table~\ref{ExtTab1}), with associated 3D Fourier map (Fig.~\ref{Fig1}\panc). We found a Flack factor of $\sim91\%$, which indicates that our samples are predominantly of a single structural chirality. \textit{Ab initio} electronic bulk band structure calculations predict that both chiral crystals exhibit a 3-fold degeneracy at $\Gamma$ near the Fermi level, \ef\ (Fig.~\ref{Fig1}\pand). This degeneracy is described by a low-energy Hamiltonian which exhibits a 3-fold fermion associated with Chern number $+2$ \cite{RhSi, CoSi}. We refer to this Chern number as a chiral charge, a usage of the term ``chiral'' which is distinct from the notion of structural chirality defined above and which also motivates our use of the term ``chiral fermion''  to describe these topological band crossings. The $R$ point hosts a 4-fold degeneracy corresponding to a 4-fold fermion with Chern number $-2$. These two higher-fold chiral fermions are pinned to opposite TRIMs and are consequently constrained to be maximally separated in momentum space (Fig.~\ref{Fig1}\pane), suggesting that CoSi might provide a near-ideal platform for accessing topological phenomena using a variety of techniques. The hole pocket at $M$ is topologically trivial at the band relevant for low-energy physics, but it is well-separated in momentum space from the $\Gamma$ and $R$ topological crossings. It is not expected to affect topological transport, such as the chiral anomaly ($\sigma_{xx}$ is not topological transport). The two higher-fold chiral fermions lead to a net Chern number of zero in the entire bulk Brillouin zone (BZ), as expected from basic considerations \cite{NielsenNinomiya1}. We see that CoSi and RhSi also satisfy a key criterion for an ideal topological conductor, namely that they have only two chiral fermions in the bulk BZ, the minimum non-zero number allowed.

The two chiral fermions remain topologically non-trivial over a wide energy range. In particular, CoSi maintains constant-energy surfaces with non-zero Chern number over an energy window of 0.85 eV, while for RhSi this window is 1.3 eV (Fig.~\ref{Fig1}\pand). This prediction suggests that $X$Si satisfies another criterion for an ideal topological conductor---a large topologically non-trivial energy window. An \textit{ab initio} calculated Fermi surface shows that the projection of the higher-fold chiral fermions to the (001) surface results in a hole (electron) pocket at $\bar{\Gamma}$ ($\bar{M}$) with Chern number $+2$ ($-2$; Fig.~\ref{Fig1}\panf; Extended Data Fig.~\ref{ExtFig2}g). As a result, we expect that $X$Si hosts Fermi arcs of length $\pi$ spanning the entire surface BZ, again suggesting that these materials may realise a near-ideal topological conductor.

\section{ARPES overview of CoSi}

%Figure 2
Using low-photon-energy ARPES, we experimentally study the (001) surface of CoSi and RhSi to reveal their surface electronic structure. For CoSi, the measured constant-energy contours show the following dominant features: two concentric contours around the $\bar{\Gamma}$ point, a faint contour at the $\bar{X}$ point, and long winding states extending along the $\bar{M}-\bar{\Gamma}-\bar{M}$ direction (Fig.~\ref{Fig2}\pana). Both the $\bar{\Gamma}$ and $\bar{X}$ pockets show a hole-like behaviour (Extended Data Fig.~\ref{ExtFig8_bulk_pockets}). The measured surface electronic structure for RhSi shows similar features (Fig.~\ref{Fig2}\panb; Extended Data Fig.~\ref{ExtFig1}). Using only our spectra, we first sketch the key features of the experimental Fermi surface for CoSi (Fig.~\ref{Fig2}\panc). Then, to better understand the $\textit{k}$-space trajectory of the long winding states in CoSi, we study Lorentzian fits to the momentum distribution curves (MDCs) of the ARPES spectrum. We plot the Lorentzian peak positions as the extracted band dispersion (Fig.~\ref{Fig2}\pand, \pane) and we find that the long winding states extend from the center of the BZ to the $\bar{M}$ pocket (Fig.~\ref{Fig2}\panf). To better understand the nature of these states, we perform an ARPES photon energy dependence and we find that the long winding states do not disperse as we vary the photon energy, suggesting that they are surface states (Extended Data Fig.~\ref{ExtFig6}). Moreover, we observe an overall agreement between the ARPES data and the $\textit{ab initio}$ calculated Fermi surface, where topological Fermi arcs connect the $\bar{\Gamma}$ and $\bar{M}$ pockets (Fig.~\ref{Fig1}\panf). Taken together, these results suggest that the long winding states observed in ARPES may be topological Fermi arcs.

% Indeed, such spectroscopic methods to determine the Chern number have become a well-accepted approach.

\section{Observation of Fermi arcs in CoSi}

Grounded in the framework of topological band theory, the bulk-boundary correspondence of chiral fermions makes it possible for ARPES (spectroscopic) measurements to determine the Chern numbers of a crystal by probing the surface state dispersion (Fig.~\ref{Fig3}\pana; Methods). Such spectroscopic methods to determine Chern numbers have become well-accepted in the field \cite{Nobel}. Using this approach, we provide two spectral signatures of Fermi arcs in CoSi. We first look at the dispersion of the candidate Fermi arcs along a pair of energy-momentum cuts on opposite sides of the $\bar{\Gamma}$ pocket, taken at fixed $+k_x$ (Cut I) and $-k_x$ (Cut II; Fig.~\ref{Fig2}\panf). In Cut I, we observe two right-moving chiral edge modes (Fig.~\ref{Fig3}\panb,\panc). Since the cut passes through two BZs (Fig.~\ref{Fig2}\panf), we associate one right-moving mode with each BZ. Next, we fit Lorentzian peaks to the MDCs and we find that the extracted dispersion again suggests two chiral edge modes, as well as a hole-like pocket near $\bar{X}$ (Fig.~\ref{Fig3}\pand). Along Cut II, we observe two left-moving chiral edge modes (Fig.~\ref{Fig3}\pane,\panf). Consequently, one chiral edge mode is observed for each measured surface BZ on Cuts I and II, but with opposite Fermi velocity direction. In this way, our ARPES spectra suggest that the number of chiral edge modes $n$ changes by $+2$ when the $k$-slice is swept from Cut I to Cut II. This again suggests that the long winding states are topological Fermi arcs. Moreover, these ARPES results imply that projected topological charge with net Chern number $+2$ lives near $\bar{\Gamma}$.

%Figure 3, part 2 (closed E-k cut)
Next we search for other Chern numbers encoded by the surface state band structure. We study an ARPES energy-momentum cut on a loop $\mathcal{P}$ enclosing $\bar{M}$ (Fig.~\ref{Fig2}\panf; Fig.~\ref{Fig4}a, inset). Again following the bulk-boundary correspondence (Methods), we aim to extract the Chern number of chiral fermions projecting on $\bar{M}$. The cut $\mathcal{P}$ shows two right-moving chiral edge modes dispersing towards \ef\ (Fig.~\ref{Fig4}\pana,\panb), suggesting a Chern number $-2$ on the associated bulk manifold. Furthermore, the \textit{ab initio} calculated surface spectral weight along $\mathcal{P}$ is consistent with our experimental results (Fig.~\ref{Fig4}\panc). Our ARPES spectra on Cut I, Cut II and $\mathcal{P}$ suggest that CoSi hosts a projected chiral charge of $+2$ at $\bar{\Gamma}$ with its partner chiral charge of $-2$ projecting on $\bar{M}$. This again provides evidence that the long winding states are a pair of topological Fermi arcs which traverse the surface BZ on a diagonal, connecting the $\bar{\Gamma}$ and $\bar{M}$ pockets. Our ARPES spectra on RhSi also provide evidence for gigantic topological Fermi arcs following a similar analysis (Extended Data Fig.~\ref{ExtFig1}). 

%Figure 4, part 1 (Helicoid Fermi arcs)
To further explore the topological properties of CoSi, we examine in greater detail the structure of the Fermi arcs near $\bar{M}$. We consider the dispersion on $\mathcal{P}$ (plotted as a magenta loop in Fig.~\ref{Fig4}d, inset) and we also extract a dispersion from Lorentzian fitting on a second, tighter circle (black loop; Extended Data Fig.~\ref{ExtFig_hel_fit}). We observe that as we decrease the binding energy (approach \ef), the extracted dispersion spirals in a clockwise fashion on both loops, suggesting that as a given $\textit{k}$ point traverses the loop, the energy of the state does not return to its initial value after a full cycle. Such a counter-intuitive electron dispersion directly signals a projected chiral charge at $\bar{M}$ (Fig.~\ref{Fig4}d). In fact, the extracted dispersion is characteristic of the helicoid structure of topological Fermi arcs as they wind around a chiral fermion (Fig.~\ref{Fig4}e), suggesting that CoSi provides a rare example of a non-compact Riemann surface in nature \cite{HelicodalFermiArcs, KramersWeyl}.

\begin{figure}
\centering
\includegraphics[width=150mm]{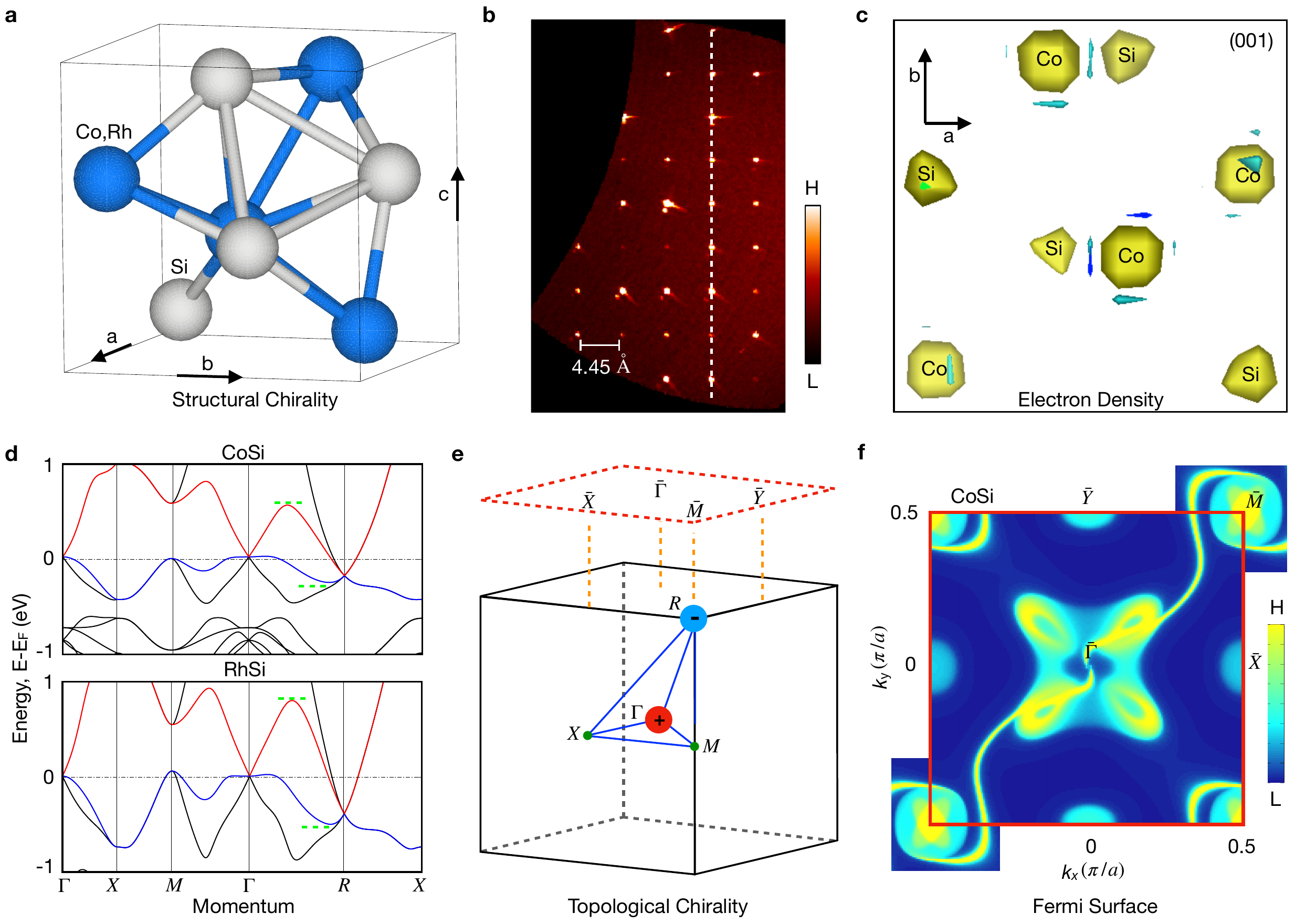}
\caption{\textbf{Structural chirality and topological chirality.}
\textbf{a}, Chiral crystal structure of $X$Si ($X=$ Co, Rh), space group $P2_{1}3$, No. 198. \textbf{b}, Single crystal X-ray diffraction precession pattern of the (0$kl$) planes of CoSi at 100 K. The resolved spots confirm space group $P2_{1}3$ with lattice constant $a=4.433(4)\ \textrm{\AA}$. \textbf{c}, Three-dimensional Fourier map showing the electron density in the $B20$ CoSi structure. \textbf{d}, \textit{Ab initio} calculation of the electronic bulk band structure along high-symmetry lines. A 3-fold degenerate topological chiral fermion is predicted at $\Gamma$ and a 4-fold topological chiral fermion at $R$; these carry Chern numbers $+2$ and $-2$, respectively. The highest valence (blue) and lowest conduction (red) bands fix a topologically non-trivial energy window (green dotted lines). \textbf{e}, Bulk Brillouin zone (BZ) and (001) surface BZ with high-symmetry points and the predicted chiral fermions (red and blue spheres) marked. \textbf{f}, \textit{Ab initio} calculation of the surface spectral weight on the (001) surface for CoSi, with the (001) surface BZ marked (red box). The predicted bulk chiral fermions project onto $\bar{\Gamma}$ and $\bar{M}$, connected by a pair of topological Fermi arcs extending diagonally across the surface BZ.
}
\label{Fig1}
\end{figure}

\begin{figure}[t]
\centering
\includegraphics[width=150mm]{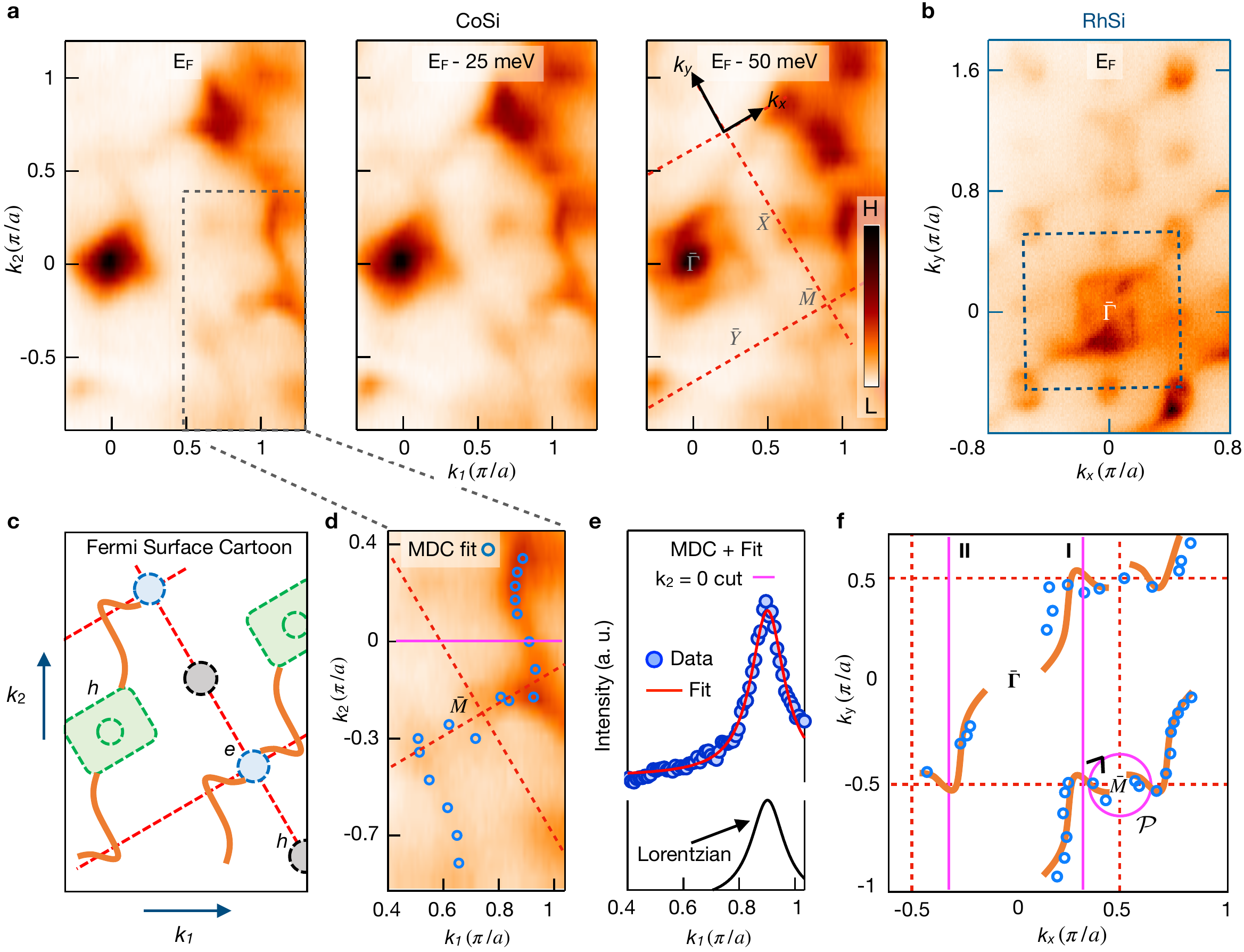}
\caption{\textbf{Fermiology of the (001) surface electronic structure in CoSi and RhSi.}
\textbf{a}, ARPES constant-energy contours for CoSi measured at incident photon energy 50 eV with the Brillouin zone (BZ) boundary marked (rightmost panel, red dotted line). We observe long winding states connecting the $\bar{\Gamma}$ and $\bar{M}$ pockets. \textbf{b}, Fermi surface for RhSi measured at incident photon energy 82 eV with BZ boundary marked (blue dotted line). Again, we observe long winding states extending diagonally across the BZ (Extended Data Fig.~\ref{ExtFig1}). \textbf{c}, Schematic of the measured Fermi surface for CoSi showing hole-like ($h$) and electron-like ($e$) bulk pockets and long winding states (orange). \textbf{d}, Zoomed-in view of the long winding states with a trajectory obtained by fitting Lorentzians to the momentum distribution curves (MDCs) of the ARPES spectrum (blue circles). \textbf{e}, Representative Lorentzian fit to the MDC along $k_1$ for $k_2 = 0$ at binding energy $E-E_\textrm{F}=-10$ meV (Extended Data Fig.~\ref{ExtFig5}). \textbf{f}, Schematic overlaid with Lorentzian peaks extracted from the MDCs passing through the long winding states. We mark two straight cuts, Cut I and II (magenta arrows), as well as a closed loop cut $\mathcal{P}$ (magenta circle). 
}
\label{Fig2}
\end{figure}

\begin{figure}[t]
\includegraphics[width=150mm]{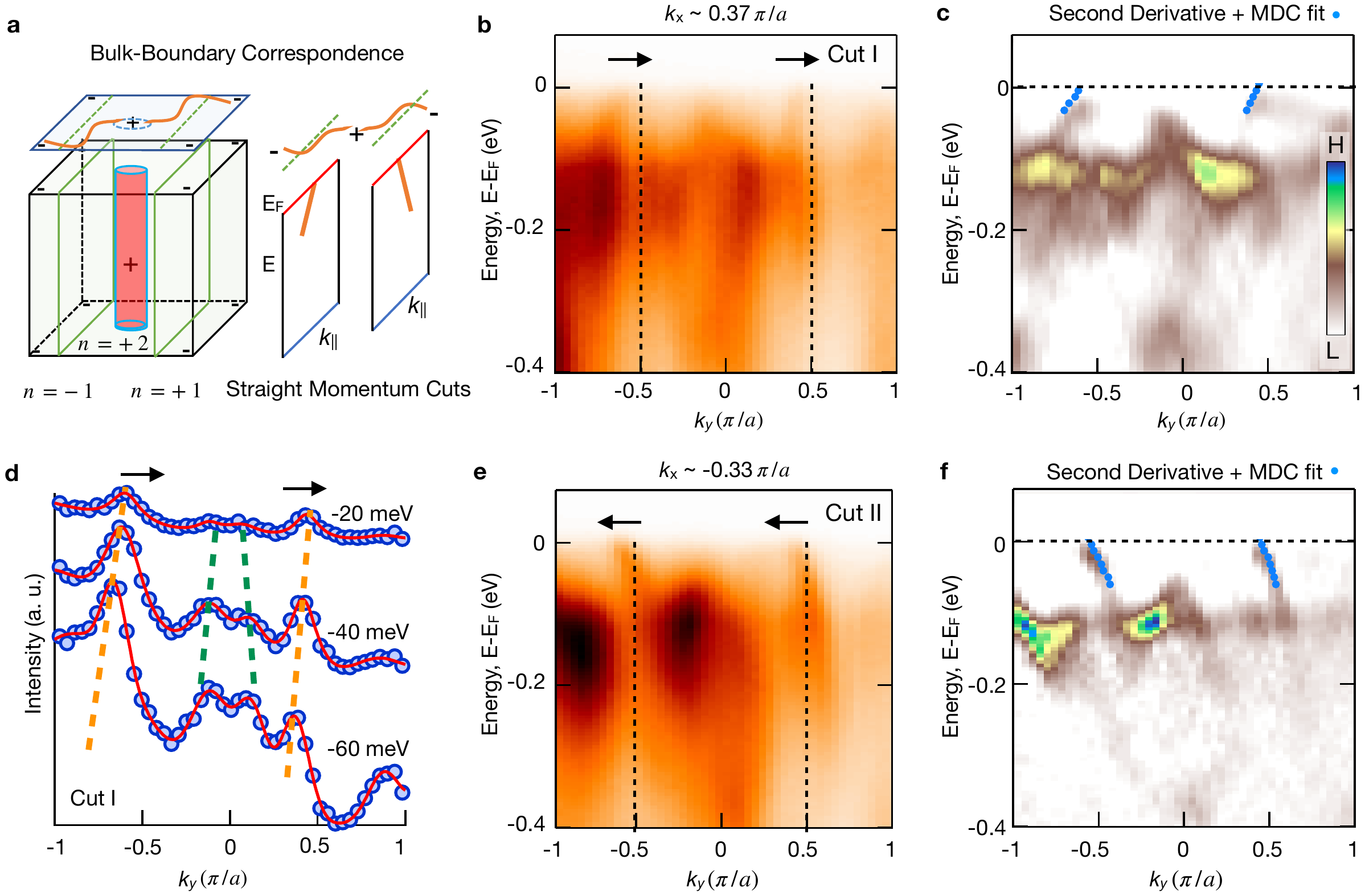}
\caption{ \textbf{Observation of chiral edge modes and a Chern number in CoSi.}
\textbf{a}, Left panel: schematic 3D bulk and 2D surface Brillouin zone (BZ) of CoSi with chiral fermions marked ($\pm{}$) and including several examples of 2D manifolds hosting a Chern number $n$ (green planes, red cylinder) \cite{arcDetect1}. Every plane in the bulk has a non-zero $n$. The cylinder enclosing the bulk chiral fermion at $\Gamma$ has $n=+2$. Right panel: Fermi arcs (orange curves) show up as chiral edge modes (orange lines). Energy-momentum cuts on opposite sides of $\bar{\Gamma}$ are expected to show chiral modes propagating in opposite directions. \textbf{b}, ARPES spectrum along Cut I (as marked in Fig.~\ref{Fig2}f), suggesting two right-moving chiral edge modes (black arrows). Vertical dotted lines mark the BZ boundaries. \textbf{c}, Second derivative plot of Cut I, with fitted Lorentzian peaks for a series of momentum distribution curves (MDCs) to track the dispersion (blue dots). \textbf{d}, MDCs and Lorentzian fits tracking the chiral edge modes (dotted orange lines) and hole-like $\bar{X}$ pocket (dotted green curve) for Cut I. \textbf{e}, Same as (b), but for Cut II, suggesting two left-moving chiral edge modes (black arrows). \textbf{f}, Same as (c), but for Cut II. The difference in the net number of chiral edge modes on Cut I and Cut II suggests a Chern number $+2$ living near $\bar{\Gamma}$.
}
\label{Fig3}
\end{figure}

To further understand these experimental results, we consider the \textit{ab initio} calculated spectral weight for the (001) surface and we observe a pair of Fermi arcs winding around the $\bar{\Gamma}$ and $\bar{M}$ pockets in a counterclockwise and clockwise manner, respectively, with decreasing binding energy (approaching \ef; Fig.~\ref{Fig4}\panf). The clockwise winding around $\bar{M}$ is consistent with our observation by ARPES of a $-2$ projected chiral charge. Moreover, from our \textit{ab initio} calculations, we predict that the $-2$ charge projecting to $\bar{M}$ arises from a 4-fold chiral fermion at the bulk $R$ point (Fig.~\ref{Fig1}d). The $+2$ chiral charge which we associate with $\bar{\Gamma}$ from ARPES (Fig.~\ref{Fig3}) is further consistent with the 3-fold chiral fermion predicted at the bulk $\Gamma$ point. By fully accounting for the predicted topological charges in experiment, our ARPES results suggest the demonstration of a topological chiral crystal in CoSi. We can similarly account for the predicted topological charges in RhSi from our ARPES data (Extended Data Fig.~\ref{ExtFig1},~\ref{ExtFig2}).

%\section{Discussion}

 %Discussion
The surface state dispersions in our ARPES spectra, taken together with the topological bulk-boundary correspondence established in theory \cite{Wan, arcDetect1}, demonstrate that CoSi is a topological chiral crystal. This experimental result is further consistent with the numerical result determined from first-principles calculations of the surface state dispersions and topological invariants. Unlike previously-reported WSMs, the Fermi arcs which we observe in CoSi and RhSi stretch diagonally across the entire (001) surface Brillouin zone, from $\bar{\Gamma}$ to $\bar{M}$. In fact, the Fermi arcs in $X$Si are longer than those found in TaAs by a factor of thirty. Our surface band structure measurements also demonstrate two well-separated Fermi pockets carrying Chern number $\pm{2}$. Lastly, we observe for the first time in an electronic material the helicoid structure of topological Fermi arcs, offering an example of a non-compact Riemann surface \cite{HelicodalFermiArcs, KramersWeyl}. Our results suggest that CoSi and RhSi are excellent candidates for studying topological phenomena distinct to chiral fermions, using a variety of techniques \cite{rev4, rev7, rev6}.

%  Although the higher-fold chiral fermion at $R$ is below \ef, it remains isolated in $\textit{k}$ space with the associated Fermi surface carrying a non-zero Chern number.

Crucial for applications, the topologically non-trivial energy window in CoSi is an order of magnitude larger than that in TaAs \cite{ARPES-TaAs1, ARPES-TaAs2}, rendering its quantum properties robust against changes in surface chemical potential and disorder. Moreover, the energy offset between the higher-fold chiral fermions at $\Gamma$ and $R$ is predicted to be $\sim225$ meV. Such an energy offset is essential for inducing the chiral magnetic effect \cite{chiralmagenticeffect1} and the quantised photogalvanic effect (optical) \cite{chiral_photogalvanic}. When coupled to a compatible superconductor, CoSi is a compelling platform for studying the superconducting pairing of Fermi surfaces with non-zero Chern numbers, which may be promising for realizing a new type of topological superconducting phase recently proposed by Li and Haldane \cite{Haldane}. CoSi further opens the door to exploring other exotic quantum phenomena when combined with the isochemical material FeSi. Fe$_{1-x}$Co$_x$Si may simultaneously host $\textit{k}$-space topological defects (chiral fermions) and real-space topological defects (skyrmions). Through our observation of a helicoid surface state dispersion by ARPES, our results suggest the discovery of the first topological chiral crystal. In this way, our work provides a much-needed new platform for further study of topological conductors.

\ \\
While the present article was in review, a related work studying the bulk dispersion in CoSi was posted on the arXiv \cite{CoSi_ARPES}.

%\ \\
%\textbf{Data availability}. The data supporting the findings of this study are available within the paper and other findings of this study are available from the corresponding author upon reasonable request.
%\clearpage

%\begin{acknowledgements}
%\textbf{Acknowledgments:} 

%Work at Princeton University and Princeton-led synchrotron based ARPES measurements were by the Gordon and Betty Moore Foundation through the EPIQS program Grant No. GBMF4547-HASAN and by the National Science Foundation under Grant No. NSF-DMR-1507585. The crystal growth was supported by National Basic Research Program in China (Grant Nos. 2014CB239302), National Natural Science Foundation of China No.11774007 and the Key Research Program of the Chinese Academy of Science (Grant No. XDPB08-1). T.-R.C. is supported by the Ministry of Science and Technology under MOST Young Scholar Fellowship: MOST Grant for the Columbus Program NO. 107-2636-M-006 -004-, National Cheng Kung University, Taiwan, and National Center for Theoretical Sciences (NCTS), Taiwan. The authors thank S. K. Mo and J. D. Denlinger for their beamline assistance at the Advanced Light Source (ALS) in Berkeley, California. We also thank G. Bian and H. Verlinde for useful discussions.
%\end{acknowledgements}

\clearpage

\clearpage

\clearpage

\begin{figure}
\includegraphics[width=155mm]{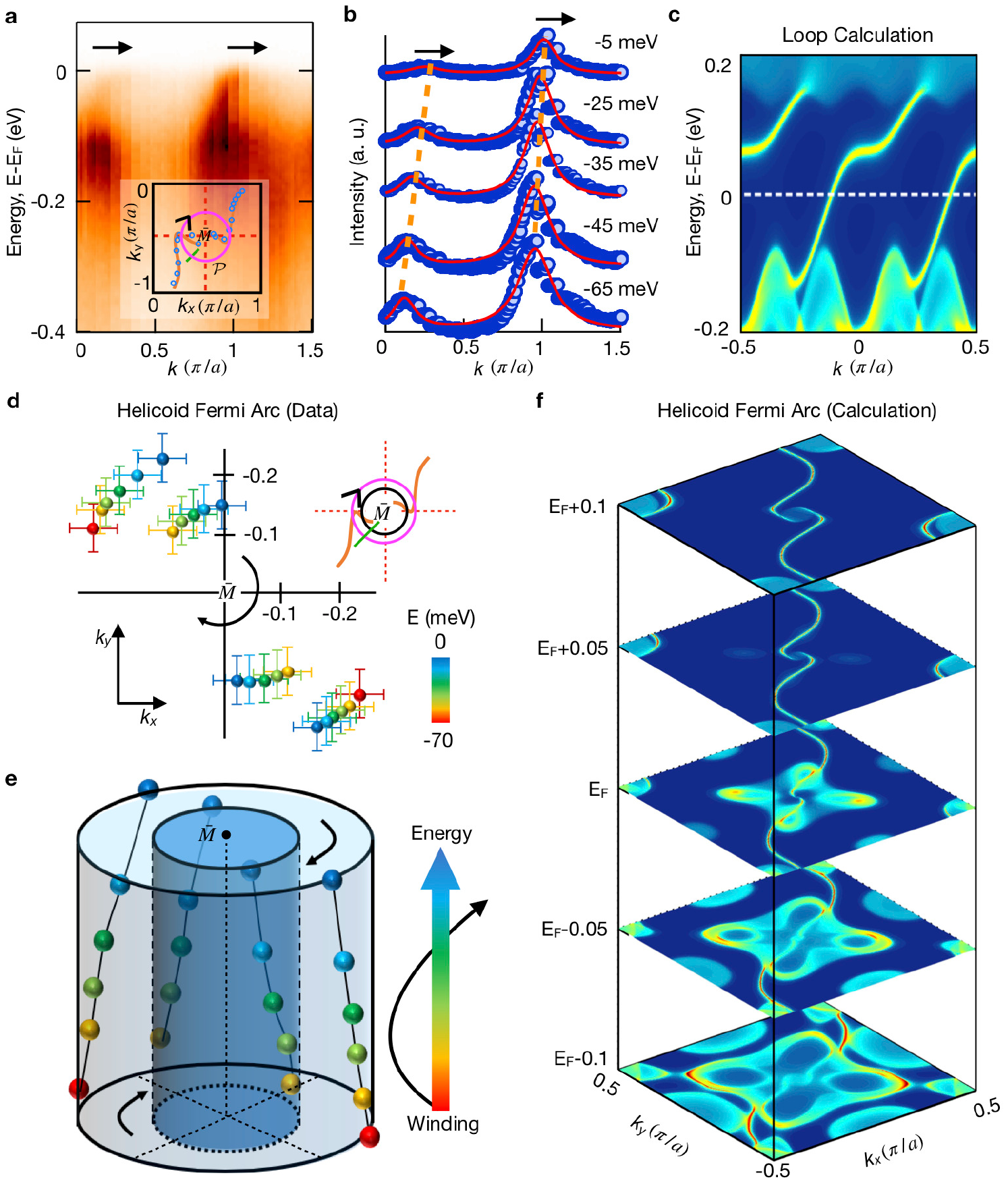}
\end{figure}
%\addtocounter{figure}{-1}
\begin{figure*}[t!]
\caption{\textbf{Observation of helicoid Fermi arcs.}
\textbf{a}, ARPES spectrum along loop $\mathcal{P}$, suggesting two right-moving chiral edge modes and a projected chiral charge of $-2$ at $\bar{M}$. Inset: definition of the loop $\mathcal{P}$, starting from the green mark and proceeding clockwise. \textbf{b}, Lorentzian fits to a series of momentum distribution curves (MDCs) along $\mathcal{P}$ to track the dispersion of the chiral edge modes. \textbf{c}, \textit{Ab initio} calculation of the dispersion along a loop around $\bar{M}$ showing two right-moving chiral edge modes, consistent with the ARPES data. \textbf{d}, Extracted dispersion of the chiral edge modes on $\mathcal{P}$ and a second inner loop from Lorentzian fits to the MDCs. Error bars correspond to the momentum resolution. Inset: definition of the second inner loop (black). We observe that the chiral edge modes spiral in a clockwise way with decreasing binding energy (approaching \ef). \textbf{e}, Perspective plot of (d), where the two loops now correspond to two concentric cylinders. The winding of the chiral modes around $\bar{M}$ as a function of binding energy suggests that the Fermi arcs have a helicoid structure \cite{HelicodalFermiArcs, KramersWeyl}. \textbf{f}, \textit{Ab initio} calculated constant-energy contours, consistent with the helicoid Fermi arc structure observed in our ARPES spectra.}
\label{Fig4}
\end{figure*}
 \clearpage

\section{Analogous observation of the Fermi arcs in RhSi}

\begin{figure}
\centering
\includegraphics[width=150mm]{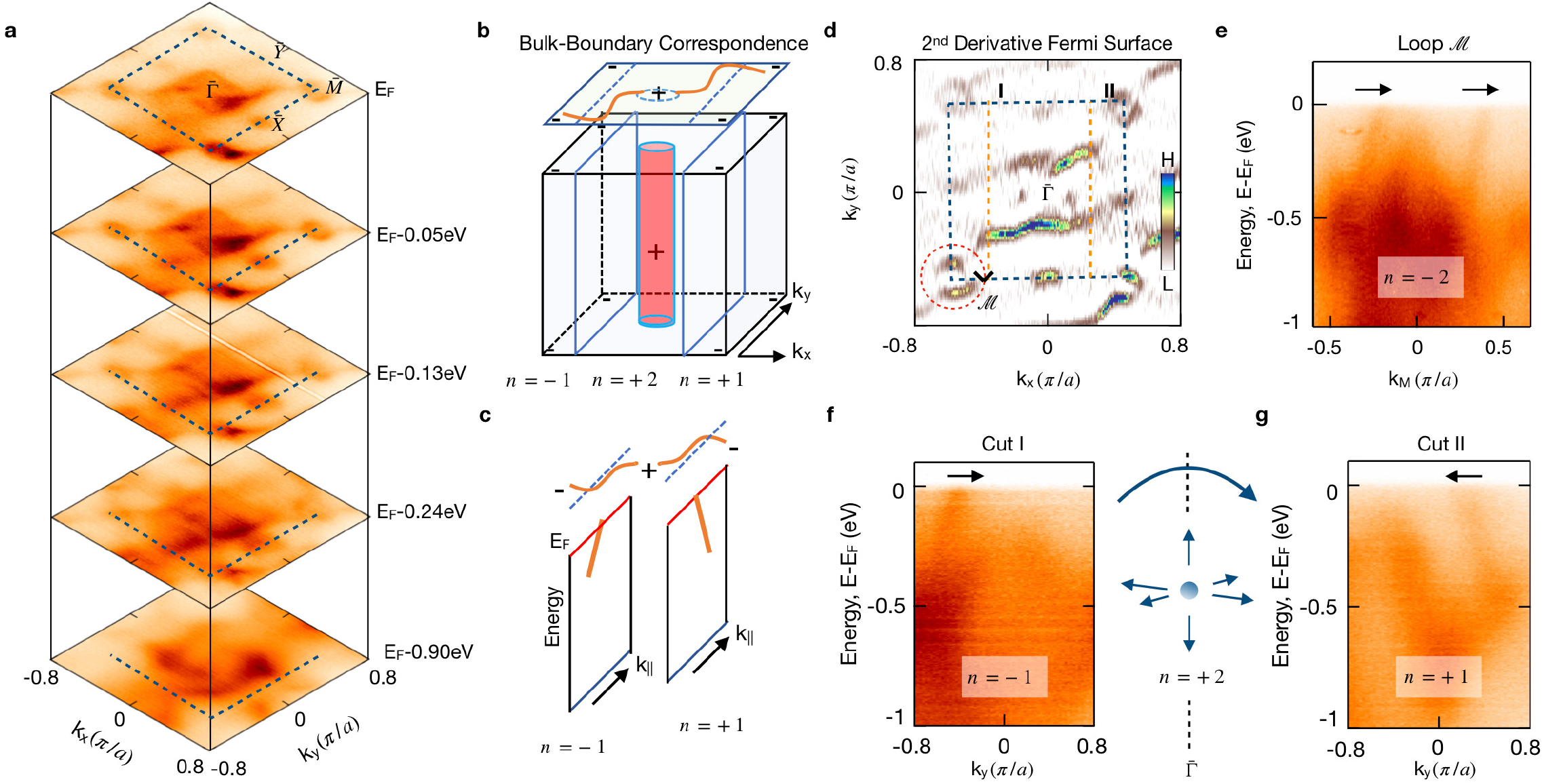}
\caption{\textbf{Long Fermi arcs in topological chiral crystal RhSi.}
\textbf{a}, ARPES measured Fermi surface and constant binding energy contours with an incident photon energy of 82eV at 10K. The Brillouin zone (BZ) boundary is marked in blue. \textbf{b}, 3D bulk and 2D surface BZ with higher-fold chiral fermions ($\pm{}$). The planes outlined in blue and the red cylinder are 2D manifolds with indicated Chern number $n$ \cite{arcDetect1}. The cylinder enclosing the bulk chiral fermion at $\Gamma$ has $n=+2$. \textbf{c}, Fermi arcs (orange) connect the projected chiral fermions. Energy dispersion cuts show that these two chiral edge modes are time-reversed partners propagating in opposite directions. \textbf{d}, Second derivative Fermi surface with the straight and loop cuts of interest marked. \textbf{e}, ARPES spectrum along a loop $\mathcal{M}$ showing two right-moving chiral edge modes, suggesting that the $4$-fold chiral fermion at $R$ carries Chern number $-2$. \textbf{f}, ARPES spectrum along Cut I showing a right-moving chiral edge mode. \textbf{g}, ARPES spectrum along Cut II on the opposite side of the $3$-fold chiral fermion at $\bar{\Gamma}$ (illustrated by the blue sphere) showing a left-moving chiral edge mode.
}
\label{ExtFig1}
\end{figure}

Motivated by an interest in establishing families of closely-related topological materials \cite{burkov, Jia, H.Weyl, TypeIIWeyl1, TypeIIWeyl2, TypeIIWeyl3, TypeIIWeyl4}, we expand our experiment to include RhSi, an isoelectronic cousin of CoSi. Rhodium silicide, RhSi, crystallises in a chiral cubic lattice, space group $P2_{1}3$, No. 198. The calculated electronic band structure is generally similar to that of CoSi (side-by-side comparison in Fig.~\ref{ExtFig1}\pand). Constant-energy contours measured by ARPES further suggest features similar to those in CoSi (Extended Data Fig.~\ref{ExtFig1}\pana). In particular, we observe contours at the $\bar{\Gamma}$, $\bar{X}$ and $\bar{Y}$ points and long winding states that extend diagonally from $\bar{\Gamma}$ to $\bar{M}$. Again taking advantage of the bulk-boundary correspondence, we focus on counting chiral edge modes to determine the topological nature of RhSi (Extended Data Fig.~\ref{ExtFig1}\panb,\panc). A second-derivative plot of the Fermi surface further suggests the presence of long states stretching diagonally across the BZ, motivating a study of energy-momentum cuts through these states (Extended Data Fig.~\ref{ExtFig1}\pand-\pang). The cuts show that the long states take the form of a right-moving chiral edge mode (Cut I) and a left-moving chiral edge mode (Cut II) on opposite sides of $\bar{\Gamma}$. The net difference in the number of right-moving chiral edge modes suggests that a Chern number of $+2$ projects to $\bar{\Gamma}$ and that the long states are topological Fermi arcs. Proceeding again by analogy to CoSi, we study the band structure along a loop $\mathcal{M}$ enclosing the $\bar{M}$ point and we observe two right-moving chiral edge modes, suggesting a Chern number of $-2$ at the $R$ point. These results suggest that RhSi, an isoelectronic cousin of CoSi, provides another example of a near-ideal topological conductor.

%\section{Helicoid structure of the Fermi arcs in RhSi}

\begin{figure}[t]
\centering
\includegraphics[width=150mm]{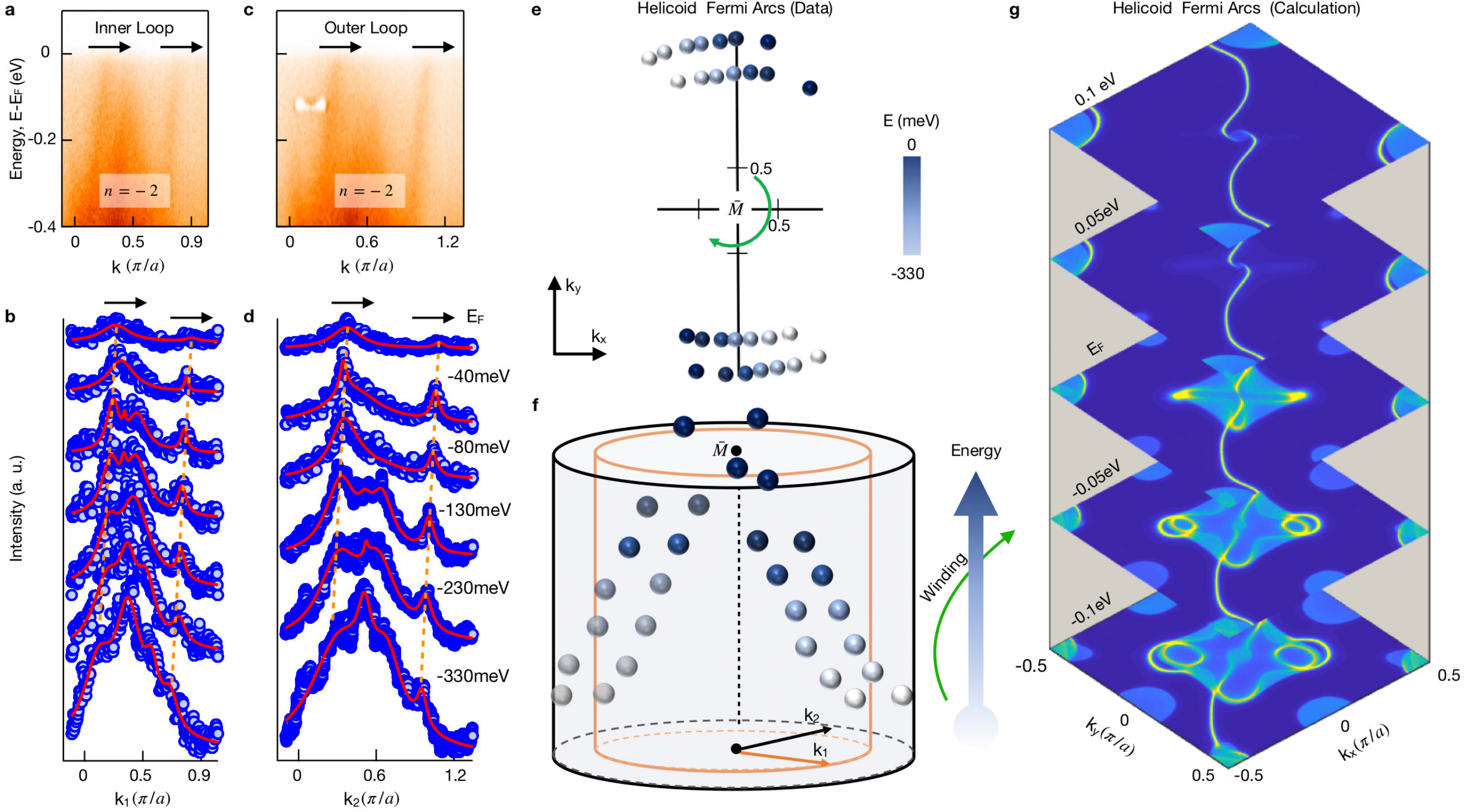}
\caption{\textbf{Fermi arc helicoid in RhSi.} \textbf{a}, Energy dispersion cut on an inner loop of radius $0.18\pi/a$ enclosing the $\bar{M}$. \textbf{b}, Lorentzian fits (red traces) to the momentum distribution curves (MDCs; blue dots) to track the observed chiral edge modes. \textbf{c}, \textbf{d}, Similar analysis to (a, b), but for an outer loop of radius $0.23\pi/a$. Black arrows show the Fermi velocity direction for the chiral edge modes. \textbf{e}, Top view and \textbf{f}, perspective view of the helicoid dispersion extracted from the MDCs, plotted on the two concentric loops, suggesting a clockwise spiral with decreasing binding energy. \textbf{g}, $\textit{Ab initio}$ calculated constant-energy contours show a consistent helicoid structure.
}
\label{ExtFig2}
\end{figure}

We study the Fermi arcs near $\bar{M}$ in RhSi, by analogy with the analysis performed for CoSi (Fig.~\ref{Fig4}). First, we consider ARPES energy-momentum cuts on an inner and outer loop enclosing the $\bar{M}$ point and we observe signatures of two right-moving chiral edge modes on each loop, suggesting an enclosed projected chiral charge of $-2$ (Extended Data Fig.~\ref{ExtFig2}\pana-\pand). We further fit Lorentzians to the MDCs and we find that the extracted dispersion exhibits a clockwise winding on both loops for decreasing binding energy (approaching \ef; Extended Data Fig.~\ref{ExtFig2}\pane,\panf). This dispersion again suggests a chiral charge $-2$ at $\bar{M}$ and is characteristic of the helicoid structure of Fermi arcs as they wind around a chiral fermion. Lastly, we study $\textit{ab initio}$ calculations of the constant-energy contours (Extended Data Fig.~\ref{ExtFig2}\pang) which exhibit the same winding pattern. These ARPES results suggest that RhSi, like CoSi, exhibits helicoid topological Fermi arcs.

\section{Theory of the bulk-boundary correspondence} By measuring the surface state band structure of a crystal, ARPES is capable of demonstrating Fermi arcs and counting Chern numbers \cite{arcDetect1}. We briefly review the details of this method. First, recall that topological invariants are typically defined for a gapped system. A topological conductor is by definition gapless and so we cannot define a topological invariant for the full bulk Brillouin zone of the three-dimensional system. However, we may be able to choose certain two-dimensional $\textit{k}$-space manifolds where the band structure is fully gapped and a two-dimensional invariant can then be defined on this slice of momentum space (Fig.~\ref{Fig3}a). We consider specifically the case of a chiral fermion in a topological conductor and we study a gapped momentum-space slice which cuts in between two chiral fermions. By definition, a chiral fermion is associated with a non-zero Chern number, so at least some of these slices will be characterized by a non-zero Chern number. Following the bulk-boundary correspondence, these slices will then contribute chiral edge modes to the surface state dispersion. If we image collecting together the chiral edge modes from all of the gapped slices, we will assemble the entire topological Fermi arc of the topological conductor. If we run the bulk-boundary correspondence in reverse, we can instead measure the surface states by ARPES and count the chiral edge modes on a particular one-dimensional slice of the surface Brillouin zone to determine the Chern number of the underlying two-dimensional slice of the bulk Brillouin zone. These Chern numbers in turn fix the chiral charges of the topological fermions.

Next we highlight two spectral signatures that can determine the chiral charges specifically in $X$Si ($X=$ Co, Rh). First, we consider chiral edge modes along straight $\textit{k}$-slices. Theoretically, the two-dimensional $\textit{k}$-slices on the two sides of the chiral fermion at $\Gamma$ are related by time-reversal symmetry and therefore should have equal and opposite Chern numbers ($n_{l}=-n_{r}$; Fig. 3b). On the other hand, because the 3-fold chiral fermion at $\Gamma$ is predicted to carry chiral charge $+2$, we expect that the difference should be $n_{l}-n_{r}=+2$, resulting in $n_{l(r)}=\pm{1}$. Therefore we expect one net left-moving chiral edge mode on one cut and one net right-moving chiral edge mode on the other. For the second signature, we study the chiral edge modes along a closed loop in the surface BZ. Any loop that encloses the projected chiral charge of $+2$ at $\bar{\Gamma}$ or $-2$ at $\bar{M}$ has this number of net chiral edge modes along this path. By taking advantage of this correspondence, we can count surface states on loops in our ARPES spectra to determine enclosed projected chiral charges.

\section{Materials and methods}

\subsection{Growth of CoSi and RhSi single crystals} Single crystals of CoSi were grown using a chemical vapor transport (CVT) technique. First, polycrystalline CoSi was prepared by arc-melting stoichiometric amounts of Co slices and Si pieces. After being crushed and ground into a powder, the sample was sealed in an evacuated silica tube with an iodine concentration of approximately $0.25$ mg/cm$^3$. The transport reaction took place at a temperature gradient from $1000\degree$C (source) to $1100\degree$C (sink) for two weeks. The resulting CoSi single crystals had a metallic luster and varied in size from $1$ mm to $2$ mm. Single crystals of RhSi were grown from a melt using the vertical Bridgman crystal growth technique at a non-stoichiometric composition. In particular, we induced a slight excess of Si to ensure a flux growth inside the Bridgman ampoule. First, a polycrystalline ingot was prepared by pre-melting the highly pure constituents under an argon atmosphere using an arc furnace. The crushed powder was poured into a custom-designed sharp-edged alumina tube and then sealed inside a tantalum tube again under an argon atmosphere. The sample was heated to $1550\degree$C and then slowly pooled to the cold zone at a rate of $0.8$ mm$/$h. Single crystals on average $\sim15$ mm in length and $\sim6$ mm in diameter were obtained.

\subsection{X-ray diffraction} 
\begin{figure}[t]
\includegraphics[width=150mm]{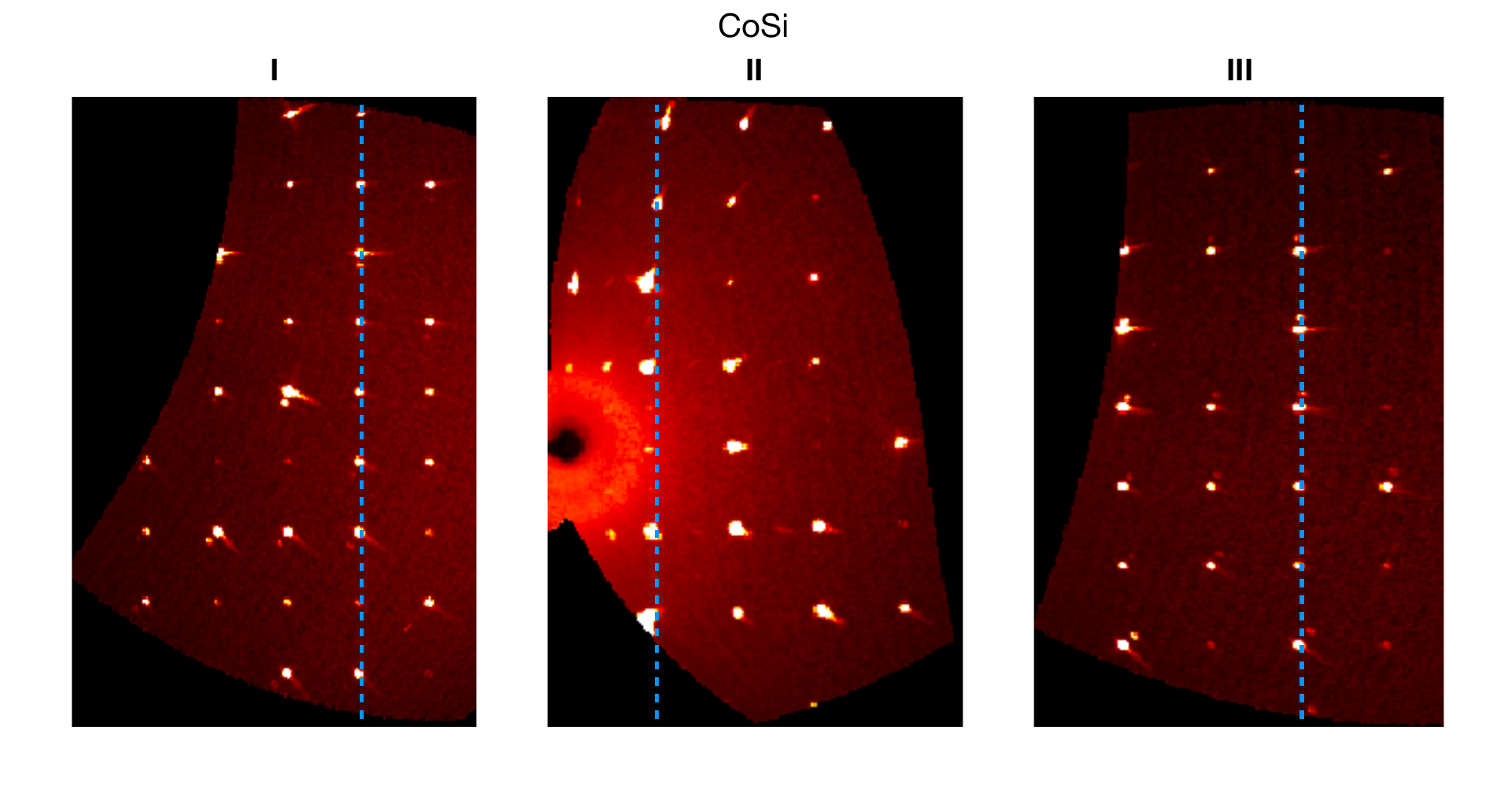}
\caption{ \textbf{X-ray diffraction.} Single crystal X-ray diffraction precession image of the ($0kl$) planes in the reciprocal lattice of CoSi. The resolved spots from scans I and III are consistent with space group $P2_13$ at 100K. This is also shown in the reflection intersection, scan II.
}
\label{ExtFig3}
\end{figure}

Single crystals of CoSi were mounted on the tips of Kapton loops. The low-temperature (100 K) intensity data was collected on a Bruker Apex II X-ray diffractometer with Mo radiation K$\alpha_{1}$ ($\lambda=0.71073\textrm{\AA}^{-1}$). Measurements were performed over a full sphere of $\textit{k}$-space with 0.5$\degree$ scans in $\omega$ with an exposure time of 10 seconds per frame (Extended Data Fig.~\ref{ExtFig3}). The SMART software was used for data acquisition. The extracted intensities were corrected for Lorentz and polarization effects with the SAINT program. Numerical absorption corrections were accomplished with XPREP, which is based on face-indexed absorption \cite{XRay1}. The twin unit cell was tested. With the SHELXTL package, the crystal structures were solved using direct methods and refined by full-matrix least-squares on F$^{2}$ \cite{XRay2}. No vacancies were observed according to the refinement and no residual electron density was detected, indicating that the CoSi crystals were of high quality.

\subsection{Sample surface preparation for ARPES} For CoSi single crystals, the surface preparation procedure followed the conventional \textit{in situ} mechanical cleaving approach. This cleaving method resulted in a low success rate and the resulting surface typically appeared rough under a microscope. We speculate that the difficulty in cleaving CoSi single crystals may be a result of its cubic structure, strong covalent bonding and lack of a preferred cleaving plane. For RhSi single crystals, their large size allowed them to be mechanically cut and polished along the (001) surface. An \textit{in situ} sputtering and annealing procedure, combined with LEED/RHEED characterization, was used to obtain a clean surface suitable for ARPES measurements. The typical spectral line-width was narrowed and the background signal was reduced for these samples as compared with the mechanically-cleaved CoSi samples.

\subsection{Angle-resolved photoemission spectroscopy} ARPES measurements were carried out at beamlines (BL) 10.0.1 and 4.0.3 at the Advanced Light Source in Berkeley, CA, USA. A Scienta R4000 electron analyser was used at BL 10.0.1 and a Scienta R8000 was used at BL 4.0.3. At both beamlines the angular resolutions was $< 0.2 \degree$ and the energy resolution was better than 20 meV. Samples were cleaved or sputtered/annealed \textit{in situ} and measured under vacuum better than $5 \times 10^{-11}$ Torr at $T < 20$ K.

\subsection{First-principles calculations} Numerical calculations of $X$Si ($X=$ Co, Rh) were performed within the density functional theory (DFT) framework using the OPENMX package and the full potential augmented plane-wave method as implemented in the package WIEN2k \cite{DFT1, DFT2, DFT3}. The generalised gradient approximation (GGA) was used \cite{DFT4}. Experimentally measured lattice constants were used in DFT calculations of material band structures \cite{xraydata}. A $\Gamma$-centered $\textit{k}$-point $10\times10\times10$ mesh was used and spin-orbit coupling (SOC) was included in self-consistent cycles. To generate the (001) surface states of CoSi and RhSi, Wannier functions were generated using the $p$ orbitals of Si and the $d$ orbitals of Co and Rh. The surface states were calculated for a semi-infinite slab by the iterative Green's function method.

\subsection{Electronic bulk band structure} 
\begin{figure}
\includegraphics[width=150mm, trim={3cm, 4cm, 2cm, 3cm}, clip]{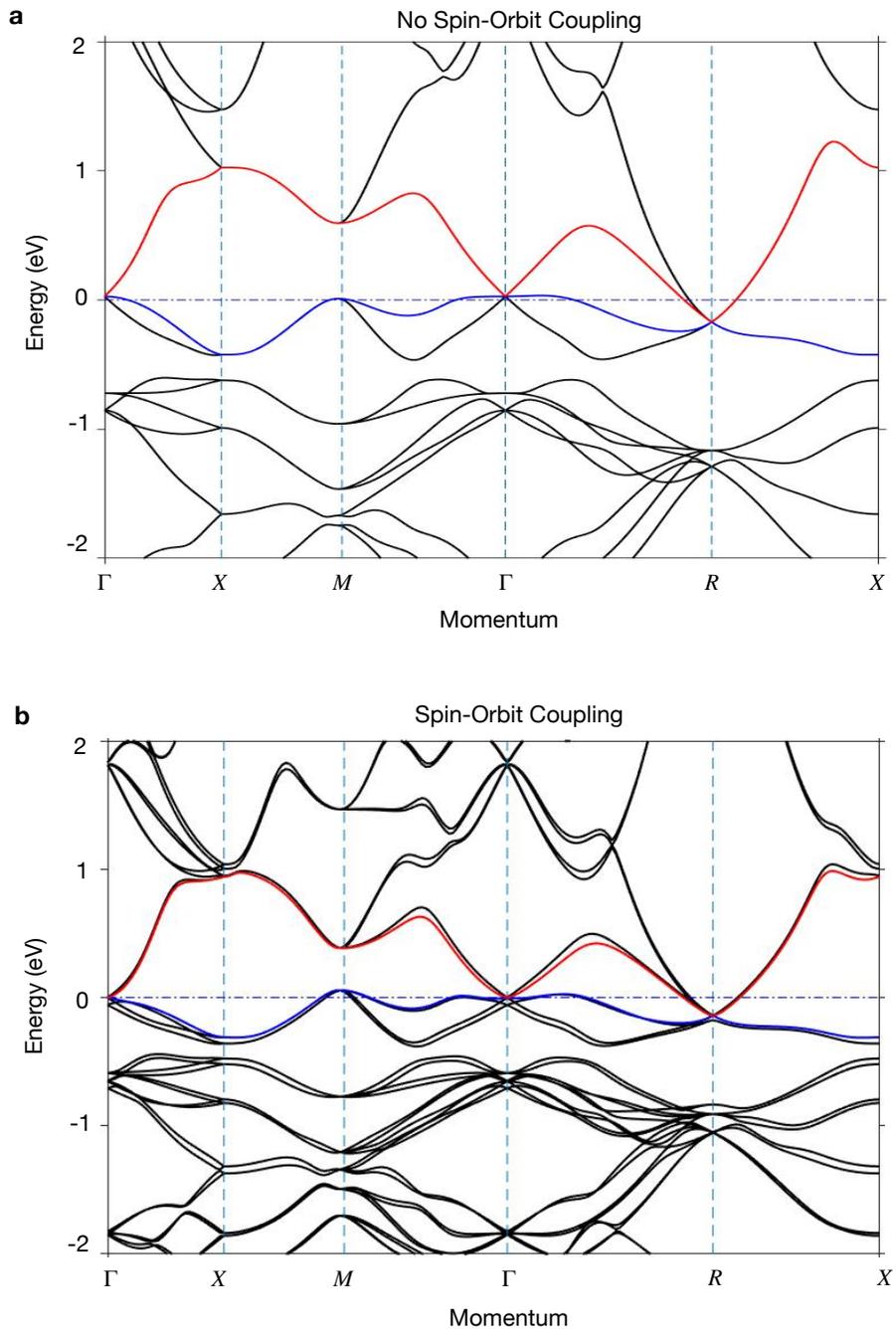}
\caption{\textbf{Electronic bulk band structure.} 
\textbf{a}, Band structure of CoSi in the absence of SOC interactions. \textbf{b}, Band structure in the presence of SOC. The highest valence and lowest conduction bands are colored in blue and red, respectively.
}

\label{ExtFig4}
\end{figure}

The electronic bulk band structure of CoSi is shown with and without spin-orbit coupling (SOC; Extended Data Fig.~\ref{ExtFig4}). The small energy splitting that arises from introducing SOC is approximately 40 meV, which is negligible compared to the $\sim$1.2 eV topologically non-trivial energy window. Our results suggest that we do not need to consider SOC, either theoretically or experimentally, to demonstrate a chiral charge in CoSi.

\subsection{Tracking the Fermi arcs in momentum distribution curves} 
\begin{figure}
\includegraphics[width=150mm]{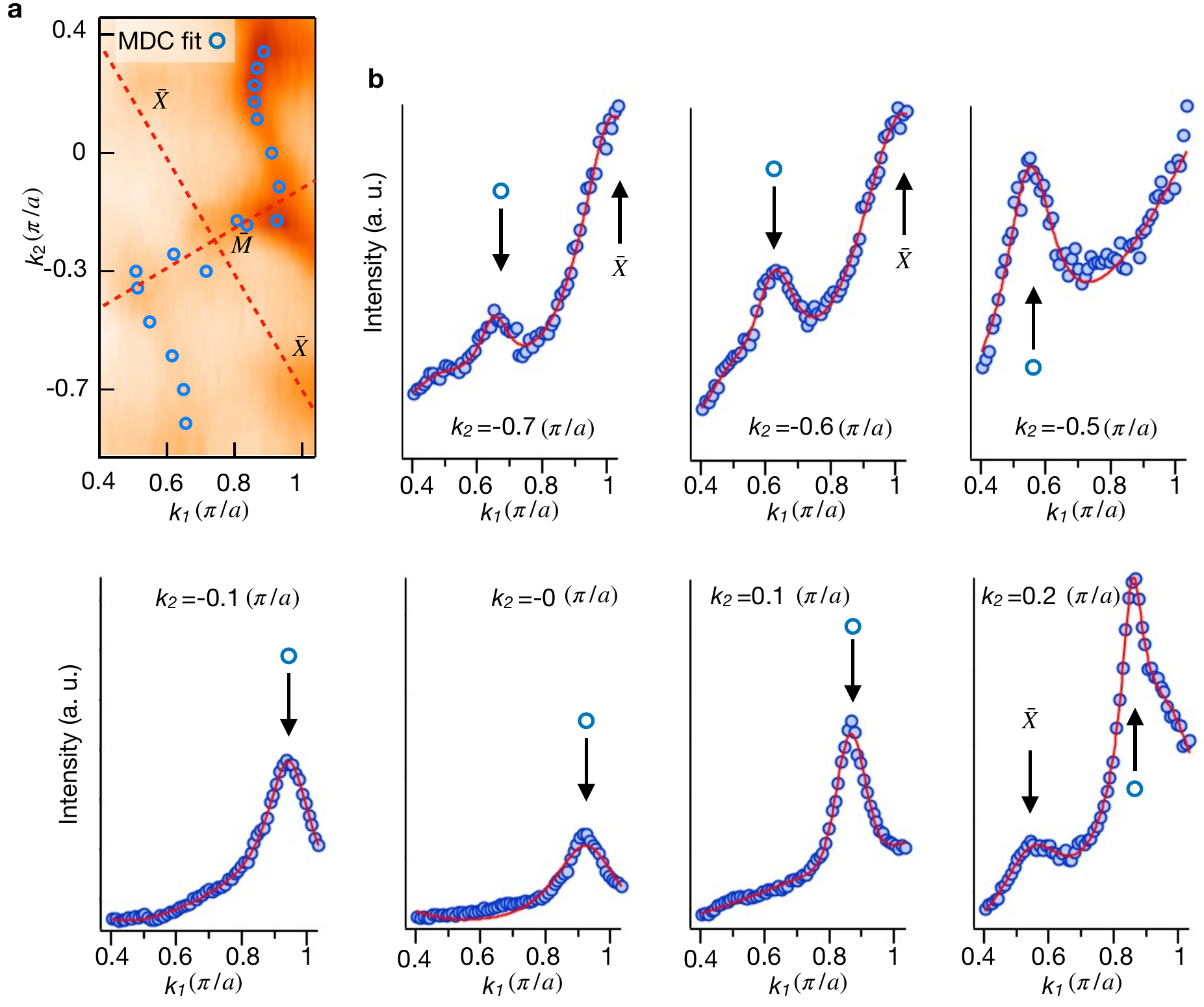}
\caption{\textbf{Tracking the Fermi arcs by fitting Lorentzians to momentum distribution curves (MDCs).} 
\textbf{a}, Zoomed-in region of the Fermi surface (Fig.~\ref{Fig2}d), with Fermi arcs tracked (blue circles) and the surface Brillouin zone marked (red dotted lines). \textbf{b}, Representative fits of Lorentzian functions (red lines) to the MDCs (filled blue circles). The peaks indicate the extracted positions of the Fermi arc (open blue circles) and the $\bar{X}$ pocket.
}

\label{ExtFig5}
\end{figure}

The following procedure was performed to track the Fermi arcs in the surface BZ of CoSi (Extended Data Fig.~\ref{ExtFig5}\pana). At $E=E_\textrm{F}$, MDCs were collected for various fixed $k_y$ values (extending along the region of interest). Each MDC was fitted with a Lorentzian function to pinpoint the $k_x$ value that corresponds to the peak maximum. The MDC fitted peak maximum was then plotted on top of the measured Fermi surface and marked with blue circles. Where necessary, the peak maximum corresponding to the $\bar{X}$ pockets is annotated on each MDC fitting panel (Extended Data Fig.~\ref{ExtFig5}\panb). For the chiral edge modes discussed in the main text and the Extended Data, a similar MDC fitting procedure was used at different binding energies to track the dispersion.

\subsection{Independence of the Fermi arcs with photon energy} 
\begin{figure}
\includegraphics[width=140mm]{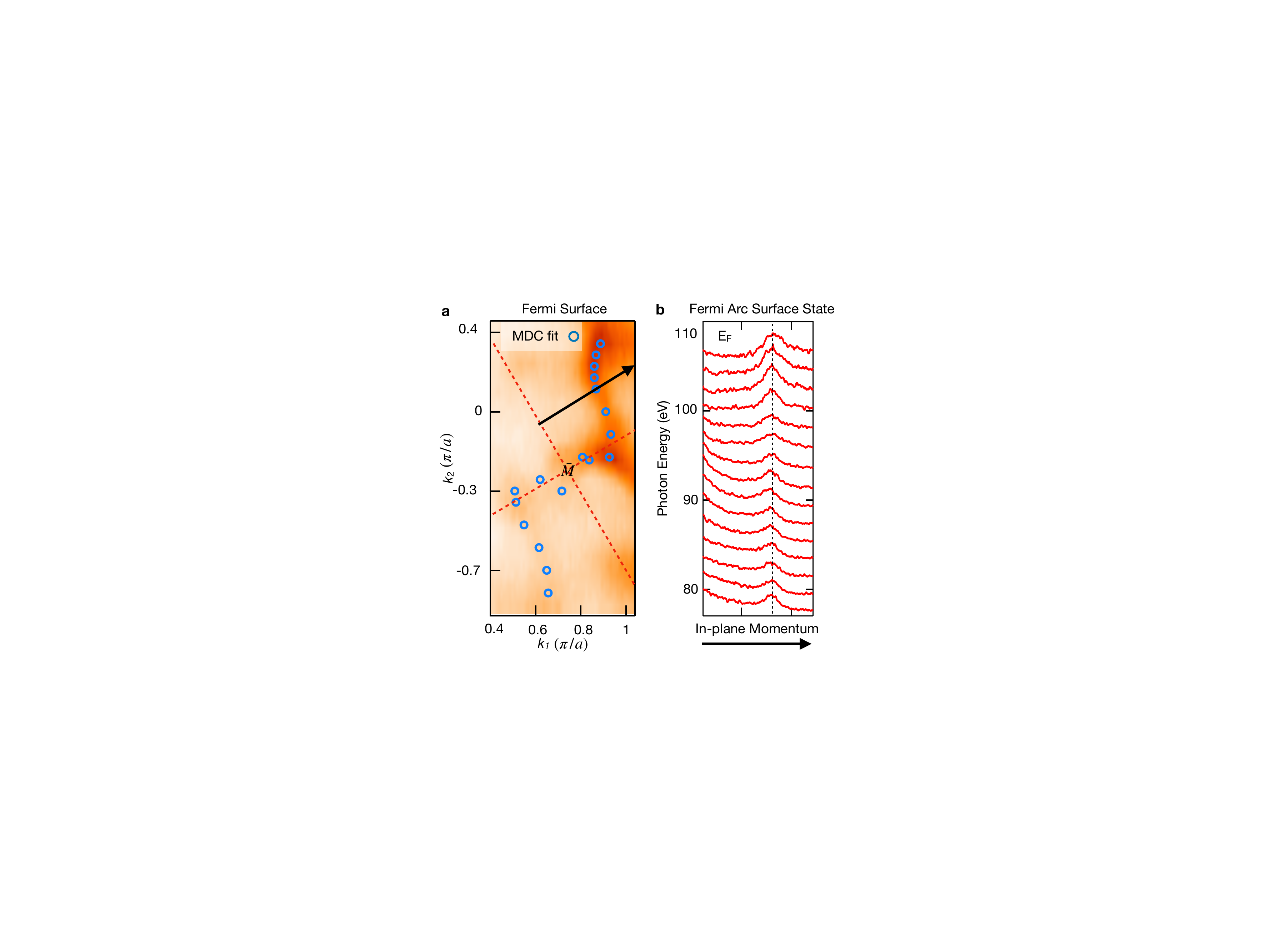}
\caption{\textbf{Photon energy dependence of the Fermi arc in CoSi.} \textbf{a}, Location of the in-plane momentum direction (black arrow) along which a photon energy dependence study was performed, plotted on the Fermi surface (Fig.~\ref{Fig2}d), with surface Brillouin zone (red dotted lines). \textbf{b}, Momentum distribution curves (MDCs) at \ef\ along the in-plane momentum direction illustrated in (a), obtained for a series of photon energies from 80 eV to 110 eV in steps of 2 eV. The peak associated with the Fermi arc does not disperse significantly in photon energy (dotted line), providing further evidence that the observed Fermi arc is indeed a surface state.
}

\label{ExtFig6}
\end{figure}

\begin{figure}
\includegraphics[width=150mm]{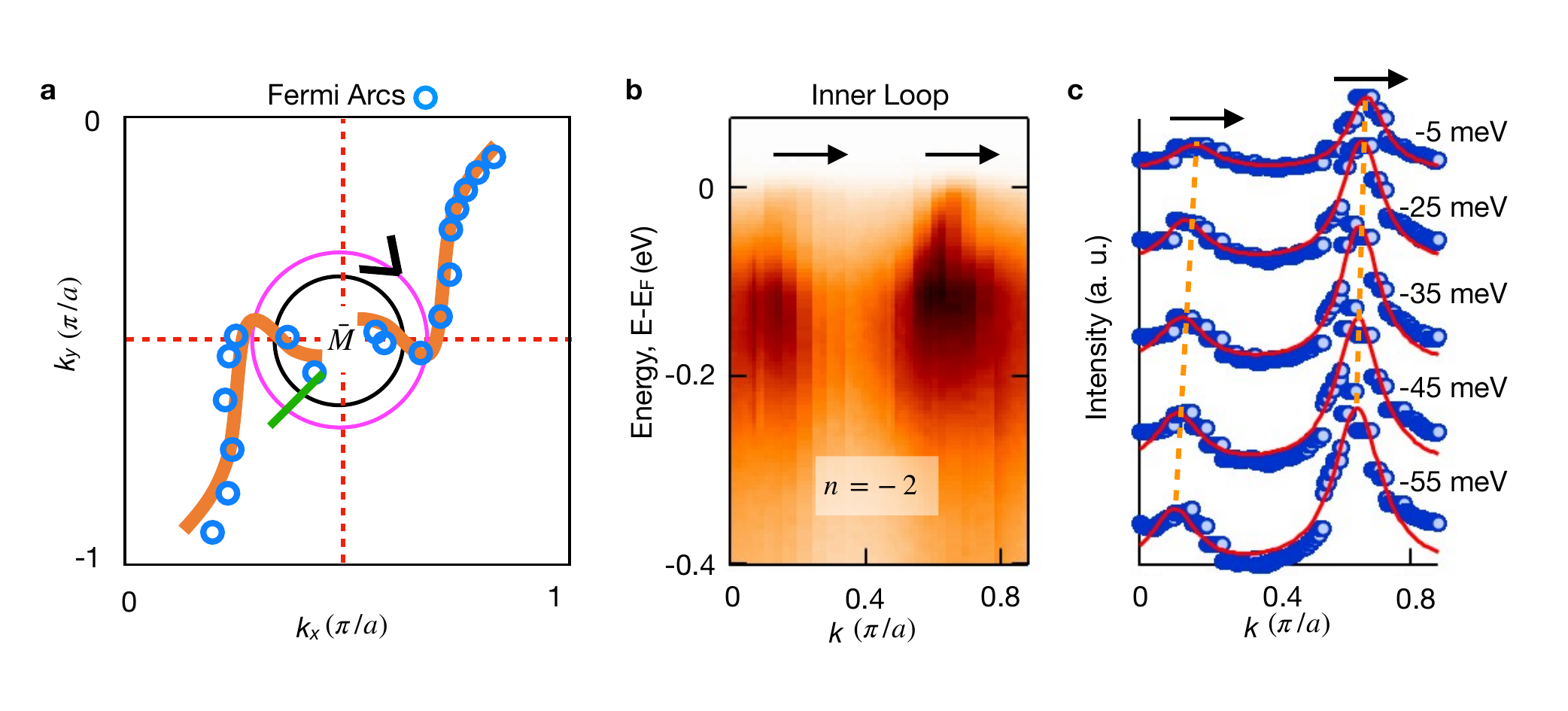}
\caption{\textbf{Systematics for the helicoid fitting in CoSi.} \textbf{a}, Fermi arc trajectory extracted from Lorentzian fits to the MDCs (blue open circles) near $\bar{M}$ with overlaid schematic of the observed features (orange lines). There are two closed contours enclosing $\bar{M}$ on which we count chiral edge modes, the outer loop (magenta, Fig.~\ref{Fig4}) and the inner loop (black). \textbf{b}, Energy-momentum cut along the inner loop, radius $0.14\pi/a$ starting from the green notch in (a) and winding clockwise, black arrow in (a). \textbf{c}, Lorentzian fits (red curves) to the MDCs (blue dots) to extract the helicoid dispersion of the Fermi arcs. We observe two right-moving chiral edge modes dispersing towards \ef, marked schematically (dotted orange lines). The corresponding bulk manifold has Chern number $n=-2$ \cite{Wan}.}
\label{ExtFig_hel_fit}
\end{figure}

\begin{figure}
\includegraphics[width=150mm]{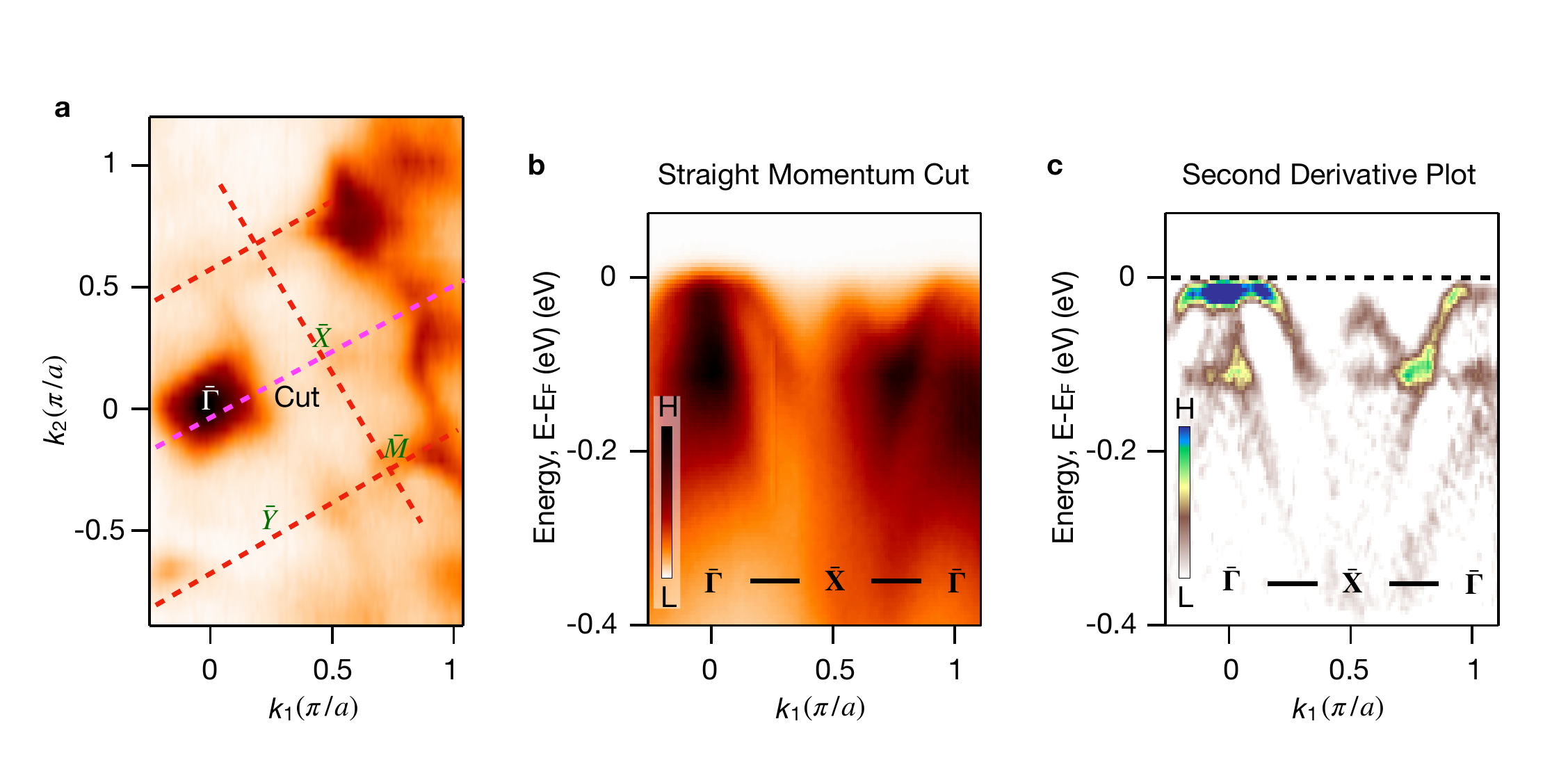}
\caption{\textbf{Surveying the $\bar{\Gamma}$ and $\bar{X}$ pockets on the (001) surface of CoSi.} 
\textbf{a}, ARPES Fermi surface with the Brillouin zone boundary (red dotted line) and the location of the energy-momentum cut of interest (magenta dotted line). \textbf{b}, $\bar{\Gamma}-\bar{X}-\bar{\Gamma}$ high-symmetry line energy-momentum cut. \textbf{c}, Second derivative plot of (b). We observe an outer and inner hole-like band at $\bar{\Gamma}$, while at $\bar{X}$ we observe signatures of a single hole-like band.}
\label{ExtFig8_bulk_pockets}
\end{figure}

We study an MDC cutting through the Fermi arc as a function of incident photon energy (Extended Data Fig.~\ref{ExtFig6}). We find that the Fermi arc does not disperse significantly from 80 eV to 110 eV, providing additional evidence that it is indeed a surface state.

\clearpage
\newpage

\clearpage
\newpage

\begin{table}
\begin{center}
\centering
\textsf{\footnotesize
\begin{tabular}{p{6cm}p{3.5cm}p{3.5cm}}
\hline
CoSi & T = 100(2) K & T = 300(2) K\\
\hline
Scan & 1 & 2 \\
F.W. (g/mol) & 87.02 & 87.02 \\
Space Group; $Z$ & $P2_{1}3$ (No.198); 4 & $P2_{1}3$ (No.198); 4\\
$a (\textrm{\AA})$ & 4.433(4) & 4.4245(16)\\
$V (\textrm{\AA}^{3}$) & 87.1(2) & 86.61(9)\\
Absorption Correction & Numerical & Numerical\\
Extinction Coefficient  & 0.39(9) & 0.5(2)\\
$\theta$ range (deg) & 19.301- 31.714 & 18.740- 31.781\\
No. Reflections; R$_{int}$ & 165; 0.0222 & 167;0.0232\\
No. Independent Reflections & 79 & 77\\
No. Parameters & 9 & 9\\
$R_1$; $wR_2$ (all \textit{I}) & 0.0223; 0.0561 & 0.0533; 0.1209\\
Goodness of fit & 1.123 & 1.142\\
Diffraction peak and hole (e$^{-}/\textrm{\AA}^3$) & 0.637; -0.736 & 1.155; -1.895\\
\hline
\end{tabular}
}
\end{center}
\caption{Single crystal crystallographic data for CoSi at 100(2) and 300(2)K.}
\label{ExtTab1}
\end{table}

%\begin{table}
%\begin{center}
%\centering
%\textsf{\footnotesize
%\begin{tabular}{p{2cm}p{2cm}p{2cm}p{2cm}p{2cm}p{2cm}p{2cm}}
%\hline
%T = 100(2) K\\
%\hline
% Atom & Wyckoff & Occupancy & x ($\textrm{\AA}$) & y ($\textrm{\AA}$) & z ($\textrm{\AA}$) & U$_{eq}$ ($\textrm{\AA}^{2}$)\\
%\hline
%Co & 4$a$ & 1 & 0.3934(1) & 0.3934(1)& 0.3934(1)& 0.0009(7)\\
%Si & 4$a$ & 1 & 0.0936(3) & 0.0936(3) & 0.0936(3) & 0.0016(7)\\
%\hline
%\end{tabular}
%\centering
%\begin{tabular}{p{2cm}p{2cm}p{2cm}p{2cm}p{2cm}p{2cm}p{2cm}}
%T = 300(2) K\\
%\hline
% Atom & Wyckoff & Occupancy & x ($\textrm{\AA}$) & y ($\textrm{\AA}$) & z ($\textrm{\AA}$) & U$_{eq}$ ($\textrm{\AA}^{2}$)\\
%\hline
%Co & 4$a$ & 1 & 0.3934(1) & 0.3934(1)& 0.3934(1)& 0.0009(7)\\
%Si & 4$a$ & 1 & 0.0936(3) & 0.0936(3) & 0.0936(3) & 0.0016(7)\\
%\hline
%\end{tabular}
%}
%\end{center}
%\caption{Atomic coordinates and equivalent isotropic displacement parameters of CoSi at 100(2) and 300(2) K. $U_{eq}$ is defined as one-third of the trace of the orthogonalized $U_{ij}$ tensor.}
%\label{ExtTab2}
%\end{table}

\clearpage
\newpage

\clearpage

\clearpage

\clearpage

\clearpage
\newpage

\clearpage
\newpage

\clearpage
\newpage

\clearpage
\newpage

\clearpage
\newpage

\clearpage
%\cleardoublepage
\ifdefined\phantomsection
  \phantomsection  % makes hyperref recognize this section properly for pdf link
\else
\fi
\addcontentsline{toc}{section}{Bibliography}

{\singlespacing

}

%\end{document} 

%% file: ch-magnet/magnet.tex
\chapter{Topological Weyl lines and drumhead surface states\\ in a room-temperature magnet}
\label{ch:magnet}

%{\singlespacing
%\begin{chapquote}{The Naked and Famous, \textit{Hearts Like Ours}}
%Could we try to reinvent?\\
%Feed the head with common sense\\
%Through the streets and avenues\\
%Climbing up the walls with you\\
%\end{chapquote}}

%{\singlespacing
%\begin{chapquote}{The Naked and Famous, \textit{What We Want}}
%Yeah we don't know\\
%What we want\\
%Just keep it trivial
%\end{chapquote}}

%{\singlespacing
%\begin{chapquote}{The Naked and Famous, \textit{What We Want}}
%All the chemicals reel in the absence of the noise\\
%We are fools in the wake of the physical\\
%I just don't know what I want\\
%You just don't know what you want\\
%Yeah we don't know what we want\\
%Just keep it trivial
%\end{chapquote}}

{\singlespacing
\begin{chapquote}{Charli XCX, \textit{Break The Rules}}
I don't wanna go to school\\
I just wanna break the rules
\end{chapquote}}

%\RequirePackage[right]{lineno}
%\setlength\linenumbersep{1.5cm}
%\documentclass[aps,prl,preprint,nopacs,superscriptaddress]{revtex4}
%\usepackage{amsmath}
%\usepackage{amssymb}
%\usepackage{graphicx}
%\usepackage{hyperref}
%\usepackage[utf8]{inputenc}
%\pagestyle{headings}
%\newcommand{\beq}{\begin{equation*}}
%\newcommand{\eeq}{\end{equation*}}

% If compiling all chapters
\newcommand{\s}{Co$_2$MnGa}
\renewcommand{\eb}{$E_\textrm{B}$}
\newcommand{\ai}{\textit{ab initio}}
\renewcommand{\pana}{A}
\renewcommand{\panb}{B}
\renewcommand{\panc}{C}
\renewcommand{\pand}{D}
\renewcommand{\pane}{E}
\renewcommand{\panf}{F}
\renewcommand{\pang}{G}
\renewcommand{\panh}{H}
\renewcommand{\pani}{I}
\renewcommand{\panj}{J}

\noindent This chapter is based on the article, \textit{Discovery of topological Weyl lines and drumhead surface states in a room-temperature magnet} by Ilya Belopolski*, Kaustuv Manna* \textit{et al}., available at \href{https://arxiv.org/abs/1712.09992}{arXiv:1712.09992}.\\

\lettrine[lines=3]{T}{opological} phases of matter are known to exhibit anomalous transport and surface state dispersion. While the quantum anomalous Hall state has been demonstrated in two-dimensional materials, a three-dimensional topological magnetic state with its anomalous electron transport and surface states remains elusive. Here we use photoemission spectroscopy, density functional theory and transport to elucidate the electronic topology of the room temperature ferromagnet Co$_2$MnGa. We observe sharp bulk Weyl line dispersions indicative of non-trivial topological invariants existing in the magnetic ground state. On the surface of the sample, we observe electronic states which take the form of drumheads, directly visualizing the topological bulk-boundary correspondence. The topological Weyl lines are further found to contribute substantially to the giant anomalous Hall response in transport. Our experimental results demonstrate the topological transport-bulk-surface correspondence of a magnet for the first time and suggest the rich interplay of magnetism and topology in quantum materials.
%\end{abstract}
%
%\date{\today}
%\maketitle

\newcommand{\mbs}{Mn:Bi$_2$Se$_3$}
\newcommand{\mbstf}{Mn(2.5\%):Bi$_2$Se$_3$}

\section{Prologue: Chern insulator state in \mbs\ accessed by spin-resolved ARPES}

Before discussing the magnetic topological semimetal \s, I first review our key earlier work demonstrating a two-dimensional Chern insulating state on the surface of $\mathbb{Z}_2$ topological insulator with magnetic impurities \cite{SYX_NP_MnBi2Se3_2012}. The magnetic impurities introduce a perturbation which breaks time-reversal symmetry $T$. Under such a perturbation, the $\mathbb{Z}_2$ topological insulator Dirac cone acquires a mass gap with continuum Hamiltonian of the form $h(k) = k_x \sigma_x + k_y \sigma_y + m \sigma_z$. In the continuum, such a Dirac cone with mass gap carries a half-integer quantum Hall conductivity, as can be shown by integrating the Berry curvature which I write down at the end of Chap. \ref{ch:theory}. In this way, magnetizing a topological insulator through magnetic impurities offers the opportunity to realize a Chern insulating state on the surface of the sample. Notably, the phase is robust in the sense that the sign of the mass gap is unimportant. This is because a change in the sign of the Dirac mass $m$ constitutes a band inversion which changes the conductivity by $1$, but in the present case this simply corresponds to a change of $\sigma_{xy} = +1/2 \rightarrow -1/2$, so both signs of $m$ essentially give rise to the same half-integer quantum Hall state. Apart from the dispersion gap, a key hallmark of this state is a hedgehog (pseudo)spin texture. These two smoking-gun signatures lead to ARPES as a natural probe to demonstrate the Chern insulating state. In Ref. \cite{SYX_NP_MnBi2Se3_2012}, we study thin films of Bi$_2$Se$_3$ doped with Mn, which acts as a magnetic impurity that breaks $T$ (Fig. \ref{MnBi2Se3}a,b). This then leads to a Dirac gap directly observed by ARPES (Fig. \ref{MnBi2Se3}c), with associated characteristic hedgehog spin texture (Fig. \ref{MnBi2Se3}d,e). Furthermore, we demonstrate through {\it in situ} NO$_2$ dosing that the Fermi  level of the surface state can be tuned into the Chern gap, producing a quantum anomalous Hall state (Fig. \ref{MnBi2Se3}f-h). The magnetic gap on the surface of \mbs, along with the hedgehog spin texture and {\it in situ} Fermi level tuning, confirms the realization of a two-dimensional Chern insulating state on the surface of a $\mathbb{Z}_2$ topological insulator with $T$-breaking perturbation. The observation of this magnetic topological insulating state sets the stage for the observation of a magnetic topological semimetal, which I now discuss.

\clearpage
\begin{figure}
\centering
\includegraphics[width=15.5cm,trim={0.5in 1in 0.5in 1in},clip]{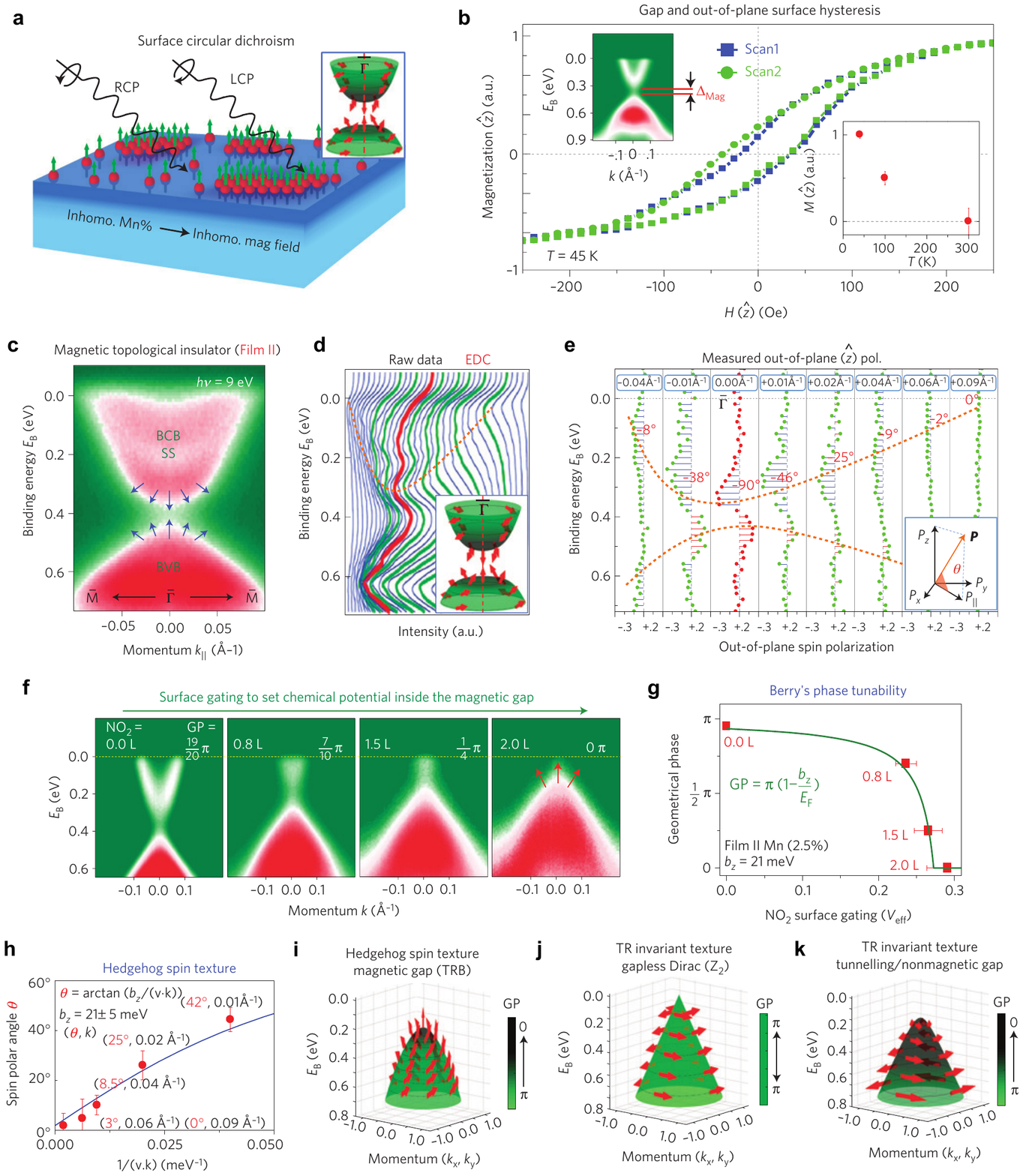}
\end{figure}
\clearpage

\begin{figure*}[h!]
\caption{\label{MnBi2Se3} {\bf A Chern gap on the surface of a magnetized $\mathbb{Z}_2$ topological insulator.} {\bf a,} The Mn atoms on the surface of the film have a finite out-of-plane magnetic moment component, serving as a local magnetic field which results in the spin texture reorientation. {\bf b,} Two independent hysteresis measurements at $T=45$ K using XMCD reveal the ferromagnetic character of the \mbs\ film surface. The lower inset shows the remanent surface magnetization as a function of temperature. The upper inset shows the gap at the Dirac point in the \mbstf\ film. {\bf c, d,} Spin-integrated dispersion and EDCs. The EDCs selected for detailed spin-resolved measurements are highlighted in green and red. {\bf e,} Measured out-of-plane spin polarization of the EDCs corresponding to {\bf d}. Inset defines the definition of the spin polarization vector $P$ and the polar angle $\theta$. The momentum value of each spin-resolved EDC is noted on the top. The polar angles $\theta$ of the spin vectors obtained from measurements are also noted. The $90^{\circ}$ polar angle observed at $\bar{\Gamma}$ suggests that the spin vector at $\bar{\Gamma}$ momenta points in the vertical direction. The spin behaviour at $\bar{\Gamma}$ and its surrounding momentum space reveals a hedgehog-like spin configuration for each Dirac band separated by the gap, which breaks TR symmetry ($E(k=0,\uparrow) \neq E(k=0,\downarrow)$), as schematically shown in the inset of {\bf d}. {\bf f,} Measured surface state dispersion on {\it in situ} NO$_2$ surface adsorption on the \mbs\ surface. The NO$_2$ dosage in Langmuir units ($1.0 L=1 \times 10^{-6}$ Torr s) and the tunable geometrical phase (GP) associated with the topological surface state are noted on the top-left and top-right corners of the panels, respectively. The red arrows depict the TR breaking out-of-plane spin texture at the gap edge based on the experimental data. {\bf g,} GP or Berry's phase associated with the spin texture on the iso-energetic contours on the \mbs\ surface as a function of effective gating voltage induced by NO$_2$ surface adsorption. Red squares represent the GP experimentally realized by NO$_2$ surface adsorption, as shown in {\bf f}. $GP=0$  (NO$_2 =2.0$ L) is the condition for axion dynamics. The error bars of the red squares are estimated based on the combination of the energy resolution of the ARPES experiments, the binding energy calibration with respect to the Fermi level, as well as the fitting uncertainty in fitting the ARPES measured surface states in {\bf f}, using the experimentally based $k\cdot p$ model. {\bf h,} The magnetic interaction strength $b_z$ (see \cite{SYX_NP_MnBi2Se3_2012} for definition), which corresponds to half of the magnetic gap magnitude, is obtained on the basis of spin-resolved data sets (polar angle $\theta$, momentum $k$) for \mbstf film II, {\bf c-e}. {\bf i,} The TR breaking spin texture features a singular hedgehog-like configuration when the chemical potential is tuned to lie within the magnetic gap, corresponding to the experimental condition presented in the last panel in {\bf f}. {\bf j,k,} Spin texture schematic based on measurements of Zn-doped Bi$_2$Se$_3$ film (60 QL), and 3 QL undoped ultrathin film. (Adapted from Ref. \cite{SYX_NP_MnBi2Se3_2012}.)}
\end{figure*}

% The out-of-plane magnetic hysteresis and ARPES gap were found to be correlated with each other.

\section{Introduction}

The discovery of topological phases of matter has led to a new paradigm in physics, which not only explores the analogs of particles relevant for high energy physics, but also offers new perspectives and pathways for the application of quantum materials \cite{news_Castelvecchi,RMPMajorana_Franz,ReviewQuantumMaterials_Nagaosa,QAH_Xue,ReviewQuantumMaterials_Hsieh,RMPTopoBandThy_Bansil,Colloquium_Zahid,ARCMP_me,RMPWeylDirac_Armitage}. To date, most topological phases have been discovered in non-magnetic materials \cite{Colloquium_Zahid,RMPWeylDirac_Armitage,ARCMP_me}, which severely limits their magnetic field tunability and electronic/magnetic functionality. Identifying and understanding electronic topology in magnetic materials will not only provide indispensable information to make their existing magnetic properties more robust, but also has the potential to lead to the discovery of novel magnetic response that can be used to explore future spintronics technology. Recently, several magnets were found to exhibit a large anomalous Hall response in transport, which has been linked to a large Berry curvature in their electronic structures \cite{Co2MnGa_Kaustuv,Co2MnGa_Nakatsuji,Co3Sn2S2_Enke,Mn3Sn_Nakatsuji,Fe3GeTe2_Kim_Pohang,Fe3Sn2_Checkelsky,Co2MnGa_Felser_Nernst,Co2MnGa_Felser_films}. However, it is largely unclear in experiment whether the Berry curvature originates from a topological band structure, such as Dirac/Weyl point or line nodes, due to the lack of spectroscopic investigation. In particular, there is no direct visualization of a topological magnetic phase demonstrating a bulk-boundary correspondence with associated anomalous transport.

Here we use angle-resolved photoemission spectroscopy (ARPES), \ai\ calculation and transport to explore the electronic topological phase of the full-Heusler ferromagnet \s\ \cite{Co2MnGa_Guoqing}. In our ARPES spectra we discover a line node in the bulk of the sample. Taken together with our \ai\ calculations, we conclude that we observe Weyl lines protected by crystalline mirror symmetry and requiring magnetic order. In ARPES we further observe drumhead surface states connecting the bulk Weyl lines, revealing a bulk-boundary correspondence in a magnet. Combining our ARPES and \ai\ calculation results with transport, we further find that Berry curvature concentrated by the Weyl lines accounts for a significant fraction of the anomalous Hall response in \s.

Weyl lines can be understood within a simple framework where one categorizes a topological phase by the dimensionality of the band touching: there are topological insulators, point node semimetals and line node semimetals \cite{ClassificationGapped_Schnyder,SemimetalReview_Vishwanath,ClassificationReflection_Schnyder,TINI_Balents}. Point nodes are often further sub-categorized as Dirac points, Weyl points and other exotic point touchings \cite{SemimetalReview_Vishwanath,RMPWeylDirac_Armitage}. Analogously, line nodes can include Dirac lines (four-fold degenerate), Weyl lines (two-fold degenerate) and possibly other one-dimensional band crossings \cite{WeylDiracLoop_Nandkishore,WeylLoopSuperconductor_Nandkishore,WeylLines_Kane}. Line nodes can be protected by crystal mirror symmetry, giving rise to a $\pi$ Berry phase topological invariant and drumhead surface states \cite{ClassificationReflection_Schnyder,PbTaSe2_Guang,NodalChain_Soluyanov,RMPClassification_ChingKai,DiracLineNodes_WeiderKane,Ca3P2_Schnyder}. \s\ takes the full-Heusler crystal structure (Fig. \ref{gacoFig1}\pana), with a cubic face-centered Bravais lattice, space group $Fm\bar{3}m$ (No. 225), indicating the presence of several mirror symmetries in the system. Moreover, the material is ferromagnetic with Curie temperature $T_\textrm{C} = 690$ K (Fig. \ref{gacoFig1}\panb) \cite{Co2MnGa_CurieTemp}, indicating intrinsically broken time-reversal symmetry. This suggests that all bands are generically singly-degenerate and that mirror symmetry may give rise to two-fold degenerate line nodes. As a result, we anticipate that \s\ favors Weyl lines in its band structure (Fig. \ref{gacoFig1}\panc).

\section{Observation of magnetic Weyl lines}

\begin{figure}
\centering
\includegraphics[width=14cm,trim={1in 4.1in 1in 1in},clip]{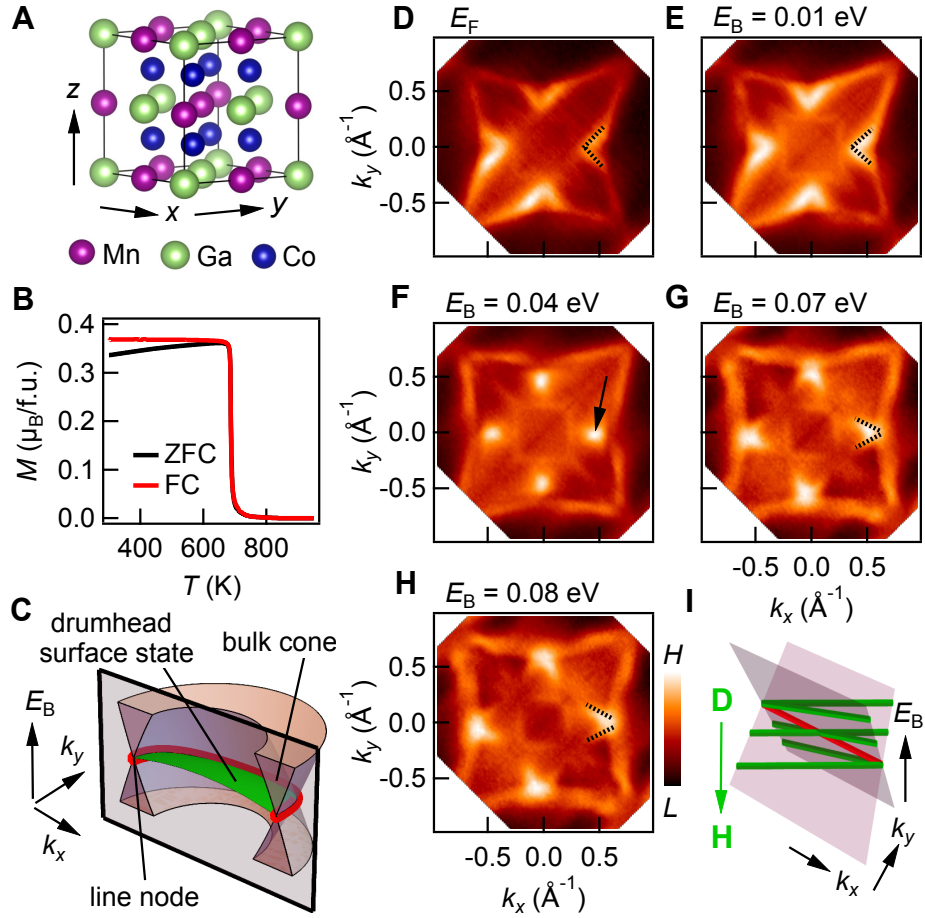}
\caption{\label{gacoFig1} {\bf Magnetic line node in \s.} ({\bf \pana}) Crystal structure of \s. ({\bf \panb}) Magnetization as a function of temperature of \s\ single crystals, in the absence of a magnetic field (zero-field-cooled, ZFC) and cooled under a constant magnetic field of $\mu_0 H = 200$ Oe oriented along the [001] crystallographic axis (field-cooled, FC). We find a Curie temperature $T_C = 690$ K. ({\bf \panc}) Schematic of a generic line node. A line node (red curve) is a band degeneracy along an entire curve in the bulk Brillouin zone. It is associated with a $\pi$ Berry phase topological invariant and a drumhead surface state stretching across the line node (green sheet). In the case of a mirror-symmetry-protected line node, the line node lives in a mirror plane of the Brillouin zone, but it is allowed to disperse in energy. ({\bf \pand-\panh}) Constant-energy surfaces of \s\ measured by ARPES at $h \nu = 50$ eV and $T = 20$ K, presented at a series of binding energies, \eb, from the Fermi level, $E_\textrm{F}$, down to \eb\ = 0.08 eV. ({\bf \pani}) Schematic of constant-energy cuts (green curves) of a line node, suggesting a correspondence with the observed ARPES dispersion.}
\end{figure}

\begin{figure}
\centering
\includegraphics[width=15cm,trim={1in 5.5in 1in 1in},clip]{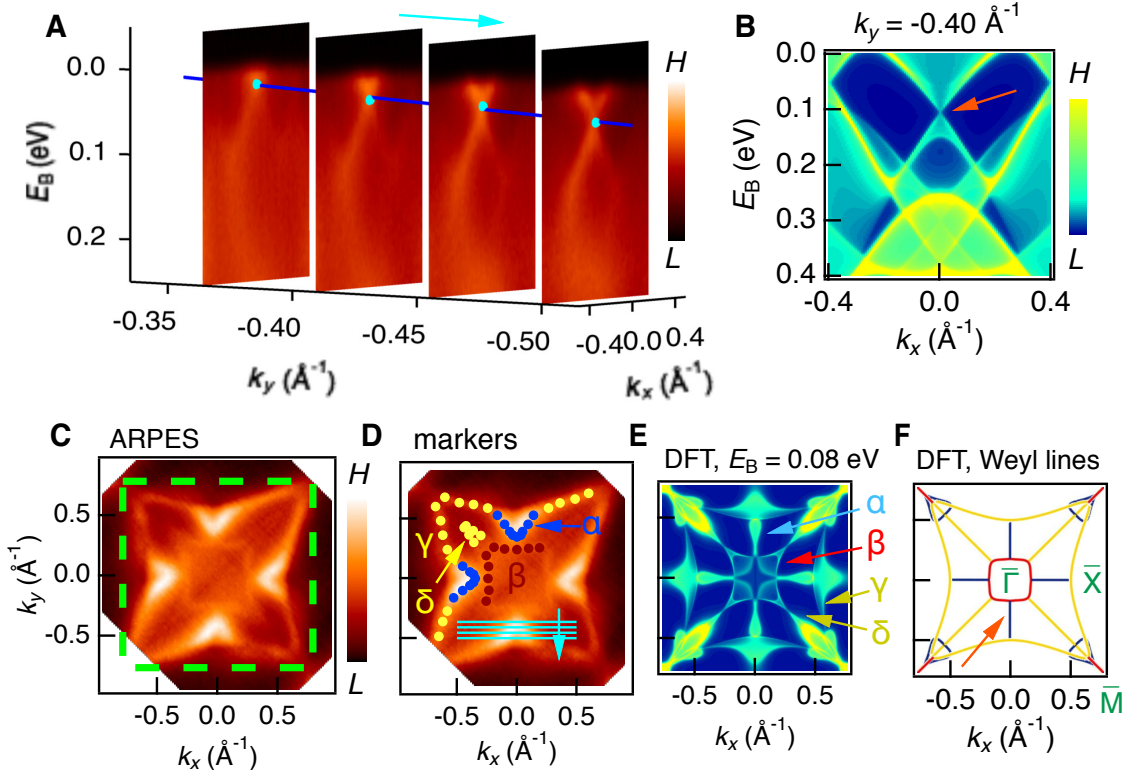}
\caption{\label{gacoFig2} {\bf Evidence for a Weyl line.} ({\bf \pana}) Series of ARPES $E_\textrm{B}-k_x$ cuts through the candidate line node, corresponding to the feature discussed in Fig. \ref{gacoFig1}. The band crossing points near $k_x = 0$ are fit with a single Lorentzian peak (cyan dots) and the train of dots is then fit with a line (blue line), the experimentally-observed line node dispersion. ({\bf \panb}) \textit{Ab initio} $E_\textrm{B}-k_x$ predicted bulk bands of \s\ in the ferromagnetic state, projected on the (001) surface, predicting a Weyl line at $k_x = 0$ (orange arrow) \cite{Co2MnGa_Guoqing}. The colors indicate the spectral weight of a given bulk state on the surface, obtained using an iterative Green's function method \cite{GreensFunction_Bryant}. ({\bf \panc}) Same as Fig. \ref{gacoFig1}\pane, with (001) surface Brillouin zone marked (green box). ({\bf \pand}) Key features of the data, obtained from analysis of the momentum and energy distribution curves of the ARPES spectrum. ({\bf \pane}) \textit{Ab initio} constant-energy surface at binding energy $E_\textrm{B} = 0.08$ eV below $E_\textrm{F}$, on the (001) surface with MnGa termination, showing qualitative agreement with the ARPES, as marked by $\alpha-\delta$. ({\bf \panf}) Projection of the predicted Weyl lines on the (001) surface, with energy axis collapsed, suggesting that the key features observed in ARPES and DFT arise from the predicted Weyl lines: $\alpha$ (blue Weyl line), $\beta$ (red), $\gamma$ (yellow), $\delta$ (another copy of the yellow Weyl line). }
\end{figure}

Motivated by these fundamental considerations, we investigate \s\ single crystals by ARPES. We focus first on the constant energy surfaces at different binding energies, \eb. We readily observe a feature which exhibits an unusual evolution from a $<$ shape (Fig. \ref{gacoFig1}\pand,\pane) to a dot (Fig. \ref{gacoFig1}\panf) to a $>$ shape (Fig. \ref{gacoFig1}\pang-\panh). This feature suggests that we observe a pair of bands which touch at a series of points in momentum space. As we shift downward in \eb, the touching point moves from left to right (black guides to the eye) and we note that at certain \eb\ (Fig. \ref{gacoFig1}\panf) the spectral weight appears to be dominated by the crossing point. This series of momentum-space patterns is characteristic of a line node (Fig. \ref{gacoFig1}\pani). For the constant-energy surfaces of a line node, as we slide down in \eb\ the touching point slides from one end of the line node to the other, gradually zipping closed an electron-like pocket (upper band) and unzipping a hole-like pocket (lower band). To better understand this result, we consider $E_\textrm{B}-k_x$ cuts passing through the line node feature (Fig. \ref{gacoFig2}\pana). On these cuts, we observe a candidate band crossing near $k_x = 0$. We further find that this crossing persists in a range of $k_y$ and moves downward in energy as we cut further from $\bar{\Gamma}$ (more negative $k_y$). We can fit the candidate band crossing with a single Lorentzian peak, suggestive of a series of touching points between the upper and lower bands (Fig. S13). Taking these fitted touching points, we can in turn fit the dispersion of the candidate line node to linear order, obtaining a slope $m = 0.079 \pm 0.018\ \textrm{eV\AA}$. Lastly, we observe that at a given $k_y$, the bands disperse linearly in energy away from the touching points. In this way, our ARPES results suggest the presence of a line node at the Fermi level in \s.

To better understand our experimental results, we compare our spectra with an \textit{ab initio} calculation of \s\ in the ferromagnetic state \cite{Co2MnGa_Guoqing}. We consider the spectral weight of bulk states on the (001) surface and we study an $E_\textrm{B}-k_x$ cut in the region of interest (Fig.\ref{gacoFig2}\panb). At $k_x = 0$ we observe a band crossing (orange arrow) which we can trace back in numerics to a line node associated with a $\pi$ Berry phase. The line node lives near the $X$ point of the bulk Brillouin zone, shown as the blue line node in Fig. S9. It is protected by the mirror symmetry of the crystal structure of \s\ \cite{Co2MnGa_Guoqing}. This line node is a Weyl line, in the sense that it is a two-fold degenerate band crossing extended along one dimension \cite{WeylDiracLoop_Nandkishore,WeylLoopSuperconductor_Nandkishore,WeylLines_Kane}. It can be viewed as a Weyl point where the point crossing has been upgraded to a line crossing by the addition of a mirror symmetry. To compare experiment and theory in greater detail, we plot the calculated dispersion of the Weyl line against the dispersion as extracted from Lorentzian fits of ARPES data (Fig. S13). We observe a hole-doping of experiment relative to theory of $E_\textrm{B} = 0.08 \pm 0.01 \ \textrm{eV}$. We speculate that this shift may be due to a chemical doping of the sample or an approximation in the way that DFT captures magnetism in this material. The correspondence between the crossing observed in \textit{ab initio} calculation and ARPES suggests that we have observed a magnetic Weyl line in \s.

Having considered the blue line node, we search for other line nodes in our data. We compare an ARPES spectrum (Fig. \ref{gacoFig2}\panc,\pand) to an \textit{ab initio} calculation of the surface spectral weight of bulk states, taking into account the observed effective hole-doping of our sample (Fig. \ref{gacoFig2}\pane). In addition to the blue Weyl line (labelled here as $\alpha$), we observe a correspondence between three features in experiment and theory: $\beta$, $\gamma$ and $\delta$. To better understand the origin of these features, we consider all of the predicted Weyl lines in \s\ \cite{Co2MnGa_Guoqing} and we plot their surface projection with the energy axis collapsed (Fig. \ref{gacoFig2}\panf). We observe a correspondence between $\beta$ in the ARPES spectrum and the red Weyl line. Similarly, we see that $\gamma$ and $\delta$ match with predicted yellow Weyl lines. To further test this correspondence, we look again at our ARPES constant-energy cuts and we find that $\delta$ exhibits a $<$ to $>$ transition suggestive of a line node (Figs. S15, S16). The comparison between ARPES and \ai\ calculation suggests that an entire network of magnetic Weyl lines live in \s.

\section{Signatures of drumhead surface states}

\begin{figure}
\centering
\includegraphics[width=15.5cm,trim={1.2in 6.6in 1in 1in},clip]{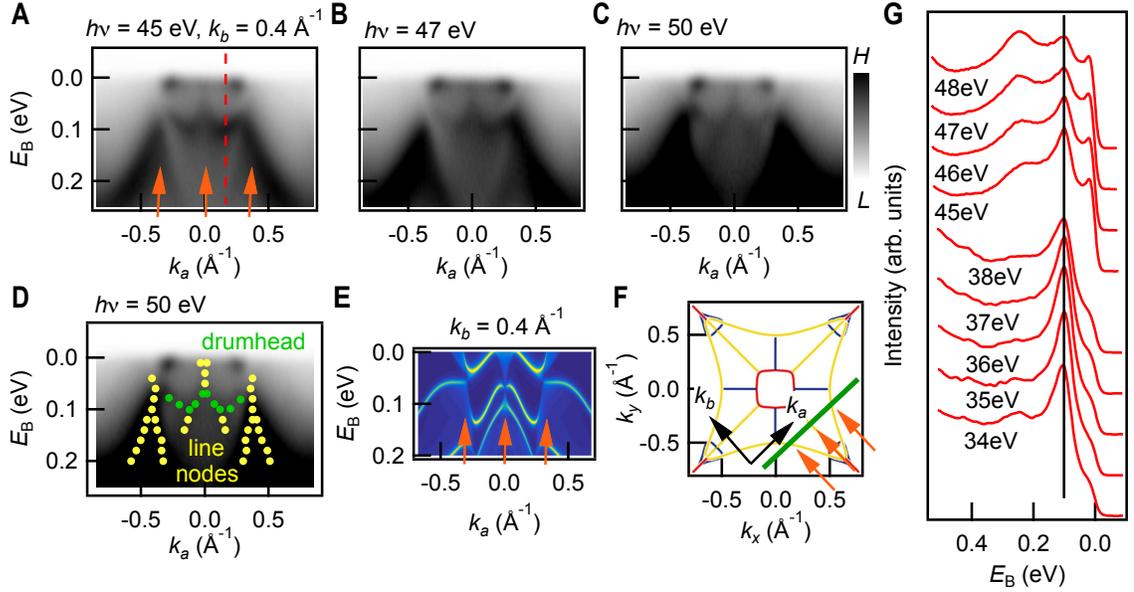}
\caption{\label{FigDrum} \textbf{Drumhead surface states in Co$_2$MnGa.} (\textbf{\pana}-\textbf{\panc}) ARPES $E_\textrm{B}-k_a$ cuts at different photon energies. We observe three cone-like features (orange arrows). (\textbf{\pand}) Key features of the data of \panc, obtained from analysis of the momentum and energy distribution curves (MDCs/EDCs). Apart from the cone-like features (yellow) there are additional states (green) connecting the cones. (\textbf{\pane}) The corresponding $E_\textrm{B}-k_a$ cut from \textit{ab initio} calculation, crossing three Weyl lines (orange arrows) connected by drumhead surface states. (\textbf{\panf}) Same as Fig. \ref{gacoFig2}\panf, marking the location of the ARPES spectra in \textbf{\pana}-\textbf{\panc} (green line) and defining the $k_{a,b}$ axes. (\textbf{\pang}) Photon energy dependence of an EDC passing through the candidate drumhead state (red dotted line in \pana). The peaks marked by the black vertical line correspond to the drumhead surface state. We observe no dispersion as a function of photon energy, providing evidence that the candidate drumhead is a surface state.}
\end{figure}

Next we explore the topological surface states. We study the ARPES spectrum along $k_a$, as marked by the green line in (Fig. \ref{FigDrum}\panf). On this cut we observe three cones (orange arrows in Fig. \ref{FigDrum}\pana) which are consistent with the yellow Weyl lines. Interestingly, we also observe a pair of states which appear to connect one cone to the next (Fig. \ref{FigDrum}\pana-\panc). Moreover, these extra states consistently terminate on the candidate yellow Weyl lines as we vary $k_b$ (Fig. S19). We further carry out a photon energy dependence and we discover that these extra states do not disperse with photon energy from $h \nu = 34$ to $48$ eV, suggestive of a surface state (Fig. \ref{FigDrum}\pang). In \ai\ calculation, we observe a similar pattern of yellow Weyl lines pinning a surface state (Fig. \ref{FigDrum}\pane). These observations suggest that we have observed a drumhead surface state stretching across Weyl lines in \s. The pinning of the surface states to the cones further points to a bulk-boundary correspondence between the bulk Weyl lines and the drumhead surface state dispersion.

\section{Weyl lines as the source of the intrinsic anomalous Hall response}

\begin{figure}
\centering
\includegraphics[width=16cm,trim={0in 3.3in 0in 0.4in},clip]{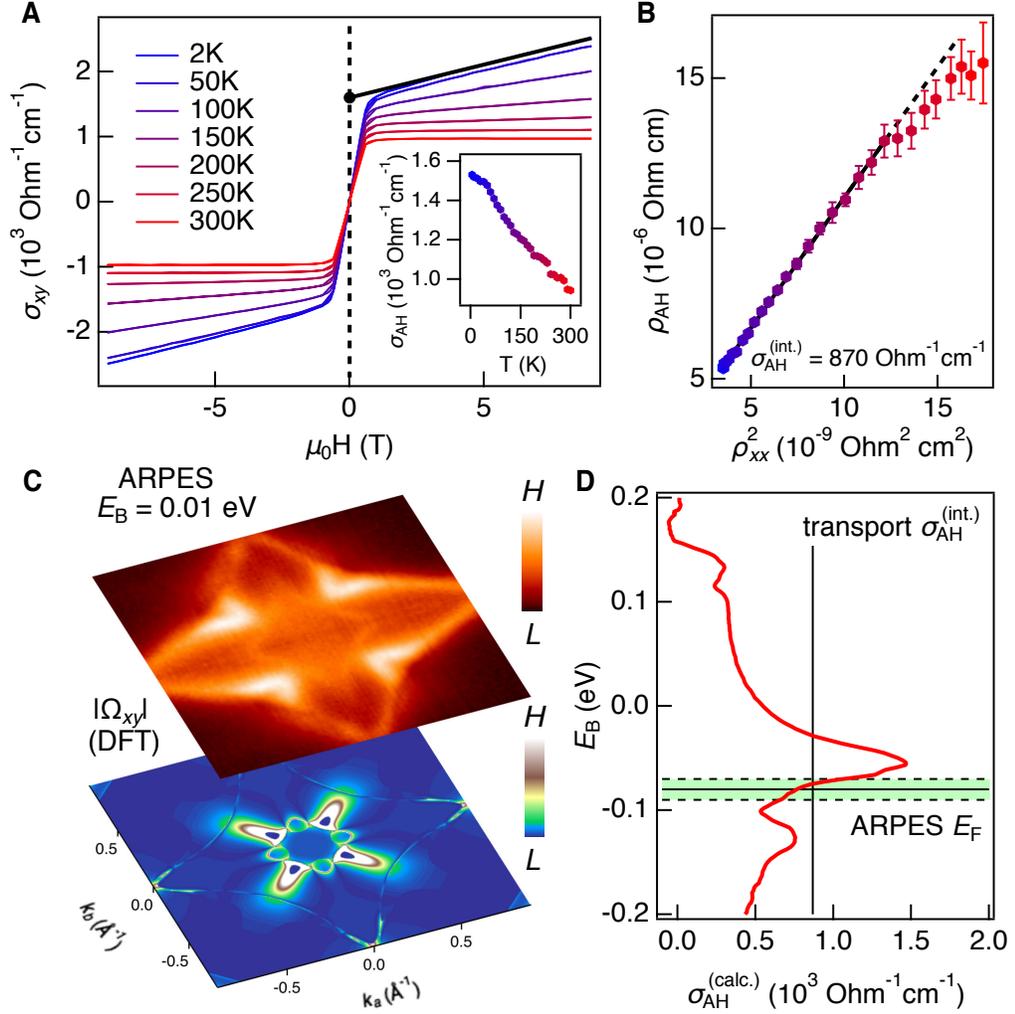}
\caption{\label{AHE} {\bf Large anomalous Hall response from topological Weyl lines.} ({\bf \pana}) The Hall conductivity $\sigma_{xy}$, measured as a function of applied magnetic field $\mu_0 H$ at several representative temperatures $T$, after two-point averaging of the raw data (Fig. S5), with $\mu_0 H$ applied along $[110]$ and current along $[001]$. Inset: the anomalous Hall conductivity, $\sigma_{\textrm{AH}}(T)$, obtained from $\sigma_{xy}$. ({\bf \panb}) The anomalous Hall resistivity $\rho_{\textrm{AH}}$ plotted against $\rho_{xx}^2$, both as functions of $T$, as indicated by the colors: blue (2 K) $\rightarrow$ red (300 K). A linear scaling relation estimates the intrinsic, Berry curvature contribution to the AHE, given by the slope of the line \cite{ScalingAHE_TianYeJin_2009,ScalingAHE_TianJinNiu_2015,ScalingAHE_TianJin_2016}. ({\bf \panc}) Bottom: $z$-component of the Berry curvature integrated up to $E_\textrm{B} = -0.09$ eV, $|\Omega_{xy}|$. Top: the ARPES constant-energy surface at the corresponding \eb\ (same as Fig. \ref{gacoFig1}\pand). The correspondence between ARPES and DFT suggests that the Berry curvature is dominated by the Weyl lines. ({\bf \pand}) 
Prediction of $\sigma^{\textrm{int}}_{\textrm{AH}}$ by integrating the Berry curvature from DFT up to a given \eb\ (red curve), with $E_\textrm{F}$ set from ARPES and compared to the estimated $\sigma^{\textrm{int}}_{\textrm{AH}}$ from transport. The ARPES/DFT prediction is consistent with transport, suggesting that the line nodes dominate the large, intrinsic AHE in \s.}
\end{figure}

Now that we have provided spectroscopic evidence for a magnetic bulk-boundary correspondence in \s, we investigate the relationship between the topological line nodes and the anomalous Hall effect (AHE). We study the Hall conductivity $\sigma_{xy}$ under magnetic field $\mu_0 H$ at different temperatures $T$ and we extract the anomalous Hall conductivity $\sigma_\textrm{AH} (T)$ (Fig. \ref{AHE}\pana). We obtain a very large AHE value of $\sigma_{\textrm{AH}} = 1530 \ \Omega^{-1} \ \textrm{cm}^{-1}$ at 2 K, consistent with earlier reports \cite{Co2MnGa_Kaustuv,Co2MnGa_Nakatsuji}. To understand the origin of the large AHE, we study the scaling relation between the anomalous Hall resistivity, $\rho_\textrm{AH}$, and the square of the longitudinal resistivity, $\rho_{xx}^2$, both considered as a function of temperature. It has been shown that under the appropriate conditions, the scaling relation takes the form,
\beq
\rho_\textrm{AH} = (\alpha \rho_{xx0} + \beta \rho^2_{xx0}) + b \rho^2_{xx},
\eeq
where $\rho_{xx0}$ is the residual longitudinal resistivity, $\alpha$ represents the contribution from skew scattering, $\beta$ represents the side-jump term and $b$ represents the intrinsic Berry curvature contribution to the AHE \cite{ScalingAHE_TianYeJin_2009,ScalingAHE_TianJinNiu_2015,ScalingAHE_TianJin_2016}. When we plot $\rho_\textrm{AH}$ against $\rho_{xx}^2$, we observe that a linear scaling appears to hold below $\sim 230$ K (Fig. \ref{AHE}\panb). We conjecture that the deviation from linearity at high temperature may signify a change in the Berry curvature distribution. From the linear fit, we find that the intrinsic Berry curvature contribution to the AHE is $b = 870 \ \Omega^{-1} \ \textrm{cm}^{-1}$. This large intrinsic AHE leads us to consider the role of the Weyl lines in producing a large Berry curvature. To explore this question, we compare the intrinsic AHE measured in transport with a prediction based on ARPES and DFT. We observe in first-principles that the Berry curvature distribution is dominated by the topological line nodes (Fig. \ref{AHE}\panc). Next we integrate the Berry curvature up to a given binding energy to predict $\sigma^\textrm{int}_{\textrm{AH}}$ as a function of the Fermi level. Then we set the Fermi level from ARPES, predicting $\sigma^\textrm{int}_{\textrm{AH}} = 770_{- 100}^{+ 130} \ \Omega^{-1} \ \textrm{cm}^{-1}$ (Fig. \ref{AHE}\pand). This is in remarkable agreement with the value extracted from transport, suggesting that the topological line nodes contribute significantly to the large AHE in \s.

\section{Conclusion}

In summary, our ARPES and corresponding transport experiments, supported by \ai\ calculation, provide evidence for magnetic Weyl lines in the room-temperature ferromagnet \s. We further find that the Weyl lines give rise to drumhead surface states and a large anomalous Hall response, providing the first demonstration of a topological magnetic bulk-boundary correspondence with associated anomalous transport. Since there are 1651 magnetic space groups and thousands of magnets in three-dimensional solids, the experimental methodology of transport-bulk-boundary exploration established here can be a valuable guideline in probing and discovering novel topological phenomena on the surfaces and the bulk of magnetic materials.

%\section{Acknowledgments}

%I.B. acknowledges the support of the US National Science Foundation GRFP and the Harold W. Dodds Fellowship of Princeton University. Work at Princeton University is supported by the Emergent Phenomena in Quantum Systems (EPiQS) Initiative of the Gordon and Betty Moore Foundation under Grant No. GBMF4547 (M.Z.H.) and by the National Science Foundation, Division of Materials Research, under Grants No. NSF-DMR-1507585 and No. NSF-DMR-1006492. H.L. acknowledges the support of the Singapore National Research Foundation under NRF Award No. NRF-NRFF2013-03. Use of the Stanford Synchrotron Radiation Lightsource (SSRL), SLAC National Accelerator Laboratory, is supported by the US Department of Energy, Office of Science, Office of Basic Energy Sciences under Contract No. DE-AC02-76SF00515. The authors thank Donghui Lu and Makoto Hashimoto for enthusiastic beamtime support at SSRL Beamlines 5-2 and 5-4. K.M., B.E. and C.F. acknowledge the financial support by the ERC Advanced Grant No. 291472 `Idea Heusler' and 742068 `TOPMAT'.

%\bibliography{ch-magnet/mybib}{}
%\bibliographystyle{ch-magnet/Science}

\clearpage

\clearpage

\clearpage

\clearpage

%\RequirePackage[right]{lineno}
%\setlength\linenumbersep{1.5cm}
%\documentclass[aps,prl,preprint,nopacs,superscriptaddress]{revtex4}
%\usepackage{amsmath}
%\usepackage{amssymb}
%\usepackage{graphicx}
%\usepackage{hyperref}
%\usepackage[utf8]{inputenc}
%\linenumbers
%\pagestyle{headings}
%\newcommand{\beq}{\begin{equation*}}
%\newcommand{\eeq}{\end{equation*}}
%\newcommand{\s}{Co$_2$MnGa}
\newcommand{\stitle}{C\lowercase{o}$_2$M\lowercase{n}G\lowercase{a}}
\newcommand{\Ai}{\textit{Ab initio}}
\newcommand{\invA}{$\textrm{\AA}^{-1}$}

\newcommand{\pank}{K}
\newcommand{\panl}{L}

\section{Materials and methods}

\subsection{Single crystal growth}

\begin{figure*}[h]
\centering
\includegraphics[width=13cm]{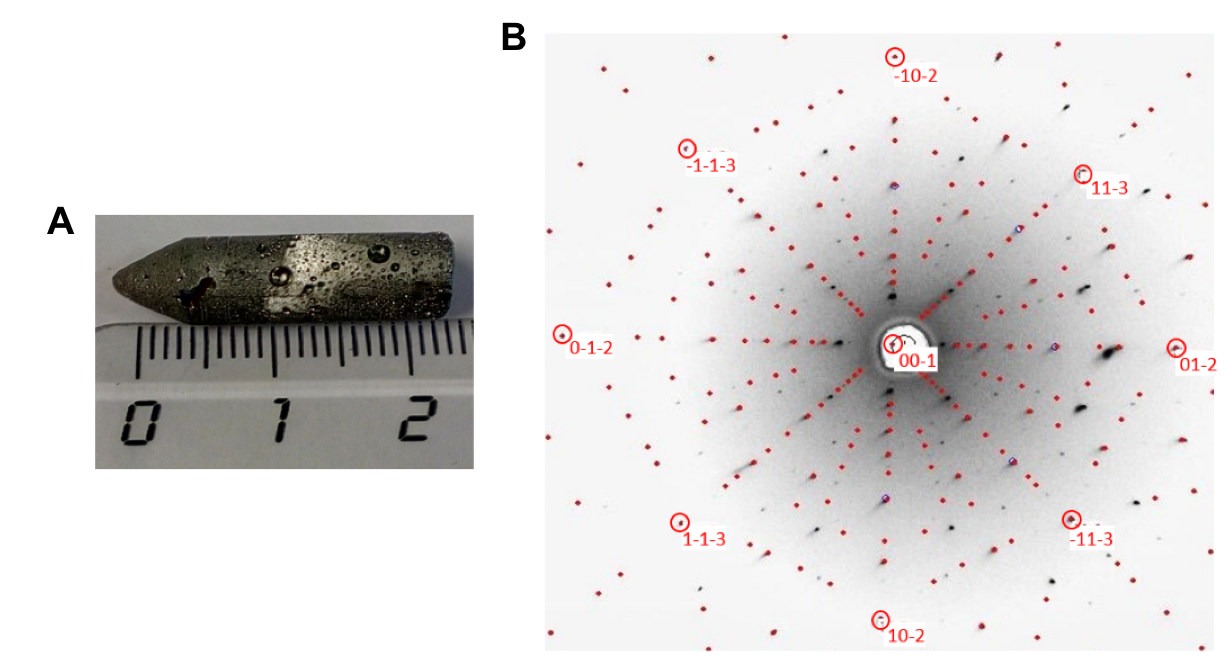}
\caption{\label{FigS1} \textbf{Crystal structure of Co$_2$MnGa.} (\textbf{\pana}) Grown single crystal of the full Heusler material Co$_2$MnGa. (\textbf{\panb}) Laue diffraction pattern of a [001] oriented crystal superposed with a theoretically simulated one.}
\end{figure*}

Single crystals of Co$_2$MnGa were grown using the Bridgman-Stockbarger crystal growth technique. First, we prepared a polycrystalline ingot using the induction melt technique with the stoichiometric mixture of Co, Mn and Ga metal pieces of 99.99\% purity. Then, we poured the powdered material into an alumina crucible and sealed it in a tantalum tube. The growth temperature was controlled with a thermocouple attached to the bottom of the crucible. For the heating cycle, the entire material was melted above 1200$^{\circ}$C and then slowly cooled below 900$^{\circ}$C (Fig. \ref{FigS1}\pana). We analyzed the crystals with white beam backscattering Laue X-ray diffraction at room temperature (Fig. \ref{FigS1}\panb). The samples show very sharp spots that can be indexed by a single pattern, suggesting excellent quality of the grown crystals without any twinning or domains. We show a representative Laue diffraction pattern of the grown Co$_2$MnGa crystal superimposed on a theoretically-simulated pattern, Fig. S1B. The crystal structure is found to be cubic $Fm\bar{3}m$ with lattice parameter $a=5.771(5)$ $\textrm{\AA}$.

\subsection{Magnetization, transport}

Magnetic measurements were performed using a Quantum Design vibrating sample magnetometer (VSM) operating in a temperature range of $2 - 950$ K with magnetic field up to 7 T. The transport experiments were performed in a Quantum Design physical property measurement system (PPMS, ACT option) in a temperature range of $2 - 350$ K with magnetic field up to 9 T. For the longitudinal and Hall resistivity measurements, we employed a 4-wire and 5-wire geometry, respectively, with a 25 $\mu$m platinum wire spot-welded on the surface of the oriented \s\ single crystals.

\subsection{Angle-resolved photoemission spectroscopy}

\begin{figure*}[h!]
\centering
\includegraphics[width=15cm,trim={1in 6.9in 1in 1in},clip]{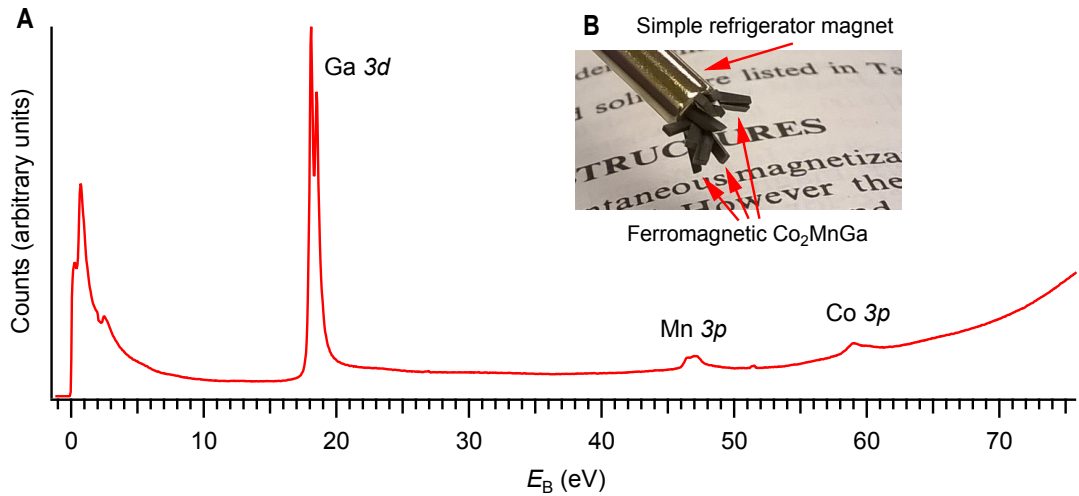}
\caption{\label{FigCore} \textbf{Core level spectrum of Co$_2$MnGa.} (\textbf{\pana}) An XPS spectrum of our Co$_2$MnGa samples clearly shows Ga $3d$, Mn $3p$ and Co $3p$ peaks, without significant irrelevant core level peaks, suggesting that our samples are of high quality. (\textbf{\panb}) The single crystal Co$_2$MnGa samples are readily picked up by an ordinary refrigerator magnet at room temperature.}
\end{figure*}

Ultraviolet ARPES measurements were carried out at Beamlines 5-2 and 5-4 of the Stanford Synchrotron Radiation Lightsource, SLAC in Menlo Park, CA, USA with a Scienta R4000 electron analyzer. The angular resolution was better than 0.2$^{\circ}$ and the energy resolution better than 20 meV, with a beam spot size of about $50 \times 40$ $\mu$m for Beamline 5-2 and $100 \times 80$ $\mu$m for Beamline 5-4. Samples were cleaved $\textit{in situ}$ and measured under vacuum better than $5 \times 10^{-11}$ Torr at temperatures $<$ 25 K. A core level spectrum of \s\ measured with 100 eV photons showed peaks consistent with the elemental composition (Fig. S\ref{FigCore}).

\subsection{First-principles calculations}

Numerical calculations of Co$_{2}$MnGa were performed within the density functional theory (DFT) framework using the projector augmented wave method  as implemented in the VASP package \cite{DFT1, DFT2,DFT3}. The generalized gradient approximation (GGA) \cite{DFT4} and a $\Gamma$-centered $k$-point $12 \times 12 \times 12$ mesh were used. Ga $s, p$ orbitals and Mn, Co $d$ orbitals were used to generate a real space tight-binding model, giving the Wannier functions. The surface states on a (001) semi-infinite slab were calculated from the Wannier functions by an iterative Green's function method.

\section{Additional systematic ARPES \& transport}

\subsection{Magnetotransport}

\begin{figure*}[h]
\centering
\includegraphics[width=15cm]{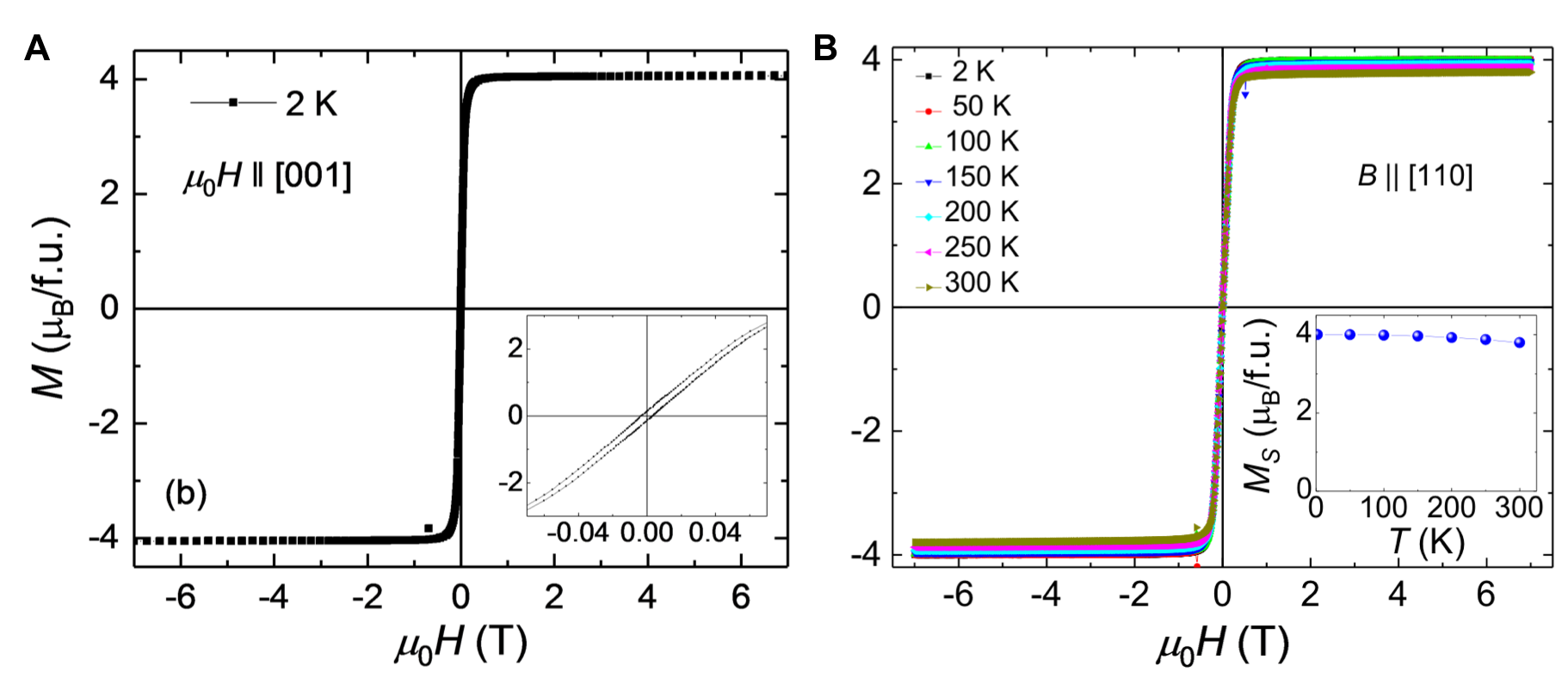}
\caption{\label{FigS2} \textbf{Magnetic hysteresis of Co$_2$MnGa.} (\textbf{\pana}) Hysteresis loop at 2 K for a [001] oriented Co$_2$MnGa crystal. Inset: zoomed-in view at low field, with hysteresis loop. (\textbf{\panb}) Hysteresis loop for various temperatures with field applied along the [110] direction. Inset: temperature dependence of the saturation magnetization, which decreases slightly from $4.0 \ \mu_B$ at 2 K to $3.8 \ \mu_B$ at 300 K.}
\end{figure*}

\begin{figure*}[h]
\centering
\includegraphics[width=9.5cm,trim={0in 0.35in 0in 0.8in},clip]{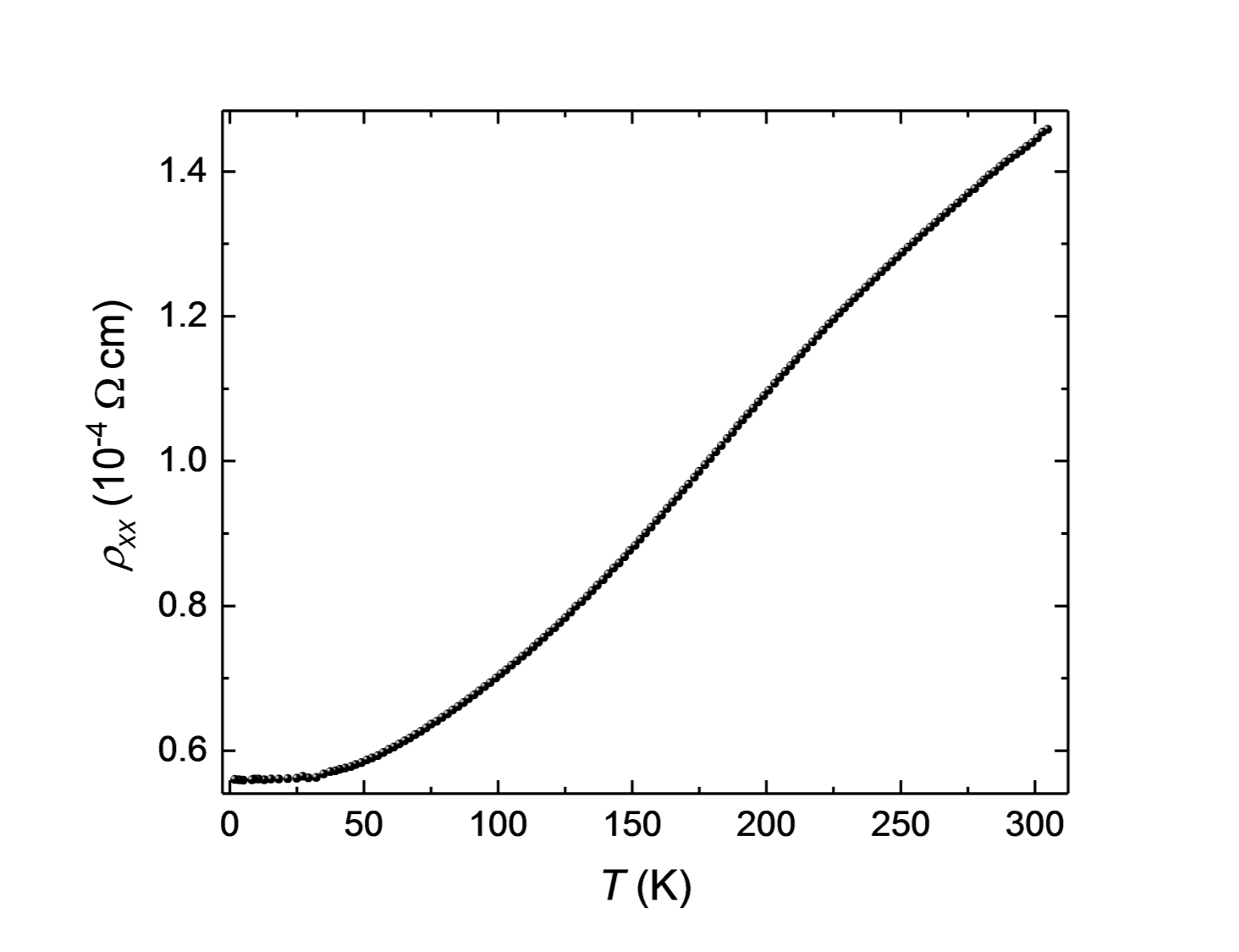}
\caption{\label{FigS3} \textbf{Longitudinal resistivity.} Resistivity as a function of temperature for a Co$_2$MnGa single crystal with current along the [001] direction.}
\end{figure*}

The magnetic hysteresis loop recorded at 2 K shows a soft ferromagnetic behavior, (Fig. \ref{FigS2}\pana). The magnetization saturates above $\sim 0.5$ T field with saturation magnetization $M_\textrm{S} \sim 4.0 \mu_{\textrm{B}}$/f.u.  Evidently, the compound follows the Slater-Pauling rule, $M_\textrm{S} = N-24$, where $N$ is the number of valence electrons, $N = 28$ for Co$_2$MnGa. We see a ferromagnetic loop opening with coercive field $\sim 35$ Oe (Fig. \ref{FigS2}\pana, inset). We observe that $M_\textrm{S}$ decreases slightly with increasing temperature (Fig. \ref{FigS2}\panb). We also study the temperature dependent longitudinal resistivity of our samples $\rho_{xx} (T) $, with zero applied magnetic field (Fig. \ref{FigS3}). We measure current along the [100] direction. Clearly the compound shows metallic behavior throughout the temperature range with very low residual resistivity: $\rho_{xx}$ $(2\textrm{K}) \sim 5.6 \times 10^{-5}$ $\Omega$ $\textrm{cm}$ and residual resistivity ratio (RRR) $\rho_{xx}$ $(300 \textrm{K})/\rho_{xx}$ $(2 \textrm{K}) = 2.6$.\\

\begin{figure*}[h]
\centering
\includegraphics[width=15cm,trim={0in 0.2in 0in 0in},clip]{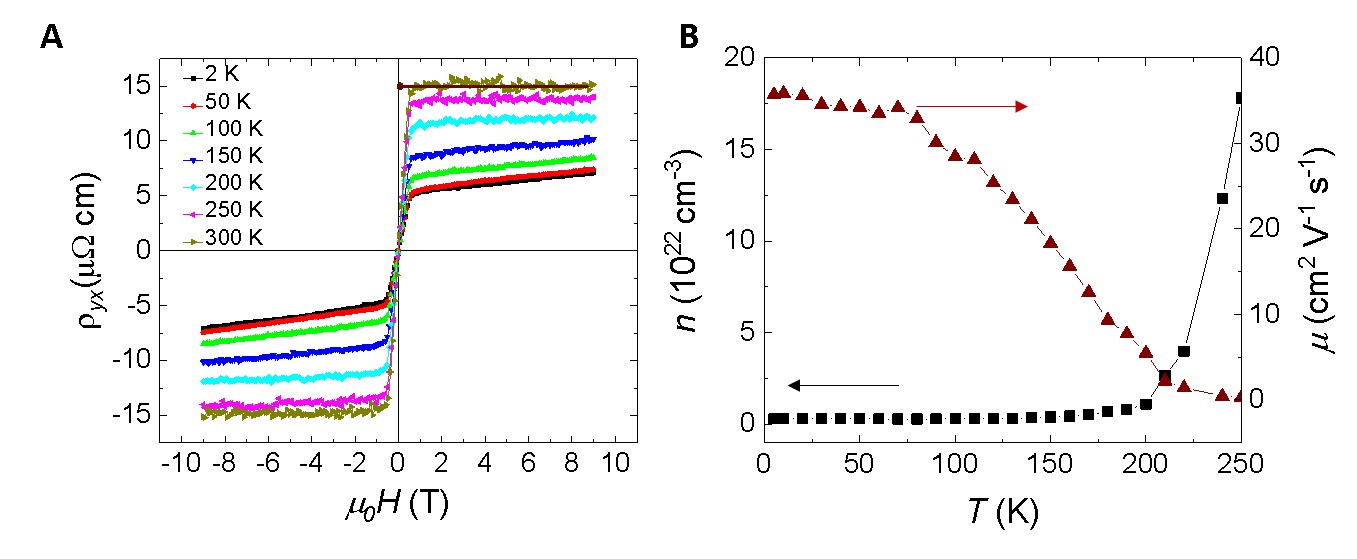}
\caption{\label{Hall} \textbf{Hall resistivity.} (\textbf{\pana}) Magnetic field dependence of the Hall resistivity at several representative temperatures. The magnetic field is applied along the [110] direction and the current along [001]. (\textbf{\panb}) Temperature dependence of carrier concentration and mobility of \s\ calculated from the Hall coefficient of the $\rho_{yx}$ data, as described in the text.}
\end{figure*}

\begin{figure}
\centering
\includegraphics[width=14cm,trim={0in 0in 0in 0in},clip]{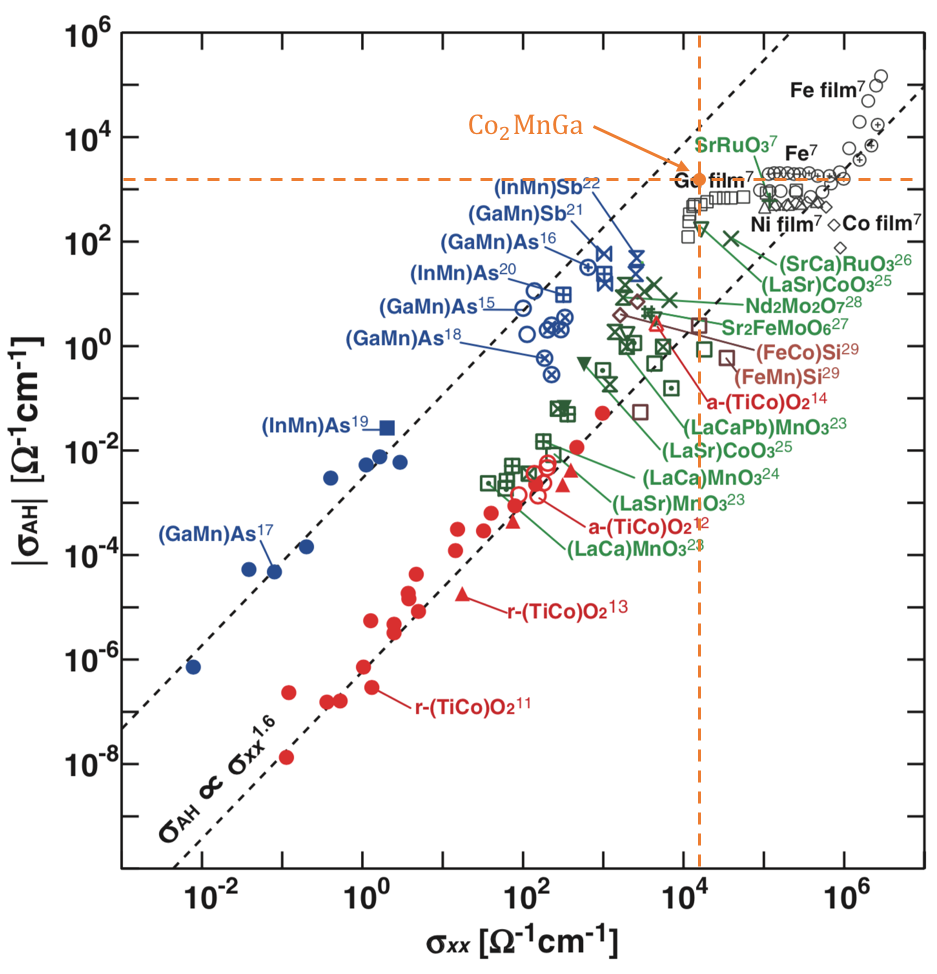}
\caption{\label{AHE_fig} The anomalous Hall response in various materials, reproduced from \href{http://iopscience.iop.org/article/10.1143/JJAP.46.L642}{\textit{Jpn. J. Appl. Phys.} {\bf 46}, L642 (2007)} (see references within). We add the measured values for \s. Only Fe (shown here) and Co$_2$MnAl (Ref. \cite{Co2MnAl_Jakob}) are known to have a larger AHE.}
\end{figure}

We provide some additional background on the Hall measurements presented in maintext Fig. 4. The Hall resistivity $\rho_{xy}$ is generally expressed as,
\beq
\rho_{xy}=R_0 \mu_0 H + \rho_{\textrm{AH}}
\eeq
where $R_0$ is known as the ordinary Hall coefficient arising from the Lorentz force and $\rho_\textrm{AH}$ is the anomalous Hall contribution \cite{ReviewAHE_Nagaosa_Ong}. The Hall conductivity $\sigma_{xy}$ is defined from the matrix inverse of $\rho$,
\beq
\sigma_{xy} = \frac{-\rho_{xy}}{\rho^2_{xy} + \rho^2_{xx}}
\eeq
% The matrix inverse formula is different in his doc.

%Here ?_xy^A at each temperature is estimated by extrapolating the high field data of ?_(xx ) (B) to the B?0 limit.

% It is well established that the extrinsic contributions contribute only at the low temperatures in the electrical transport1,2.

% Therefore, we conjecture that this deviation might be connected with the anomalous Hall angle (?=?_xy^A/?_(xx )) which sharply increases upto 300 K to a value of ~ 12\% as reported by Manna et al3.

where $\rho_{xx}$ is the longitudinal resistivity. Therefore, we can obtain $\sigma_{\textrm{AH}}$ (or, in a similar way, $\rho_{\textrm{AH}}$) by extrapolating the high field $\sigma_{xy}$ value back to $\mu_0 H = 0$ to find the $y$ intercept, as shown in main text Fig. 4\pana\ and Fig. \ref{Hall}\pana. We find $\sigma_{\textrm{AH}} = 1530 \ \Omega^{-1} \ \textrm{cm}^{-1}$ at 2 K, the largest anomalous Hall response in any known material except Fe and Co$_2$MnAl, Fig. \ref{AHE_fig} and Ref. \cite{Co2MnAl_Jakob}. Further, from the high field slope of $\rho_{yx}$ as a function of $\mu_0 H$, we find an ordinary Hall coefficient $R_0 = 2.76 \times 10^{-3} \ \textrm{cm}^3/\textrm{C}$ at 2 K and $9.98 \times 10^{-5} \ \textrm{cm}^3/\textrm{C}$ at 300 K. The positive sign of $R_0$ suggests that the charge carriers in \s\ are majority hole type through the full temperature range. We estimate the carrier concentration as $n=1/(eR_0)$ and the carrier mobility as $\mu = R_0/\rho_{xx}$, where $e$ is the electron charge, Fig. \ref{Hall}\panb.

\subsection{Survey band structure calculation}

\begin{figure*}[h]
\centering
\includegraphics[width=12cm,trim={0in 0.5in 0in 0in},clip]{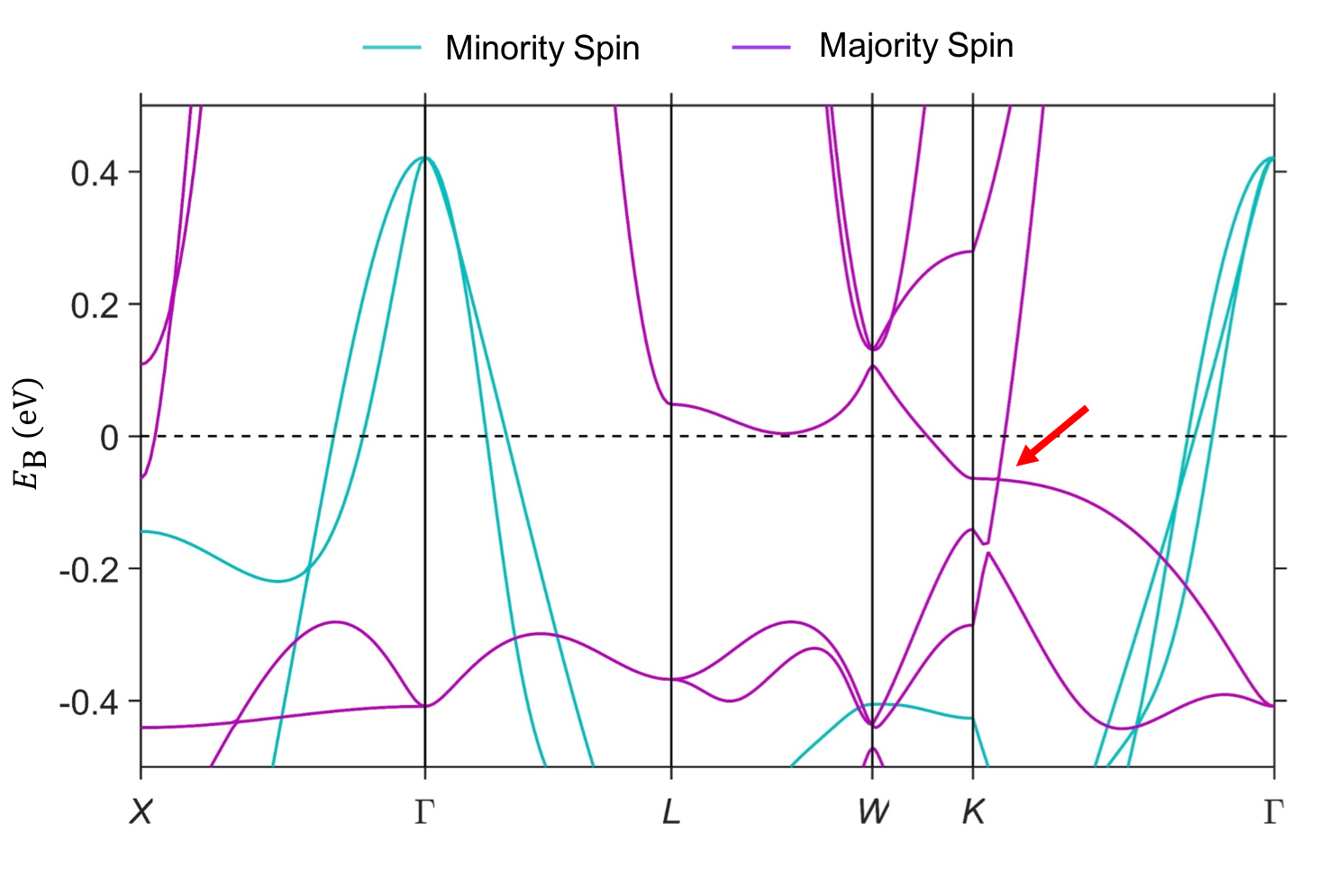}
\caption{\label{FigSurvey} \textbf{Survey band structure of Co$_2$MnGa.} \textit{Ab initio} band structure of \s\ in the ferromagnetic state. Two majority spin bands near the Fermi level form Weyl lines (orange arrow).}
\end{figure*}

\begin{figure*}[h]
\centering
\includegraphics[width=13cm,trim={0in 0in 0in 0in},clip]{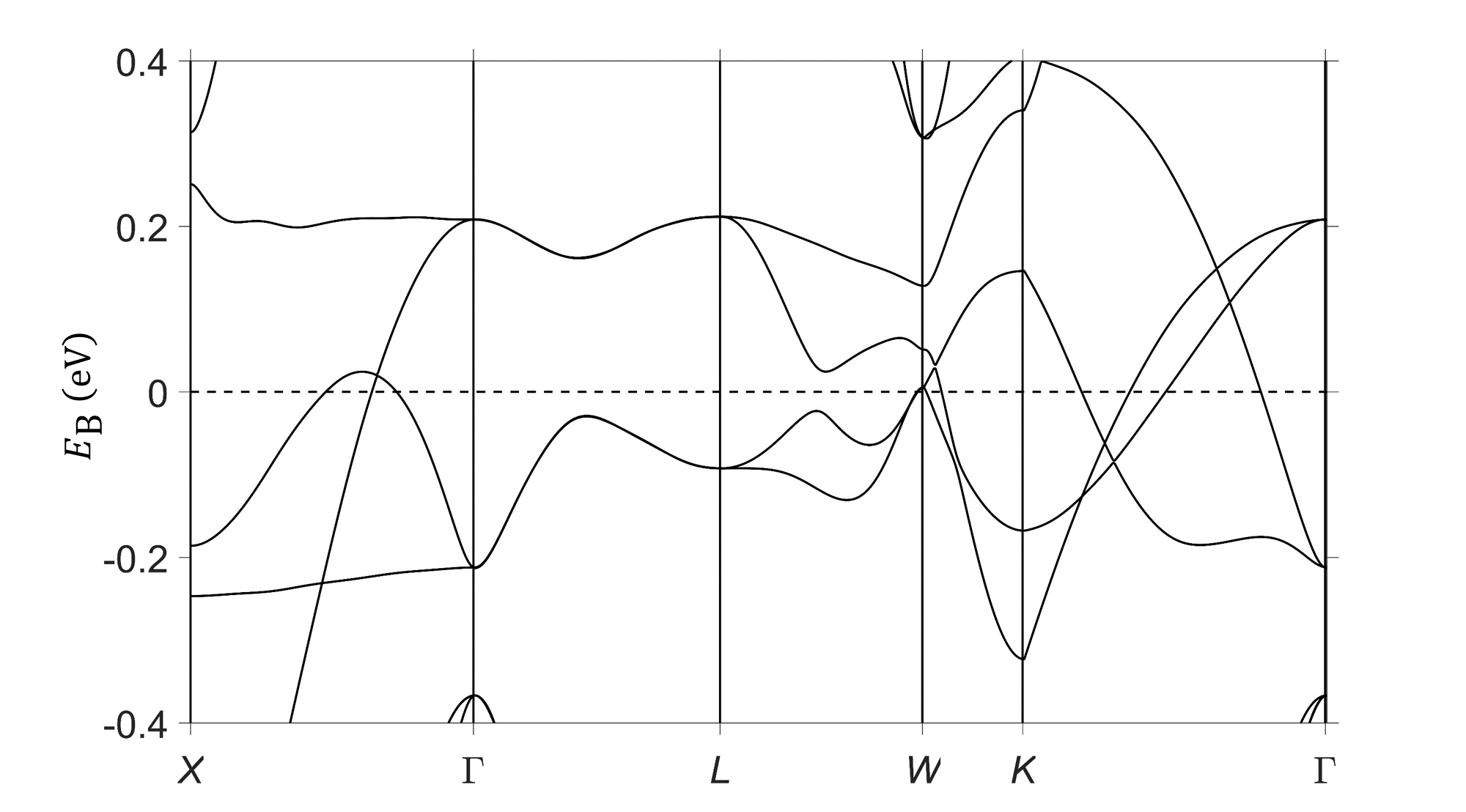}
\caption{\label{FigNonMag} \textbf{Non-magnetic survey band structure.} \textit{Ab initio} band structure, ignoring ferromagnetism. The magnetic Weyl lines disappear.}
\end{figure*}

We consider a bird's eye view of the \ai\ bulk band structure in the ferromagnetic state (Fig. \ref{FigSurvey}). We observe two majority spin bands near the Fermi level---these are the bands which form the Weyl lines \cite{Co2MnGa_Guoqing}. There is also a large, irrelevant minority spin pocket around $\Gamma$ which we experimentally suppress by judicious choice of photon energy in ARPES. For completeness, we present ARPES measurements on this minority spin pocket below (Fig. \ref{spinmin}). Without magnet order, the \ai\ band structure changes drastically (Fig. \ref{FigNonMag}). This provides additional evidence that in ARPES we access the magnetic state of \s. Moreover, this result suggests that the Weyl lines we observe are essentially magnetic in the sense that they disappear if we remove the magnetic order.

\begin{figure*}[h]
\centering
\includegraphics[width=16cm,trim={1in 5.4in 1in 1.6in},clip]{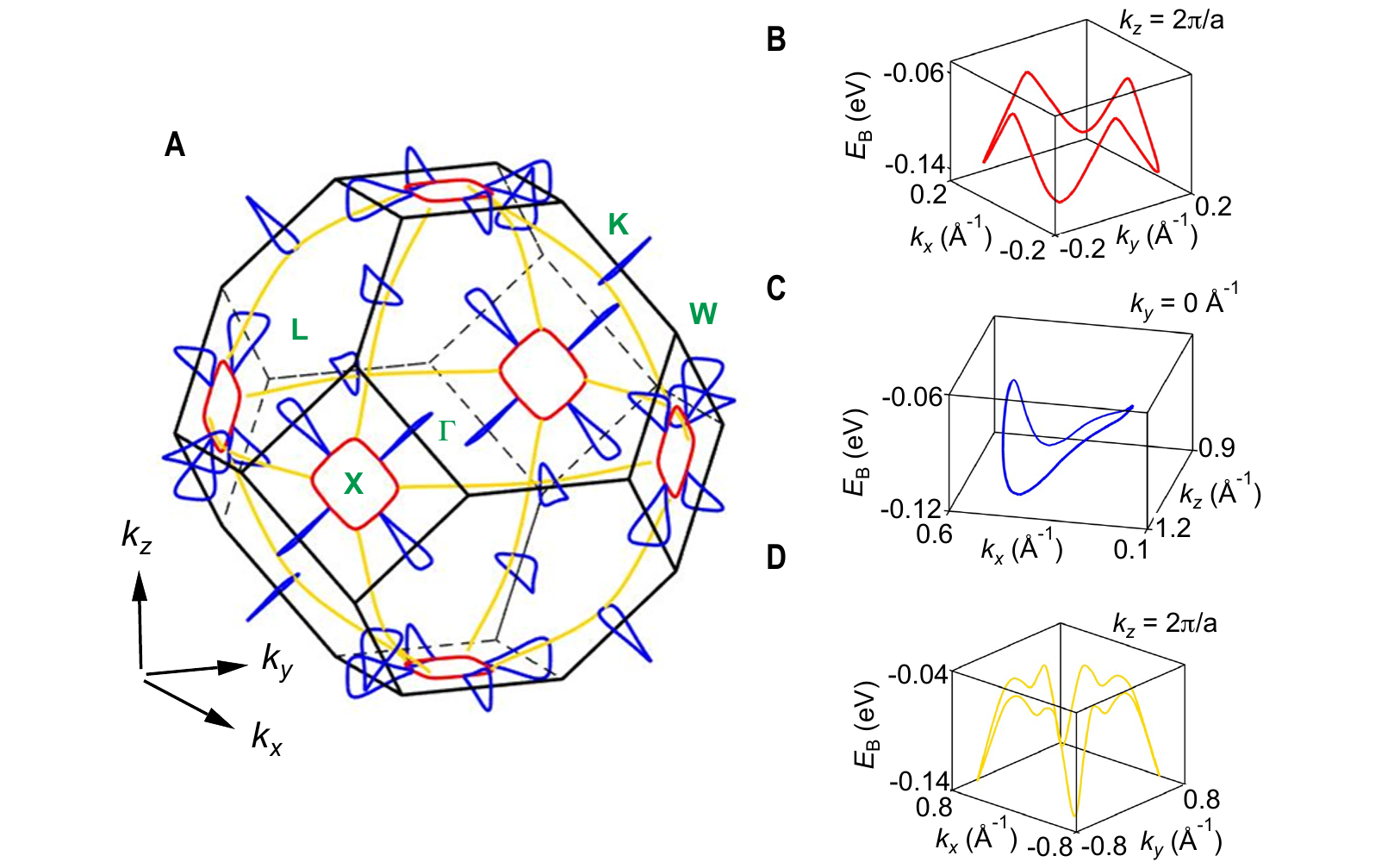}
\caption{\label{FigCalc} \textbf{Predicted Weyl lines in Co$_2$MnGa.} (\textbf{\pana}) \textit{Ab initio} prediction of Weyl lines in Co$_2$MnGa, with the $E_\textrm{B}$ axis collapsed. (\textbf{\panb}-\textbf{\pand}) Theory predicts three unique Weyl lines, which we call red, blue and yellow. Although they are locked on a mirror plane, the line nodes disperse in energy. Each one is copied many times through the Brillouin zone by the symmetries of the lattice.}
\end{figure*}

\subsection{Weyl lines from calculation}

Here we give a more systematic introduction to the full network of Weyl lines in \s\ as predicted by \ai\ \cite{Co2MnGa_Guoqing}. Recall that a Weyl line is a one-dimensional crossing between a pair of singly-degenerate bands. In \s, the Weyl lines are contained in the mirror planes of the bulk Brillouin zone but they are allowed to disperse in energy. As a result, it is instructive to plot each Weyl line as a function of $k_x, k_y$ and $E_\textrm{B}$, where $k_x$ and $k_y$ without loss of generality are the two momentum axes of the mirror plane. \textit{Ab initio} predicts three independent Weyl lines in Co$_2$MnGa, which we denote the red, blue and yellow Weyl lines (Fig. \ref{FigCalc}\panb-\pand). The energies are marked with respect to the Fermi level observed in numerics. Since we find in experiment that the Fermi level is at $-0.08 \pm 0.01$ eV relative to calculation, the experimental Fermi level cuts through all of the Weyl lines. To better view the full pattern of line nodes throughout the bulk Brillouin zone, we collapse the energy axis and plot the line nodes in $k_x, k_y, k_z$ (Fig. \ref{FigCalc}\pana). Although we start with three independent line nodes, these are each copied many times by the symmetries of the crystal lattice, giving rise to a rich line node network.\\

Since we are studying the (001) surface, it's useful to consider more carefully how the line nodes project onto the surface Brillouin zone---in other words, to see how Fig. 2\panf\ arises from Fig. \ref{FigCalc}\pana. To start, the center red line node will project straight up on its face around $\bar{\Gamma}$. By contrast, the adjacent blue line nodes are ``standing up'' and will project on their side, so that in fact there will be a ``double'' cone in the surface projection. Moreover, the blue line node projection will form an open line node segment. The other red line nodes will also project on their side, forming ``double" red cones along an open line segment. The yellow line nodes also projects in two distinct ways. The yellow line nodes in the $k_z = 0$ plane will produce a single yellow line node projection, while the yellow line nodes in the $k_x = 0$ and $k_y = 0$ planes are standing up, so they will produce a double yellow line node in an open segment. Band folding associated with the (001) surface projection further sends the $k_z = 0$ yellow line node towards $\bar{\Gamma}$.

\begin{figure*}[h!]
\centering
\includegraphics[width=16cm,trim={1.4in 4.8in 1.4in 1.1in},clip]{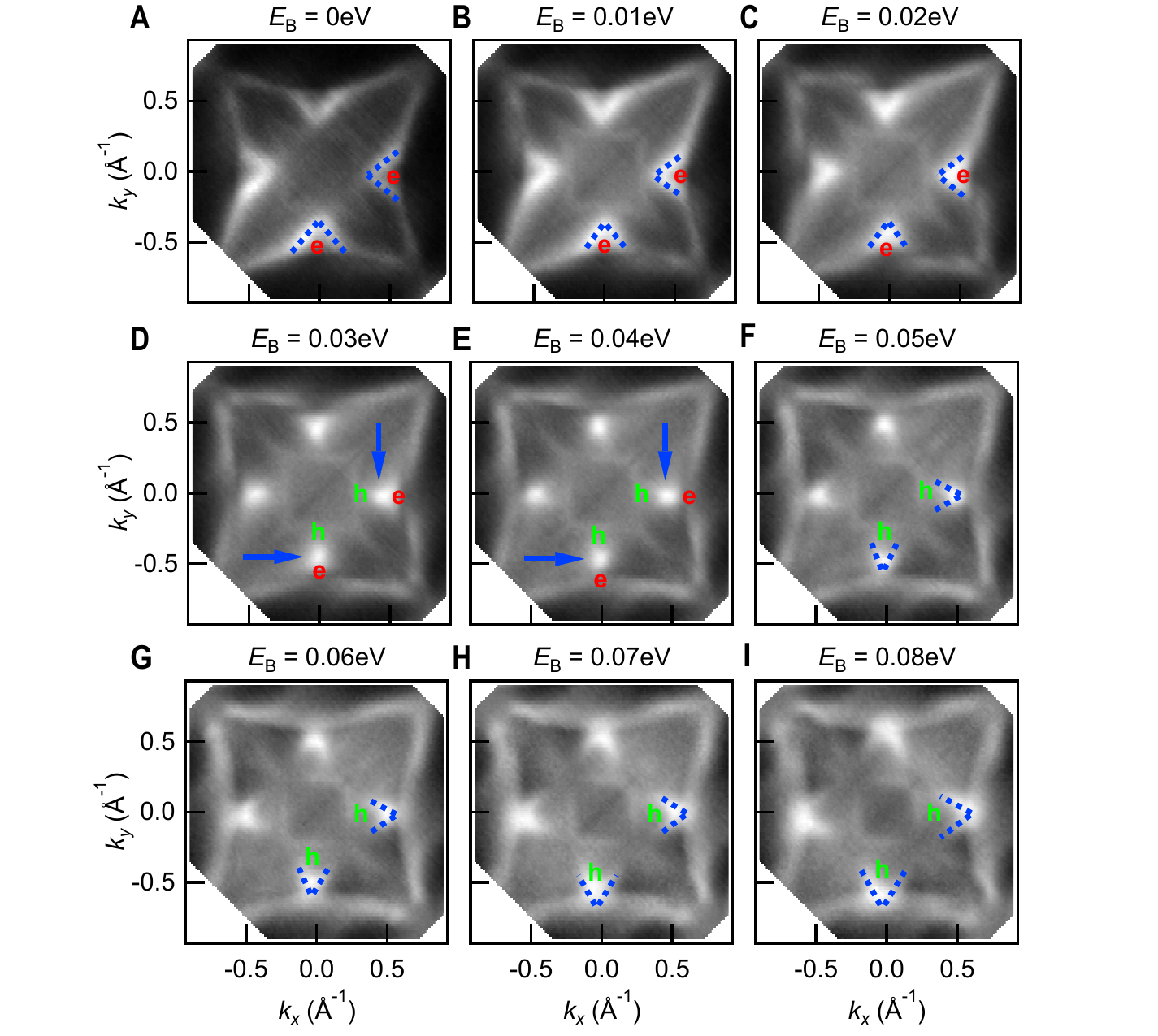}
\caption{\label{BlueFS} \textbf{Constant energy surfaces of Co$_2$MnGa.} (\textbf{\pana}-\textbf{\pani}) We expand on the dataset shown in main text Fig. 1\pand-\panh\ by plotting constant energy surfaces at additional \eb. The evolution is characteristic of a line node dispersion.}
\end{figure*}

\subsection{ARPES systematics on the blue Weyl line}

\begin{figure*}[h!]
\centering
\includegraphics[width=16cm,trim={1in 6.6in 1in 1.2in},clip]{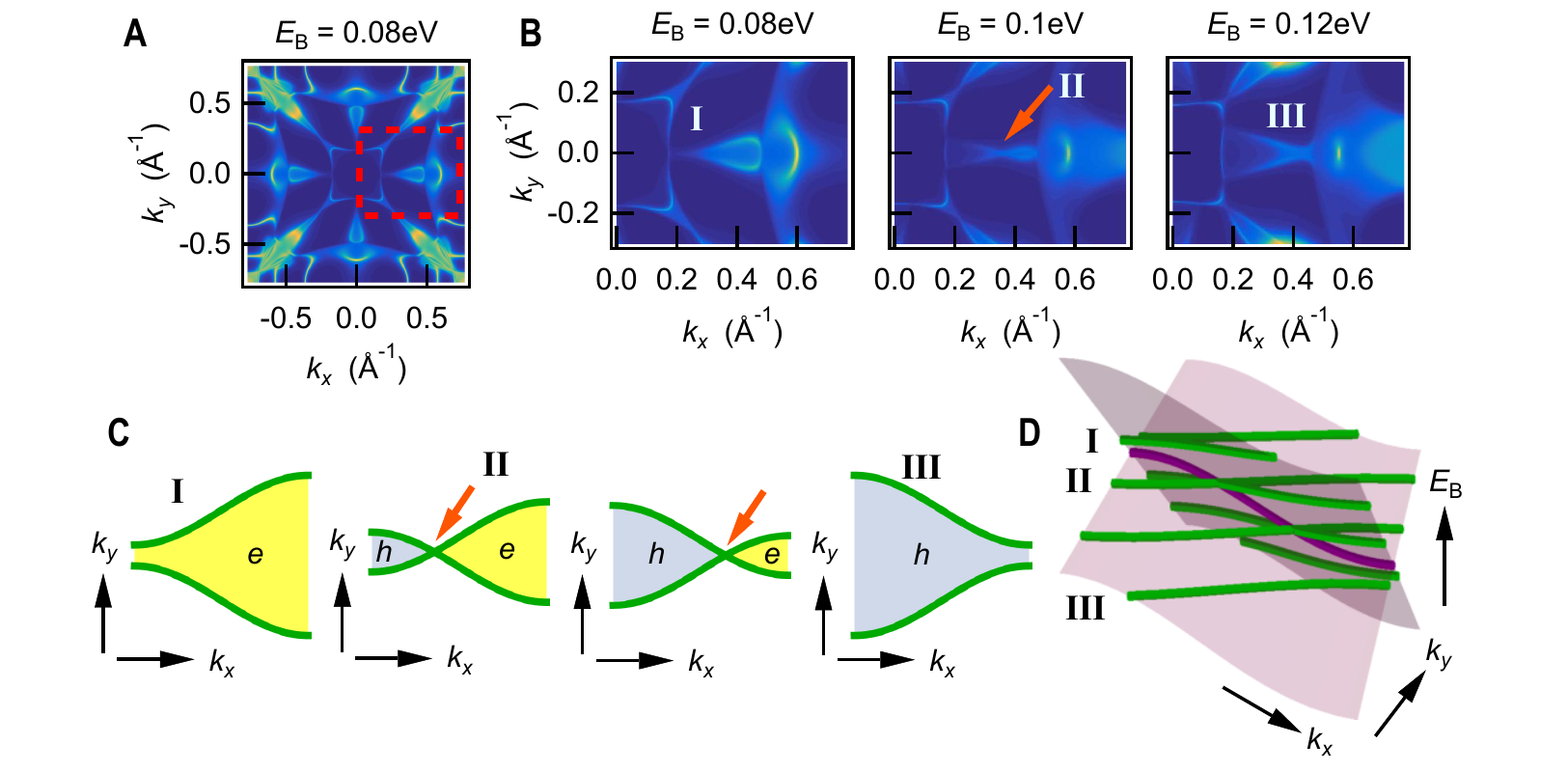}
\caption{\label{BlueFSCalc} {\bf \Ai\ evolution of the blue Weyl line in \eb.} ({\bf \pana}) Constant-energy surface with red boxed region marking ({\bf \panb}) the zoomed-in slices at different \eb. ({\bf \panc}, {\bf \pand}) Cartoon of the constant-energy cuts through a generic line node. Scanning from shallower to deeper \eb, the bands evolve from {\bf I}, an electron pocket, to {\bf II}, a hole and electron pocket touching at a point (orange arrows), to {\bf III}, a hole pocket. The touching point at each energy is a point on the Weyl line.}
\end{figure*}

We present an extended dataset for main text Fig. 1\pand-\panh\ (Fig. \ref{BlueFS}), with a finer \eb\ sampling. We observe in greater detail the evolution of the dispersion from a $<$ to a dot to a $>$. For instance, we find that the crossing point moves systematically away from $\bar{\Gamma}$ with deeper \eb, consistent with a line node. In \ai, we observe a similar evolution for the blue Weyl line (Fig. \ref{BlueFSCalc}\pana,\panb). We can better understand this evolution by considering the constant-energy surfaces for a generic line node (Fig. \ref{BlueFSCalc}\panc,\pand). For \eb\ above the line node, the slice intersects only the upper cone, giving {\bf I}. For \eb\ which cross the line node, we find electron and hole pockets intersecting at a point, as in {\bf II}. As we continue to move downward the intersection point traces out the line node, shifting from left to right. Comparing to our ARPES spectra, we observe that the photoemission cross-section appears to be dominated by the intersection point for this range of \eb. Lastly, as we scan below the line node, the intersection point completely zips closed the electron pocket and zips open the hole pocket, as in {\bf III}. A detailed study of the $E_\textrm{B}$ dependence of the constant energy surface in ARPES again suggests a line node in \s.\\

%  As we scan downward through \eb, the intersection point zips the electron pock away and unzips a hole pocket.

\begin{figure*}
\centering
\includegraphics[width=15cm,trim={1.2in 4.8in 1.2in 1in},clip]{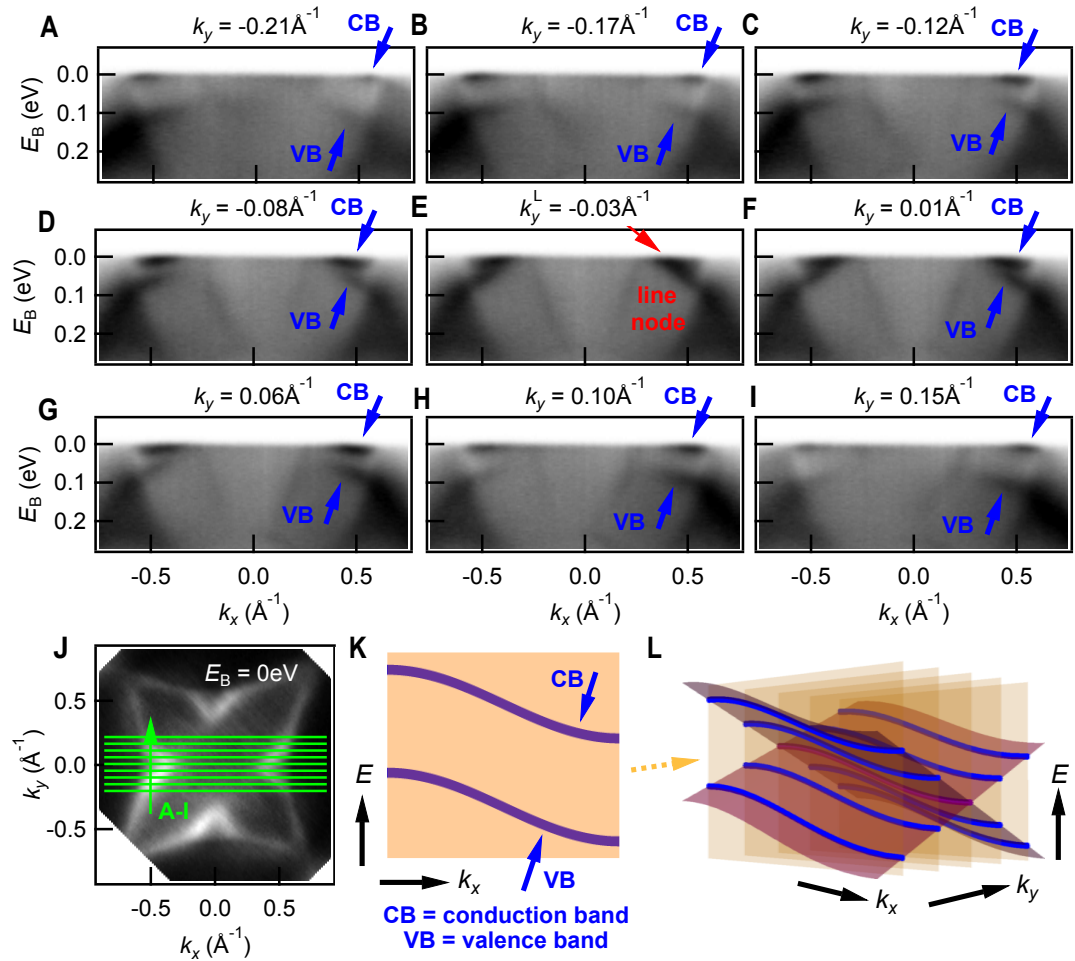}
\caption{\label{FigBluePar} \textbf{Cutting parallel to the bulk Weyl line.} (\textbf{\pana}-\textbf{\pani}) $E_\textrm{B}-k_x$ cuts sweeping through the Weyl line in $k_y$, as indicated in (\textbf{\panj}). The valence and conduction bands appear to meet at a single $k_y^\textrm{L}$ (red arrow). Note that $k_y^\textrm{L}$ is slightly away from $k_y = 0$ $\textrm{\AA}^{-1}$, probably due to a small misalignment. (\textbf{\pank}) A generic cut parallel to a Weyl line and (\textbf{\panl}) its evolution in $k_y$.}
\end{figure*}

\begin{figure*}
\centering
\includegraphics[width=15cm,trim={1.2in 2.0in 1.2in 1.0in},clip]{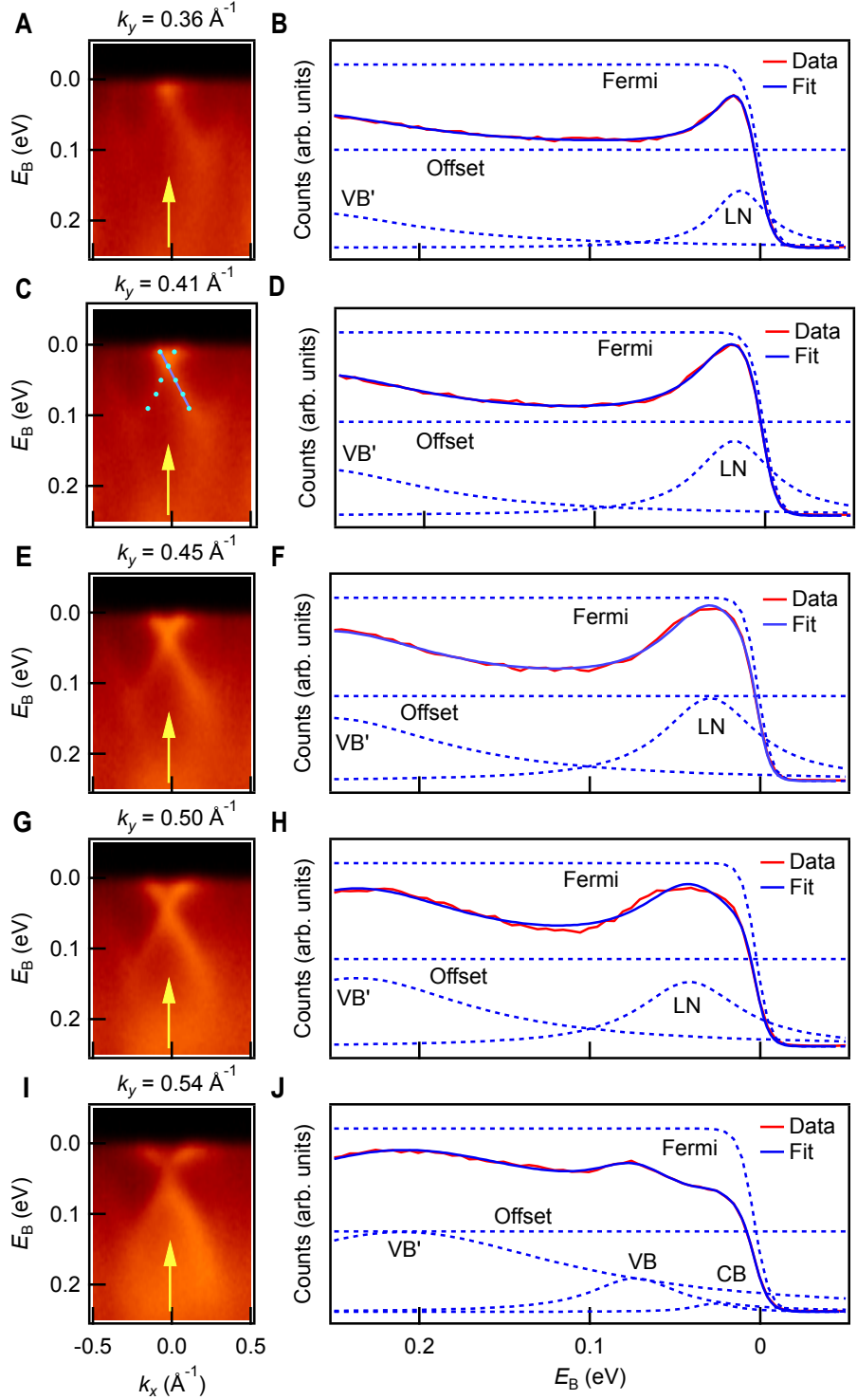}
\caption{\label{FigBlueFit} \textbf{Lorentzian fitting of the blue Weyl line.} (\textbf{\pana}, \textbf{\panc}, \textbf{\pane}, \textbf{\pang}) $E_\textrm{B}-k_x$ cuts through the line node, same  as those in main text Fig. 2\pana. The yellow arrows mark the center EDC. Band dispersion obtained from analysis of the spectra (cyan dots, \panc) with linear fit (purple line, \panc). (\textbf{\pani}) Additional cut further away from $\bar{\Gamma}$, past the predicted end of the blue Weyl line. (\textbf{\panb}, \textbf{\pand}, \textbf{\panf}, \textbf{\panh}, \textbf{\panj}) Fits of the center EDC (fitting form discussed in the text), suggesting a band crossing.}
\end{figure*}

\begin{figure*}
\centering
\includegraphics[width=15cm,trim={1.2in 7in 1.2in 1in},clip]{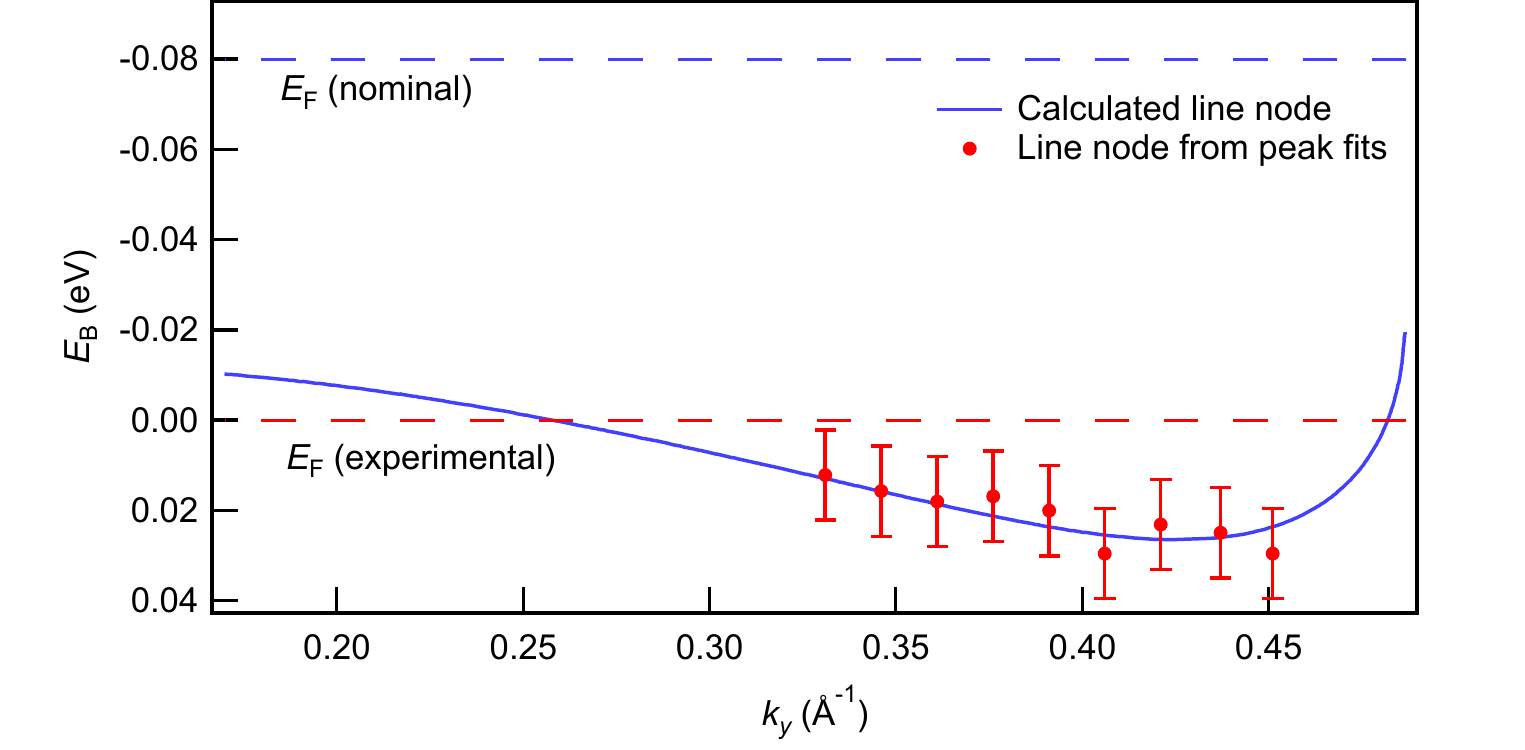}
\caption{\label{FigBlueTrack} \textbf{Comparing Lorentzian peaks to the calculated Weyl line.} Line node dispersion as obtained from Lorentzian peak fits, superimposed on the calculated Weyl line with a $0.08$ eV experimental shift. We observe a match between Lorentzian fits and \textit{ab initio}. The error bars reflect the experimental energy resolution $\delta E = 0.02$ eV.}
\end{figure*}

To provide another perspective, we cut parallel to the blue Weyl line (Fig. \ref{FigBluePar}). In contrast to main text Figs. 1,2, here we cut \textit{along} the Weyl line. Sweeping in $k_y$, we see the conduction and valence bands approach (Fig. \ref{FigBluePar}\pana-\pand), touch each other at fixed $k_y^\textrm{L} = - 0.03\ \textrm{\AA}^{-1}$ along a finite range of $k_x$ (Fig. \ref{FigBluePar}\pane) and then move apart again (Fig. \ref{FigBluePar}\panf-\pani). These parallel $E_\textrm{B} - k_x$ cuts again suggest a line node dispersion.\\

Next, we perform a Lorentzian peak fitting of the blue Weyl line. We begin with the $E_\textrm{B}-k_x$ cuts discussed in main text Fig. 2 and we choose the energy distribution curve (EDC) passing through the crossing point (Fig. \ref{FigBlueFit}). We fit these EDCs to the following form,

\beq
I(x) = (C + L_1(x) + L_2(x))f(x)
\eeq

\beq
L_i(x) = \frac{A_i^2}{(x - B_i)^2 + C_i^2}  \hspace{1cm}    f(x) = (\exp(\beta(x - \mu)) + 1)^{-1}
\eeq\\
We include two Lorentzian peaks $L_1(x)$ and $L_2(x)$, where the first peak corresponds to the line node crossing LN, while the second peak corresponds to a deeper valence band VB' which is useful for improving the fit. We also include the Fermi-Dirac distribution $f(x)$ and a constant offset $C$ which we interpret as a background spectral weight approximately constant within the energy range of the fit. We find a high-quality fit close to $\bar{\Gamma}$ (Fig. \ref{FigBlueFit}\pana-\pand) using a single LN peak. Away from the crossing point, the peaks are well-described by a linear dispersion, further suggesting a band crossing (Fig. \ref{FigBlueFit}\panc). At $k_y = 0.45$ \invA, we observe that the fit begins to deviate from the data, and at $k_y = 0.5$ \invA there is an even more noticeable error (Fig. \ref{FigBlueFit}\pane-\panh). We speculate that this deviation may arise due to our finite $k_y$ resolution/linewidth as well as the fact that these spectra cut near the extremum of the Weyl line, producing a smeared-out energy gap. Another explanation considers the detailed dispersion of the blue Weyl line, which exhibits a rapid upward dispersion at its extremum (Fig. \ref{FigCalc}\panc). Due to broadening along $k_y$, we may capture LN peaks from a range of $k_y$, smearing out this rapid dispersion and producing a plateau structure in the EDC. For $k_y = 0.54$ \invA, we clearly observe two peaks on the EDC, so we fit with an additional Lorentzian $L_3 (x)$ (Fig. \ref{FigBlueFit}\pani-\panj). This gives VB and CB, the peaks corresponding to the conduction and valence bands of the line node. This interpretation is consistent with \ai, which predicts that the blue Weyl line ends at $k_y = 0.5$ \invA.

Lastly, we take the results of our peak fitting and compare them with the calculated blue line node dispersion. We plot the LN peak maxima and the standard deviation of the peak positions, Fig. \ref{FigBlueTrack}. We ignore EDCs at $k_y > 0.45$ \invA\ because the plateau shape in the EDC is poorly described by a single Lorentzian, as discussed above. We compare these numerical fitting results with a first-principles calculation of the blue Weyl line dispersion, shifted by $0.08$ eV. We find a reasonable quantitative agreement between the fit and calculated dispersion. Note that there is some expected contribution to the error from the \textit{ab initio} calculation as well as corrections to the Lorentzian fitting form. These results support our observation of a line node in our ARPES spectra on Co$_2$MnGa.

% We estimate an error of $\sim 5$ meV by eye from the $E_\textrm{B}-k_y$ cuts of the line node. This estimate of the error is larger than the na\''ive error of the fit, which we find to be $< 2$ meV in all cases, which assumes only a normal distribution of noise.

\begin{figure*}
\centering
\includegraphics[width=16cm,trim={1.2in 4.2in 1.2in 1in},clip]{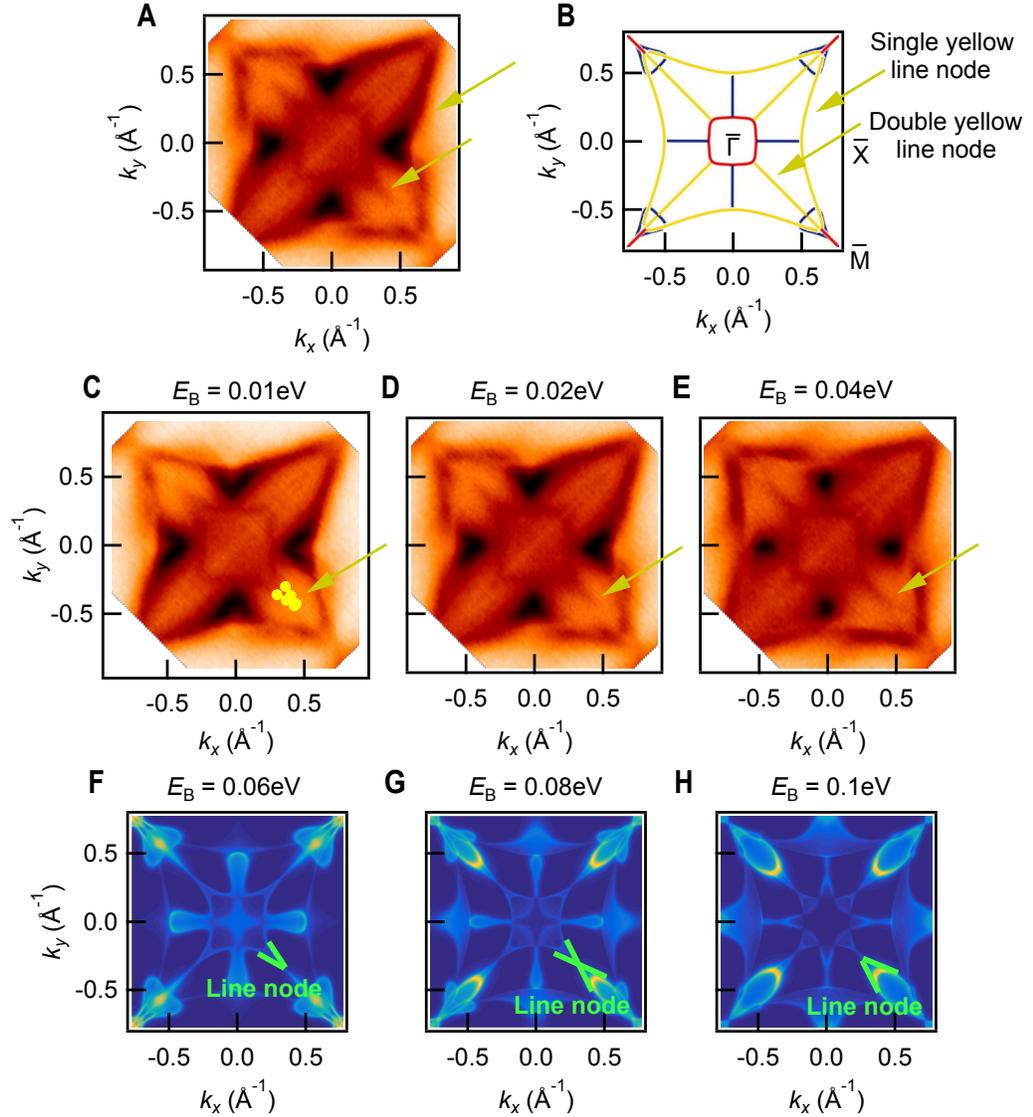}
\caption{\label{FigYellow1} \textbf{Signatures of the yellow Weyl line.} ({\bf \pana}) Constant-energy surface from ARPES compared with ({\bf \panb}) the (001) projection of the line nodes from calculation, emphasizing signatures of the yellow Weyl lines (yellow arrows). Note that the Weyl lines project both ``standing up'' so that the band crossings project in pairs into the surface Brillouin zone (double yellow Weyl line). The remaining yellow line node forms a large ring around the entire surface Brillouin zone, projecting simply ``face up'' (single yellow Weyl line). (\textbf{\panc}-\textbf{\pane}) Constant-energy surfaces showing the characteristic $<$ to $>$ transition (yellow arrows). (\textbf{\panf}-\textbf{\panh}) Corresponding \ai\ cuts showing the double yellow Weyl line (green guides to the eye).}
\end{figure*}

\begin{figure*}
\centering
\includegraphics[width=16cm,trim={1.2in 3.1in 1.2in 1.3in},clip]{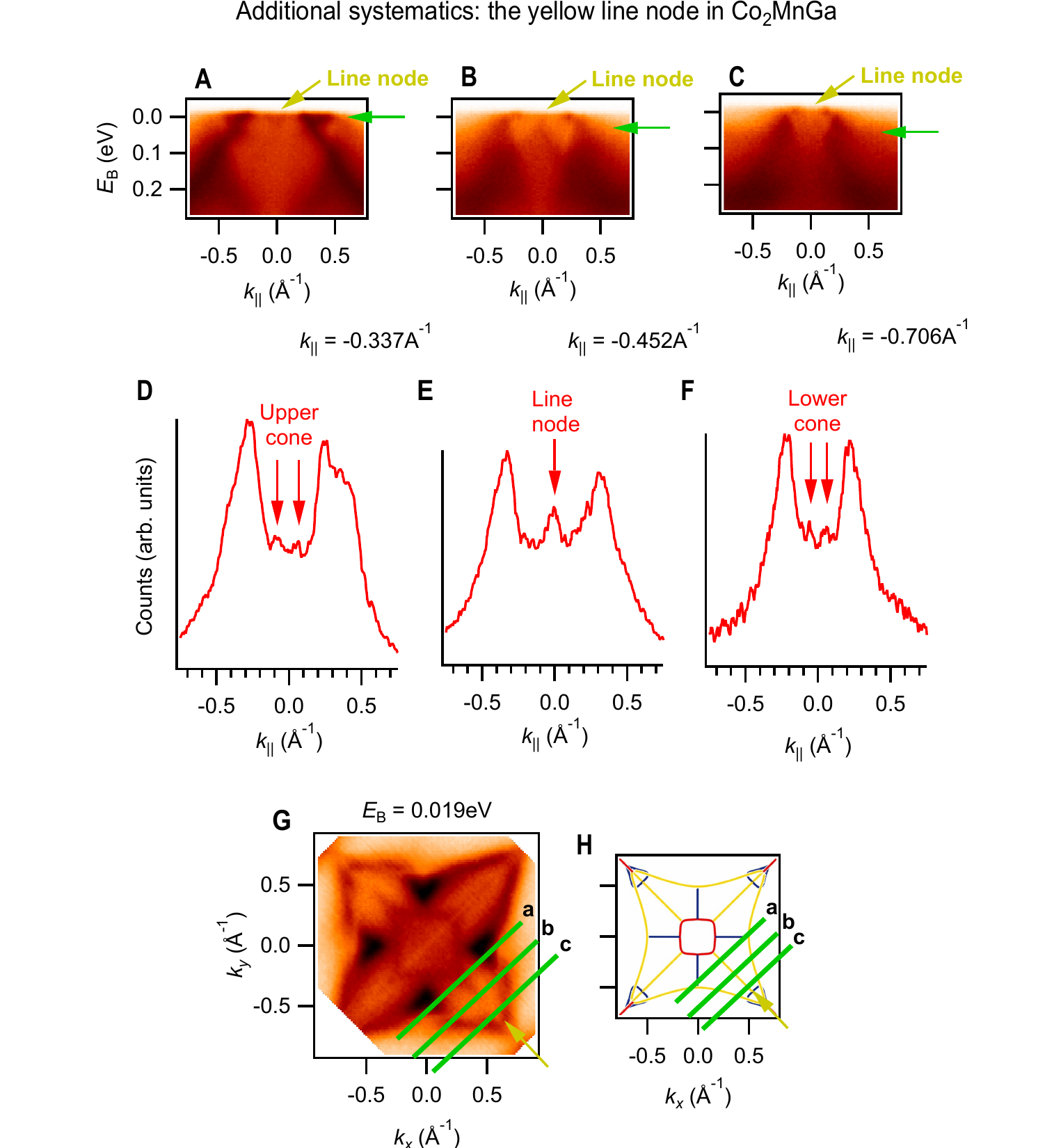}
\caption{\label{FigYellow2} \textbf{Yellow Weyl line cones.} (\textbf{\pana}-\textbf{\panc}) $E_\textrm{B}-k_{||}$ cuts sweeping perpendicular to the line node. In \pana\ we observe the upper cone associated with the double yellow line node; in \panb\ we see a line node crossing and the lower cone; in \panc\ we find a weak signature remaining from the lower cone. (\textbf{\pand}-\textbf{\pane}) MDCs taken from the $E_\textrm{B}-k_{||}$ cuts, as indicated by the green arrows in \pana-\panc. The weak peaks associated with the double yellow line node are marked by the red arrows. (\textbf{\pang}, \textbf{\panh}) The locations of the cuts in \pana-\panc, as marked. Our MDC analysis provides additional evidence for the yellow line node in Co$_2$MnGa.}
\end{figure*}

\subsection{ARPES study of the yellow Weyl line}

We can also observe signatures of yellow Weyl lines in our ARPES data, at incident photon energy $h\nu = 50$. We can identify a candidate yellow Weyl line by comparing an ARPES constant-energy surface and the projected nodal lines (Fig. \ref{FigYellow1}\pana, \panb). We reiterate that there are two different ways in which the yellow Weyl lines can project on the (001) surface. In particular, the four yellow Weyl lines along $\bar{\Gamma} - \bar{M}$ are ``standing up'', so two crossings project onto the same point in the surface Brillouin zone, similar to the blue Weyl line we discussed above. By contrast, the outer yellow Weyl line runs in a single large loop around the entire surface Brillouin zone. It projects ``lying down'', with single crossing projections. Here we focus on the double yellow line node. We study constant-energy surfaces at various binding energies (Fig. \ref{FigYellow1}\panc-\pane) and we observe the same $<$ to $>$ switch that we discussed in the case of the blue Weyl line. We see similar behavior on an \textit{ab initio} constant energy surface (Fig. \ref{FigYellow1}\pani-\pank). Note crucially that the electron-to-hole transition occurs in the same direction in the ARPES spectra and calculation, suggesting that the line node dispersion has the same slope in experiment and theory. This provides additional evidence for the yellow Weyl line. Next we search for signatures of the yellow Weyl line on a series of $E_\textrm{B}-k_a$ cuts (green lines in Fig. \ref{FigYellow2}\pang, \panh). We observe the upper cone near the center of the cut for $k_a$ closest to $\bar{\Gamma}$ (Fig. \ref{FigYellow2}\pana), corresponding to the yellow markers in Fig. \ref{FigYellow1}\panc. As we slide away from $\bar{\Gamma}$, we observe the band crossing and lower cone (Fig. \ref{FigYellow2}\panb, \panc). To pinpoint the cone, we study MDCs through the line node. We observe twin peaks corresponding to the upper and lower cone (\ref{FigYellow2}\pand,\panf), as well as a single peak when we cut through the line node (\ref{FigYellow2}\pane). In this way, we observe signatures of the ``double'' yellow Weyl line of Co$_2$MnGa in our ARPES spectra.\\

Lastly, we search for signatures of the other, ``single'' yellow Weyl line. The outer features in Fig. \ref{FigYellow2}\panb, \panc, as well as the large off-center peaks in Fig. \ref{FigYellow2}\pane, \panf, correspond well to the predicted locations of the single yellow Weyl line. The valence band further shows a cone shape. However, we note that the conduction band appears to be shifted/offset in momentum relative to the valence band, for instance near the Fermi level at $k_a \sim 0.25$ $\textrm{\AA}^{-1}$ (Fig. \ref{FigYellow2}\panb). This suggests that we are perhaps observing a surface state or surface resonance which partly traces out the line node. This explanation appears to be consistent with our calculations, compare main text Fig. 3\pane, which show a similar surface resonance. In summary, we provide evidence for the yellow Weyl lines in our ARPES spectra.

\begin{figure*}
\centering
\includegraphics[width=15cm,trim={1.1in 4.3in 1.1in 1in},clip]{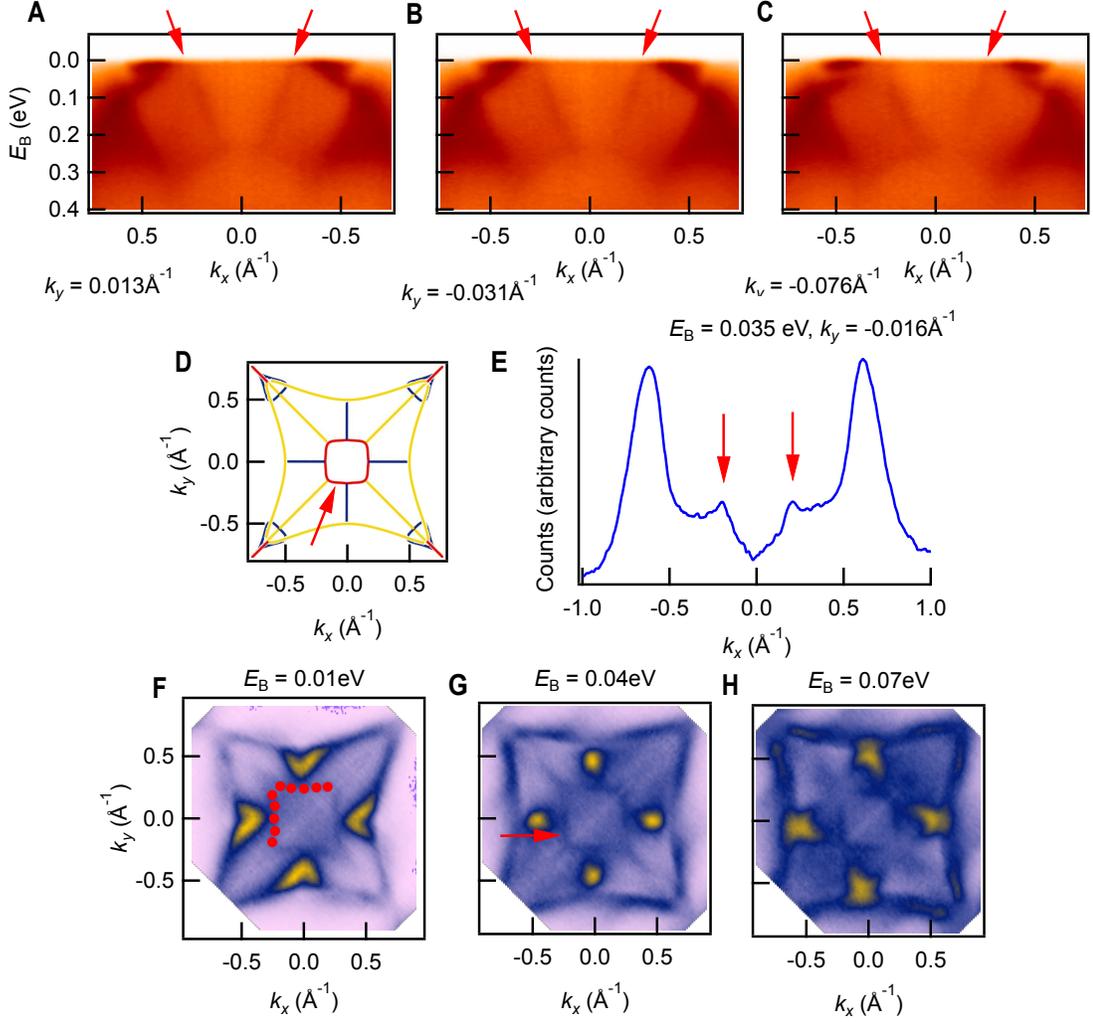}
\caption{\label{FigRed} \textbf{Signatures of the red Weyl line.} (\textbf{\pana}-\textbf{\panc}) $E_\textrm{B}-k_x$ cuts passing through the red line node near $\bar{\Gamma}$. We see two bands dispersing away from $\bar{\Gamma}$ as we approach $E_\textrm{F}$, consistent with (\textbf{\pand}) the red line node in calculation. (\textbf{\pane}) The two bands on an MDC (taken at the red arrow in {\pang}). (\textbf{\panf}-\textbf{\panh}) Constant-energy surfaces, emphasizing a square feature around $\bar{\Gamma}$ in agreement with the predicted red Weyl line. The square feature (red dots) appears to arise from the red Weyl line.}
\end{figure*}

\subsection{ARPES study of the red Weyl line}

Having discussed the blue and yellow Weyl lines, we search for ARPES signatures of the red Weyl line. On the constant-energy surfaces, we observe a square feature around $\bar{\Gamma}$ (Fig. \ref{FigRed}\panf-\panh). This corresponds well to the predicted red Weyl line (Fig. \ref{FigRed}\pand). Next, we study a series of $E_\textrm{B}-k_x$ cuts passing through the center square feature. We see two clear branches dispersing away from $k_x = 0\ \textrm{\AA}^{-1}$ as we approach the Fermi level (red arrows, Fig. \ref{FigRed}\pana-\panc). We can mark these features on an MDC (Fig. \ref{FigRed}\pane). However, we observe no cone or crossing. We speculate that this may be because the other branch of the Weyl line has low photoemission cross-section under these measurement conditions.

%It may also be that we are actually observing a surface state or a resonance state which is pushed out of the bulk gap against the bulk projection. If there is one such state, then we can expect that it will be associated with one branch of the bulk line node, so that we see no crossing. Alternatively, the absence of a surface state in the red line node could be understood within a na\"ive topological theory. First, note that we do observe a drumhead surface state in the region bounded by the single, outer yellow line node, as discussed in the main text. These two observations, (1) a drumhead inside the outer yellow line node projection and (2) no drumhead inside the red line node projection, can be understood by considering the Berry phase topological invariant associated with the line nodes. In particular, the red line node and the outer yellow line node project on top of each other, so the region bounded by the red line node actually has two line nodes projecting together. If we view each line node as contributing a $\pi$ Berry phase, the result is that the two line nodes cancel each other out and we expect no surface state within the red line node projection.

%In the following section, we comment further on the relationship of the drumhead surface state to the topological invariants of the line nodes. Here, we only conclude that we observe clear signatures of the red line node in that we see a square feature with excellent match between experiment and calculation.

\begin{figure*}[h!]
\centering
\includegraphics[width=15cm,trim={1in 5.4in 1in 1.35in},clip]{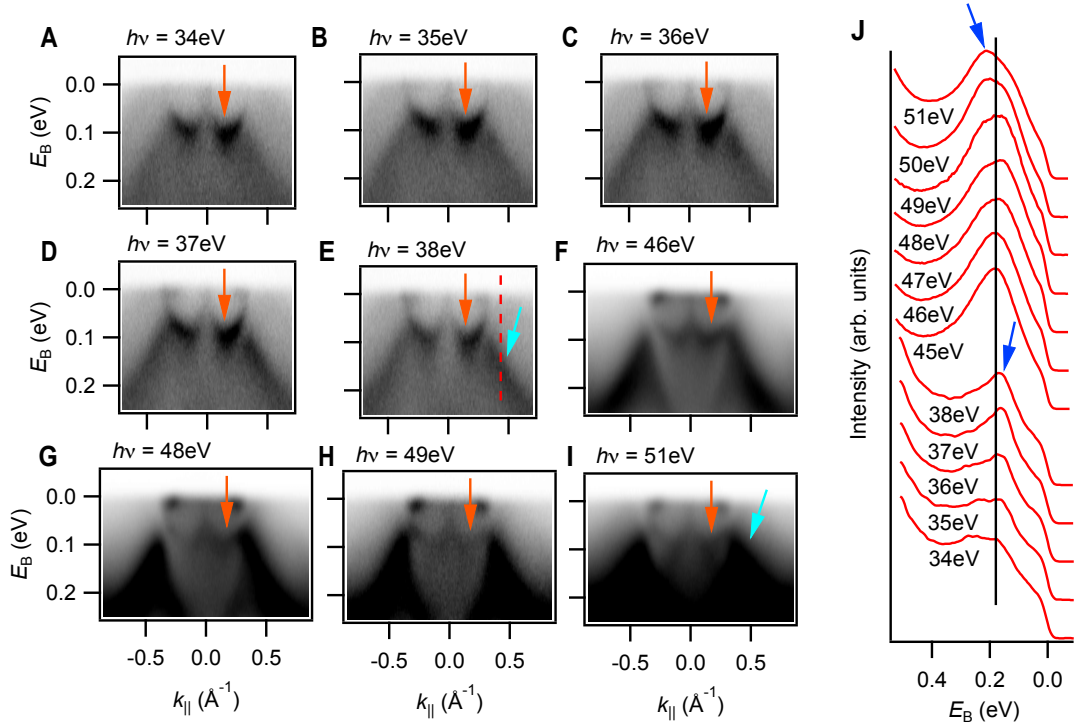}
\caption{\label{DH1} \textbf{Evidence for $k_z$ dispersion of the yellow Weyl line.} (\textbf{\pana-\pani}) $E_\textrm{B}-k_{||}$ cuts analagous to main text Fig. 3\pana-\panc, but at more photon energies. (\textbf{\panj}) Stack of EDCs as a function of $h\nu$, analogous to main text Fig. 3\pang, but instead of cutting through the drumhead surface state, the EDC cuts through the yellow line node, at $k_{||} = 0.45$ $\textrm{\AA}^{-1}$ (dotted red line in \pane). We clearly observe the drumhead surface state in all cuts (orange arrows). Recall that the drumhead showed no photon energy dependence, suggesting that it is a surface state. Here, by contrast, we see a photon energy dependence (blue arrows in \panj), associated with the yellow Weyl line (cyan arrows in \pane, \pani), suggesting a $k_z$ dispersion for the yellow line node cone.}
\end{figure*}

\subsection{Photon energy dependence of the drumhead surface state}

We present an extended dataset to accompany main text Fig. 3, showing the drumhead surface states. In main text Fig. 3, we presented an energy distribution curve (EDC) stack cutting through the drumhead surface state at different photon energies. Here we present the full $E_\textrm{B}-k_x$ cut for each photon energy included in the stack. We observe the drumhead surface state consistently at all energies (orange arrow, Fig. \ref{DH1}\pana-\pani). Additionally, we show an EDC stack at a different momentum, $k_{||} = 0.45 \ \textrm{\AA}^{-1}$, which cuts not through the drumhead surface state but the yellow line node cone (Fig. \ref{DH1}\panj) analogous to main text Fig. 3\pang. When cutting through the candidate drumhead, the EDC stack showed no dispersion in the peak energy as a function of photon energy, indicating no $k_z$ dispersion and suggesting a surface state. Here, by contrast, we can observe that the peak positions shift with photon energy (blue arrows). This shift suggests that the yellow line node lives in the bulk.

\subsection{In-plane dispersion of the drumhead surface state}

\begin{figure*}[h]
\centering
\includegraphics[width=15cm,trim={1in 6.7in 1in 1.35in},clip]{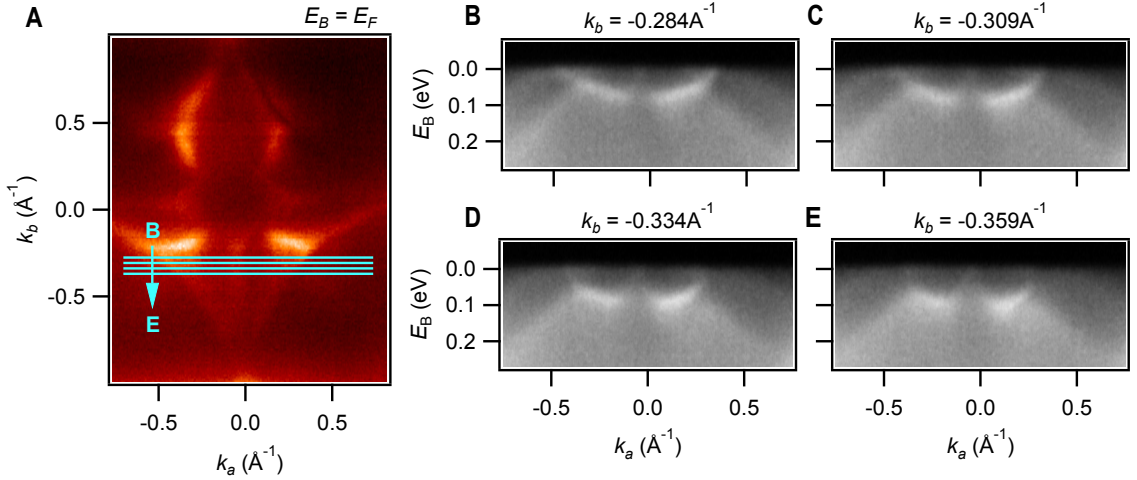}
\caption{\label{DH2} \textbf{In-plane dispersion of the drumhead.} (\textbf{\pana}) Fermi surface at $h\nu = 35$ eV and (\textbf{\panb-\pane}) $E_\textrm{B}-k_{||}$ cuts showing the drumhead surface state, with locations as marked in \pana\ (cyan lines). We observe a weak dispersion of the surface state doward in energy as we move away from $\bar{\Gamma}$. So we a observe an in-plane dispersion of the drumhead, but not an out-of-plane dispersion (main text Fig. 4), demonstrating a surface state.}
\end{figure*}

We briefly study the in-plane dispersion of the drumhead surface states. At $h\nu = 35$ eV, with Fermi surface as shown in Fig. \ref{DH2}\pana, we study a sequence of $E_\textrm{B}-k_x$ cuts scanning through the drumhead surface state, Fig. \ref{DH2}\panb-\pane. We observe that the surface state disperses slightly downward in energy as we scan away from $\bar{\Gamma}$ and that it narrows in $k_{||}$, as expected because the line node cones move towards each other as we approach the corner of the Brillouin zone. The dispersion of the candidate drumhead in-plane (but not out-of-plane) is consistent with the behavior of a surface state.

\subsection{ARPES and \ai\ study of the minority spin pocket}

\begin{figure*}[h]
\centering
\includegraphics[width=15cm,trim={1.7in 7.9in 2in 1.05in},clip]{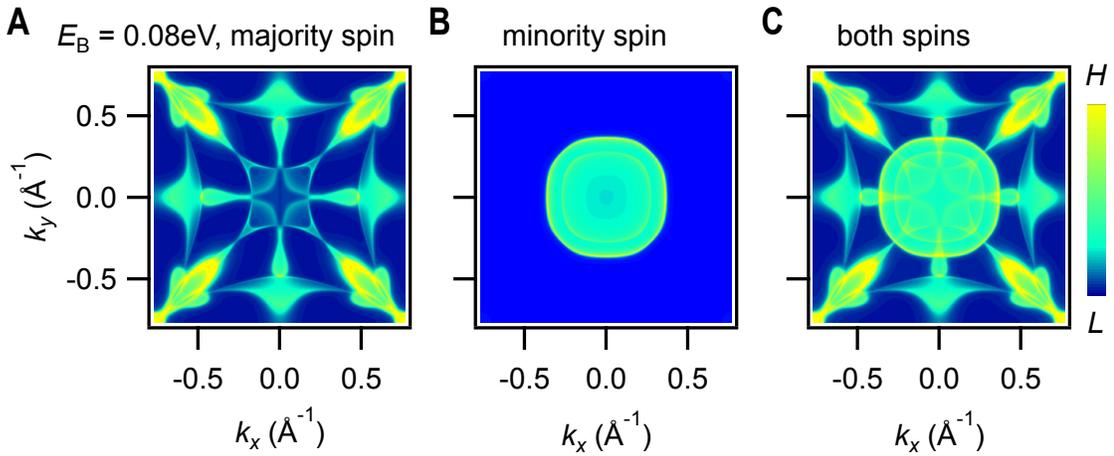}
\caption{\label{spinmin_calc} \textbf{Minority spin pocket in calculation.} ({\bf \pana}) Bulk projection of the majority spin states, same as main text Fig. 2\pane. ({\bf \panb}) Bulk projection of the minority spin at the same energy. ({\bf \panc}) The sum of \pana and \panb, the bulk projection of all states at the given binding energy.}
\end{figure*}

\begin{figure*}[h!]
\centering
\includegraphics[width=10.5cm,trim={2.2in 2.1in 2.2in 1.35in},clip]{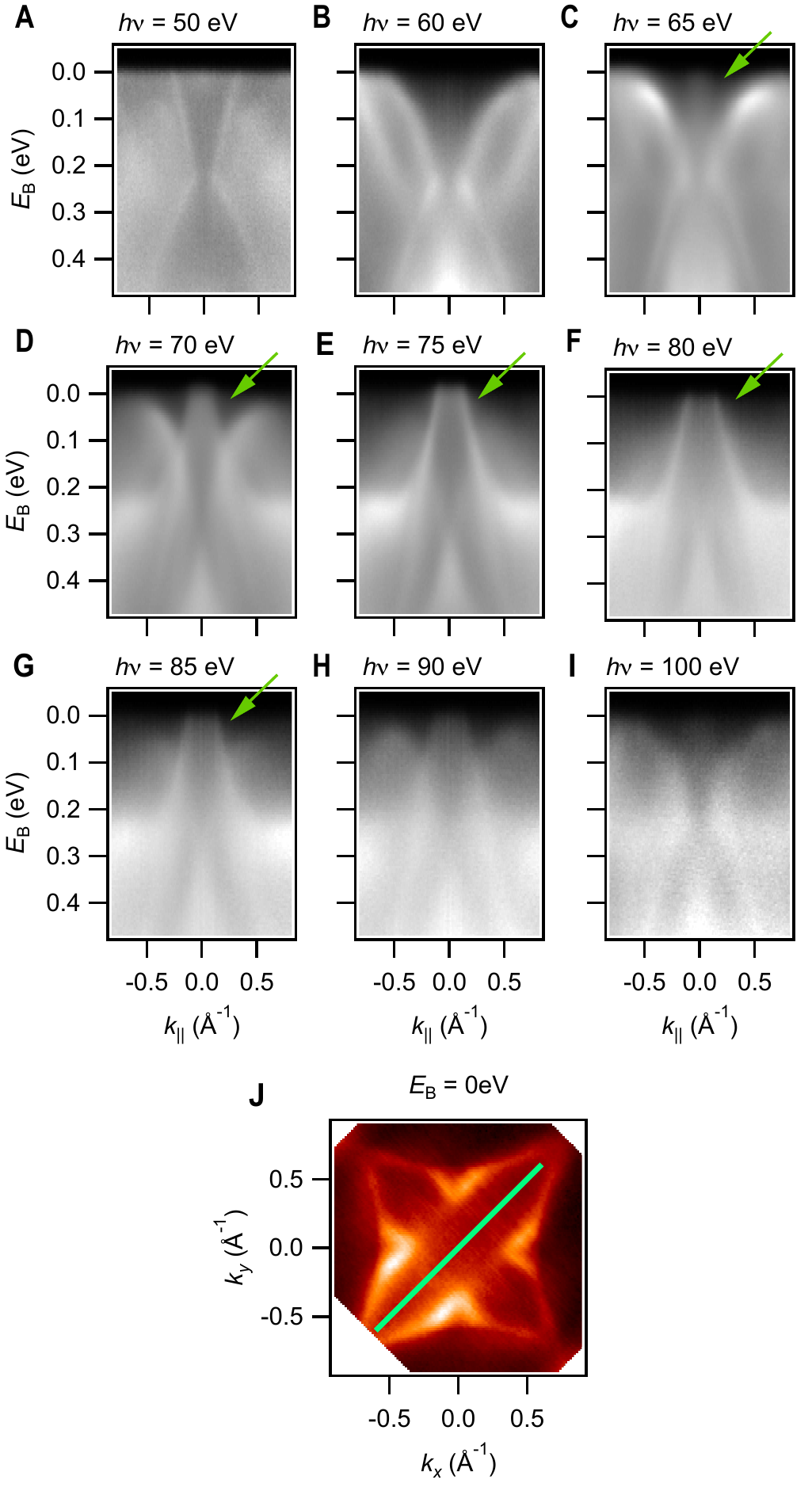}
\caption{\label{spinmin} \textbf{The minority spin pocket in ARPES.} (\textbf{\pana-\pani}) $E_\textrm{B}-k_a$ cuts through $\bar{\Gamma}$ at different photon energies, with the location of the cut shown in (\textbf{\panj}) by the green line. At  $h \nu > 65$ eV a large hole pocket appears (green arrows), consistent with the minority spin pocket seen in calculation.}
\end{figure*}

We briefly noted that Co$_2$MnGa has a large minority spin pocket around the $\Gamma$ point (Fig. \ref{FigSurvey}). We can omit this pocket from our ARPES measurements by an appropriate choice of photon energy $h\nu$, which then corresponds to a $k_z$ away from $\Gamma$. In particular, at $h \nu = 50$ eV, main text Fig. 2, we find that we cut near the top of the Brillouin zone (near the $X$ point) and far from the $\Gamma$ point. As a result, we then compare our data with the majority spin bands from \textit{ab initio}, as in main text Fig. 2\pane. However, to further compare ARPES and DFT, it is useful to search for this irrelevant pocket in ARPES and better understand why it does not compete with the line node and drumhead states in photoemission. In the DFT bulk projection, the minority spin projects onto a large pocket around $\bar{\Gamma}$, see Fig. \ref{spinmin_calc}\pana-\panc. Experimentally, we performed a photon energy dependence on a cut passing through $\bar{\Gamma}$. For $h \nu > 50$ eV we find that the red Weyl line disappears and a large hole pocket appears near $\bar{\Gamma}$ (Fig. \ref{spinmin}\panb-\panj). This pocket matches well with the minority spin pocket in calculation. This photon energy dependence suggests that the minority spin pocket does not interfere with our measurements of the Weyl lines because at $h\nu = 50$ eV we cut near the top of the bulk Brillouin zone in $k_z$ (near the $X$ point).

\clearpage
\newpage
\clearpage

%\cleardoublepage
\ifdefined\phantomsection
  \phantomsection  % makes hyperref recognize this section properly for pdf link
\else
\fi
\addcontentsline{toc}{section}{Bibliography}

{\singlespacing

} %single spacing

%\bibliography{ch-magnet/mybib}{}
%\bibliographystyle{../../Science}

%\end{document}

%\end{document}

%% file: ch-mainpubs/ch-mainpubs.tex
\chapter{Principal publications}

\begin{enumerate}
\item {\bf Ilya Belopolski} {\it et al}. A three-dimensional magnetic topological phase. Submitted to {\it Science} (2018), arXiv:1712.09992.

\item	Daniel S. Sanchez*, {\bf Ilya Belopolski}* {\it et al}. Topological chiral crystals with helicoid-arc quantum states. {\it Nature} {\bf 567}, 500 (2019).

\item	{\bf Ilya Belopolski} {\it et al}. A novel condensed matter lattice and a new platform for one-dimensional topological phases. {\it Sci. Adv.} {\bf 3}, e1501692 (2017).

\item	{\bf Ilya Belopolski} {\it et al}. Discovery of a new type of topological Weyl fermion semimetal state in Mo$_x$W$_{1-x}$Te$_2$. {\it Nat. Commun.} {\bf 7}, 13643 (2016).

\item	{\bf Ilya Belopolski} {\it et al}. Signatures of a time-reversal symmetric Weyl semimetal with only four Weyl points. {\it Nat. Commun.} {\bf 8}, 942 (2017).

\item	M. Zahid Hasan, Su-Yang Xu, {\bf Ilya Belopolski} and Shin-Ming Huang. Discovery of Weyl Fermion Semimetals and Topological Fermi Arc States. {\it Ann. Rev. Cond. Mat. Phys.} {\bf 8}, 289 (2017).

\item	Su-Yang Xu*, {\bf Ilya Belopolski}* {\it et al}. Discovery of a Weyl Fermion Semimetal and Topological Fermi Arcs. {\it Science} {\bf 349}, 6248 (2015). [First Weyl semimetal: experiment]

\item	Shin-Ming Huang, Su-Yang Xu, {\bf Ilya Belopolski} {\it et al}. A Weyl Fermion semimetal with surface Fermi arcs in the transition metal monopnictide TaAs class. {\it Nat. Commun.} {\bf 6}, 7373 (2015). [First Weyl semimetal: prediction]

\item	{\bf Ilya Belopolski} {\it et al}. Criteria for directly detecting topological Fermi arcs in Weyl semimetals. {\it Phys. Rev. Lett.} {\bf 116}, 066802 (2016).

\item	Su-Yang Xu, {\bf Ilya Belopolski} {\it et al}. Experimental discovery of a topological Weyl semimetal state in TaP. {\it Sci. Adv.} {\bf 1}, e1501092 (2015).

\item	Su-Yang Xu, Nasser Alidoust, {\bf Ilya Belopolski} {\it et al}. Discovery of a Weyl fermion state with Fermi arcs in niobium arsenide. {\it Nat. Phys.} {\bf 11}, 748 (2015).

\item	Chenglong Zhang, Su-Yang Xu, {\bf Ilya Belopolski} {\it et al}. Signatures of the Adler-Bell-Jackiw chiral anomaly in a Weyl Fermion semimetal. {\it Nat. Commun.} {\bf 7}, 10735 (2016).
 
\item	{\bf Ilya Belopolski} {\it et al}. Measuring Chern numbers above the Fermi level in the Type II Weyl semimetal Mo$_x$W$_{1-x}$Te$_2$. {\it Phys. Rev. B} {\bf 94}, 085127 (2016).

\item	Su-Yang Xu, {\bf Ilya Belopolski} {\it et al}. Spin polarization and texture of the Fermi arcs in the Weyl Fermion semimetal TaAs. {\it Phys. Rev. Lett.} {\bf 116}, 096801 (2016).

\end{enumerate}

%% file: ch-pubs/ch-pubs.tex
\chapter{Complete list of publications}

\begin{enumerate}

\item {\bf Ilya Belopolski} {\it et al.} A three-dimensional magnetic topological phase. Submitted to {\it Science} (2018), arXiv:1712.09992.

\item Xitong Xu, Xirui Wang, Tyler A. Cochran, Daniel S. Sanchez, {\bf Ilya Belopolski} {\it et al.} Crystal growth and quantum oscillations in the topological chiral semimetal CoSi. arXiv:1904.00630.

\item Jia-Xin Yin, Songtian S. Zhang, Guoqing Chang, Qi Wang, Stepan Tsirkin, Zurab Guguchia, Biao Lian, Huibin Zhou, Kun Jiang, {\bf Ilya Belopolski} {\it et al.} Negative flat band magnetism in a spin-orbit coupled correlated kagome magnet. {\it Nat. Phys.} (2019). 

\item Daniel S. Sanchez*, {\bf Ilya Belopolski}* {\it et al.} Topological chiral crystals with helicoid-arc quantum states. {\it Nature} 567, 500 (2019).

\item Guoqing Chang, Benjamin J. Wieder, Frank Schindler, Daniel S. Sanchez, {\bf Ilya Belopolski} {\it et al.} Topological quantum properties of chiral crystals. {\it Nat. Mat.} (2018).

\item Jia-Xin Yin, Songtian S. Zhang, Hang Li, Kun Jiang, Guoqing Chang, Bingjing Zhang, Biao Lian, Cheng Xiang, {\bf Ilya Belopolski} {\it et al.} Giant and anisotropic many-body spin-orbit tunability in a strongly correlated kagome magnet. {\it Nature} 562, 91-95 (2018).

\item Jiaxin Yin, Songtian S. Zhang, Guangyang Dai, Hao Zheng, Guoqing Chang, {\bf Ilya Belopolski} {\it et al.} Vector field based emergent phase diagram in a correlated superconductor. arXiv:1802.10059.

\item M. Mofazzel Hosen, Klauss Dimitri, Alex Aperis, Pablo Maldonado, {\bf Ilya Belopolski} {\it et al.} Observation of Gapless Dirac Surface States in ZrGeTe. {\it Phys. Rev. B} 97, 121103 (2018).

\item Hao Zheng, Guoqing Chang, Shin-Ming Huang, Cheng Guo, Xiao Zhang, Songtian Zhang, Jiaxin Yin, Su-Yang Xu, {\bf Ilya Belopolski} {\it et al.} Mirror protected Dirac fermions on a Weyl semimetal NbP surface. {\it Phys. Rev. Lett.} 119, 196403 (2017).

\item Guoqing Chang, Su-Yang Xu, Xiaoting Zhou, Shin-Ming Huang, Bahadur Singh, Baokai Wang, {\bf Ilya Belopolski} {\it et al.} Topological Hopf and chain link semimetal states and their application to Co$_2$MnGa. {\it Phys. Rev. Lett.} 119, 156401 (2017).

\item Guoqing Chang, Su-Yang Xu, Benjamin J. Wieder, Daniel S. Sanchez, Shin-Ming Huang, {\bf Ilya Belopolski} {\it et al.} Large Fermi Arcs in Unconventional Weyl Semimetal RhSi. {\it Phys. Rev. Lett.} 119, 206401 (2017).

\item Q. R. Zhang, B. Zeng, D. Rhodes, S. Memaran, T. Besara, R. Sankar, F. Chou, N. Alidoust, S.-Y. Xu, {\bf Ilya Belopolski} {\it et al.} Magnetic field-induced electronic and topological phase transitions in Weyl type-I semi-metals. arXiv:1705.00920

\item {\bf Ilya Belopolski} {\it et al.} A novel condensed matter lattice and a new platform for one-dimensional topological phases. {\it Sci. Adv.} 3, e1501692 (2017).

\item M. Zahid Hasan, Su-Yang Xu, {\bf Ilya Belopolski} and Shin-Ming Huang. Discovery of Weyl Fermion Semimetals and Topological Fermi Arc States. {\it Ann. Rev. Cond. Mat. Phys.} 8, 289 (2017).

\item {\bf Ilya Belopolski} {\it et al.} Topological Weyl phase transition in Mo$_x$W$_{1-x}$Te$_2$. arXiv:1612.07793

\item M. Mofazzel Hosen, Klauss Dimitri, {\bf Ilya Belopolski} {\it et al.} Tunability of the topological nodal-line semimetal phase in ZrSi$X$-type materials ($X =$ S, Se, Te). {\it Phys. Rev. B}. 95, 161101 (2017).

\item {\bf Ilya Belopolski}, {\it et al.} Discovery of a new type of topological Weyl fermion semimetal state in Mo$_x$W$_{1-x}$Te$_2$. {\it Nat. Commun.} 7, 13643 (2016).

\item Hao Zheng, Guang Bian, Guoqing Chang, Hong Lu, Su-Yang Xu, Guangqiang Wang, Tay-Rong Chang, Songtian Zhang, {\bf Ilya Belopolski}, {\it et al.} Atomic-Scale Visualization of Quasiparticle Interference on a Type-II Weyl Semimetal Surface. {\it Phys. Rev. Lett.} 117, 266804 (2016).

\item Guoqing Chang, Daniel S. Sanchez, Benjamin J. Wieder, Su-Yang Xu, Frank Schindler, {\bf Ilya Belopolski}, {\it et al.} Kramers theorem-enforced Weyl fermions: Theory and Materials Predictions (Ag$_3$BO$_3$, TlTe$_2$O$_6$ and Ag$_2$Se related families). arXiv:1611.07925

\item Madhab Neupane, Nasser Alidoust, M. Mofazzel Hosen, Jian-Xin Zhu, Klauss Dimitri, Su-Yang Xu, Nagendra Dhakal, Raman Sankar, {\bf Ilya Belopolski}, {\it et al.} Observation of the spin-polarized surface state in a noncentrosymmetric superconductor BiPd. {\it Nat. Commun.} 7, 13315 (2016).

\item {\bf Ilya Belopolski}, {\it et al.} A minimal, ``hydrogen atom" version of an inversion-breaking Weyl semimetal. {\it Nat. Commun.} (2017) \& arXiv:1610.02013

\item Tay-Rong Chang, Su-Yang Xu, Daniel S. Sanchez, Shin-Ming Huang, Guoqing Chang, Chuang-Han Hsu, Guang Bian, {\bf Ilya Belopolski}, {\it et al.} Type-II Topological Dirac Semimetals: Theory and Materials Prediction (VAl$_3$ family). arXiv:1606.07555

\item Guoqing Chang, Su-Yang Xu, Shin-Ming Huang, Daniel S. Sanchez, Chuang-Han Hsu, Guang Bian, Zhi-Ming Yu, {\bf Ilya Belopolski}, {\it et al.} Nexus fermions in topological symmorphic crystalline metals. {\it Sci. Rep.} 7, 1688 (2017).

\item Madhab Neupane, M Mofazzel Hosen, {\bf Ilya Belopolski}, {\it et al.} Observation of Dirac-like semi-metallic phase in NdSb. {\it J. Phys: Cond. Mat.} 28, 23LT02 (2016).

\item Su-Yang Xu, Nasser Alidoust, Guoqing Chang, Hong Lu, Bahadur Singh, {\bf Ilya Belopolski}, {\it et al.} Discovery of Lorentz-violating Weyl fermion semimetal state in LaAlGe materials. arXiv:1603.07318.

\item {\bf Ilya Belopolski}, {\it et al.} Criteria for directly detecting topological Fermi arcs in Weyl semimetals. {\it Phys. Rev. Lett.} 116, 066802 (2016).

\item Chenglong Zhang, Su-Yang Xu, {\bf Ilya Belopolski}, {\it et al.} Signatures of the Adler-Bell-Jackiw chiral anomaly in a Weyl Fermion semimetal. {\it Nat. Commun.} 7, 10735 (2016).

\item Nasser Alidoust, A. Alexandradinata, Su-Yang Xu, {\bf Ilya Belopolski}, {\it et al.} A new form of (unexpected) Dirac fermions in the strongly-correlated cerium monopnictides. arXiv: 1604.08571

\item {\bf Ilya Belopolski}, {\it et al.} Measuring Chern numbers above the Fermi level in the Type II Weyl semimetal Mo$_x$W$_{1-x}$Te$_2$. {\it Phys. Rev. B}. 94, 085127 (2016).

\item Guoqing Chang, Bahadur Singh, Su-Yang Xu, Guang Bian, Shin-Ming Huang, Chuang-Han Hsu, {\bf Ilya Belopolski}, {\it et al.} Theoretical prediction of magnetic and noncentrosymmetric Weyl fermion semimetal states in the $R$-Al-$X$ family of compounds ($R =$ rare earth, Al $=$ aluminium, $X =$ Si, Ge). arXiv:1604.02124

\item Madhab Neupane, {\bf Ilya Belopolski}, {\it et al.} Observation of topological nodal fermion semimetal phase in ZrSiS. {\it Phys. Rev. B}. 93, 201104 (2016).

\item Su-Yang Xu, Nasser Alidoust, Guoqing Chang, Hong Lu, Bahadur Singh, {\bf Ilya Belopolski}, {\it et al.} Discovery of Lorentz-violating type II Weyl fermions in LaAlGe. {\it Sci. Adv.} 3, e1603266 (2017).

\item Guoqing Chang, Su-Yang Xu, Hao Zheng, Bahadur Singh, Chuang-Han Hsu, Guang Bian, Nasser Alidoust, {\bf Ilya Belopolski}, {\it et al.} Room-temperature magnetic topological Weyl fermion and nodal line semimetal states in half-metallic Heusler Co$_2$Ti$X$ ($X =$ Si, Ge, or Sn). Sci. Rep. 6, 38839 (2016).

\item Cheng-Long Zhang, Su-Yang Xu, {\bf Ilya Belopolski}, {\it et al.} Signatures of the Adler-Bell-Jackiw chiral anomaly in a Weyl fermion semimetal. {\it Nat. Commun.} 7, 10735 (2016).

\item Guoqing Chang, Su-Yang Xu, Daniel S. Sanchez, Shin-Ming Huang, Chi-Cheng Lee, Tay-Rong Chang, Hao Zheng, Guang Bian, {\bf Ilya Belopolski}, {\it et al.} A strongly robust Weyl fermion semimetal state in Ta$_3$S$_2$. {\it Sci. Adv.} 2, e1600295 (2016).

\item Tay-Rong Chang, Peng-Jen Chen, Guang Bian, Titus Neupert, Raman Sankar, Su-Yang Xu, {\bf Ilya Belopolski}, {\it et al.} Topological Dirac States and Pairing Correlations in the Non-Centrosymmetric Superconductor PbTaSe$_2$. {\it Phys. Rev. B}. 93, 245130 (2016)

\item Guoqing Chang, Su-Yang Xu, Hao Zheng, Chi-Cheng Lee, Shin-Ming Huang, {\bf Ilya Belopolski}, {\it et al.} Quasi-particle interferences of the Weyl semimetals TaAs and NbP. {\it Phys. Rev. Lett.} 116, 066601 (2016).

\item Guang Bian, Ting-Fung Chung, Chang Liu, Chaoyu Chen, Tay-Rong Chang, Tailung Wu, {\bf Ilya Belopolski}, {\it et al.} Experimental observation of two massless Dirac fermion gases in graphene-topological insulator heterostructure. {\it 2D Mat.} 3, 021009 (2016).

\item Hao Zheng, Su-Yang Xu, Guang Bian, Cheng Guo, Guoqing Chang, Daniel S. Sanchez, {\bf Ilya Belopolski}, {\it et al.} Atomic Scale Visualization of Quantum Interference on a Weyl Semimetal Surface by Scanning Tunneling Microscopy/Spectroscopy. ACS Nano 10, 1378 (2016).

\item Madhab Neupane, Yukiaki Ishida, Raman Sankar, Jian-Xin Zhu, Daniel S. Sanchez, {\bf Ilya Belopolski}, {\it et al.} Electronic structure and relaxation dynamics in a superconducting topological material. {\it Sci. Rep.} 6, 22557 (2016).

\item Su-Yang Xu, {\bf Ilya Belopolski}, {\it et al.} Spin polarization and texture of the Fermi arcs in the Weyl Fermion semimetal TaAs. {\it Phys. Rev. Lett.} 116, 096801 (2016).

\item Guang Bian, Tay-Rong Chang, Hao Zheng, Saavanth Velury, Su-Yang Xu, Titus Neupert, Ching-Kai Chiu, Daniel S. Sanchez, {\bf Ilya Belopolski}, {\it et al.} Drumhead Surface States and Topological Nodal-Line Fermions in TlTaSe$_2$. Accepted in {\it Phys. Rev. B}. (2016).

\item Tay-Rong Chang, Su-Yang Xu, Guoqing Chang, Chi-Cheng Lee, Shin-Ming Huang, BaoKai Wang, Guang Bian, Hao Zheng, Daniel S. Sanchez, {\bf Ilya Belopolski}, {\it et al.} Prediction of an arc-tunable Weyl fermion metallic state in Mo$_x$W$_{1-x}$Te$_2$. {\it Nat. Commun.} 7, 10639 (2016).

\item Madhab Neupane, Su-Yang Xu, Yukiaki Ishida, Shuang Jia, Benjamin M. Fregoso, Chang Liu, {\bf Ilya Belopolski}, {\it et al.} Gigantic surface life-time of an intrinsic topological insulator revealed via time-resolved (pump-probe) ARPES. {\it Phys. Rev. Lett.} 115, 116801 (2015). 
 
\item Chi-Cheng Lee, Su-Yang Xu, Shin-Ming Huang, Daniel S. Sanchez, {\bf Ilya Belopolski}, {\it et al.} Fermi arc topology and interconnectivity in Weyl fermion semimetals TaAs, TaP, NbAs and NbP. {\it Phys. Rev. B} 92, 235104 (2015).

\item Madhab Neupane, Su-Yang Xu, R. Sankar, Q. Gibson, Y. J. Wang, {\bf Ilya Belopolski}, {\it et al.} Topological phase diagram and saddle point singularity in a tunable topological crystalline insulator. {\it Phys. Rev. B} 92, 075131 (2015).

\item Su-Yang Xu, {\bf Ilya Belopolski}, {\it et al.} Experimental discovery of a topological Weyl semimetal state in TaP. {\it Sci. Adv.} 1, e1501092 (2015).

\item Madhab Neupane, Nasser Alidoust, Su-Yang Xu, {\bf Ilya Belopolski}, {\it et al.} Discovery of the topological surface state in a noncentrosymmetric superconductor BiPd. arXiv:1505.03466.

\item Guang Bian, Tay-Rong Chang, Raman Sankar, Su-Yang Xu, Hao Zheng, Titus Neupert, Ching-Kai Chiu, Shin-Ming Huang, Guoqing Chang, {\bf Ilya Belopolski}, {\it et al.} Topological nodal-line fermions in spin-orbit metal PbTaSe$_2$. {\it Nat. Commun.} 7, 10556 (2016).

\item Su-Yang Xu, Nasser Alidoust, {\bf Ilya Belopolski}, {\it et al.} Discovery of a Weyl fermion state with Fermi arcs in niobium arsenide. {\it Nat. Phys.} 11, 748 (2015).

\item Su-Yang Xu, Madhab Neupane, {\bf Ilya Belopolski}, {\it et al.} Unconventional transformation of spin Dirac phase across a topological quantum phase transition. {\it Nat. Commun.} 6, 6870 (2015).

\item Shin-Ming Huang, Su-Yang Xu, {\bf Ilya Belopolski}, {\it et al.} New type of Weyl semimetal with quadratic double Weyl fermions. {\it PNAS} 5, 113 (2016). 

\item Su-Yang Xu, Chang Liu, {\bf Ilya Belopolski}, {\it et al.} Lifshitz transition and van Hove singularity in a Topological Dirac Semimetal. {\it Phys. Rev. B} 92, 075115 (2015).

\item Su-Yang Xu*, {\bf Ilya Belopolski}*, {\it et al.} Discovery of a Weyl Fermion Semimetal and Topological Fermi Arcs. {\it Science} 349, 6248 (2015).

\item Satya K. Kushwaha, Jason W. Krizan, Benjamin E. Feldman, Andras Gyenis, Mallika T. Randeria, Jun Xiong, Su-Yang Xu, Nasser Alidoust, {\bf Ilya Belopolski}, {\it et al.} Bulk crystal growth and electronic characterization of the 3D Dirac Semimetal Na$_3$Bi. {\it APL Mater.} 3, 041504 (2015).

\item Su-Yang Xu, Chang Liu, Satya K. Kushwaha, Raman Sankar, Jason W. Krizan, {\bf Ilya Belopolski}, {\it et al.} Observation of Fermi arc surface states in a topological metal. {\it Science} 347, 6219 (2015).

\item Shin-Ming Huang, Su-Yang Xu, {\bf Ilya Belopolski}, {\it et al.} A Weyl Fermion semimetal with surface Fermi arcs in the transition metal monopnictide TaAs class. {\it Nat. Commun.} 6, 7373 (2015).

\item Madhab Neupane, Su-Yang Xu, Nasser Alidoust, Raman Sankar, {\bf Ilya Belopolski}, {\it et al.} Surface Versus Bulk Dirac States Tuning in a Three-Dimensional Topological Dirac Semimetal. {\it Phys. Rev. B} 91, 241114 (2015).

\item Madhab Neupane, Nasser Alidoust, Guang Bian, Su-Yang Xu, {\bf Ilya Belopolski}, {\it et al.} Fermi Surface Topology and Hotspots Distribution in Kondo Lattice System CeB$_6$. {\it Phys. Rev. B} 92, 104420 (2015).

\item Nasser Alidoust, Chang Liu, Su-Yang Xu, {\bf Ilya Belopolski}, {\it et al.} Observation of metallic surface states in the strongly correlated Kitaev-Heisenberg candidate Na$_2$IrO$_3$. arXiv:1410.6389.

\item Su-Yang Xu, Nasser Alidoust, {\bf Ilya Belopolski}, {\it et al.} Momentum-space imaging of Cooper pairing in a half-Dirac-gas topological superconductor. {\it Nat. Phys.} 10, 943 (2014).

\item Madhab Neupane, Su-Yang Xu, Nasser Alidoust, Guang Bian, Dae-Jeong Kim, Chang Liu, {\bf Ilya Belopolski}, {\it et al.} Non-Kondo-like Electronic Structure in the Correlated Rare-Earth Hexaboride YbB$_6$. {\it Phys. Rev. Lett.} 114, 016403 (2015).

\item Madhab Neupane, Anthony Richardella, Jaime Sanchez-Barriga, Su-Yang Xu, Nasser Alidoust, {\bf Ilya Belopolski}, {\it et al.} Observation of quantum-tunneling modulated spin texture in ultrathin topological insulator Bi$_2$Se$_3$ films. {\it Nat. Commun.} 5, 3841 (2014).

\item Chang Liu, Su-Yang Xu, Nasser Alidoust, Tay-Rong Chang, Hsin Lin, Chetan Dhital, Sovit Khadka, Madhab Neupane, {\bf Ilya Belopolski}, {\it et al.} Spin-correlated electronic state on the surface of a spin-orbit Mott system. {\it Phys. Rev. B} 90, 045127 (2014).

\item Su-Yang Xu, Chang Liu, Anthony Richardella, {\bf Ilya Belopolski}, {\it et al.} Fermi-level electronic structure of a topological-insulator/cuprate-superconductor in the superconducting proximity effect regime. {\it Phys. Rev. B} 90, 085128 (2014).

\item Nasser Alidoust, Guang Bian, Su-Yang Xu, Raman Sankar, Madhab Neupane, Chang Liu, {\bf Ilya Belopolski}, {\it et al.} Observation of monolayer valence band spin-orbit effect and induced quantum well states in Mo$X_2$. {\it Nat. Commun.} 5, 4673 (2014).

\item Su-Yang Xu, Chang Liu, S. K. Kushwaha, T.-R. Chang, J. W. Krizan, R. Sankar, C. M. Polley, J. Adell, T. Balasubramanian, K. Miyamoto, N. Alidoust, Guang Bian, M. Neupane, {\bf Ilya Belopolski}, {\it et al.} Observation of a bulk 3D Dirac multiplet, Lifshitz transition, and nestled spin states in Na$_3$Bi. arXiv:1312.7624.

\item Madhab Neupane, Nasser Alidoust, SuYang Xu, Takeshi Kondo, Yukiaki Ishida, Dae-Jeong Kim, Chang Liu, {\bf Ilya Belopolski}, {\it et al.} Surface electronic structure of the topological Kondo insulator candidate correlated electron system SmB$_6$. {\it Nat. Commun.} 4, 2991 (2013).

\item M. Neupane, N. Alidoust, Suyang Xu, Chang Liu, {\bf Ilya Belopolski}, {\it et al.} An experimental algorithm for identifying the topological nature of Kondo and mixed valence insulators. arXiv:1311.7111.

\item Madhab Neupane, Su-Yang Xu, R. Sankar, N. Alidoust, G. Bian, Chang Liu, {\bf Ilya Belopolski}, {\it et al.} Observation of a three-dimensional topological Dirac semimetal phase in high-mobility Cd$_3$As$_2$. {\it Nat. Commun.} 5, 3786 (2014).

\item M. Neupane, S. Basak, N. Alidoust, S.-Y. Xu, Chang Liu, {\bf Ilya Belopolski}, {\it et al.} Oscillatory surface dichroism of an insulating topological insulator Bi$_2$Te$_2$Se. {\it Phys. Rev. B}. 88, 165129 (2013).

\item Su-Yang Xu, Chang Liu, Nasser Alidoust, M. Neupane, D. Qian, {\bf Ilya Belopolski}, {\it et al.} Observation of a topological crystalline insulator phase and topological phase transition in Pb$_{1-x}$Sn$_x$Te. {\it Nat. Commun.} 3, 1192 (2012).

\end{enumerate}

%% file: thesis.bbl
\begin{thebibliography}{10}

\bibitem{TKNN}
D.~J. Thouless, M. Kohmoto, M.~P. Nightingale, M. den Nijs, Quantized {H}all {C}onductance in a {T}wo-{D}imensional {P}eriodic {P}otential. {\it Phys. Rev. Lett.} {\bf 49}, 405 (1982).

\bibitem{Nobel_FDMHaldane}
F.~D.~M. Haldane, {T}opological quantum matter. {\it Rev. Mod. Phys.\/} {\bf 89}, 040502 (2017).

\bibitem{Review_HasanKane}
M.~Z. Hasan, C.~L. Kane, Colloquium: {T}opological insulators. {\it Rev. Mod. Phys.\/} {\bf 82}, 3045 (2010).

\bibitem{ReviewTopology_XiaoGangWen}
X.-G. Wen, Choreographed entanglement dances: {T}opological states of quantum matter. {\it Science\/} {\bf 363}, 834 (2019).

\bibitem{Review_Kitaev}
A. Kitaev, Periodic table for topological insulators and superconductors. {\it AIP Conf. Proc.} {\bf 1134}, 22 (2009).

\bibitem{ReviewSSH_Kane}
C.~L. Kane, {\it Topological Insulators\/}, Contemporary Concepts of Condensed
  Matter Science (Elsevier, 2013), Chap.~1.

\bibitem{Review_QiZhang}
X.-L.~Qi, S.-C. Zhang, Topological insulators and superconductors. {\it Rev. Mod. Phys.} {\bf 83}, 1057 (2011).

\bibitem{Topo_Andrei}
A.~Bernevig, {\it Topological {I}nsulators and {T}opological
  {S}uperconductors\/} (Princeton University Press, 2013).

\bibitem{TaAs_Suyang}
S.-Y. Xu, I. Belopolski {\it et~al.\/}, Discovery of a {W}eyl fermion semimetal and
  topological {F}ermi arcs. {\it Science\/} {\bf 349}, 613 (2015).

\bibitem{NbAs_Suyang}
S.-Y. Xu, N. Alidoust, I. Belopolski {\it et~al.\/}, Discovery of a {W}eyl fermion state with {F}ermi arcs
  in niobium arsenide. {\it Nat. Phys.\/} {\bf 11}, 748 (2015).

\bibitem{TaP_Suyang}
S.-Y. Xu, I. Belopolski {\it et~al.\/}, Experimental discovery of a topological {W}eyl
  semimetal state in {T}a{P}. {\it Sci. Adv.\/} {\bf 1}, e1501092 (2015).

\bibitem{NbP_me}
I.~Belopolski, {\it et~al.\/}, Criteria for directly detecting topological
  {F}ermi arcs in {W}eyl semimetals. {\it Phys. Rev. Lett.\/} {\bf 116}, 066802
  (2016).

\bibitem{MoWTe2_me}
I.~Belopolski {\it et~al.\/}, Discovery of a new type of topological weyl
  fermion semimetal state in {Mo$_x$W$_{1-x}$Te$_2$}. {\it Nat. Commun.\/} {\bf
  7}, 13643 (2016).

\bibitem{MoWTe2_me2}
I.~Belopolski {\it et~al.\/}, Fermi arc electronic structure and {C}hern
  numbers in the type-{II} weyl semimetal candidate {Mo$_x$W$_{1-x}$Te$_2$}.
  {\it Phys. Rev. B\/} {\bf 94}, 085127 (2016).

\bibitem{TaIrTe4_me}
I.~Belopolski {\it et~al.\/}, Signatures of a time-reversal symmetric weyl
  semimetal with only four weyl points. {\it Nat. Commun.\/} {\bf 8}, 942
  (2017).

\bibitem{ARCMP_me}
M.~Z. Hasan, S.-Y. Xu, I.~Belopolski, S.-M. Huang, Discovery of {W}eyl fermion
  semimetals and topological {F}ermi arc states. {\it Ann. Rev. Cond. Matt.
  Phys.\/} {\bf 2017}, 289 (8).

\bibitem{LineNode_ChenFang}
C.~Fang, Y.~Chen, H.-Y. Kee, L.~Fu, Topological nodal line semimetals with and
  without spin-orbital coupling. {\it Phys. Rev. B\/} {\bf 92}, 081201 (2015).

\bibitem{Ca3P2_Schnyder_2016}
Y.-H. Chan, C.-K. Chiu, M.~Y. Chou, A.~P. Schnyder, {Ca$_3$P$_2$} and other
  topological semimetals with line nodes and drumhead surface states. {\it
  Phys. Rev. B\/} {\bf 93}, 205132 (2016).

\bibitem{TINI_Balents}
A.~A. Burkov, M.~D. Hook, L.~Balents, Topological nodal semimetals. {\it Phys.
  Rev. B\/} {\bf 84}, 235126 (2011).
  
\bibitem{Review_HasanXuBian}
M.~Z. Hasan, S.-Y. Xu, G. Bian, Topological insulators, topological superconductors and Weyl fermion semimetals: discoveries, perspectives and outlooks. {\it Phys. Scr.\/} {\bf 2015}, 014001 (2015).

\end{thebibliography}

\begin{thebibliography}{1}

\bibitem{Hertz} H.~Hertz, Ueber einen Einfluss des ultravioletten Lichtes auf die electrische Entladung, {\it Annalen der Physik} {\bf 267}, 983 (1887).

\bibitem{Einstein} A.~Einstein, Ueber einen die Erzeugung und Verwandlung des Lichtes betreffenden heuristischen Gesichtspunkt. {\it Annalen der Physik} {\bf 17}, 132 (1905).

\bibitem{ARPES_Hufner}
S.~H\"ufner, {\it Photoelectron Spectroscopy.} Springer, New York (1996).

\bibitem{ARPES_Damascelli}
A.~Damascelli, Z.~Hussain, Z.-X. Shen, Angle-resolved photoemission studies of
  the cuprate superconductors. {\it Rev. Mod. Phys.\/} {\bf 75}, 473 (2003).

\bibitem{ARPES_Damascelli2}
A.~Damascelli, Probing the Electronic Structure of Complex Systems by ARPES. {\it Phys. Scr.\/} {\bf 2004}, 61 (2004).

% FULL AUTHOR LIST; full author list for all papers in Photoemission chapter
\bibitem{BL4_a}
R.~Reininger, J.~Bozek, Y.-D.~Chuang, M.~Howells, N.~Kelez, S.~Prestemon, S.~Marks, T.~Warwick, C.~Jozwiak, A.~Lanzara, M.~Z.~Hasan, Z.~Hussain, MERLIN - A meV Resolution Beamline at the ALS (for ARPES and RIXS techniques). {\it AIP Conf. Proc.\/} {\bf 879}, 509 (2007).

\bibitem{BL4_b}
Y.-D.~Chuang, J.~Pepper, W.~McKinney, Z.~Hussain, E.~Gullikson, P.~Batson, D.~Qian, M.~Z.~Hasan, High-resolution soft X-ray emission spectrograph at the Advanced Light Source. {\it J. Phys. Chem. Solids\/} {\bf 66}, 2173 (2005).

%\bibitem{laser_xenon_Harter}
%J.~W. Harter, {\it et~al.\/}, A tunable low-energy photon source for
%  high-resolution angle-resolved photoemission spectroscopy. {\it Rev. Sci.
%  Inst.\/} {\bf 83}, 113103 (2012).

\bibitem{ADRESS_Strocov}
V.~N. Strocov, T.~Schmitt, U.~Flechsig, T.~Schmidt, A.~Imhof, Q.~Chen, J.~Raabe, R.~Betemps, D.~Zimoch, J.~Krempasky, X.~Wang, M.~Grioni, A.~Piazzalunga, L.~Pattheya, High-resolution soft {X}-ray beamline {ADRESS} at the {S}wiss {L}ight {S}ource for resonant inelastic {X}-ray scattering and angle-resolved photoelectron spectroscopies. {\it J. Synch. Rad.\/} {\bf 17}, 631 (2010).

%\bibitem{DA30} P.~Karlsson, ``DA30-L, Advanced Training'', ScientaOmicron (2018).

\bibitem{Kondo_analyzer}
K.~Kuroda, K.~Yaji, A.~Harasawa, R.~Noguchi, T.~Kondo, F.~Komori, S.~Shin, Experimental Methods for Spin- and Angle-Resolved Photoemission Spectroscopy Combined with Polarization-Variable Laser. {\it J. Vis. Exp.\/} {\bf 136}, e57090, doi:10.3791/57090 (2018).

\bibitem{Ishida_gap_functions}
Y.~Ishida, S.~Shin, Functions to map photoelectron distributions in a variety of setups in angle-resolved photoemission spectroscopy. {\it Rev. Sci. Inst.\/} {\bf 89}, 043903 (2018).

\end{thebibliography}

\begin{thebibliography}{21}

\bibitem{Weyl} H. Weyl, Elektron und gravitation. I. \textit{Z. Phys.} $\mathbf{56}$, 330-352 (1929).
\bibitem{Balents_viewpoint} L. Balents, Weyl electrons kiss. \textit{Physics} \textbf{4}, 36 (2011).
\bibitem{Wilczek} F. Wilczek, Why are there Analogies between Condensed Matter and Particle Theory? \textit{Phys. Today} $\mathbf{51}$, 11 (1998).

%\bibitem{Ashvin_Review} A. M. Turner, A. Vishwanath, Beyond band insulators: topology of semi-metals and interacting phases. http://arxiv.org/abs/1301.0330 (2013).

%\bibitem{Haldane} F. D. M. Haldane, Attachment of surface "Fermi arcs" to the bulk Fermi surface: "Fermi-level plumbing" in topological metals. http://arxiv.org/abs/1401.0529 (2014).
\bibitem{TI_book_2014} M. Z. Hasan, S.-Y. Xu, M. Neupane, Topological Insulators, Topological Crystalline Insulators, Topological Kondo Insulators, and Topological Semimetals.  in {\em Topological Insulators: Fundamentals and Perspectives} edited by F. Ortmann, S. Roche, S. O. Valenzuela (John Wiley \& Sons, 2015).

\bibitem{Nielsen1983} H. B. Nielsen, M. Ninomiya, The Adler-Bell-Jackiw anomaly and Weyl fermions in a crystal. \textit{Phys. Lett. B} \textbf{130}, 389-396 (1983).
\bibitem{Wan2011} X. Wan, A. M. Turner, A. Vishwanath, S. Y. Savrasov, Topological Semimetal and Fermi-arc surface states in the electronic structure of pyrochlore iridates. \textit{Phys. Rev. B} \textbf{83}, 205101 (2011).

%\bibitem{Burkov2011} A. A. Burkov, L. Balents, Weyl semimetal in a topological insulator multilayer. \textit{Phys. Rev. Lett.} \textbf{107},127205 (2011).

%\bibitem{HgCrSe} G. Xu \textit{et al.,} Chern semi-metal and quantized anomalous Hall effect in HgCr$_2$Se$_4$. \textit{Phys. Rev. Lett.} \textbf{107}, 186806 (2011).
\bibitem {Thallium} B. Singh, A. Sharma, H. Lin, M. Z. Hasan, R. Prasad, A Bansil, Topological electronic structure and Weyl semimetal in the TlBiSe$_2$ class of semiconductors. \textit{Phys. Rev. B} $\mathbf{86}$, 115208 (2012).

\bibitem{Suyang} S.-Y. Xu \textit{et al.,} Topological phase transition and texture inversion in a tunable topological insulator, TlBiSe$_2$. \textit{Science} $\mathbf{332}$, 560-564 (2011).

\bibitem{Hsin_TaAs} S.-M. Huang, S.-Y. Xu, I. Belopolski \textit{et al}., A Weyl Fermion semimetal with surface Fermi arcs in the transition metal monopnictide TaAs class. http://arxiv.org/abs/1501.00755 \& {\it Nat. Commun.} {\bf 6}, 7373 (2015).

\bibitem{Dai_TaAs} H. Weng, C. Fang, Z. Fang, B. A. Bernevig, X. Dai, Weyl Semimetal Phase in Noncentrosymmetric Transition-Metal Monophosphides. http://arxiv.org/abs/1501.00060 \& {\it Phys. Rev. X} {\bf 5}, 011209 (2015).

\bibitem{Hsin_TaAs_PRB} C.-C. Lee, S.-Y. Xu, S.-M. Huang, D. S. Sanchez, I. Belopolski {\it et al}., Fermi surface interconnectivity and topology in Weyl fermion semimetals TaAs, TaP, NbAs, and NbP. {\it Phys. Rev. B} {\bf 92}, 235104 (2015).


\bibitem{Vanderbilt} J. Liu, D. Vanderbilt, Weyl semimetals from noncentrosymmetric topological insulators. \textit{Phys. Rev. B} $\mathbf{90}$, 155316 (2014).



\bibitem{Chiral} A. A. Zyuzin, A. A. Burkov, Topological response in Weyl semimetals and the chiral anomaly. \textit{Phys. Rev. B} $\mathbf{86}$, 115133 (2012).
\bibitem{NMR} D. T. Son, B. Z. Spivak, Chiral anomaly and classical negative magnetoresistance of Weyl metals. \textit{Phys. Rev. B} $\mathbf{88}$, 104412 (2013).
\bibitem{Chiral_Qi} C.-X. Liu, P. Ye, X.-L. Qi, Chiral gauge field and axial anomaly in a Weyl semimetal. \textit{Phys. Rev. B} $\mathbf{87}$, 235306 (2013).

%\bibitem{Franz2013} M. M. Vazifeh, M. Franz, Electromagnetic response of Weyl semimetals. \textit{Phys. Rev. Lett.} \textbf{111},027201 (2013).




\bibitem{Aji2012} H. Wei, S.-P. Chao, V. Aji, Excitonic phases from Weyl semimetals. \textit{Phys. Rev. Lett.} $\mathbf{109}$, 196403 (2012).
\bibitem{Carbotte2013} P. E. C. Ashby, J. P. Carbotte, Magneto-optical conductivity of Weyl semimetals. \textit{Phys. Rev. B} \textbf{87}, 245131 (2013).

%\bibitem{SC} G. Y. Cho, J. H. Bardarson, Y.-M. Lu, J. E. Moore, Superconductivity of doped Weyl semimetals: Finite momentum pairing and electronic analog of the $^3$He-A phase. \textit{Phys. Rev. B} $\mathbf{86}$, 214514 (2012).
\bibitem{Hosor} P. Hosur, Friedel oscillations due to Fermi arcs in Weyl semimetals. \textit{Phys. Rev. B} $\mathbf{86}$, 195102 (2012).
\bibitem{Ojanen} T. Ojanen, Helical Fermi arcs and surface states in time-reversal invariant Weyl semimetals. \textit{Phys. Rev. B} $\mathbf{87}$, 245112 (2013).
\bibitem{Ashvin2} A. C. Potter, I. Kimchi, A. Vishwanath, Quantum oscillations from surface Fermi arcs in Weyl and Dirac semimetals. \textit{Nat. Commun.} $\mathbf{5}$, 5161 (2014).


%\bibitem{Burkov2014} I. Panfilov, A. A. Burkov, D. A. Pesin, Density response in Weyl metals. \textit{Phys. Rev. B} \textbf{89}, 245103 (2014).

%\bibitem{Yee2014} G. Basar, D. E. Kharzeev, H.-U. Yee, Triangle anomaly in Weyl semimetals. \textit{Phys. Rev. B} \textbf{89}, 035142 (2014).

\bibitem{Murakami2007} S. Murakami, Phase transition between the quantum spin Hall and insulator phases in 3D: emergence of a topological gapless phase. \textit{New J. Phys.} \textbf{9}, 356 (2007).


\bibitem{Hasan2010} M. Z. Hasan, C. L. Kane, \textit{Colloquium}: topological insulators. \textit{Rev. Mod. Phys.} \textbf{82}, 3045-3067 (2010).

\bibitem{Qi2011} X.-L. Qi, S.-C. Zhang, Topological insulators and superconductors. \textit{Rev. Mod. Phys.} \textbf{83}, 1057-1110 (2011).

\bibitem{Shuang_TaAs} C.-L. Zhang \textit{et al.,} Observation of the Adler-Bell-Jackiw chiral anomaly in a Weyl semimetal. http://arxiv.org/abs/1503.02630 (2015).
\bibitem{TaAs_Crystal_1} J. J. Murray, J. B. Taylor, L. D. Calvert, Y. Wang, E. J. Gabe, J. G. Despault, Phase relationships and thermodynamics of refractory metal pnictides: The metal-rich tantalum arsenides. \textit{J. Less Common Met.} $\mathbf{46}$, 311-320 (1976).

%\bibitem{TaAs_Crystal_2} H. Boller, E. Parth, The transposition structure of NbAs and of similar monophosphides and arsenides of niobium and tantalum. \textit{Acta Crystallogr.} $\mathbf{16}$, 1095-1101 (1963).
\bibitem{TaAs_Crystal_3} S. Furuseth, K. Selte, A. Kjekshus, On the Arsenides and Antimonides of Tantalum. \textit{Acta Chem. Scand.} $\mathbf{19}$, 95-106 (1965).
%\bibitem{TaAs_Crystal_4} J. O. Willerstr\"om, Stacking disorder in NbP, TaP, NbAs and TaAs. \textit{J. Less Common Met.} $\mathbf{99}$, 273-283 (1984).
%\bibitem{TaAs_Transport} B. Saparov, J. E. Mitchell, A. S. Sefat. Properties of binary transition-metal arsenides. \textit{Supercond. Sci. Technol.} $\mathbf{25}$, 084016 (2012).



%\bibitem{Graphene} A. K. Geim, K. S. Novoselov, The rise of graphene. \textit{Nat. Mater.} $\mathbf{6}$, 183-191 (2007).

\bibitem{Hasan_Na3Bi} S.-Y. Xu \textit{et al.}, Observation of Fermi arc surface states in a topological metal. \textit{Science} $\mathbf{347}$, 294-298 (2015).

\bibitem{Nagaosa} B.-J. Yang, N. Nagaosa, Classification of stable three-dimensional Dirac semimetals with nontrivial topology. \textit{Nat. Commun.} $\mathbf{5}$, 4898 (2014).
\bibitem{SXARPES} V. N. Strocov \textit{et al.}, Soft-X-ray ARPES facility at the ADRESS beamline of the SLS: Concepts, technical realisation and scientific applications. \textit{J. Synchrotron Rad.} $\mathbf{21}$, 32-44 (2014).

\bibitem{Perdew} J. P. Perdew, K. Burke, M. Ernzerhof, Generalized Gradient Approximation Made Simple. \textit{Phys. Rev. Lett.} $\mathbf{77}$, 3865-3868 (1996).
\bibitem{Weng} H. Weng, T. Ozaki, K. Terakura, Revisiting magnetic coupling in transition-metal-benzene complexes with maximally localized Wannier functions. \textit{Phys. Rev. B} $\mathbf{79}$, 235118 (2009).

\bibitem{Review_HasanXuBian_2015}
M.~Z. Hasan, S.-Y. Xu, G. Bian, Topological insulators, topological superconductors and Weyl fermion semimetals: discoveries, perspectives and outlooks. {\it Phys. Scr.\/} {\bf 2015}, 014001 (2015).

\end{thebibliography}

\begin{thebibliography}{99}
% 1
\bibitem{ColdAtoms} G. Jotzu, \textit{et al}. Experimental realization of the topological Haldane model with ultracold fermions. \textit{Nature} {\bf 515}, 237 (2014).
% 2
\bibitem{PhotonicWeyl} L. Lu, \textit{et al}. Experimental observation of Weyl points. \textit{Science} {\bf 349} 6248 (2015).
% 3
% 4
\bibitem{Mele} E. J. Mele. The winding road to topological insulators. \textit{Phys. Scr.} {\bf 2015}, 014004 (2015).
% 5
\bibitem{HasanKane} M. Z. Hasan, C. L. Kane, Topological insulators. \textit{Rev. Mod. Phys.} {\bf 82} 3045-3067 (2010).
\bibitem{BurkovBalents} A. Burkov \& L. Balents. Weyl semimetal in a topological insulator multilayer. \textit{Phy. Rev. Lett.} {\bf 107}, 127205 (2011).
% 6
\bibitem{Nikesh} M. Brahlek, \textit{et al}. Tunable inverse topological heterostructure utilizing (Bi$_{1-x}$In$_x$)$_2$Se$_3$. arXiv:1603.00683.
% 7
\bibitem{HKU} Z. Y. Wang, \textit{et al}. Superlattices of Bi$_2$Se$_3$/In$_2$Se$_3$: growth characteristics and structural properties. \textit{App. Phys. Lett.} {\bf 99}, 023112 (2011).
\bibitem{transport} Y. Zhao, \textit{et al}. Crossover from 3D to 2D Quantum Transport in Bi$_2$Se$_3$/In$_2$Se$_3$ Superlattices. \textit{Nano Lett.} {\bf 14}, 5244-5249 (2014).
% 8
\bibitem{MatthewBiSe} Y. Xia, \textit{et al}., Observation of a large-gap topological-insulator class with a single Dirac cone on the surface. \textit{Nature Phys.} {\bf 5}, 398-402 (2009).
% 9
\bibitem{ZhangBiSe} H. J. Zhang, \textit{et al}., Topological insulators in Bi$_2$Se$_3$, Bi$_2$Te$_3$ and Sb$_2$Te$_3$ with a single Dirac cone on the surface. \textit{Nature Phys.} {\bf 5}, 438-442 (2009).
% 10
\bibitem{BrahlekTPT} M. Brahlek, N. Bansal, N. Koirala, S.-Y. Xu, M. Neupane, C. Liu, M. Z. Hasan, S. Oh, Topological-metal to band-insulator transition in (Bi$_{1-x}$In$_x$)$_2$Se$_3$ thin films. \textit{Phys. Rev. Lett.} {\bf 109}, 186403 (2012).
% 11
\bibitem{QiZhang} X. L. Qi, S. C. Zhang, Topological insulators and superconductors. \textit{Rev. Mod. Phys.} {\bf 83}, 1057-1110 (2011).
% 12
\bibitem{BAB} B. A. Bernevig, \textit{Topological insulators and topological superconductors} (Princeton, 2013).
% 13
\bibitem{HasanMoore} M. Z. Hasan, J. E. Moore, Three-dimensional topological insulators. \textit{Ann. Rev. Cond. Mat. Phys.} {\bf 2}, 55-78 (2011).
% 14
\bibitem{HasanSusu} M. Z. Hasan, S.-Y. Xu \& G. Bian, Topological insulators, topological superconductors and Weyl fermion semimetals: discoveries, perspectives and outlooks. \textit{Phys. Scr.} {\bf 2015}, 014001 (2015).
% 15
\bibitem{MadhabQL} M. Neupane, \textit{et al}., Observation of quantum-tunneling modulated spin texture in ultrathin topological insulator Bi$_2$Se$_3$ films. \textit{Nature Commun.} {\bf 5}, 3841 (2014).
% 16
\bibitem{XueQikunQL} Y. Zhang, \textit{et al}., Crossover of the three-dimensional topological insulator Bi$_2$Se$_3$ to the two-dimensional limit. \textit{Nature Phys.} {\bf 6}, 584-588 (2010).
% 17
\bibitem{YuCardona} P. Yu, M. Cardona, \textit{Fundamentals of Semiconductors}, Chap. 9 (Springer, 2001).
% 18
\bibitem{EsakiTsu} L. Esaki, R. Tsu, Superlattice and negative differential conductivity in semiconductors. \textit{IBM J. Res. Develop.} {\bf 14}, 61-65 (1970).
% 19
\bibitem{SSHOriginalPaper} W. P. Su, J. R. Schrieffer, A. J. Heeger, Solitons in polyacetylene. \textit{Phys. Rev. Lett.} {\bf 42}, 1698-1701 (1979).
% 20
\bibitem{SSHK} A. J. Heeger, S. Kivelson, J. R. Schrieffer, W. P. Su. Solitons in conducting polymers. \textit{Rev. Mod. Phys.} {\bf 60}, 781-850 (1988).
% 21
\bibitem{PeriodicTable} A. P. Schnyder, \textit{et al}. Classification of topological insulators and superconductors in three spatial dimensions. \textit{Phys. Rev. B} {\bf 78}, 195125 (2008).
% 22
\bibitem{MBE} N. Bansal, \textit{et al.} Epitaxial growth of topological insulator \topo\ film on Si(111) with atomically sharp interface. \textit{Thin Solid Films} {\bf 520}, 224 (2011).
% 23
\bibitem{ADRESS} V. N. Strocov, \textit{et al}., Soft-X-ray ARPES facility at the ADRESS beamline of the SLS: Concepts, technical realisation and scientific applications. \textit{J. Synchrotron Rad.} {\bf 21}, 32 (2014).
% 24
\bibitem{STM} N. Koirala, \textit{et al}., Record Surface State Mobility and Quantum Hall Effect in Topological Insulator Thin Films via Interface Engineering. \textit{Nano Lett.} {\bf 15}, 8245 (2015).
\end{thebibliography}

\begin{thebibliography}{99}

\bibitem{Weyl} H. Weyl. {\it Z. Phys.} {\bf 56}, 330 (1929).
\bibitem{Herring} C. Herring. {\it Phys. Rev.} {\bf 52}, 365 (1937).
\bibitem{Abrikosov} A. A. Abrikosov \& S. D. Beneslavskii. {\it J. Low Temperature Physics} {\bf 5}, 141 (1971).
\bibitem{Nielsen} H. B. Nielsen \& M. Ninomiya. {\it Phys. Lett. B} {\bf 130}, 389 (1983).
\bibitem{Volovik} G. E. Volovik, \textit{The Universe in a Helium Droplet} (Clarendon Press, Oxford, 2003).
\bibitem{Murakami} S. Murakami. {\it New Journal of Physics} {\bf 9}, 356 (2007).

\bibitem{Multilayer} A. A. Burkov \& L. Balents. {\it Phys. Rev. Lett.} {\bf 107}, 127205 (2011).

\bibitem{Pyrochlore} X. Wan, A. M. Turner, A. Vishwanath \& S. Y. Savrasov. {\it Phys. Rev. B} {\bf 83}, 205101 (2011).

\bibitem{Vish} A. Turner \& A. Vishwanath. {\it Topological Insulators\/}, Contemporary Concepts of Condensed Matter Science (Elsevier, 2013), Chap.~11.

\bibitem{Hosur} P. Hosur \& X. Qi. {\it Comp. Rend. Phy.} {\bf 14}, 857 (2013).

\bibitem{TaAsThyUs} S.-M. Huang, S.-Y. Xu, I. Belopolski, C.-C. Lee, G. Chang, B. K. Wang, N. Alidoust, G. Bian, M. Neupane, C. Zhang, S. Jia, A. Bansil, H. Lin \& M. Z. Hasan. {\it Nat. Commun.} {\bf 6}, 7373 (2015).
\bibitem{TaAsUs} S.-Y. Xu, I. Belopolski, N. Alidoust, M. Neupane, G. Bian, C. Zhang, R. Sankar, G. Chang, Z. Yuan, C.-C. Lee, S.-M. Huang, H. Zheng, J. Ma, D. S. Sanchez, B. K. Wang, A. Bansil, F. C. Chou, P. P. Shibayev, H. Lin, S. Jia \& M. Z. Hasan. {\it Science} {\bf 349}, 613 (2015).
\bibitem{NbAs} S.-Y. Xu, N. Alidoust, I. Belopolski, Z. Yuan, G. Bian, T.-R. Chang, H. Zheng, V. N. Strocov, D. S. Sanchez, G. Chang, C. Zhang, D. Mou, Y. Wu, L. Huang, C.-C. Lee, S.-M. Huang, B. K. Wang, A. Bansil, H.-T. Jeng, T. Neupert, A. Kaminski, H. Lin, S. Jia \& M. Z. Hasan. {\it Nat. Phys.} {\bf 11}, 748 (2015).
\bibitem{TaPUs} S.-Y. Xu, I. Belopolski, D. S. Sanchez, C. Zhang, G. Chang, C. Guo, G. Bian, Z. Yuan, H. Lu, T.-R. Chang, P. P. Shibayev, M. L. Prokopovych, N. Alidoust, H. Zheng, C.-C. Lee, S.-M. Huang, R. Sankar, F. C. Chou, C.-H. Hsu, H.-T. Jeng, A. Bansil, T. Neupert, V. N. Strocov, H. Lin, S. Jia \& M. Z. Hasan. {\it Sci. Adv.} {\bf 1}, 10 (2015).
\bibitem{HaoNbP} H. Zheng, S.-Y. Xu, G. Bian, C. Guo, G. Chang, D. S. Sanchez, I. Belopolski, C.-C. Lee, S.-M. Huang, X. Zhang, R. Sankar, N. Alidoust, T.-R. Chang, F. Wu, T. Neupert, F. C. Chou, H.-T. Jeng, N. Yao, A. Bansil, S. Jia, H. Lin \& M. Z. Hasan. {\it ACS Nano.} {\bf 10}, 1378 (2016).
\bibitem{LingLu} L. Lu, Z. Wang, D. Ye, L. Ran, L. Fu, J. D. Joannopoulos \& M. Solja\v{c}i\'{c}. {\it Science} {\bf 349}, 622 (2015).
\bibitem{TaAsThyThem} H. Weng, C. Fang, Z. Fang, B. A. Bernevig \& X. Dai. {\it Phys. Rev. X} {\bf 5}, 011029 (2015).
\bibitem{TaAsThem} B. Q. Lv, H. M. Weng, B. B. Fu, X. P. Wang, H. Miao, J. Ma, P. Richard, X. C. Huang, L. X. Zhao, G. F. Chen, Z. Fang, X. Dai, T. Qian \& H. Ding. {\it Phys. Rev. X} {\bf 5}, 031013 (2015).
\bibitem{TaAsNodesDing} B. Q. Lv, N. Xu, H. M. Weng, J. Z. Ma, P. Richard, X. C. Huang, L. X. Zhao, G. F. Chen, C. E. Matt, F. Bisti, V. N. Strocov, J. Mesot, Z. Fang, X. Dai, T. Qian, M. Shi \& H. Ding. {\it Nat. Phys.} {\bf 11}, 724 (2015).
\bibitem{TaPThem} N. Xu, H. M. Weng, B. Q. Lv, C. Matt, J. Park, F. Bisti, V. N. Strocov, D. Gawryluk, E. Pomjakushina, K. Conder, N. C. Plumb, M. Radovic, G. Aut\`{e}s, O. V. Yazyev, Z. Fang, X. Dai, G. Aeppli, T. Qian, J. Mesot, H. Ding \& M. Shi. {\it Nat. Commun.} {\bf 7}, 11006 (2016).
\bibitem{TaAsChen} L. X. Yang, Z. K. Liu, Y. Sun, H. Peng, H. F. Yang, T. Zhang, B. Zhou, Y. Zhang, Y. F. Guo, M. Rahn, D. Prabhakaran, Z. Hussain, S.-K. Mo, C. Felser, B. Yan \& Y. L. Chen. {\it Nat. Phys.} {\bf 11}, 728 (2015).
\bibitem{Bernevig} A. Bernevig \& T. Hughes, \textit{Topological Insulators and Topological Superconductors} (Princeton University Press, Princeton, 2013).
\bibitem{Na3Bi} S.-Y. Xu, C. Liu, S. K. Kushwaha, R. Sankar, J. W. Krizan, I. Belopolski, M. Neupane, G. Bian, N. Alidoust, T.-R. Chang, H.-T. Jeng, C.-Y. Huang, W.-F. Tsai, H. Lin, P. P. Shibayev, F. C. Chou, R. J. Cava \& M. Z. Hasan. {\it Science} {\bf 347}, 294 (2015).
\bibitem{NbPThem} D.-F. Xu, Y.-P. Du, Z. Wang, Y.-P. Li, X.-H. Niu, Q. Yao, P. Dudin, Z.-A. Xu, X.-G. Wan \& D.-L. Feng. {\it Chin. Phys. Lett.} {\bf 32}, 10 (2015).
\bibitem{NbPAndo} S. Souma, Z. Wang, H. Kotaka, T. Sato, K. Nakayama, Y. Tanaka, H. Kimizuka, T. Takahashi, K. Yamauchi, T. Oguchi, K. Segawa \& Y. Ando. {\it Phys. Rev. B} {\bf 93}, 161112 (2016).
\bibitem{NbPChen} Z. K. Liu, L. X. Yang, Y. Sun, T. Zhang, H. Peng, H. F. Yang, C. Chen, Y. Zhang, Y. F. Guo, D. Prabhakaran, M. Schmidt, Z. Hussain, S.-K. Mo, C. Felser, B. Yan \& Y. L. Chen. {\it Nat. Mat.} {\bf 15}, 27 (2016).
\bibitem{OpenMX} T. Ozaki. {\it Phys. Rev. B} {\bf 67}, 155108 (2003).
\bibitem{Perdew} J. P. Perdew, K. Burke \& M. Ernzerhof. {\it Phys. Rev. Lett.} {\bf 77}, 3865 (1996).
\bibitem{Crystal1} S. Rundqvist. {\it Nature} {\bf 211}, 847 (1966).
\bibitem{Crystal2} J. O. Willerstr\"{o}m. {\it J. Less Common Metals} {\bf 99}, 273 (1984).
\bibitem{Crystal3} J. Xu, M. Greenblatt, T. Emge \& P. H\"{o}hn. {\it Inorgan. Chem.} {\bf 35}, 845 (1996).
\bibitem{Wannier1} N. Marzari \& D. Vanderbilt. {\it Phys. Rev. B} {\bf 56}, 12847 (1997).
\bibitem{Wannier2} A. A. Mostofi, J. R. Yates, Y.-S. Lee, I. Souza, D. Vanderbilt \& N. Marzari. {\it Comp. Phys. Commun.} {\bf 178}, 685 (2008).
\bibitem{FourCompounds} C.-C. Lee, S.-Y. Xu, S.-M. Huang, D. S. Sanchez, I. Belopolski, G. Chang, G. Bian, N. Alidoust, H. Zheng, M. Neupane, B. K. Wang, A. Bansil, M. Z. Hasan \& H. Lin. {\it Phys. Rev. B} {\bf 92}, 235104 (2015).

\bibitem{Crystal1} S. Rundqvist. New metal-rich phosphides of niobium, tantalum and tungsten. \textit{Nature} {\bf 211}, 847 (1966).
\bibitem{Crystal2} J. O. Willerstr\"{o}m. Stacking disorder in NbP, TaP, NbAs and TaAs. \textit{J. Less Common Metals} {\bf 99}, 273 (1984).
\bibitem{Crystal3} J. Xu, M. Greenblatt, T. Emge \& P. H\"{o}hn. Crystal structure, electrical transport and magnetic properties of niobium monophosphide. \textit{Inorgan. Chem.} {\bf 35}, 845 (1996).
\bibitem{FourCompounds} C.-C. Lee, S.-Y. Xu, S.-M. Huang, D. S. Sanchez, I. Belopolski, G. Chang, G. Bian, N. Alidoust, H. Zheng, M. Neupane, B. K. Wang, A. Bansil, M. Z. Hasan \& H. Lin. Fermi surface interconnectivity and topology in Weyl fermion semimetals TaAs, TaP, NbAs and NbP. \textit{Phys. Rev. B} {\bf 92}, 235104 (2015).
\bibitem{TaAsUs} S.-Y. Xu, I. Belopolski, N. Alidoust, M. Neupane, G. Bian, C. Zhang, R. Sankar, G. Chang, Z. Yuan, C.-C. Lee, S.-M. Huang, H. Zheng, J. Ma, D. S. Sanchez, B. K. Wang, A. Bansil, F. C. Chou, P. P. Shibayev, H. Lin, S. Jia \& M. Z. Hasan. Discovery of a Weyl fermion semimetal and topological Fermi arcs. \textit{Science} {\bf 349}, 613 (2015).
\bibitem{TaPUs} S.-Y. Xu, I. Belopolski, D. S. Sanchez, C. Zhang, G. Chang, C. Guo, G. Bian, Z. Yuan, H. Lu, T.-R. Chang, P. P. Shibayev, M. L. Prokopovych, N. Alidoust, H. Zheng, C.-C. Lee, S.-M. Huang, R. Sankar, F. C. Chou, C.-H. Hsu, H.-T. Jeng, A. Bansil, T. Neupert, V. N. Strocov, H. Lin, S. Jia \& M. Z. Hasan. Experimental discovery of a topological Weyl semimetal state in TaP. \textit{Sci. Adv.} {\bf 1}, 10 (2015).

\end{thebibliography}

\begin{thebibliography}{99}

\bibitem{Weyl} H. Weyl. Elektron und Gravitation. Z. Phys. {\bf 56} 330 (1929).
\bibitem{Peskin} M. Peskin \& D. Schroeder. \textit{An Introduction to Quantum Field Theory} (Perseus Books, Reading, MA, 1995)

\bibitem{AndreiNature} Soluyanov, A. \textit{et al}. Type II Weyl semimetals. \textit{Nature} {\bf 527}, 495 (2015).
\bibitem{Grushin} Grushin, A. G. Consequences of a condensed matter realization of Lorentz violating QED in
Weyl semi-metals. \textit{Phys. Rev. D} {\bf 86} 045001 (2012).
\bibitem{Bergholtz} Bergholtz, E. J. \textit{et al}. Topology and Interactions in a Frustrated Slab: Tuning from Weyl Semimetals to $C > 1$ Fractional Chern Insulators. \textit{Phys. Rev. Lett.} {\bf 114}, 016806 (2015).
\bibitem{Trescher} Trescher, M. \textit{et al}. Quantum transport in Dirac materials: Signatures of tilted and anisotropic Dirac and Weyl cones. \textit{Phys. Rev. B} {\bf 91}, 115135 (2015).
\bibitem{Beenakker} Beenakker, C. Tipping the Weyl cone. \textit{Journal Club for Condensed Matter Physics}, posted August, 2015.
\bibitem{Zyuzin} Zyuzin, A. A. \& Tiwari, R. P. Intrinsic anomalous Hall effect in type-II Weyl semimetals. {\it JETP Lett.} {\bf 103}, 717 (2016).
\bibitem{Isobe} Isobe, H. \& Nagaosa, N. Coulomb interaction effect in tilted Weyl fermion in two dimensions.
\textit{Phys. Rev. Lett.} {\bf 116}, 116803 (2016).

\bibitem{TayRong} T.-R. Chang \textit{et al}. Prediction of an arc-tunable Weyl Fermion metallic state in Mo$_x$W$_{1-x}$Te$_2$. {\it Nat. Commun.} {\bf 7}, 10639 (2016).

\bibitem{Binghai} Y. Sun \textit{et al}. Prediction of Weyl semimetal in orthorhombic MoTe$_2$. {\it Phys. Rev. B} {\bf 92}, 161107 (2015).

\bibitem{Zhijun} Z. J. Wang \textit{et al}. MoTe$_2$: A Type-II Weyl Topological Metal. {\it Phys. Rev. Lett.} {\bf 117}, 056805 (2016).

\bibitem{Ta3S2} Chang, G. \textit{et al}. A strongly robust type II Weyl fermion semimetal state in Ta$_3$S$_2$. {\it Sci. Adv.} {\bf 2}, e1600295 (2016).

\bibitem{Koepernik} Koepernik, K. \textit{et al}. TaIrTe$_4$: A ternary type-II Weyl semimetal. {\it Phys. Rev. B} {\bf 93}, 201101 (2016).

\bibitem{Autes} Aut\'es, G. \textit{et al}. Robust Type-II Weyl Semimetal Phase in Transition Metal Diphosphides $X$P$_2$ ($X = $Mo, W). {\it Phys. Rev. Lett.} {\bf 117}, 066402 (2016).

\bibitem{Adam1} L. Huang, T. M. McCormick, M. Ochi, Z. Zhao, M. Suzuki, R. Arita, Y. Wu, D. Mou, H. Cao, J. Yan, N. Trivedi, A. Kaminski. Spectroscopic evidence for a type II Weyl semimetallic state in MoTe$_2$. {\it Nat. Mat.} {\bf 15}, 1155 (2016).
\bibitem{Adam2} Y. Wu, D. Mou, N. H. Jo, K. Sun, L. Huang, S. L. Bud'ko, P. C. Canfield, A. Kaminski. Observation of Fermi arcs in the type-II Weyl semimetal candidate WTe$_2$. {\it Phys. Rev. B} {\bf 94}, 121113 (2016).
\bibitem{Xinjiang2} C. Wang, Y. Zhang, J. Huang, S. Nie, G. Liu, A. Liang, Y. Zhang, B. Shen, J. Liu, C. Hu, Y. Ding, D. Liu, Y. Hu, S. He, L. Zhao, L. Yu, J. Hu, J. Wei, Z. Mao, Y. Shi, X. Jia, F. Zhang, S. Zhang, F. Yang, Z. Wang, Q. Peng, H. Weng, X. Dai, Z. Fang, Z. Xu, C. Chen, X. J. Zhou. Observation of Fermi arc and its connection with bulk states in the candidate type-II Weyl semimetal WTe$_2$. {\it Phys. Rev. B} {\bf 94}, 241119 (2016).

\bibitem{Shuyun} K. Deng, G. Wan, P. Deng, K. Zhang, S. Ding, E. Wang, M. Yan, H. Huang, H. Zhang, Z. Xu, J. Denlinger, A. Fedorov, H. Yang, W. Duan, H. Yao, Y. Wu, S. Fan, H. Zhang, X. Chen, S. Zhou. Experimental observation of topological Fermi arcs in type-II Weyl semimetal MoTe$_2$. {\it Nat. Phys.} {\bf 12}, 1105 (2016).
\bibitem{Chen} J. Jiang, Z. K. Liu, Y. Sun, H. F. Yang, R. Rajamathi, Y. P. Qi, L. X. Yang, C. Chen, H. Peng, C.-C. Hwang, S. Z. Sun, S.-K. Mo, I. Vobornik, J. Fujii, S. S. P. Parkin, C. Felser, B. H. Yan, Y. L. Chen. Observation of the Type-II Weyl Semimetal Phase in MoTe$_2$. {\it Nat. Commun.} {\bf 8}, 13973 (2017).
\bibitem{Xinjiang} A. Liang, J. Huang, S. Nie, Y. Ding, Q. Gao, C. Hu, S. He, Y. Zhang, C. Wang, B. Shen, J. Liu, P. Ai, L. Yu, X. Sun, W. Zhao, S. Lv, D. Liu, C. Li, Y. Zhang, Y. Hu, Y. Xu, L. Zhao, G. Liu, Z. Mao, X. Jia, F. Zhang, S. Zhang, F. Yang, Z. Wang, Q. Peng, H. Weng, X. Dai, Z. Fang, Z. Xu, C. Chen, X. J. Zhou. Electronic Evidence for Type II Weyl Semimetal State in MoTe$_2$, arXiv:1604.01706.
\bibitem{HongDing} N. Xu, Z. J. Wang, A. P. Weber, A. Magrez, P. Bugnon, H. Berger, C. E. Matt, J. Z. Ma, B. B. Fu, B. Q. Lv, N. C. Plumb, M. Radovic, E. Pomjakushina, K. Conder, T. Qian, J. H. Dil, J. Mesot, H. Ding, M. Shi. Discovery of Weyl semimetal state violating Lorentz invariance in MoTe$_2$, arXiv:1604.02116.

\bibitem{Baumberger} F. Y. Bruno, A. Tamai, Q. S. Wu, I. Cucchi, C. Barreteau, A. de la Torre, S. McKeown Walker, S. Ricco, Z. Wang, T. K. Kim, M. Hoesch, M. Shi, N. C. Plumb, E. Giannini, A. A. Soluyanov, F. Baumberger. Observation of large topologically trivial Fermi arcs in the candidate type-II Weyl semimetal WTe$_2$. {\it Phys. Rev. B} {\bf 94}, 121112 (2016).
%Surface states and bulk electronic structure in the candidate type-II Weyl semimetal WTe$_2$. arXiv:1604.02411.

\bibitem{LaAlGe} Xu, S.-Y. \textit{et al}. Discovery of Lorentz-violating type II Weyl fermions in LaAlGe. {\it Sci. Adv.} {\bf 3}, e1603266 (2017).

\bibitem{RAlX} Chang, G. \textit{et al}. Magnetic and noncentrosymmetric Weyl fermion semimetals in the $R$AlGe
 family of compounds ($R = $ rare earth). {\it Phys. Rev. B} {\bf 97}, 041104 (2018).

% Theoretical prediction of magnetic and noncentrosymmetric Weyl fermion semimetal states in the $R$-Al-$X$ family of compounds ($R =$ rare Earth, Al $ = $ aluminium, $X =$ Si, Ge). arXiv:1604.02124.

\bibitem{NbPme} Belopolski, I. \textit{et al}. Criteria for Directly Detecting Topological Fermi Arcs in Weyl Semimetals. \textit{Phys. Rev. Lett.} {\bf 116}, 066802 (2016).

\bibitem{SM} See the Supplemental Material for additional ARPES data on the electronic structure of \comp\ and the methods.

\bibitem{Trivedi} T. M. McCormik, I. Kimchi \& N. Trivedi. Minimal models for topological Weyl semimetals. {\it Phys. Rev. B} {\bf 95}, 075133 (2017).

\bibitem{IshidaMethods} Y. Ishida \textit{et al}. Time-resolved photoemission apparatus achieving sub-20-meV energy resolution and high stability. {\it Rev. Sci. Instr.} {\bf 85}, 123904 (2014).

\bibitem{MoTe2WTe2} B. E. Brown. The crystal structures of WTe$_2$ and high-temperature MoTe$_2$. {\it Acta. Cryst.} {\bf 20} 268 (1966).
\bibitem{KaminskiMethods} R. Jiang \textit{et al}. Tunable vacuum ultraviolet laser based spectrometer for angle-resolved photoemission spectroscopy. {\it Rev. Sci. Instr.} {\bf 85}, 033902 (2014).

\bibitem{XRD1} G. M. Sheldrick. \textit{SADABS} (University of G\"ottingen, G\"ottingen, Germany, 1996).
\bibitem{XRD2} \textit{CrystalClear, v1.3.5} (Rigaku Corp., The Woodlands, TX, 1999).
\bibitem{XRD3} G. M Sheldrick. \textit{SHELXTL, v5.1} (Bruker-AXS, Madison, WI, 1998).

\bibitem{PAW1} P. E. Bl\"ochl. Projector augmented-wave method. {\it Phys. Rev. B} {\bf 50}, 17953 (1994).
\bibitem{PAW2} G. Kresse \& J. Joubert. From ultrasoft pseudopotentials to the projector augmented-wave method. {\it Phys. Rev. B} {\bf 59}, 1758 (1999).
\bibitem{TransitionMetals} G. Kresse \& J. Hafner. \textit{Ab initio} molecular dynamics for open-shell transition metals. {\it Phys. Rev. B} {\bf 48}, 13115 (1993).
\bibitem{PlaneWaves1} G. Kresse \& J. Furthm\"uller. Efficiency of \textit{ab initio} total energy calculations for metals and semiconductors using a plane-wave basis set. {\it Comput. Mater. Sci} {\bf 6}, 15 (1996).
\bibitem{PlaneWaves2} G. Kresse \& J. Furthm\"uller. Efficient iterative schemes for \textit{ab initio} total energy calculations using a plane-wave basis set. {\it Phys. Rev. B} {\bf 54}, 11169 (1996).
\bibitem{GGA} J. P. Perdew, K. Burke \& M. Ernzerhof. Generalized gradient approximation made simple. {\it Phys. Rev. Lett} {\bf 77}, 3865 (1996).
\bibitem{MLWF1} N. Marzari \& D. Vanderbilt. Maximally localized generalized Wannier functions for composite energy bands. {\it Phys. Rev. B} {\bf 56}, 12847 (1997).
\bibitem{MLWF2} I. Souza, N. Marzari \& D.Vanderbilt. Maximally localized Wannier functions for entangled energy bands. {\it Phys. Rev. B} {\bf 65}, 035109 (2001).
\bibitem{Wannier90} A. A. Mostofi \textit{et al}. Wannier90: a tool for obtaining maximally-localized Wannier functions. {\it Comp. Phys. Commun.} {\bf 178}, 685 (2008).
\bibitem{MLWF3} C. Franchini \textit{et al}. Maximally localized Wannier functions in LaMnO$_3$ within PBE$ + $U, hybrid functionals and partially self-consistent GW: an efficient route to construct \textit{ab initio} tight-binding parameters for $e_g$ perovskites. {\it J. Phys. Cond. Mat.} {\bf 24}, 235602 (2012).
\bibitem{Green} H. J. Zhang \textit{et al}. Topological insulators in Bi$_2$Se$_3$, Bi$_2$Te$_3$ and Sb$_2$Te$_3$ with a single Dirac cone on the surface. {\it Nat. Phys.} {\bf 5}, 438 (2009).


%\bibitem{MoTe2WTe2} B. E. Brown. The crystal structures of WTe$_2$ and high-temperature MoTe$_2$. Acta. Cryst. {\bf 20} 268 (1966).
%\bibitem{TayRong} T.-R. Chang \textit{et al}. Arc-tunable Weyl fermion metallic state in Mo$_x$W$_{1-x}$Te$_2$. arXiv:1508.06723.
%\bibitem{KaminskiMethods} R. Jiang \textit{et al}. Tunable vacuum ultraviolet laser based spectrometer for angle-resolved photoemission spectroscopy. Rev. Sci. Instr. {\bf 85}, 033902 (2014).

%\bibitem{IshidaMethods} Y. Ishida \textit{et al}. Time-resolved photoemission apparatus achieving sub-20-meV energy resolution and high stability. Rev. Sci. Instr. {\bf 85}, 123904 (2014).

%\bibitem{XRD1} G. M. Sheldrick. \textit{SADABS} (University of G\"ottingen, G\"ottingen, Germany, 1996).
%\bibitem{XRD2} \textit{CrystalClear, v1.3.5} (Rigaku Corp., The Woodlands, TX, 1999).
%\bibitem{XRD3} G. M Sheldrick. \textit{SHELXTL, v5.1} (Bruker-AXS, Madison, WI, 1998).

%\bibitem{PAW1} P. E. Bl\"ochl. Projector augmented-wave method. Phys. Rev. B. {\bf 50}, 17953 (1994).
%\bibitem{PAW2} G. Kresse \& J. Joubert. From ultrasoft pseudopotentials to the projector augmented-wave method. Phys. Rev. B. {\bf 59}, 1758 (1999).
%\bibitem{TransitionMetals} G. Kresse \& J. Hafner. \textit{Ab initio} molecular dynamics for open-shell transition metals. Phys. Rev. B. {\bf 48}, 13115 (1993).
%\bibitem{PlaneWaves1} G. Kresse \& J. Furthm\"uller. Efficiency of \textit{ab initio} total energy calculations for metals and semiconductors using a plane-wave basis set. Comput. Mater. Sci. {\bf 6}, 15 (1996).
%\bibitem{PlaneWaves2} G. Kresse \& J. Furthm\"uller. Efficient iterative schemes for \textit{ab initio} total energy calculations using a plane-wave basis set. Phys. Rev. B. {\bf 54}, 11169 (1996).
%\bibitem{GGA} J. P. Perdew, K. Burke \& M. Ernzerhof. Generalized gradient approximation made simple. Phys. Rev. Lett. {\bf 77}, 3865 (1996).
%\bibitem{MLWF1} N. Marzari \& D. Vanderbilt. Maximally localized generalized Wannier functions for composite energy bands. Phys. Rev. B {\bf 56}, 12847 (1997).
%\bibitem{MLWF2} I. Souza, N. Marzari \& D.Vanderbilt. Maximally localized Wannier functions for entangled energy bands. Phys. Rev. B {\bf 65}, 035109 (2001).
%\bibitem{Wannier90} A. A. Mostofi \textit{et al}. Wannier90: a tool for obtaining maximally-localized Wannier functions. Comp. Phys. Commun. {\bf 178}, 685 (2008).
%\bibitem{MLWF3} C. Franchini \textit{et al}. Maximally localized Wannier functions in LaMnO$_3$ within PBE$ + $U, hybrid functionals and partially self-consistent GW: an efficient route to construct \textit{ab initio} tight-binding parameters for $e_g$ perovskites. J. Phys. Cond. Mat. {\bf 24}, 235602 (2012).
%\bibitem{Green} H. J. Zhang \textit{et al}. Topological insulators in Bi$_2$Se$_3$, Bi$_2$Te$_3$ and Sb$_2$Te$_3$ with a single Dirac cone on the surface. Nat. Phys. {\bf 5}, 438 (2009).

\end{thebibliography}

\begin{thebibliography}{99}

\bibitem{TPT} Belopolski, I. {\it et al}., Topological Weyl phase transition in Mo$_x$W$_{1-x}$Te$_2$. arXiv:1612.07793 (2016).
\bibitem{TaAsUs} Xu, S.-Y. \textit{et al}. Discovery of a Weyl fermion semimetal and topological Fermi arcs. \textit{Science} {\bf 349}, 613-617 (2015).
\bibitem{LingLu} Lu, L. \textit{et al}. Experimental observation of Weyl points. \textit{Science} {\bf 349}, 622-624 (2015).
\bibitem{TaAsThem} Lv, B. Q. \textit{et al}. Experimental discovery of Weyl semimetal TaAs. \textit{Phys. Rev. X} {\bf 5}, 031013 (2015).
\bibitem{TaAsThyUs} Huang, S.-M. \textit{et al}. A Weyl Fermion semimetal with surface Fermi arcs in the transition metal monopnictide TaAs class. \textit{Nat. Commun.} {\bf 6}, 7373 (2015).
\bibitem{TaAsThyThem} Weng, H., Fang, C., Fang, Z., Bernevig, B. A. \& Dai, X. Weyl semimetal phase in noncentrosymmetric transition-metal monophosphides. \textit{Phys. Rev. X} {\bf 5}, 011029 (2015).

\bibitem{Weyl} Weyl, H. Elektron und gravitation. \textit{Z. Phys.} {\bf 56}, 330-352 (1929).
\bibitem{Peskin} Peskin, M. \& Schroeder, D. \textit{An Introduction to Quantum Field Theory} (Perseus Books Publishing, 1995).
\bibitem{Herring} Herring, C. Accidental degeneracy in the energy bands of crystals. \textit{Phys. Rev.} {\bf 52}, 365-373 (1937).
\bibitem{Abrikosov} Abrikosov, A. A. \& Beneslavskii, S. D. Some properties of gapless semiconductors of the second kind. \textit{J. Low Temp. Phys.} {\bf 5}, 141-154 (1971).
\bibitem{Nielsen} Nielsen, H. B. \& Ninomiya, M. The Adler-Bell-Jackiw anomaly and Weyl fermions in a crystal. \textit{Phys. Lett. B} {\bf 130}, 389-396 (1983).
\bibitem{Volovik} Volovik, G. E. \textit{The Universe in a Helium Droplet} (Oxford University Press, 2003).
\bibitem{Murakami} Murakami, S. Phase transition between the quantum spin Hall and insulator phases in 3D: emergence of a topological gapless phase. \textit{New J. Phys.} {\bf 9}, 356 (2007).
\bibitem{Multilayer} Burkov, A. A. \& Balents, L. Weyl semimetal in a topological insulator multilayer. \textit{Phys. Rev. Lett.} {\bf 107}, 127205 (2011).
\bibitem{Pyrochlore} Wan, X., Turner, A. M., Vishwanath, A. \& Savrasov, S. Y. Topological semimetal and Fermi-arc surface states in the electronic structure of pyrochlore iridates. \textit{Phys. Rev. B} {\bf 83}, 205101 (2011).
\bibitem{Vish} Turner, A. \& Vishwanath, A. Beyond band insulators: topology of semimetals and interacting phases. {\it Topological Insulators\/}, Contemporary Concepts of Condensed Matter Science (Elsevier, 2013), Chap.~11.

\bibitem{Param} Parameswaran, S. A., Grover, T., Abanin, D. A., Pesin, D. A. \& Vishwanath, A. Probing the Chiral Anomaly with Nonlocal Transport in Three-Dimensional Topological Semimetals. \textit{Phys. Rev. X} {\bf 4}, 031035 (2014).
\bibitem{Hosur} Hosur, P. \& Qi, X. Recent developments in transport phenomena in Weyl semimetals. \textit{Comp. Rend. Phys.} {\bf 14}, 857-870 (2013).
\bibitem{Potter} Potter, A. C., Kimchi, I. \& Vishwanath, A. Quantum oscillations from surface Fermi arcs in Weyl and Dirac semimetals. \textit{Nat. Commun.} {\bf 5}, 5161 (2014).

\bibitem{ZhangC} Zhang, C. \textit{et al}. Signatures of the Adler-Bell-Jackiw chiral anomaly in a Weyl semimetal. \textit{Nat. Commun.} {\bf 7}, 10735 (2016).
\bibitem{HuangXiao} Huang, X. \textit{et al}. Observation of the Chiral-Anomaly-Induced Negative Magnetoresistance in 3D Weyl Semimetal TaAs. \textit{Phys. Rev. X} {\bf 5}, 031023 (2015).
\bibitem{NbPTransport} Shekhar, C. \textit{et al}. Extremely large magnetoresistance and ultrahigh mobility in the topological Weyl semimetal candidate NbP. \textit{Nat. Phys.} {\bf 11}, 645-649 (2015).

\bibitem{TR} Chang, T.-R. \textit{et al}. Prediction of an arc-tunable Weyl Fermion metallic state in \mowte. \textit{Nat. Commun.} {\bf 7}, 10639 (2016).
\bibitem{AndreiNature} Soluyanov, A. \textit{et al}. Type II Weyl semimetals. \textit{Nature} {\bf 527}, 495-498 (2015).
\bibitem{Zhijun} Wang, Z. J. \textit{et al}. MoTe$_2$: A Type-II Weyl Topological Metal. \textit{Phys. Rev. Lett.} {\bf 117}, 056805 (2016).
\bibitem{Binghai} Sun, Y., Wu, S.-C., Ali, M. N., Felser, C. \& Yan, B. Prediction of Weyl semimetal in orthorhombic MoTe$_2$. \textit{Phys. Rev. B} {\bf 92}, 161107 (2015).

\bibitem{ReviewTMDC} Mak, K. F. \& Shan, J. Photonics and optoelectronics of 2D semiconductor transition metal dichalcogenides. \textit{Nat. Photon.} {\bf 10}, 216-226 (2016).
\bibitem{LiangFu} Qian, X. F., Liu, J., Fu, L. \& Li, J. Quantum spin Hall effect in two-dimensional transition metal dichalcogenides. \textit{Science} {\bf 346}, 1344-1347 (2014). 
\bibitem{LaserPattern} Cho, S. \textit{et al}. Phase patterning for ohmic homojunction contact in MoTe$_2$. \textit{Science} {\bf 349}, 625-628 (2015).
\bibitem{Heinz} Mak, K. F., Lee, C., Hone, J., Shan, J. \& Heinz, T. F. Atomically Thin MoS$_2$: A New Direct-Gap Semiconductor. \textit{Phys. Rev. Lett.} {\bf 105}, 136805 (2010).
\bibitem{Valley} Cao, T. \textit{et al}. Valley-selective circular dichroism of monolayer molybdenum disulphide. \textit{Nat. Commun.} {\bf 3}, 887 (2012).

\bibitem{Grushin} Grushin, A. G. Consequences of a condensed matter realization of Lorentz-violating QED in
Weyl semi-metals. \textit{Phys. Rev. D} {\bf 86}, 045001 (2012).
\bibitem{Bergholtz} Bergholtz, E. J., Liu, Z., Trescher, M., Moessner, R. \& Udagawa, M. Topology and Interactions in a Frustrated Slab: Tuning from Weyl Semimetals to $C > 1$ Fractional Chern Insulators. \textit{Phys. Rev. Lett.} {\bf 114}, 016806 (2015).
\bibitem{Trescher} Trescher, M., Sbierski, B., Brouwer, P. W. \& Bergholtz, E. J. Quantum transport in Dirac materials: Signatures of tilted and anisotropic Dirac and Weyl cones. \textit{Phys. Rev. B} {\bf 91}, 115135 (2015).
\bibitem{Beenakker} Beenakker, C. Tipping the Weyl cone. \textit{Journal Club for Condensed Matter Physics}. Posted at http://www.condmatjournalclub.org/?p=2644 (2015).
\bibitem{Ta3S2} Chang, G. \textit{et al}. A strongly robust type II Weyl fermion semimetal state in Ta$_3$S$_2$. \textit{Sci. Adv.} {\bf 2}, e1600295 (2016).
\bibitem{Koepernik} Koepernik, K. \textit{et al}. TaIrTe$_4$: A ternary type-II Weyl semimetal. \textit{Phys. Rev. B} {\bf 93}, 201101 (2016).
\bibitem{Autes} Aut\'es, G., Gresch, D., Troyer, M., Soluyanov, A. A. \& Yazyev O. V. Robust Type-II Weyl Semimetal Phase in Transition Metal Diphosphides $X$P$_2$ ($X = $Mo, W). \textit{Phys. Rev. Lett.} {\bf 117}, 066402 (2016).
\bibitem{Suyang} Xu, S.-Y. \textit{et al}. Discovery of Lorentz-violating type II Weyl fermions in LaAlGe. {\it Sci. Adv.} {\bf 3}, e1603266 (2017).
\bibitem{Zyuzin} Zyuzin, A. A. \& Tiwari, R. P. Intrinsic anomalous Hall effect in type-II Weyl semimetals. \textit{JETP} {\bf 103}, 717-722 (2016).
\bibitem{Isobe} Isobe, H. \& Nagaosa, N. Coulomb interaction effect in tilted Weyl fermion in two dimensions.
\textit{Phys. Rev. Lett.} {\bf 116}, 116803 (2016).

\bibitem{myothermowte} Belopolski, I. \textit{et al}. Fermi arc electronic structure and Chern numbers in the type-II Weyl semimetal candidate \mowte. \textit{Phys. Rev. B} {\bf 94}, 085127 (2016).
\bibitem{NbPme} Belopolski, I. \textit{et al}. Criteria for Directly Detecting Topological Fermi Arcs in Weyl Semimetals. \textit{Phys. Rev. Lett.} {\bf 116}, 066802 (2016).

\bibitem{MoTe2WTe2} Brown, B. E. The crystal structures of WTe$_2$ and high-temperature MoTe$_2$. \textit{Acta. Cryst.} {\bf 20} 268-274 (1966).

\bibitem{Adam} Huang, L. \textit{et al}. Spectroscopic evidence for type II Weyl semimetallic state in MoTe$_2$. {\it Nat. Mat.} {\bf 15}, 1155 (2016).
\bibitem{Shuyun} Deng, K. \textit{et al}. Experimental observation of topological Fermi arcs in type-II Weyl semimetal MoTe$_2$. {\it Nat. Phys.} {\bf 12}, 1105 (2016).
\bibitem{Baumberger} Tamai, A. \textit{et al}. Fermi Arcs and Their Topological Character in the Candidate Type-II Weyl Semimetal MoTe$_2$. \textit{Phys. Rev. X} {\bf 6}, 031021 (2016).
\bibitem{Adam2} Wu, Y. \textit{et al}. Observation of Fermi arcs in the type-II Weyl semimetal candidate WTe$_2$. \textit{Phys. Rev. B} {\bf 94}, 121113 (2016).
\bibitem{Baumberger2} Bruno, F. Y., \textit{et al}. Observation of large topologically trivial Fermi arcs in the candidate type-II Weyl semimetal WTe$_2$. \textit{Phys. Rev. B} {\bf 94}, 121112 (2016).
\bibitem{Chen} Jiang, J. \textit{et al}. Observation of the Type-II Weyl Semimetal Phase in MoTe$_2$. {\it Nat. Commun.} {\bf 8}, 13973 (2017).
\bibitem{Xinjiang} Liang, A. \textit{et al}. Electronic Evidence for Type II Weyl Semimetal State in MoTe$_2$. Preprint at https://arxiv.org/abs/1604.01706 (2016).
\bibitem{Xinjiang2} Wang, C. \textit{et al}. Spectroscopic Evidence of Type II Weyl Semimetal State in WTe$_2$. Preprint at https://arxiv.org/abs/1604.04218 (2016).
\bibitem{HongDing} Xu, N. \textit{et al}. Discovery of Weyl semimetal state violating Lorentz invariance in MoTe$_2$. Preprint at https://arxiv.org/abs/1604.02116 (2016).

\bibitem{IshidaMethods} Ishida, Y. \textit{et al}. Time-resolved photoemission apparatus achieving sub-20-meV energy resolution and high stability. \textit{Rev. Sci. Instr.} {\bf 85}, 123904 (2014).

\bibitem{GGA} Perdew, J. P., Burke, K. \&  Ernzerhof, M. Generalized gradient approximation made simple. \textit{Phys. Rev. Lett.} {\bf 77}, 3865 (1996).
\bibitem{PAW1} Bl\"ochl, P. E. Projector augmented-wave method. \textit{Phys. Rev. B.} {\bf 50}, 17953 (1994).
\bibitem{PAW2} Kresse, G. \& Joubert, J. From ultrasoft pseudopotentials to the projector augmented-wave method. \textit{Phys. Rev. B.} {\bf 59}, 1758 (1999).
\bibitem{PlaneWaves1} Kresse, G. \& Furthm\"uller, J. Efficiency of \textit{ab initio} total energy calculations for metals and semiconductors using a plane-wave basis set. \textit{Comput. Mater. Sci.} {\bf 6}, 15-50 (1996).
\bibitem{Bonding} Mar, A., Jobic, S. \& Ibers, J. A. Metal-metal vs. tellurium-tellurium bonding in WTe$_2$ and its ternary variants TaIrTe$_4$ and NbIrTe$_4$. \textit{J. Am. Chem. Soc.} {\bf 114}, 8963 -8971 (1992).
\bibitem{MLWF1} Marzari, N. \& Vanderbilt, D. Maximally localized generalized Wannier functions for composite energy bands. \textit{Phys. Rev. B} {\bf 56}, 12847 (1997).
\bibitem{MLWF2} Souza, I., Marzari, N. \& Vanderbilt, D. Maximally localized Wannier functions for entangled energy bands. \textit{Phys. Rev. B} {\bf 65}, 035109 (2001).
\bibitem{Wannier90} Mostofi, A. A., Yates, J. R., Lee, Y.-S., Souza, I., Vanderbilt, D. \& Marzari, N. Wannier90: a tool for obtaining maximally-localized Wannier functions. \textit{Comp. Phys. Commun.} {\bf 178}, 685-699 (2008).
\bibitem{MLWF3} Franchini, C. \textit{et al}. Maximally localized Wannier functions in LaMnO$_3$ within PBE$ + $U, hybrid functionals and partially self-consistent GW: an efficient route to construct \textit{ab initio} tight-binding parameters for $e_g$ perovskites. \textit{J. Phys. Cond. Mat.} {\bf 24}, 235602 (2012).
\bibitem{Green2} Xia, Y. \textit{et al}. Observation of a large-gap topological-insulator class with a single Dirac cone on the surface. \textit{Nat. Phys.} {\bf 5}, 398-402 (2009).
\bibitem{Green} Zhang, H., Liu, C.-X., Qi X.-L., Dai, X., Fang, Z. \& Zhang, S.-C. Topological insulators in Bi$_2$Se$_3$, Bi$_2$Te$_3$ and Sb$_2$Te$_3$ with a single Dirac cone on the surface. \textit{Nat. Phys.} {\bf 5}, 438-442 (2009).

\end{thebibliography}

\begin{thebibliography}{99}
\bibitem{Weyl} Weyl, H. Elektron und Gravitation. \textit{Z. Phys.} {\bf 56} 330-352 (1929).
\bibitem{Peskin} Peskin, M. \& Schroeder, D. \textit{An Introduction to Quantum Field Theory} (Perseus Books, Reading, MA, 1995)
\bibitem{Abrikosov} Abrikosov, A. A. \& Beneslavskii, S. D. Some properties of gapless semiconductors of the second kind. \textit{J. Low Temp. Phys.} {\bf 5}, 141-154 (1971).
\bibitem{Nielsen} Nielsen, H. B. \& Ninomiya, M. The Adler-Bell-Jackiw anomaly and Weyl fermions in a crystal. \textit{Phys. Lett. B} {\bf 130}, 389-396 (1983).
\bibitem{Volovik} Volovik, G. E. \textit{The Universe in a Helium Droplet} (Clarendon Press, Oxford, 2003).

\bibitem{Murakami} Murakami, S. Phase transition between the quantum spin Hall and insulator phases in 3D: Emergence of a topological gapless phase. \textit{New J. Phys.} {\bf 9}, 356 (2007).
\bibitem{Pyrochlore} Wan, X., Turner, A. M., Vishwanath, A. \& Savrasov, S. Y. Topological semimetal and Fermi-arc surface states in the electronic structure of pyrochlore iridates \textit{Phys. Rev. B} {\bf 83}, 205101 (2011).
\bibitem{ARCMP} Hasan, M. Z., Xu, S.-Y., Belopolski, I., and Huang, S.-M. Discovery of Weyl Fermion Semimetals and Topological Fermi Arc States \textit{Ann. Rev. Cond. Matt. Phys.} {\bf 8} 289-309 (2017).
%\bibitem{Vish} Turner, A. \& Vishwanath, A. Beyond Band Insulators: Topology of Semimetals and Interacting Phases, in \textit{Contemporary Concepts of Condensed Matter Science}, M. Franz and L. Molenkamp, Eds., Vol. 6, Topological Insulators, 293-324 (2013).
\bibitem{Hosur} Hosur, P. \& Qi, X. Recent developments in transport phenomena in Weyl semimetals. \textit{Comp. Rend. Phy.} {\bf 14}, 857-870 (2013).

\bibitem{TaAsThyUs} Huang, S.-M. \textit{et al}. A Weyl Fermion semimetal with surface Fermi arcs in the transition metal monopnictide TaAs class. \textit{Nat. Commun.} {\bf 6}, 7373 (2015).
\bibitem{TaAsUs} Xu, S.-Y., Belopolski, I. \textit{et al}. Discovery of a Weyl fermion semimetal and topological Fermi arcs. \textit{Science} {\bf 349}, 613-617 (2015).


%\bibitem{TaAsNodesDing} B. Q. Lv, N. Xu, H. M. Weng, J. Z. Ma, P. Richard, X. C. Huang, L. X. Zhao, G. F. Chen, C. E. Matt, F. Bisti, V. N. Strocov, J. Mesot, Z. Fang, X. Dai, T. Qian, M. Shi \& H. Ding. Nat. Phys. {\bf 11}, 724 (2015).
\bibitem{NbAs} Xu, S.-Y., Alidoust, N., Belopolski, I. \textit{et al.} Discovery of a Weyl fermion state with Fermi arcs in niobium arsenide. \textit{Nat. Phys.} {\bf 11}, 748-754 (2015).
% N. Alidoust, I. Belopolski, Z. Yuan, G. Bian, T.-R. Chang, H. Zheng, V. N. Strocov, D. S. Sanchez, G. Chang, C. Zhang, D. Mou, Y. Wu, L. Huang, C.-C. Lee, S.-M. Huang, B. K. Wang, A. Bansil, H.-T. Jeng, T. Neupert, A. Kaminski, H. Lin, S. Jia \& M. Z. Hasan.
\bibitem{TaPUs} S.-Y. Xu, I. Belopolski \textit{et al.} Experimental discovery of a topological Weyl semimetal state in TaP. \textit{Sci. Adv.} {\bf 1}, 10 (2015).
% I. Belopolski, D. S. Sanchez, C. Zhang, G. Chang, C. Guo, G. Bian, Z. Yuan, H. Lu, T.-R. Chang, P. P. Shibayev, M. L. Prokopovych, N. Alidoust, H. Zheng, C.-C. Lee, S.-M. Huang, R. Sankar, F. C. Chou, C.-H. Hsu, H.-T. Jeng, A. Bansil, T. Neupert, V. N. Strocov, H. Lin, S. Jia \& M. Z. Hasan.
%\bibitem{TaPThem} N. Xu, H. M. Weng, B. Q. Lv, C. Matt, J. Park, F. Bisti, V. N. Strocov, D. Gawryluk, E. Pomjakushina, K. Conder, N. C. Plumb, M. Radovic, G. Aut\`{e}s, O. V. Yazyev, Z. Fang, X. Dai, G. Aeppli, T. Qian, J. Mesot, H. Ding \& M. Shi. Nat Commun. {\bf 7}, 11006 (2016).
\bibitem{HaoNbP} Zheng, H. \textit{et al.} Atomic-Scale Visualization of Quantum Interference on a Weyl Semimetal Surface by Scanning Tunneling Microscopy. \textit{ACS Nano.} {\bf 10}, 1378 (2016).
%1378-1385.
% S.-Y. Xu, G. Bian, C. Guo, G. Chang, D. S. Sanchez, I. Belopolski, C.-C. Lee, S.-M. Huang, X. Zhang, R. Sankar, N. Alidoust, T.-R. Chang, F. Wu, T. Neupert, F. C. Chou, H.-T. Jeng, N. Yao, A. Bansil, S. Jia, H. Lin \& M. Z. Hasan. 

\bibitem{TaAsThyThem} Weng, H. \textit{et al}. Weyl semimetal phase in noncentrosymmetric transition-metal monophosphides. \textit{Phys. Rev. X} {\bf 5}, 011029 (2015).
\bibitem{TaAsThem} Lv, B. Q. \textit{et al}. Experimental discovery of Weyl semimetal TaAs. \textit{Phys. Rev. X} {\bf 5}, 031013 (2015).

\bibitem{TaAsChen} Yang, L. X. \textit{et al.} Weyl semimetal phase in the non-centrosymmetric compound TaAs. \textit{Nat. Phys.} {\bf 11}, 728-732 (2015).
% Z. K. Liu, Y. Sun, H. Peng, H. F. Yang, T. Zhang, B. Zhou, Y. Zhang, Y. F. Guo, M. Rahn, D. Prabhakaran, Z. Hussain, S.-K. Mo, C. Felser, B. Yan \& Y. L. Chen.
\bibitem{AndreiNature} Soluyanov, A. \textit{et al}. Type II Weyl semimetals. \textit{Nature} {\bf 527}, 495-498 (2015).
\bibitem{TayRong} Chang, T.-R. \textit{et al}. Prediction of an arc-tunable Weyl Fermion metallic state in \mwt. \textit{Nat. Commun.} {\bf 7}, 10639 (2016).
\bibitem{Binghai} Sun, Y. \textit{et al}. Prediction of Weyl semimetal in orthorhombic MoTe$_2$. \textit{Phys. Rev. B} {\bf 92}, 161107 (2015).
\bibitem{Zhijun} Wang, Z. J. \textit{et al}. MoTe$_2$: A Type-II Weyl Topological Metal. \textit{Phys. Rev. Lett.} {\bf 117}, 056805 (2016).

\bibitem{myothermowte} Belopolski, I. \textit{et al.} Fermi arc electronic structure and Chern numbers in the type-II Weyl semimetal candidate \mwt. \textit{Phys. Rev. B} {\bf 94}, 085127 (2016).
\bibitem{Adam1} Huang, L. \textit{et al.} Spectroscopic evidence for type II Weyl semimetallic state in MoTe$_2$. \textit{Nat. Mat.} {\bf 15}, 1155-1160 (2016).
% T. M. McCormick, M. Ochi, Z. Zhao, M. Suzuki, R. Arita, Y. Wu, D. Mou, H. Cao, J. Yan, N. Trivedi, A. Kaminski.
\bibitem{Shuyun} Deng, K. \textit{et al.} Experimental observation of topological Fermi arcs in type-II Weyl semimetal MoTe$_2$. \textit{Nat. Phys.} {\bf 12}, 1105-1110 (2016).
% G. Wan, P. Deng, K. Zhang, S. Ding, E. Wang, M. Yan, H. Huang, H. Zhang, Z. Xu, J. Denlinger, A. Fedorov, H. Yang, W. Duan, H. Yao, Y. Wu, S. Fan, H. Zhang, X. Chen, S. Zhou
%\bibitem{Adam2} Y. Wu, N. H. Jo, D. Mou, L. Huang, S. L. Bud'ko, P. C. Canfield, A. Kaminski. Observation of Fermi arcs in the type-II Weyl semimetal candidate WTe$_2$. \textit{Phys. Rev. B} {\bf 94}, 121113 (2016).
\bibitem{Baumberger} Tamai, A. \textit{et al}. Fermi Arcs and Their Topological Character in the Candidate Type-II Weyl Semimetal MoTe$_2$. \textit{Phys. Rev. X} {\bf 6}, 031021 (2016).
%\bibitem{Baumberger2} F. Y. Bruno, A. Tamai, Q. S. Wu, I. Cucchi, C. Barreteau, A. de la Torre, S. McKeown Walker, S. Riccò, Z. Wang, T. K. Kim, M. Hoesch, M. Shi, N. C. Plumb, E. Giannini, A. A. Soluyanov, F. Baumberger. Observation of large topologically trivial Fermi arcs in the candidate type-II Weyl semimetal WTe$_2$. \textit{Phys. Rev. B} {\bf 94}, 121112 (2016).
%\bibitem{Xinjiang2} C. Wang, Y. Zhang, J. Huang, S. Nie, G. Liu, A. Liang, Y. Zhang, B. Shen, J. Liu, C. Hu, Y. Ding, D. Liu, Y. Hu, S. He, L. Zhao, L. Yu, J. Hu, J. Wei, Z. Mao, Y. Shi, X. Jia, F. Zhang, S. Zhang, F. Yang, Z. Wang, Q. Peng, H. Weng, X. Dai, Z. Fang, Z. Xu, C. Chen, X. J. Zhou. Spectroscopic Evidence of Type II Weyl Semimetal State in WTe$_2$. arXiv:1604.04218.
%\bibitem{Chen} J. Jiang, Z. K. Liu, Y. Sun, H. F. Yang, R. Rajamathi, Y. P. Qi, L. X. Yang, C. Chen, H. Peng, C.-C. Hwang, S. Z. Sun, S.-K. Mo, I. Vobornik, J. Fujii, S. S. P. Parkin, C. Felser, B. H. Yan, Y. L. Chen. Observation of the Type-II Weyl Semimetal Phase in MoTe$_2$. arXiv:1604.00139.
%\bibitem{Xinjiang} A. Liang, J. Huang, S. Nie, Y. Ding, Q. Gao, C. Hu, S. He, Y. Zhang, C. Wang, B. Shen, J. Liu, P. Ai, L. Yu, X. Sun, W. Zhao, S. Lv, D. Liu, C. Li, Y. Zhang, Y. Hu, Y. Xu, L. Zhao, G. Liu, Z. Mao, X. Jia, F. Zhang, S. Zhang, F. Yang, Z. Wang, Q. Peng, H. Weng, X. Dai, Z. Fang, Z. Xu, C. Chen, X. J. Zhou. Electronic Evidence for Type II Weyl Semimetal State in MoTe$_2$, arXiv:1604.01706.
%\bibitem{HongDing} N. Xu, Z. J. Wang, A. P. Weber, A. Magrez, P. Bugnon, H. Berger, C. E. Matt, J. Z. Ma, B. B. Fu, B. Q. Lv, N. C. Plumb, M. Radovic, E. Pomjakushina, K. Conder, T. Qian, J. H. Dil, J. Mesot, H. Ding, M. Shi. Discovery of Weyl semimetal state violating Lorentz invariance in MoTe$_2$, arXiv:1604.02116.

\bibitem{SrSi2} Huang, S.-M. \textit{et al}. New type of Weyl semimetal with quadratic double Weyl fermions. \textit{PNAS} {\bf 113}, 1180-1185 (2015).

\bibitem{Koepernik} Koepernik, K. \textit{et al}. TaIrTe$_4$ a ternary type-II Weyl semimetal. \textit{Phys. Rev. B} {\bf 93}, 201101 (2016).

%\bibitem{LaAlGe} Xu, S.-Y. \textit{et al}. Discovery of Lorentz-violating Weyl fermion semimetal state in LaAlGe materials. arXiv:1603.07318.
%\bibitem{RAlX} Chang, G. \textit{et al}. Theoretical prediction of magnetic and noncentrosymmetric Weyl fermion semimetal states in the $R$-Al-$X$ family of compounds ($R =$ rare Earth, Al $ = $ aluminium, $X =$ Si, Ge). arXiv:1604.02124.

\bibitem{IshidaMethods} Ishida, Y. \textit{et al}. Time-resolved photoemission apparatus achieving sub-20-meV energy resolution and high stability. {\it Rev. Sci. Instr.} {\bf 85}, 123904 (2014).

\bibitem{NbPme} Belopolski, I. \textit{et al}. Criteria for Directly Detecting Topological Fermi Arcs in Weyl Semimetals. \textit{Phys. Rev. Lett.} {\bf 116}, 066802 (2016).

\bibitem{PS} Hasan, M. Z., Xu, S.-Y. and Bian, G. Topological insulators, topological superconductors and Weyl fermion semimetals: discoveries, perspectives and outlooks. \textit{Phys. Scrip.} {\bf 2015}, T164 (2015).

\bibitem{IshidaSciRepCuprates} Ishida, Y. \textit{et al}. Quasi-particles ultrafastly releasing kink bosons to form Fermi arcs in a cuprate superconductor. {\it Sci. Rep.} {\bf 6}, 18747 (2016).

%\bibitem{SM} See the Supplemental Material for additional ARPES data on the electronic structure of \comp\ and the methods.

%\bibitem{Trivedi} T. M. McCormik, I. Kimchi \& N. Trivedi. Minimal models for topological Weyl semimetals. arXiv:1604.03096.

%\bibitem{MoTe2WTe2} B. E. Brown. The crystal structures of WTe$_2$ and high-temperature MoTe$_2$. Acta. Cryst. {\bf 20} 268 (1966).

%\bibitem{KaminskiMethods} Jiang, R. \textit{et al}. Tunable vacuum ultraviolet laser based spectrometer for angle-resolved photoemission spectroscopy. Rev. Sci. Instr. {\bf 85}, 033902 (2014).

%\bibitem{XRD1} G. M. Sheldrick. \textit{SADABS} (University of G\"ottingen, G\"ottingen, Germany, 1996).
%\bibitem{XRD2} \textit{CrystalClear, v1.3.5} (Rigaku Corp., The Woodlands, TX, 1999).
%\bibitem{XRD3} G. M Sheldrick. \textit{SHELXTL, v5.1} (Bruker-AXS, Madison, WI, 1998).

\bibitem{PAW1} Bl\"ochl, P. E. Projector augmented-wave method. {\it Phys. Rev. B} {\bf 50}, 17953 (1994).
\bibitem{PAW2} Kresse, G. \& Joubert, J. From ultrasoft pseudopotentials to the projector augmented-wave method. \textit{Phys. Rev. B} {\bf 59}, 1758 (1999).
\bibitem{TransitionMetals} Kresse, G. \& Hafner, J. \textit{Ab initio} molecular dynamics for open-shell transition metals. \textit{Phys. Rev. B} {\bf 48}, 13115 (1993).
\bibitem{PlaneWaves1} Kresse, G. \& Furthm\"uller, J. Efficiency of \textit{ab initio} total energy calculations for metals and semiconductors using a plane-wave basis set. \textit{Comput. Mater. Sci.} {\bf 6}, 15-50 (1996).
\bibitem{PlaneWaves2} Kresse, G. \& Furthm\"uller, J. Efficient iterative schemes for \textit{ab initio} total energy calculations using a plane-wave basis set. \textit{Phys. Rev. B.} {\bf 54}, 11169 (1996).
\bibitem{GGA} Perdew, J. P., Burke, K. \& Ernzerhof, M. Generalized gradient approximation made simple. \textit{Phys. Rev. Lett.} {\bf 77}, 3865 (1996).
\bibitem{TaIrTe4Structure} Mar, A., Jobic, S. \& Ibers, J. A. Metal-metal vs tellurium-tellurium bonding in WTe$_2$ and its ternary variants TaIrTe$_4$ and NbIrTe$_4$. \textit{J. Am. Chem. Soc.} {\bf 114}, 8963-8971 (1992).
%8963-8971
\bibitem{MLWF1}  Marzari, N. \& Vanderbilt, D. Maximally localized generalized Wannier functions for composite energy bands. \textit{Phys. Rev. B} {\bf 56}, 12847 (1997).
\bibitem{MLWF2} Souza, I., Marzari, N. \& Vanderbilt, D. Maximally localized Wannier functions for entangled energy bands. \textit{Phys. Rev. B} {\bf 65}, 035109 (2001).
\bibitem{Wannier90} Mostofi, A. A. \textit{et al}. Wannier90: a tool for obtaining maximally-localized Wannier functions. \textit{Comp. Phys. Commun.} {\bf 178}, 685-699 (2008).
\bibitem{MLWF3} Franchini, C. \textit{et al}. Maximally localized Wannier functions in LaMnO$_3$ within PBE$ + $U, hybrid functionals and partially self-consistent GW: an efficient route to construct \textit{ab initio} tight-binding parameters for $e_g$ perovskites. \textit{J. Phys. Cond. Mat.} {\bf 24}, 235602 (2012).

\end{thebibliography}

\begin{thebibliography}{99}

\bibitem{natnews} 
Castelvecchi, D. ``The strange topology that is reshaping physics." \textit{Nature} \textbf{547}, 272-274  (2017).

\bibitem{kmoore} 
Keimer, B. \& Moore, J. E. The physics of quantum materials. \textit{Nat. Phys.} \textbf{13}, 1045-1055  (2017).

\bibitem{revQZ} 
Qi, X. L. \& Zhang, S. C. Topological insulators and superconductors. \textit{Rev. Mod. Phys.} \textbf{83}, 6045 (2011).

\bibitem{revHK} 
Hasan, M. Z. \& Kane, C. L.  Topological insulators \textit{Rev. Mod. Phys.} \textbf{82}, 3045 (2010).

\bibitem {rev4} 
Yan, B. \& Felser, C. Topological materials: Weyl semimetals. \textit{Ann. Rev. Cond. Mat. Phys.} $\textbf{8}$, 337-354 (2017).

\bibitem {rev7}
Vafek, O. \& Vishwanath, A. Dirac fermions in solids: From high-Tc cuprates and graphene to topological insulators and Weyl semimetals. \textit{Ann. Rev. Cond. Mat. Phys.} \textbf{5}, 83-112 (2014).

\bibitem{rev6}
Burkov, A. A. Weyl metals. \textit{Ann. Rev. Cond. Mat. Phys.} \textbf{9}, 359-378  (2018).

\bibitem{NielsenNinomiya1}
Nielsen, H. B. \& Ninomiya, M. The Adler-Bell-Jackiw anomaly and Weyl fermions in a crystal. $\textit{Phys. Lett. B}$ $\textbf{130B}$, 389-396 (1983).

\bibitem{UniverseinHe}
Volovik, G. E. \textit{The Universe in a Helium Droplet} (Clarendon, Oxford, 2003).

\bibitem{Topo.gapless.phase}
Murakami, S. Phase transition between the quantum spin hall and insulator phases in 3D: Emergence of a topological gapless phase. \textit{New J. Phys.} \textbf{9}, 356 (2007).

\bibitem{Wan}
Wan, X. et al. Topological semimetal and Fermi-arc surface states in the electronic structure of pyrochlore iridates. \textit{Phys. Rev. B} \textbf{83}, 205101 (2011).

\bibitem{Weyl-Multilayer}
Burkov, A. A. \& Balents, L. Weyl semimetal in a topological insulator multilayer. \textit{Phys. Rev. Lett.} \textbf{107}, 127205 (2011).

\bibitem{Chiral-anomaly}
Son, D. T. \& Spivak, B. Z.  Chiral anomaly and classical negative magnetoresistance of Weyl metals. \textit{Phys. Rev. B} \textbf{88}, 104412 (2013).

\bibitem{HgCr2Se4}
Xu, G. et al. Chern semimetal and the quantized anomalous Hall effect in HgCr$_2$Se$_4$. \textit{Phys. Rev. Lett.} $\mathbf{107}$, 186806 (2011).

\bibitem{typeIIWeyl} 
Soluyanov, A. A. et al. Type-II Weyl semimetals. \textit{Nature} $\textbf{527}$, 495-498 (2015).

\bibitem{unconventionalWeyl}
Bradlyn, B. et al. Beyond Dirac and Weyl fermions: Unconventional quasiparticles in conventional crystals. \textsl{Science} \textbf{353}, aaf5037 (2017).

\bibitem{RhSi}
Chang, G. et al. Unconventional chiral fermions and large topological Fermi arcs in RhSi. \textit{Phys. Rev. Lett.} $\mathbf{119}$, 206401 (2017).

\bibitem{CoSi}
Tang, P. et al. Multiple types of topological termions in transition metal silicides. \textit{Phys. Rev. Lett.} $\mathbf{119}$, 206402 (2017).

\bibitem{TaAs1}
Weng H. et al. Weyl semimetal phase in non-centrosymmetric transition metal monophosphides. \textit{Phys. Rev. X} \textbf{5}, 011029 (2015).

\bibitem{TaAs2}
Huang, S.-M. et al.  A Weyl fermion semimetal with surface Fermi arcs in the transition metal monopnictide TaAs class. \textit{Nat. Commun.} \textbf{6}, 7373 (2015).

\bibitem{ARPES-TaAs1}
Xu, S.-Y., Belopolski, I. et al. Discovery of a Weyl fermion semimetal and topological Fermi arcs. \textsl{Science} \textbf{349}, 613-617 (2015).

\bibitem{ARPES-TaAs2}
Lv, B. Q. et al. Experimental discovery of Weyl semimetal TaAs. \textit{Phys. Rev. X} \textbf{5}, 031013 (2015).

\bibitem{filling_constraint1}
Watanabe, H. et al. Filling constraints for spin-orbit coupled insulators in symmorphic and nonsymmorphic crystals. \textit{Proc. Natl. Acad. Sci.} $\textbf{112}$, 14551-14556 (2015).

\bibitem{Ben1}
Wieder, B. J. et al. Spin-orbit semimetals in the layer groups. \textit{Phys. Rev. B} $\textbf{94}$, 155108 (2016).

\bibitem{KramersWeyl}
Chang, G. et al. Topological quantum properties of chiral crystals. \textit{Nat. Mat.} \textbf{17}, 978-985 (2018).

\bibitem{Manes}
Ma\~{n}es, J. L. Existence of bulk chiral fermions and crystal symmetry. \textit{Phys. Rev. B} $\textbf{85}$, 155118 (2012).

%\bibitem{idealphotonicWeyl}
%Yang, B. et al. Discovery of ideal Weyl points \textsl{Science} \textbf{11}, eaaq1221 (2018).

\bibitem{Nobel}
Royal Swedish Academy of Sciences. Topological phase transitions and topological phases of matter. Nobel Media AB (2016). \href{https://www.nobelprize.org/prizes/physics/2016/advanced-information/}{https://www.nobelprize.org/prizes/physics/2016/advanced-information/}.

\bibitem{HelicodalFermiArcs}
Fang, C. et al. Topological semimetals with helicoid surface states. \textit{Nat. Phys.} $\textbf{12}$, 936-941 (2016).

\bibitem{arcDetect1}
Belopolski, I. et al. Criteria for directly detecting topological Fermi arcs in Weyl semimetals. \textit{Phys. Rev. Lett.} $\mathbf{116}$, 066802 (2016).

\bibitem{chiralmagenticeffect1}
Zyuzin, A. A. et al. Weyl semimetal with broken time reversal and inversion symmetries. \textit{Phys. Rev. B} $\textbf{85}$, 165110 (2012).

\bibitem{chiral_photogalvanic}
Juan, F. de et al. Quantized circular photogalvanic effect in Weyl semimetals. \textit{Nat. Commun.} $\textbf{8}$, 15995 (2017).

\bibitem{Haldane} 
Li, Y. \& Haldane, F. D. M. Topological nodal Cooper pairing in doped Weyl metals. \textit{Phys. Rev. B} \textbf{120}, 067003 (2018).

\bibitem{CoSi_ARPES}
Takane, D. et al. Observation of chiral fermions with a large topological charge in CoSi. arXiv:1809.01312 (2018).


% Extended Data References

\bibitem{burkov} 
Burkov, A. A. Topological semimetals. \textit{Nat. Mat.} \textbf{15}, 1145-1148  (2016).

\bibitem{Jia} 
Jia, S. et al. Weyl semimetals, Fermi arcs and chiral anomaly \textit{Nat. Mat.} \textbf{15}, 1140-1144 (2016).

\bibitem{H.Weyl} 
Weyl, H.  Elektron und gravitation. $\textit{I. Z. Phys.}$ $\textbf{56}$, 330-352 (1929).

\bibitem{TypeIIWeyl1} 
Huang, L. et al. Spectroscopic evidence for a type II Weyl semimetallic state in MoTe$_2$. \textit{Nat. Mater.} \textbf{15}, 1155-1160 (2016).

\bibitem{TypeIIWeyl2} 
Deng, K. et al. Experimental observation of topological Fermi arcs in type-II Weyl semimetal MoTe$_2$. \textit{Nat. Phys.} $\textbf{12}$, 1105-1110 (2016).

\bibitem{TypeIIWeyl3} 
Wu, Y. et al. Observation of Fermi arcs in the type-II Weyl semimetal candidate WTe$_2$. \textit{Phys. Rev. B} $\textbf{94}$, 121113(R) (2016).

\bibitem{TypeIIWeyl4}
Belopolski, I. et al. Fermi arc electronic structure and Chern numbers in Mo$_x$W$_{1-x}$Te$_2$. \textit{Phys. Rev. B} $\mathbf{94}$, 085127 (2016).

\bibitem{XRay1}
\textit{SHELXTL}, version 6.10, Bruker (2000), Bruker AXS Inc., Madison, Wisconsin, USA.

\bibitem{XRay2}
Sheldrick, G. M. A short history of SHELX. \textit{Acta Cryst.} $\textbf{A64}$, 112-122 (2008).

\bibitem{DFT1}
Ozaki, T. et al. http://www.openmx-square.org/.

\bibitem{DFT2}
Kresse, G. \& Furthmueller, G. Efficient Iterative Schemes for \textit{ab initio} Total-energy Calculations using a Plane-wave Basis Set. \textit{Phys. Rev. B} $\textbf{54}$, 11169 (1996).

\bibitem{DFT3}
Kresse, G. \& Joubert, D. From ultrasoft pseudopotentials to the projector augmented-wave method. \textit{Phys. Rev. B} $\textbf{59}$, 1758 (1999).

\bibitem{DFT4}
Perdew, J. P.  et al. Generalized gradient approximation made simple. \textit{Phys. Rev. Lett.} $\textbf{77}$, 3865 (1996).

\bibitem{xraydata}
Demchenko, P. et al. Single crystal investigation of the new phase Er$_{0.85}$Co$_{4.31}$Si and of CoSi. \textit{Chem. Met. Alloys} $\textbf{1}$, 50-53 (2008).

\end{thebibliography}

\begin{thebibliography}{10}

\bibitem{SYX_NP_MnBi2Se3_2012}
S.-Y.~Xu {\it et al}., Hedgehog spin texture and Berry's phase tuning in a magnetic topological insulator {\it Nat. Phys.\/} {\bf 8}, 616 (2012).

\bibitem{news_Castelvecchi}
D.~Castelvecchi, The strange topology that is reshaping physics. {\it Nature\/}
  {\bf 547}, 272 (2017).

\bibitem{RMPMajorana_Franz}
S.~R. Elliot, M.~Franz, Colloquium: Majorana fermions in nuclear, particle, and
  solid-state physics. {\it Rev. Mod. Phys.\/} {\bf 87}, 137 (2015).

\bibitem{ReviewQuantumMaterials_Nagaosa}
Y.~Tokura, M.~Kawasaki, N.~Nagaosa, Emergent functions of quantum materials.
  {\it Nat. Phys.\/} {\bf 13}, 1056 (2017).

\bibitem{QAH_Xue}
C.-Z. Chang, {\it et~al.\/}, Experimental observation of the quantum anomalous
  {H}all effect in a magnetic topological insulator. {\it Science\/} {\bf 340},
  167 (2013).

\bibitem{ReviewQuantumMaterials_Hsieh}
D.~N. Basov, R.~D. Averitt, D.~Hsieh, Towards properties on demand in quantum
  materials. {\it Nat. Mat.\/} {\bf 16}, 1077 (2017).

\bibitem{RMPTopoBandThy_Bansil}
A.~Bansil, H.~Lin, T.~Das, Topological band theory. {\it Rev. Mod. Phys.\/}
  {\bf 88}, 021004 (2016).

\bibitem{Colloquium_Zahid}
M.~Z. Hasan, C.~L. Kane, Topological insulators. {\it Rev. Mod. Phys.\/} {\bf
  82}, 3045 (2010).

\bibitem{ARCMP_me}
M.~Z. Hasan, S.-Y. Xu, I.~Belopolski, S.-M. Huang, Discovery of {W}eyl fermion
  semimetals and topological {F}ermi arc states. {\it Ann. Rev. Cond. Matt.
  Phys.\/} {\bf 2017}, 289 (8).

\bibitem{RMPWeylDirac_Armitage}
N.~P. Armitage, E.~J. Mele, A.~Vishwanath, {W}eyl and {D}irac semimetals in
  three-dimensional solids. {\it Rev. Mod. Phys.\/} {\bf 90}, 015001 (2018).

\bibitem{Co2MnGa_Kaustuv}
K.~Manna, {\it et~al.\/}, From colossal to zero: Controlling the anomalous
  {H}all effect in magnetic {H}eusler compounds via {B}erry curvature design
  (2017). \MakeLowercase{a}rXiv:1712.10174.

\bibitem{Co2MnGa_Nakatsuji}
A.~Sakai, {\it et~al.\/}, Giant anomalous {N}ernst effect and quantum-critical
  scaling in a ferromagnetic semimetal. {\it Nat. Phys.\/}  (2018).
  DOI:10.1038/s41567-018-0225-6.

\bibitem{Co3Sn2S2_Enke}
E.~Liu, {\it et~al.\/}, Giant anomalous {H}all effect in a ferromagnetic
  kagome-lattice semimetal. {\it Nat. Phys.\/}  (2018).
  DOI:10.1038/s41567-018-0234-5.

\bibitem{Mn3Sn_Nakatsuji}
K.~Kuroda, {\it et~al.\/}, Evidence for magnetic {W}eyl fermions in a
  correlated metal. {\it Nat. Mat.\/} {\bf 16}, 1090 (2017).

\bibitem{Fe3GeTe2_Kim_Pohang}
K.~Kim, {\it et~al.\/}, Large anomalous {H}all current induced by topological
  nodal lines in a ferromagnetic van der {W}aals semimetal. {\it Nat. Mat.\/}
  (2018). DOI:10.1038/s41563-018-0132-3.

\bibitem{Fe3Sn2_Checkelsky}
L.~Ye, {\it et~al.\/}, Massive {D}irac fermions in a ferromagnetic kagome
  metal. {\it Nature\/} {\bf 555}, 638 (2018).

\bibitem{Co2MnGa_Felser_Nernst}
S.~N. Guin, {\it et~al.\/}, Anomalous {N}ernst effect beyond the magnetization
  scaling relation in the ferromagnetic {H}eusler compound {Co$_2$MnGa} (2018).
  \MakeLowercase{a}rXiv:1806.06753.

\bibitem{Co2MnGa_Felser_films}
H.~Reichlova, {\it et~al.\/}, Large anomalous {N}ernst effect in thin films of
  the {W}eyl semimetal {Co$_2$MnGa} (2018). \MakeLowercase{a}rXiv:1807.06487.

\bibitem{Co2MnGa_Guoqing}
G.~Chang, {\it et~al.\/}, Topological {H}opf and chain link semimetal states
  and their application to {Co$_2$MnGa}. {\it Phys. Rev. Lett.\/} {\bf 119},
  156401 (2017).

\bibitem{ClassificationGapped_Schnyder}
A.~P. Schnyder, S.~Ryu, A.~Furusaki, A.~W. Ludwig, Classification of
  topological insulators and superconductors in three spatial dimensions. {\it
  Phys. Rev. B\/} {\bf 78}, 195125 (2008).

\bibitem{SemimetalReview_Vishwanath}
A.~Vishwanath, A.~M. Turner, {\it Topological Insulators\/}, Contemporary
  Concepts of Condensed Matter Science (Elsevier, 2013), Chap.~11.

\bibitem{ClassificationReflection_Schnyder}
C.-K. Chiu, A.~P. Schnyder, Classification of reflection-symmetry-protected
  topological semimetals and nodal superconductors. {\it Phys. Rev. B\/} {\bf
  90}, 205136 (2014).

\bibitem{TINI_Balents}
A.~A. Burkov, M.~D. Hook, L.~Balents, Topological nodal semimetals. {\it Phys.
  Rev. B\/} {\bf 84}, 235126 (2011).

\bibitem{WeylDiracLoop_Nandkishore}
R.~Nandkishore, {W}eyl and {D}irac loop superconductors. {\it Phys. Rev. B\/}
  {\bf 93}, 020506 (2016).

\bibitem{WeylLoopSuperconductor_Nandkishore}
Y.~Wang, R.~Nandkishore, Topological surface superconductivity in doped {W}eyl
  loop materials. {\it Phys. Rev. B\/} {\bf 95}, 060506 (2017).

\bibitem{WeylLines_Kane}
O.~Stenull, C.~L. Kane, T.~C. Lubensky, Topological phonons and {W}eyl lines in
  three dimensions. {\it Phys. Rev. Lett.\/} {\bf 117}, 068001 (2016).

\bibitem{PbTaSe2_Guang}
G.~Bian, {\it et~al.\/}, Topological nodal-line fermions in spin-orbit metal
  {PbTaSe$_2$}. {\it Nat. Commun.\/} {\bf 7}, 10556 (2016).

\bibitem{NodalChain_Soluyanov}
T.~Bzdu{\u{s}}ek, Q.~Wu, A.~R{\"u}egg, M.~Sigrist, A.~A. Soluyanov, Nodal-chain
  metals. {\it Nature\/} {\bf 538}, 75 (2016).

\bibitem{RMPClassification_ChingKai}
C.-K. Chiu, J.~C.~Y. Teo, A.~P. Schnyder, S.~Ryu, Classification of topological
  quantum matter with symmetries. {\it Rev. Mod. Phys.\/} {\bf 88}, 035005
  (2016).

\bibitem{DiracLineNodes_WeiderKane}
Y.~Kim, B.~J. Wieder, C.~L. Kane, A.~M. Rappe, {D}irac line nodes in
  inversion-symmetric crystals. {\it Phys. Rev. Lett.\/} {\bf 115}, 036806
  (2015).

\bibitem{Ca3P2_Schnyder}
Y.-H. Chan, C.-K. Chiu, M.~Y. Chou, A.~P. Schnyder, {Ca$_3$P$_2$} and other
  topological semimetals with line nodes and drumhead surface states. {\it
  Phys. Rev. B\/} {\bf 93}, 205132 (2016).

\bibitem{Co2MnGa_CurieTemp}
H.~Ido, S.~Yasuda, Magnetic properties of {C}o-{H}eusler and related mixed
  alloys. {\it J. de Physique\/} {\bf 49}, C8 (1988).

\bibitem{ScalingAHE_TianYeJin_2009}
Y.~Tian, L.~Ye, X.~Jin, Proper scaling of the anomalous {H}all effect. {\it
  Phys. Rev. Lett.\/} {\bf 103}, 087206 (2009).

\bibitem{ScalingAHE_TianJinNiu_2015}
D.~Hou, {\it et~al.\/}, Multivariable scaling for the anomalous {H}all effect.
  {\it Phys. Rev. Lett.\/} {\bf 114}, 217203 (2015).

\bibitem{ScalingAHE_TianJin_2016}
L.~Wu, K.~Zhu, D.~Yue, Y.~Tian, X.~Jin, Anomalous {H}all effect in localization
  regime. {\it Phys. Rev. B\/} {\bf 93}, 214418 (2016).

\bibitem{GreensFunction_Bryant}
G.~Bryant, Surface states of ternary semiconductor alloys: Effect of alloy
  fluctuations in one-dimensional models with realistic atoms. {\it Phys. Rev.
  B\/} {\bf 31}, 5166 (1985).

\bibitem{DFT1}
T.~Ozaki, http://www.openmx-square.org/.

\bibitem{DFT2}
G.~Kresse, J.~Furthmueller, Efficient iterative schemes for \textit{ab initio}
  total-energy calculations using a plane-wave basis set. {\it Phys. Rev. B\/}
  {\bf 54}, 11169 (1996).

\bibitem{DFT3}
G.~Kresse, D.~Joubert, From ultrasoft pseudopotentials to the projector
  augmented-wave method. {\it Phys. Rev. B\/} {\bf 59}, 1758 (1999).

\bibitem{DFT4}
J.~P. Perdew, K.~Burke, M.~Ernzerhof, Generalized gradient approximation made
  simple. {\it Phys. Rev. Lett.\/} {\bf 77}, 3865 (1996).

\bibitem{Co2MnAl_Jakob}
E.~V. Vidal, G.~Stryganyuk, H.~Schneider, C.~Felser, G.~Jakob, Exploring
  {Co$_2$MnAl} {H}eusler compound for anomalous {H}all effect sensors. {\it
  App. Phys. Lett.\/} {\bf 99}, 132509 (2011).

\bibitem{ReviewAHE_Nagaosa_Ong}
N.~Nagaosa, J.~Sinova, S.~Onoda, A.~H. MacDonald, N.~P. Ong, Anomalous {H}all
  effect. {\it Rev. of Mod. Phys.\/} {\bf 82}, 1539 (2010).

\end{thebibliography}
